\documentclass[journal,twoside,final,letterpaper,twocolumn]{IEEEtran}

\usepackage{multirow}
\usepackage{tablefootnote}

\ifCLASSINFOpdf
   \usepackage[pdftex]{graphicx}
\else

\fi

\usepackage{epstopdf}
\usepackage{epsfig}
\usepackage{float}
\usepackage{makecell}

\usepackage{amssymb,amsbsy,subfigure,amsmath,amsfonts,multirow,color,url,bm,cite,cases,eqparbox}
\usepackage{algorithmic}
\usepackage{algorithm}
\usepackage[mathscr]{eucal}

\newcommand{\tablesizetwo}{\fontsize{7.5pt}{11pt}\selectfont}

\def\0{\mathbf{0}}
\def\1{\mathbf{1}}

\def\Gcal{\mathcal{G}}

\def\Ecal{\mathcal{E}}

\def\A{\mathbf{A}}

\def\C{\mathbf{C}}
\def\D{\mathbf{D}}

\def\S{\mathbf{S}}

\def\U{\mathbf{U}}
\def\V{\mathbf{V}}
\def\W{\mathbf{W}}

\def\Y{\mathbf{Y}}

\def\Trm{\mathrm{T}}
\def\a{\mathbf{a}}

\def\v{\mathbf{v}}

\def\y{\mathbf{y}}

\def\LAmbda{\boldsymbol{\Lambda}}

\def\Hcal{\mathcal{H}}
\def\I{\mathbf{I}}

\def\Bcal{\mathcal{B}}
\def\Ncal{\mathcal{N}}

\def\Hcal{\mathcal{H}}

\def\Rbb{\mathbb{R}}

\def\Xcal{\mathcal{X}}

\def\Ycal{\mathcal{Y}}

\def\Rcal{\mathcal{R}}

\begin{document}
\title{SMDS-Net: Model Guided Spectral-Spatial Network for Hyperspectral Image Denoising}

\author{Fengchao Xiong,~\IEEEmembership{Member,~IEEE}, Jun Zhou,~\IEEEmembership{Senior Member,~IEEE},  Shuyin Tao, Jianfeng Lu,~\IEEEmembership{Member,~IEEE}, Jiantao Zhou,~\IEEEmembership{Senior Member,~IEEE} and Yuntao Qian,~\IEEEmembership{Senior Member,~IEEE}
\thanks {This work was supported in part by Jiangsu Provincial Natural Science Foundation of China under Grant BK20200466, and the National Natural Science Foundation of China under Grant 62002169, 61905114, 61971476 and 62071421, the National Key Research and Development Program of China under Grant  2018YFB0505000 and 2018AAA0100500, Macao Young Scholars Program (No. AM2021020) and Macau Science and Technology Development Fund under 077/2018/A2. (Corresponding author: Fengchao Xiong.)}
\thanks{Fengchao Xiong is  with the School of Computer Science and Engineering, Nanjing University of Science and Technology, Nanjing 210094,  China and also  with the State Key Laboratory of Internet of Things for Smart City, Department of Computer and Information Science, University of Macau,  Macau 999078, China (e-mail: fcxiong@njust.edu.cn). }
\thanks{Jun Zhou is with the School of Information and Communication Technology, Griffith University, Nathan, Australia (e-mail: jun.zhou@griffith.edu.au).}
\thanks{Shuyin Tao and Jianfeng Lu are with the School of Computer Science and Engineering, Nanjing University of Science and Technology, Nanjing 210094, China (e-mail: taoshuyin@njust.edu.cn, lujf@njust.edu.cn). }
\thanks{Jiantao Zhou is with the State Key Laboratory of Internet of Things for
Smart City, Department of Computer and Information Science, University of
Macau, Macau 999078, China (e-mail: jtzhou@um.edu.mo).}

\thanks{Yuntao Qian  is with the College of Computer Science, Zhejiang University, Hangzhou 310027, China (e-mail: ytqian@zju.edu.cn).}
}
 \maketitle

\begin{abstract}

Deep learning (DL) based hyperspectral images (HSIs) denoising approaches directly learn the nonlinear mapping between  noisy  and  clean HSI pairs.  They normally do not consider the physical characteristics of HSIs. This drawback makes the models  lack  interpretability that is key to understand their denoising mechanism and also limits their denoising ability.   To tackle this problem, we introduce a novel model guided interpretable network for HSI denoising.  Fully considering the spatial redundancy, spectral low-rankness and spectral-spatial correlations of HSIs, we first establish a subspace based multidimensional sparse (SMDS) model under the umbrella of tensor notation.    After that, the model is unfolded into an end-to-end network named  as  SMDS-Net whose fundamental modules are seamlessly connected with the denoising procedure and optimization of  the SMDS model. This makes SMDS-Net convey clear physical meanings, i.e., learning the low-rankness and sparsity of HSIs.  Finally, all key variables are obtained by an end-to-end training. Extensive experiments and comprehensive analysis  on  synthetic and real-world HSIs confirm the strong denoising ability, learning capability and high interpretability of SMDS-Net   against the state-of-the-art HSI denoising methods.  
\end{abstract}

\IEEEpeerreviewmaketitle
\section{Introduction}

Hyperspectral images (HSIs) have enabled many practical applications such as pedestrian detection~\cite{Li2019,Zhang2019}, object tracking~\cite{Xiong2020}, medical diagnosis~\cite{Zhou2014} and more~\cite{Akhtar2020} thanks to its material identification ability provided by numerous light-wavelength indexed bands. Because of narrow bands, limited sensitivity of sensors or their defects, and the interference of the imaging environment, the captured HSIs tend to be corrupted by noises. The degradation of HSIs brings many drawbacks in the exploitation of HSIs and significantly decreases their utility as well. Therefore, HSI denoising is often needed as a critical preprocessing step to enhance the quality of HSIs.

  \begin{figure}[!htbp]
  \centering{
\includegraphics[width=0.115\textwidth,clip=true]{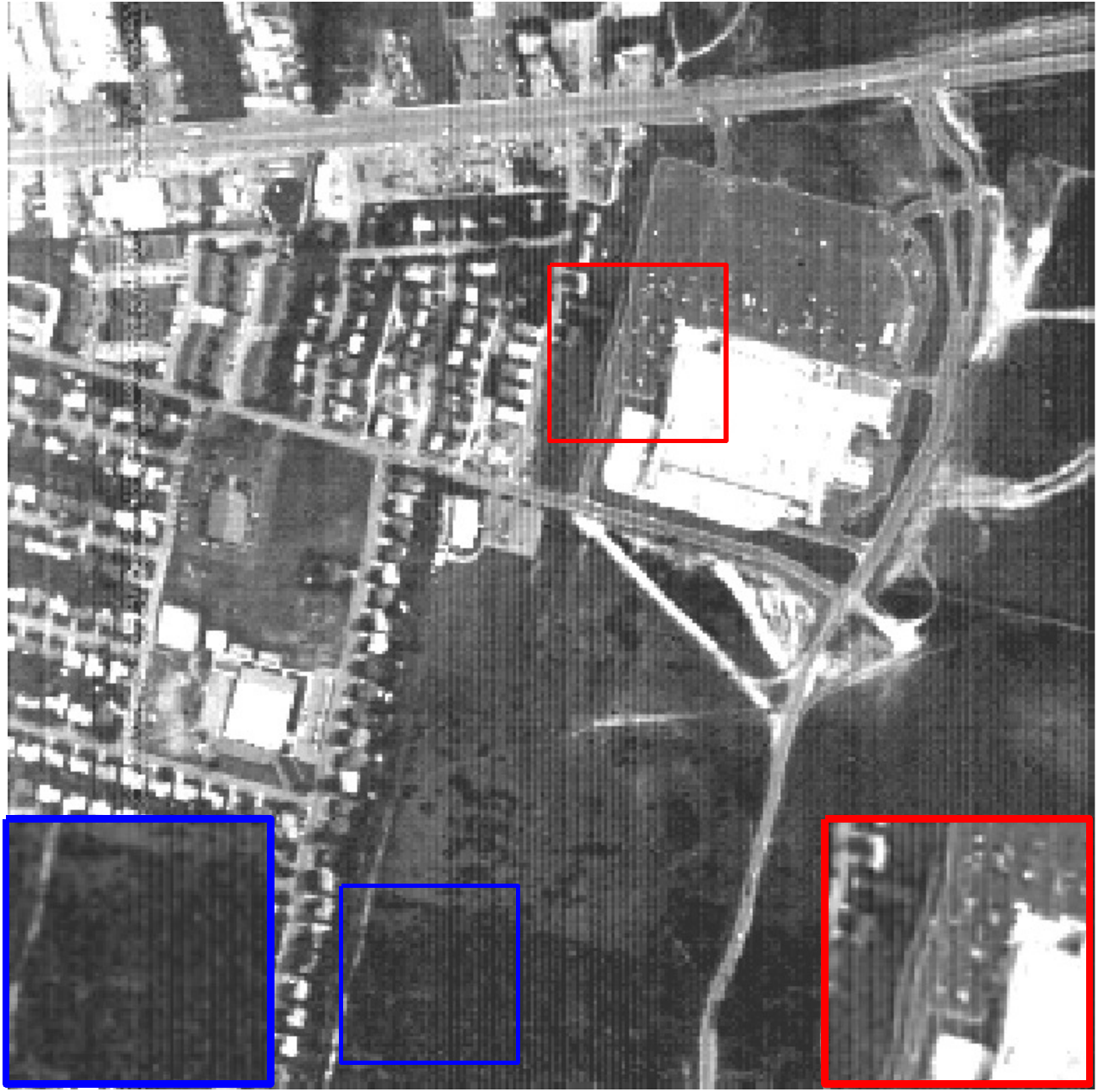}
\includegraphics[width=0.115\textwidth,clip=true]{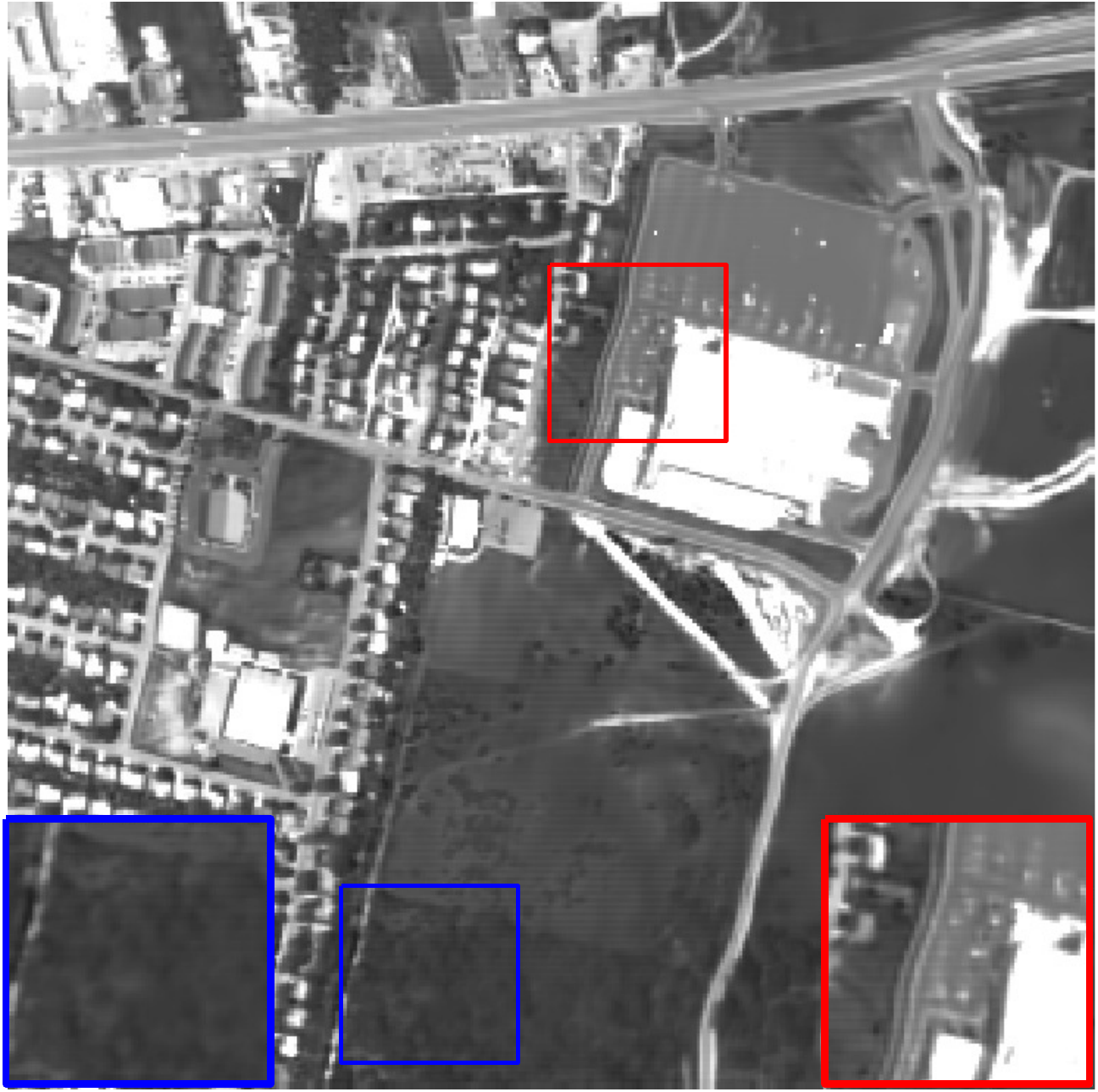}
\includegraphics[width=0.115\textwidth,clip=true]{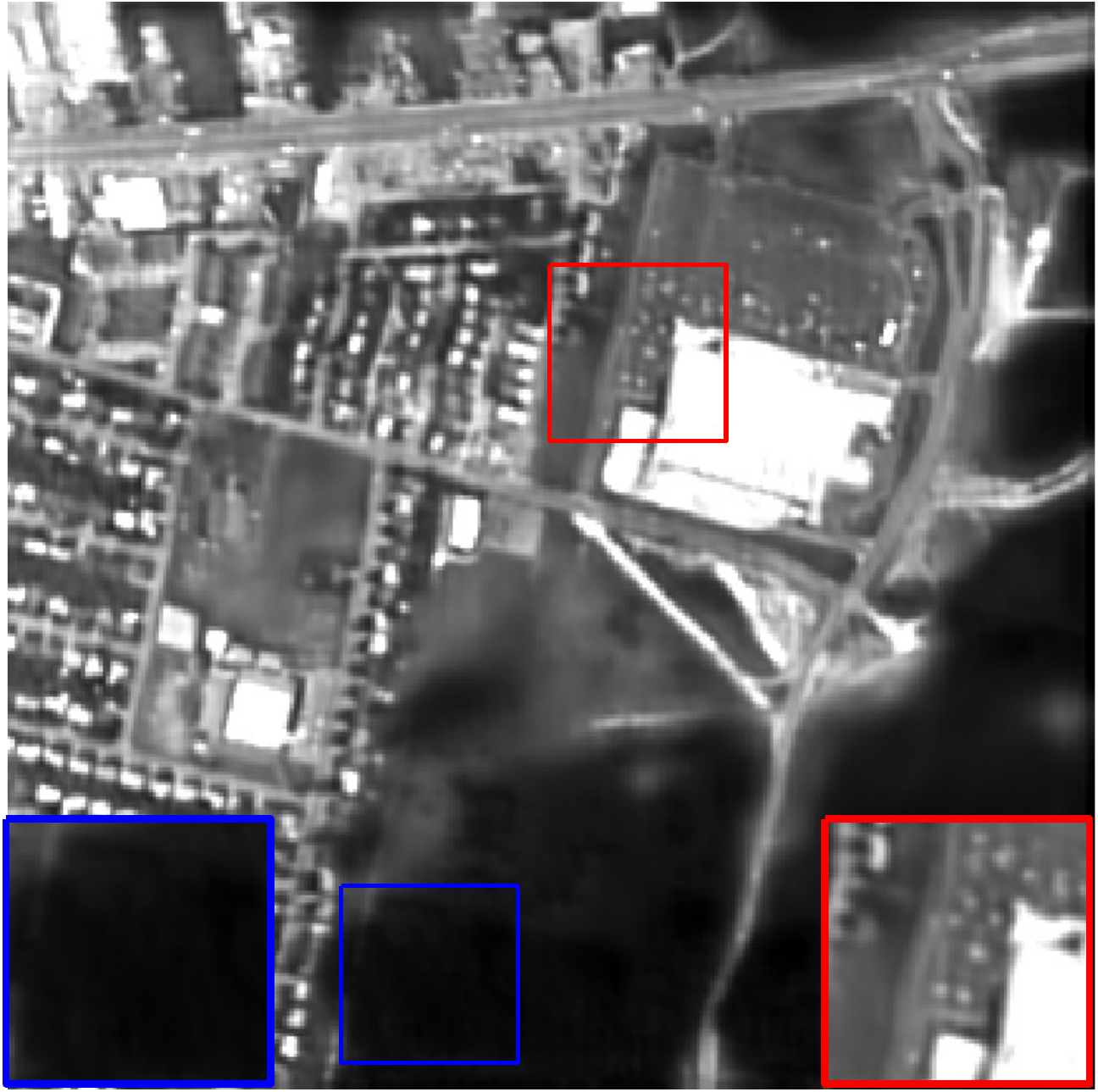}
\includegraphics[width=0.115\textwidth,clip=true]{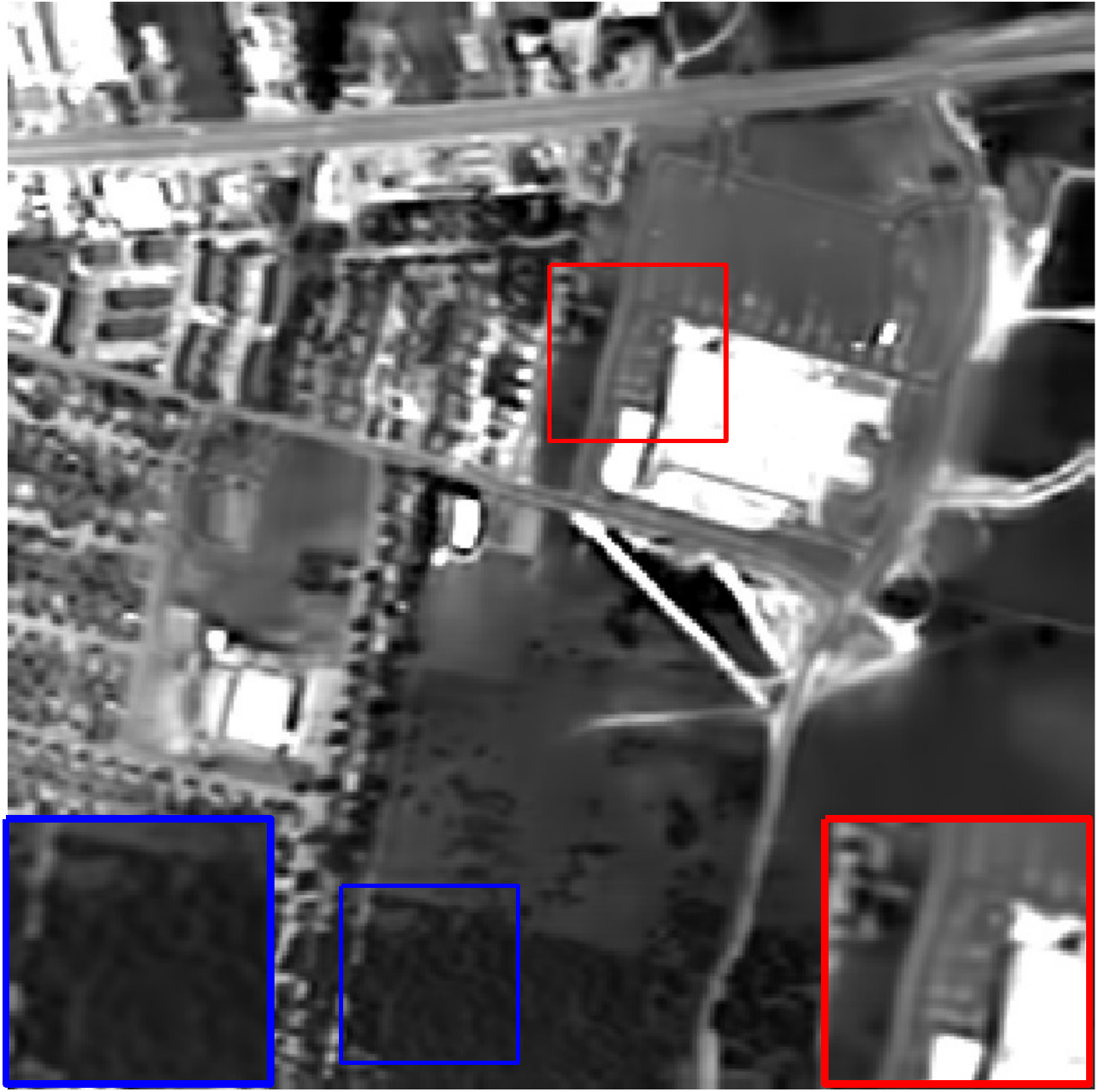}\\
\includegraphics[width=0.115\textwidth,clip=true]{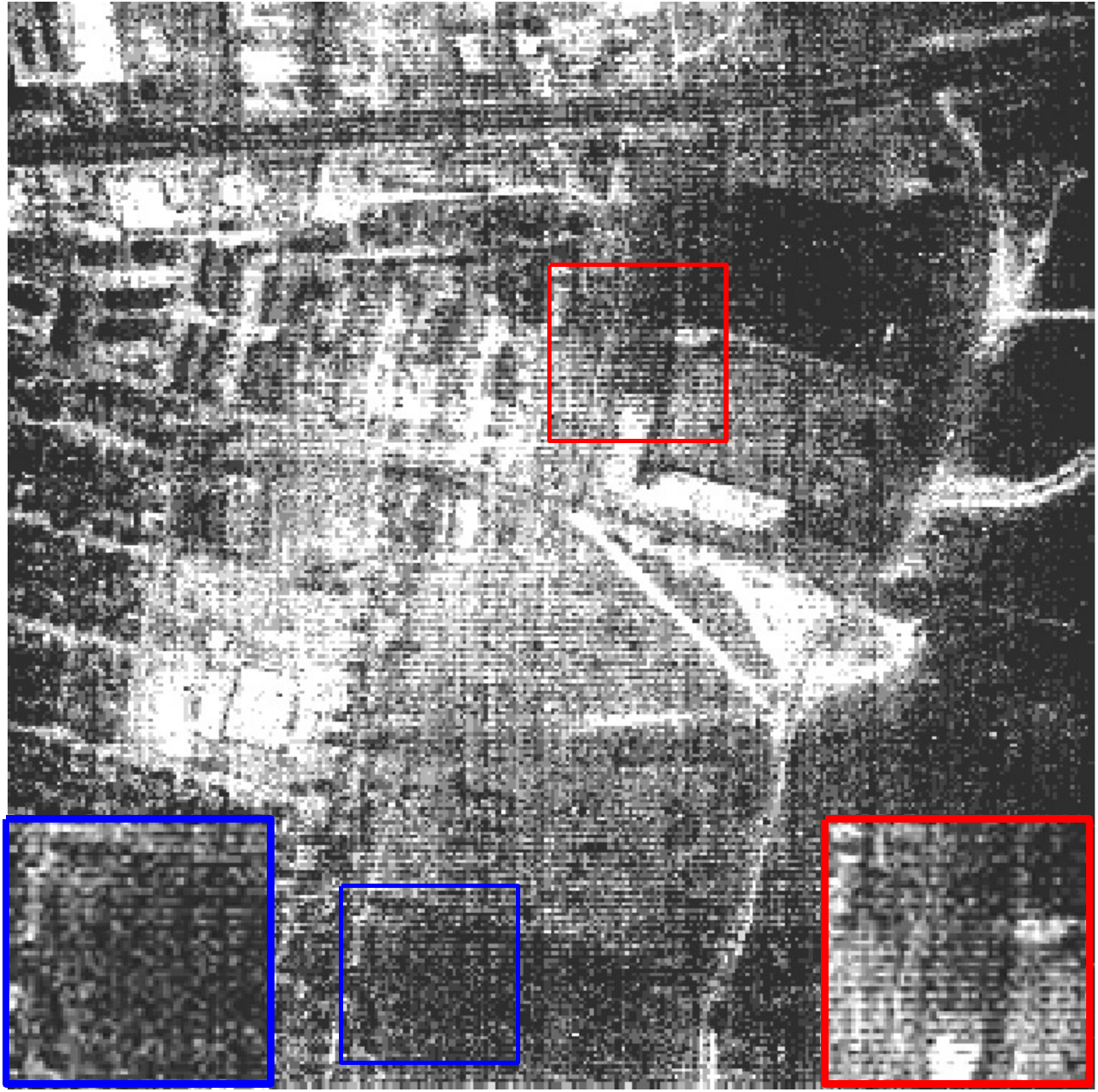}
\includegraphics[width=0.115\textwidth,clip=true]{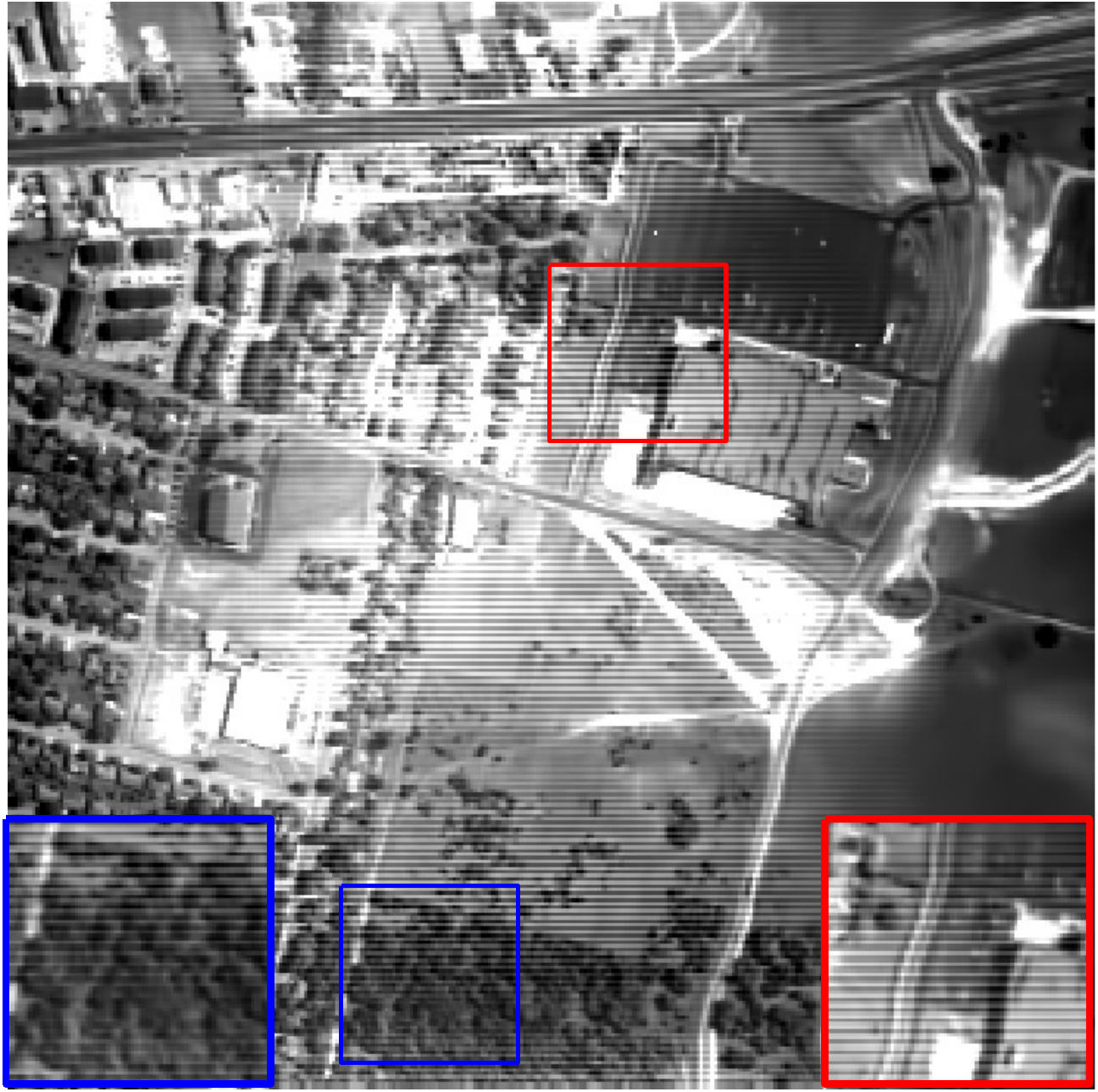}
\includegraphics[width=0.115\textwidth,clip=true]{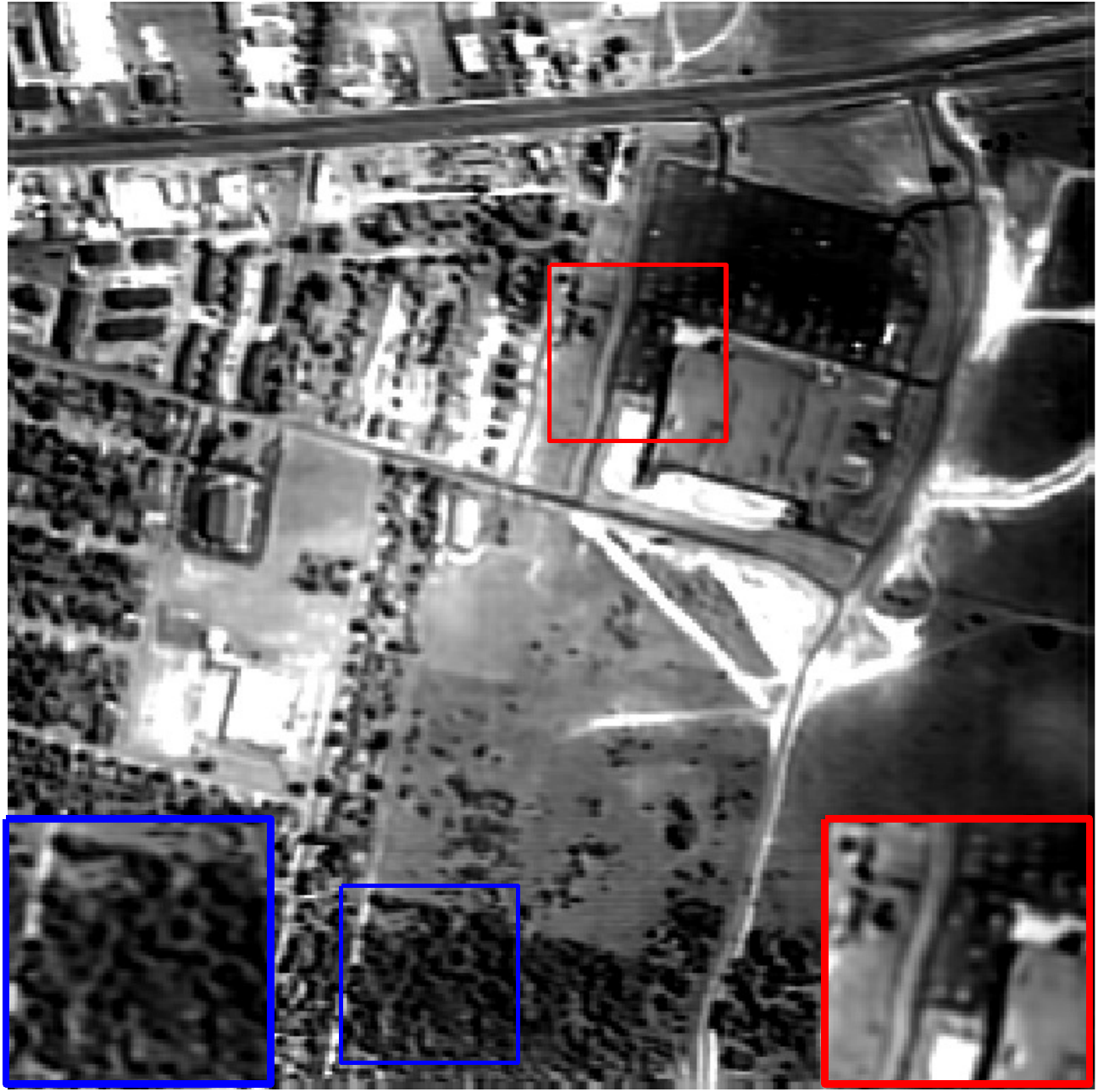}
\includegraphics[width=0.115\textwidth,clip=true]{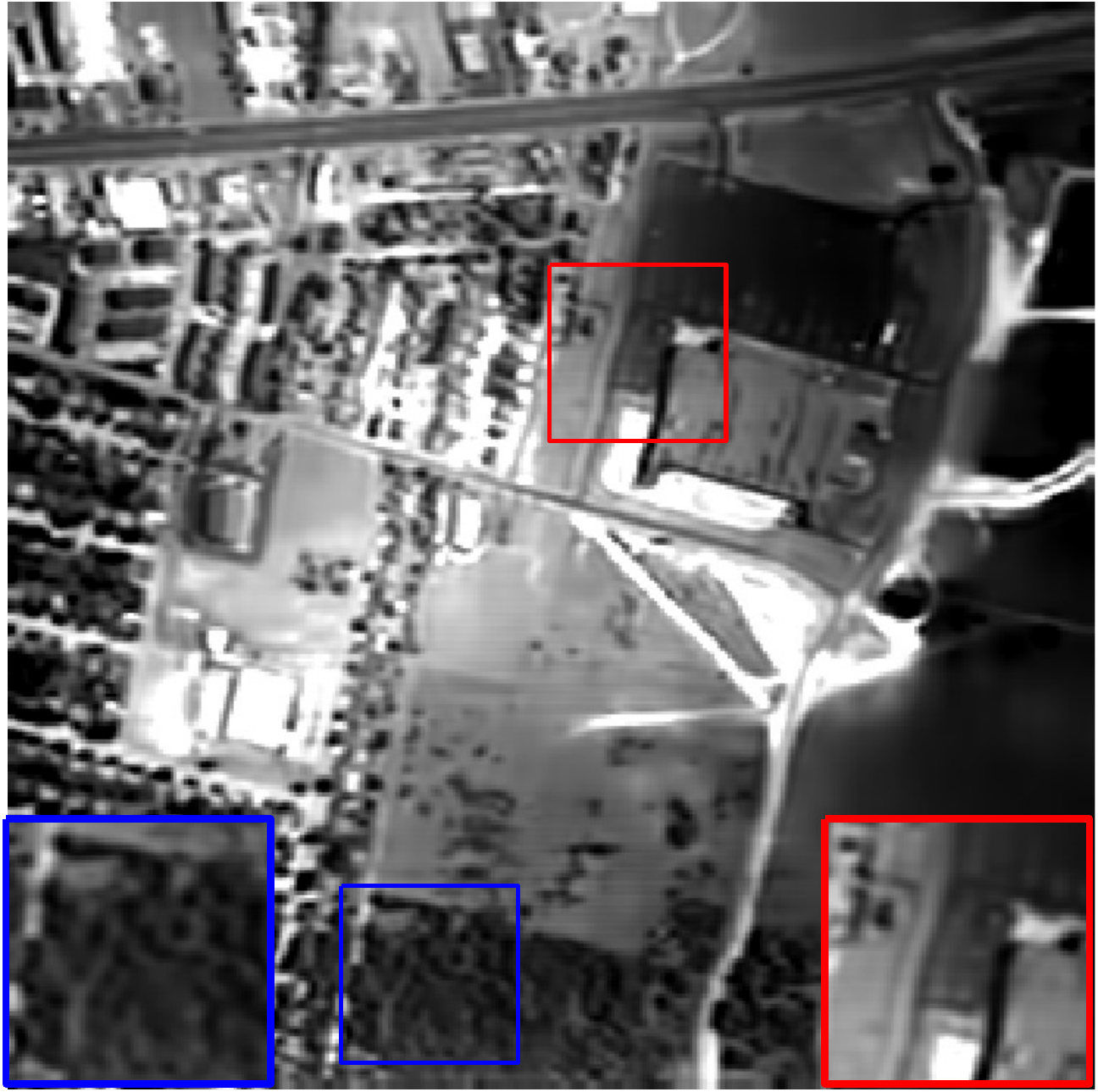}
\subfigure[\scriptsize{Noisy}]{\includegraphics[width=0.115\textwidth,clip=true]{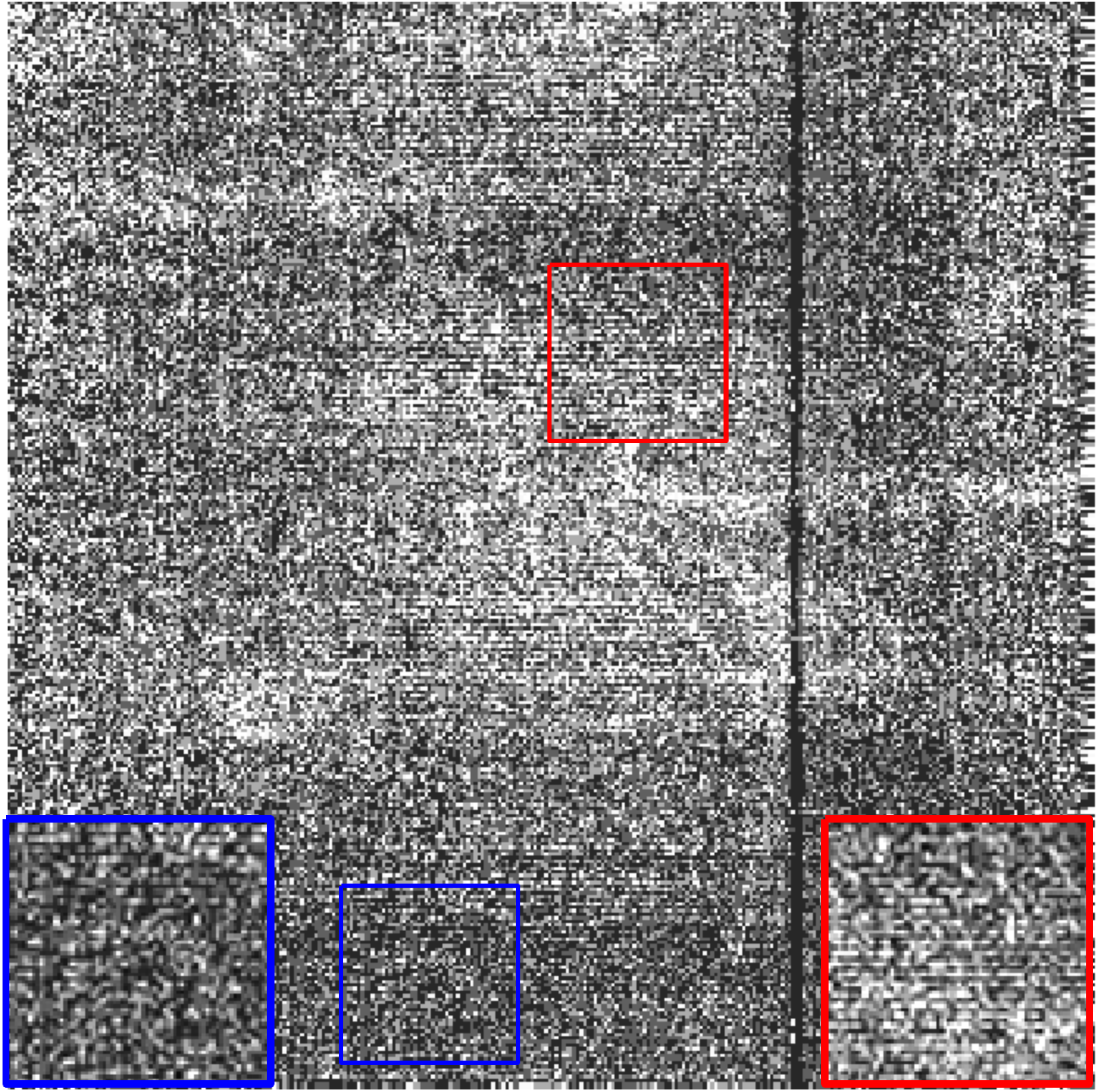}}
\subfigure[\scriptsize{NGMeet~\cite{He2019}}]{\includegraphics[width=0.115\textwidth,clip=true]{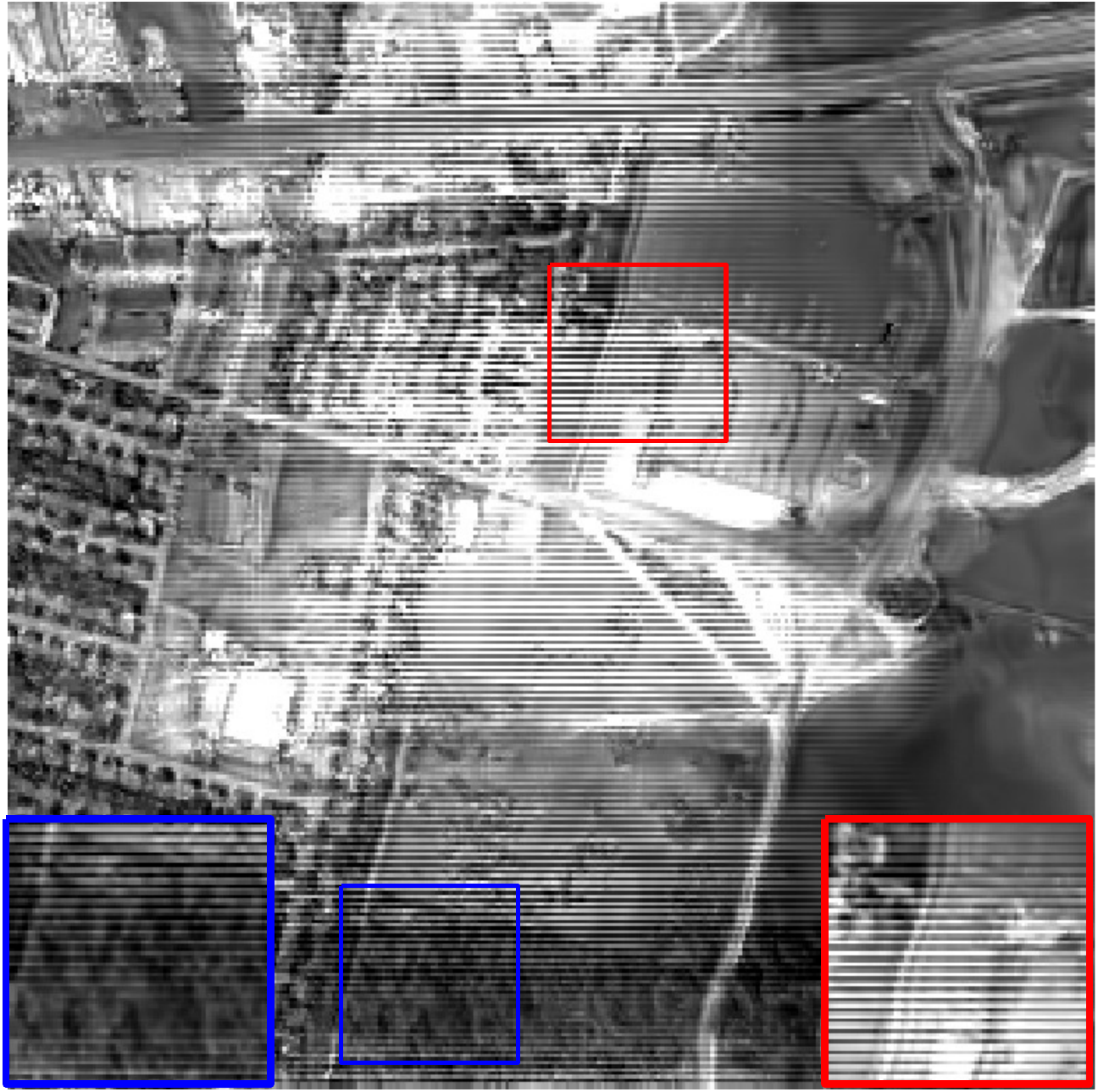}}
\subfigure[\scriptsize{QRNN3D~\cite{Wei2020}}]{\includegraphics[width=0.115\textwidth,clip=true]{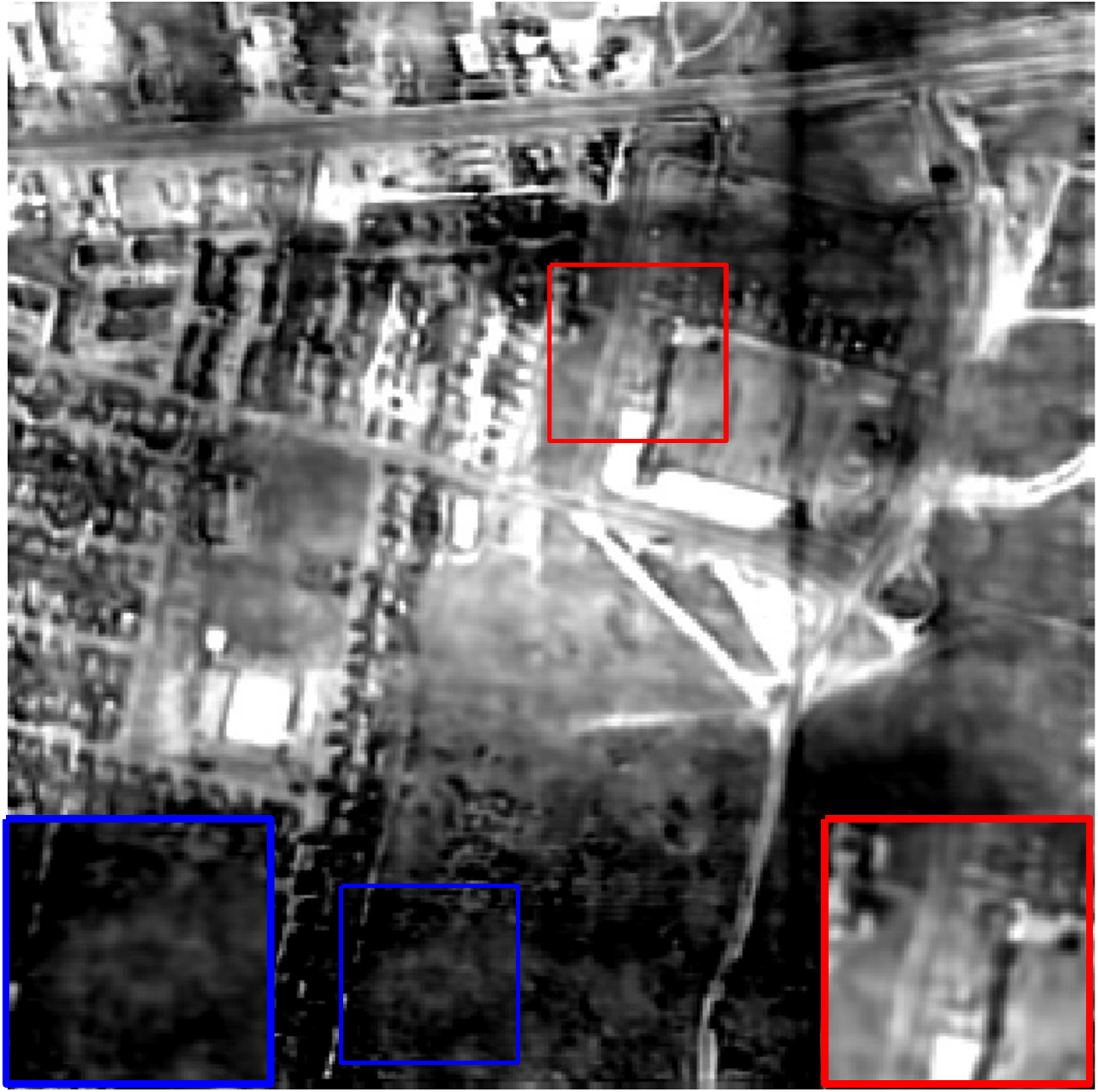}}
\subfigure[\scriptsize{Ours}]{\includegraphics[width=0.115\textwidth,clip=true]{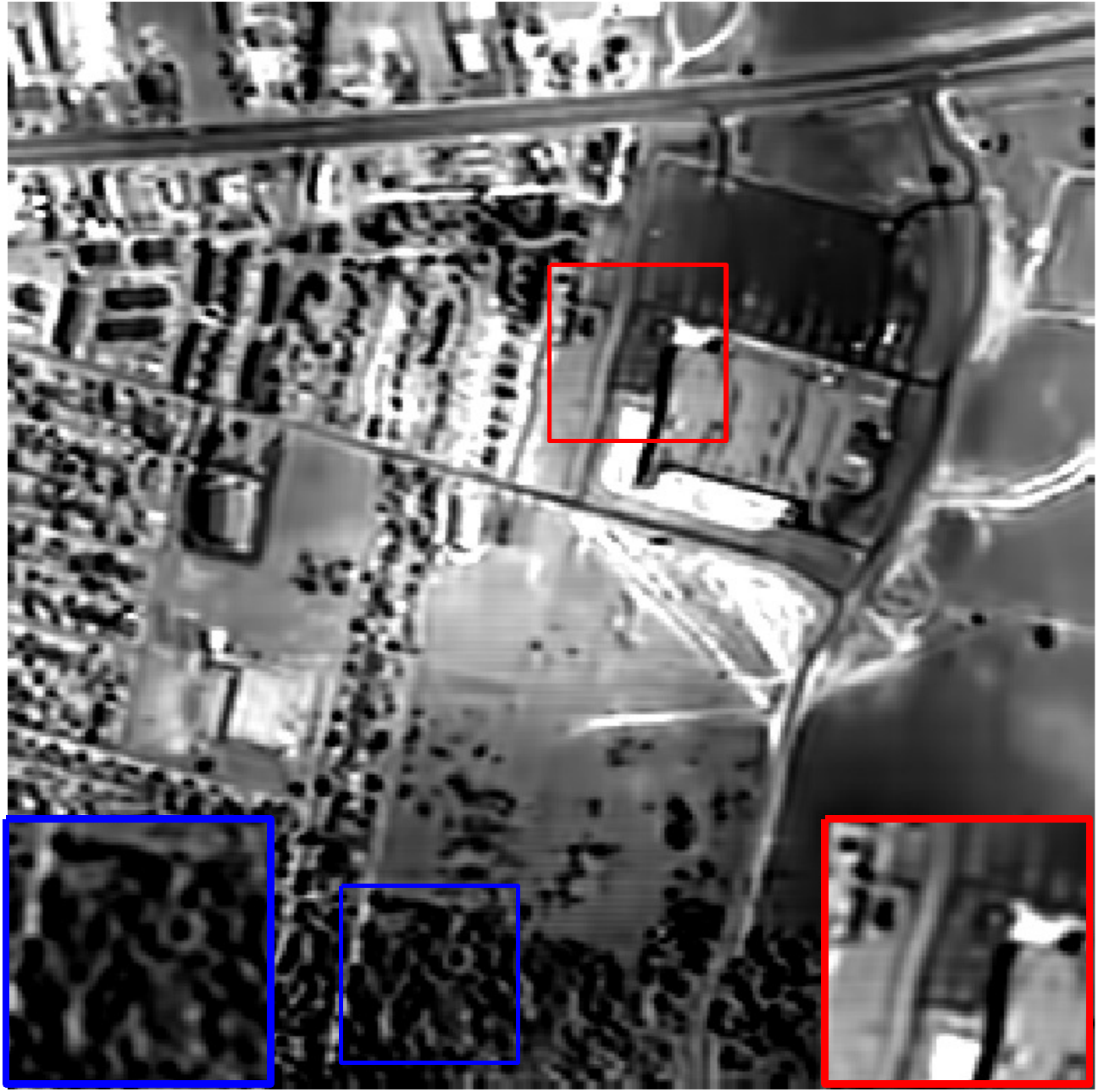}}}
\caption{Denoising results on \textbf{Real-world Urban HSI} with respect to bands 1, 109, and 208. Our SMDS-Net obtains the best qualitative  results. (\textbf{Best view on screen with zoom}) } \label{fig:icvl}
 \end{figure}

HSI denoising  methods can be divided into two  categories, i.e., model-based and learning-based. Model-based methods make explicit hypotheses on the underlying physical characteristics of HSIs and introduce various prior structures such as spectral low-rankness~\cite{Zhang2014,He2019,Liu2012,Wang2021a,Chang2017}, spatial redundancy~\cite{Ye2015,Li2016,Lu2016,Wei2017,Fu2015,Qi2018} and nonlocal similarity~\cite{Peng2014,Chen2020}.    These model-based methods are  physically interpretable   but heavily depend on the effectiveness of hand-crafted  priors, computational demanding iterative optimization, and exhaustive hyperparameter tuning.

\begin{figure*}[htbp]
    \centering
    \graphicspath{{figure/}}
    \includegraphics[width=0.96\linewidth]{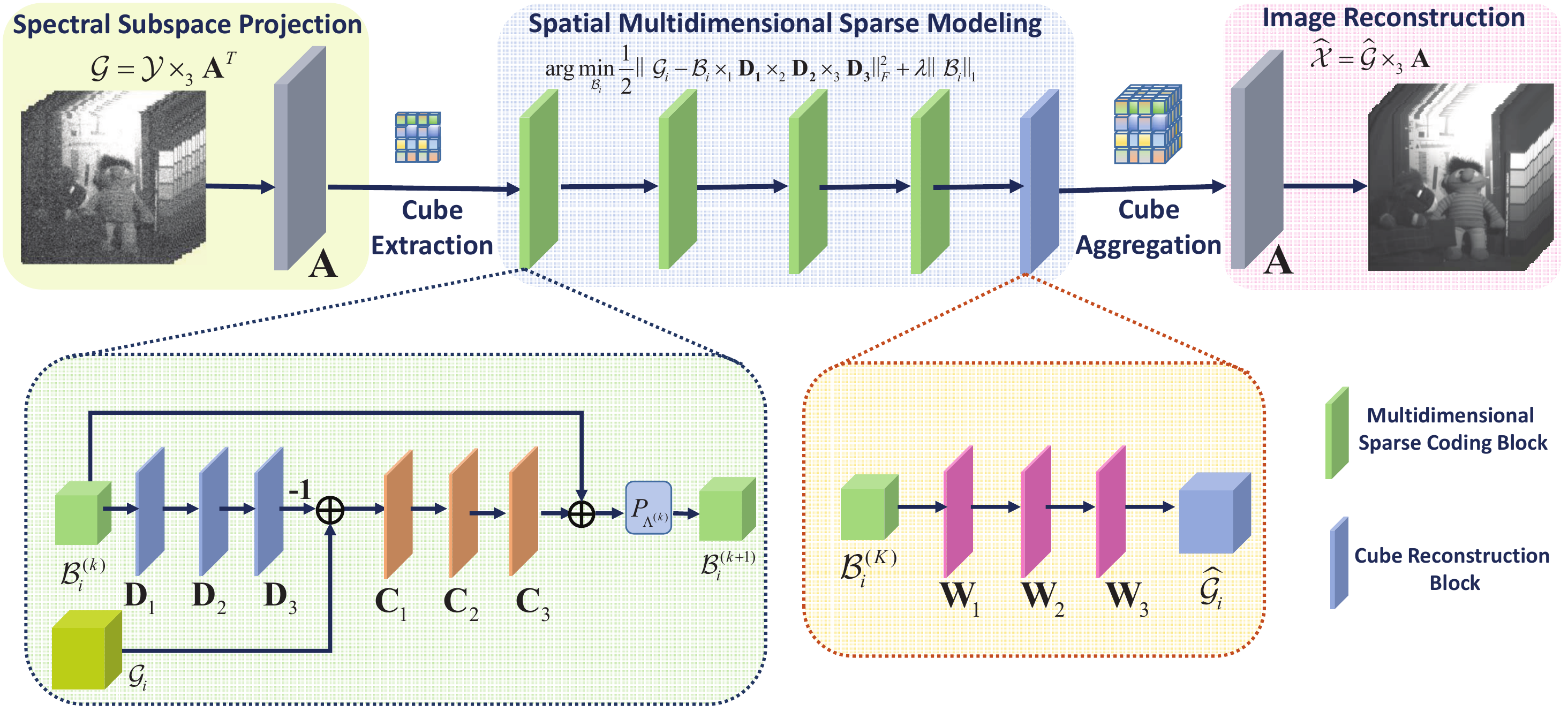}
    \caption{Illustration of the proposed method.  The observed noisy HSI $\Ycal$ is first projected into a low-dimensional image $\Gcal$ with respect to learned subspace $\A$ to depict the spectral low-rankness.  A number of overlapping cubes are then extracted from $\Gcal$ and sparsely encoded by a number of multidimensional sparsity coding blocks and a cube reconstruction block to model the spatial redundancy. After that, All the resulted cubes are aggregated and reconstructed with $\A$ to yield  the denoised image $\widehat{\Xcal}$.}\label{fig:framework}
\end{figure*}


As an alternative paradigm, deep learning (DL)-based methods directly learn the nonlinear mapping between noisy-clean HSI pairs for  denoising~\cite{Maffei2020,Lin2020,Chang2019,Yuan2019,Dong2019,Zhao2020,Wei2020} rather than making subjective physical assumptions on the underlying HSIs.  Attributing to their strong representation ability, DL-based methods provide competitive performance when trained with a large number of noisy-clean HSI pairs.  However, the high cost of acquiring hyperspectral data can dramatically limit their generalization ability and make the learning infeasible. Moreover,  complicated network architectures also hinder deep analysis of the functionality of different modules and  comprehensive understanding of the rationality behind their denoising mechanisms. 

Given the above discussion, it is preferable to design a  HSI denoising method that combines the advantages of both model-based and  DL-based methods, i.e., of high interpretability while supporting discriminative learning.  The consideration of spectral-spatial correlation  facilities information sharing among bands and improves denoising performance~\cite{Fu2016,He2019,Chen2020,Li2016}. Spectral low-rankness and spatial redundancy priors are beneficial to better model underlying clean HSI, helping more effectively remove noises~\cite{Gong2020,Ye2015,Xie2018,Lu2016}.  Therefore, to reach the above goal, the spectral-spatial correlation, spectral low-rankness, and spatial redundancy structure priors and end-to-end learning are very important factors to be considered.

%

 In this paper, a model guided network is introduced for HSI denoising,  which simultaneously considers all the above factors in a unified network whose framework is shown in Fig.~\ref{fig:framework}.  Specifically,  we first build a  subspace based multidimensional sparse (SMDS) model under the umbrella of multidimensional tensor sparse representation, which projects the high-dimensional HSIs into a low-dimensional spectral subspace and exploits the spatial redundancy property on the projected image.  SMDS has higher capability of removing noises because it considers both spectral and spatial structure powered by tensors. After that, an end-to-end deep neural network named SMDS-Net is obtained by unfolding the denoising procedure and optimization algorithm of SMDS. SMDS-Net includes five stages, \emph{i.e.} spectral subspace projection, cube extraction, spatial multidimensional sparse modeling, cube aggregation, and image reconstruction. Each stage has a clear physical meaning, implying the interpretability of our network.


 To the best of our knowledge, this is the first time that low-rankness, sparsity, and spectral-spatial structural priors were simultaneously taken into consideration in a unified end-to-end network for HSI denoising. The main contribution of this paper can be summarized as follows: 
 \begin{enumerate}
 	\item  We propose a novel subspace based multidimensional sparse  model which captures  the spectral-spatial correlations, spectral low-rankness and spatial sparsity  of HSIs under the umbrella of tensors, making the model highly capable of  removing noises.
 	\item  We construct an  end-to-end deep neural network by unfolding the denoising procedure and optimization algorithm of SMDS.  SMDS-Net combines the hybrid advantages of  model-based and DL-based methods,  such as clear interpretability, higher representation capability, and strong learning ability. 
 	\item Extensive experiments confirm the advantages of the SMDS-Net over other state-of-the-art model-based and DL-based methods.   Moreover, a comprehensive analysis of the network is carried out to fully explain its denoising mechanism and learning ability. 
 \end{enumerate}





This paper is organized as follows: Section~\ref{sec:rea} briefly reviews recent works on model-based and DL-based HSI denoising as well as deep unfolding.  In Section~\ref{sec:model}, we  present the subspace based multidimensional sparse  model and its optimization procedure. Section~\ref{sec:net} shows how to unfold the SMDS model into an end-to-end neural networks as well as the implementation details.   Section~\ref{sec:exp} reports the experimental results on both synthetic and real-world data and  compare them against several other competing approaches.  This paper concludes in  Section~\ref{sec:con}.
\section{Related work} \label{sec:rea}
This section briefly introduces the recent works concerning model-based and deep learning-based HSI denoising methods and also deep unfolding.
\subsection{Model Based Approaches}


HSIs can be considered as a stack of 2D images along the spectral dimension. Therefore, traditional 2D image denoising methods such as  Non-local means (NLM)~\cite{Buades2005}, BM3D~\cite{Dabov2007} and low-rank/sparse representation~\cite{Zha2020,Zha2020a} can be directly applied for HSI denoising in a band-wise manner. Band-wise methods ignore the spectral-spatial correlation in HSIs, leading to unsatisfactory denoising performance.  Alternatively, spectral-spatial methods  jointly leverage the information in different bands and simultaneously model HSIs in both spectral and spatial domains, thereby achieve more promising performance. Typical examples include BM4D~\cite{Maggioni2013} and 3D NLM~\cite{Qian2012}, which model HSIs at 3D cube level rather than patch level.

Following this line,  sparse representation based methods   encode the observed HSIs using a few atoms in predefined fixed  dictionaries~\cite{Qian2013} or data-driven adaptive dictionaries~\cite{Fu2015,Lu2016,Ye2015}.  Besides sparsity, the spectral correlation among bands and spatial correlation among  pixels also imply there exist low-dimensional structures  in HSIs that can be depicted by matrix rank minimization~\cite{He2016,Sarkar2021} or low-rank decomposition~\cite{Azimpour2021,Zhang2021}.  Moreover, structural sparse/low-rank representation also demonstrate their effectiveness in HSI denoising by taking the local and nonlocal similarity structures of HSIs into sparse/low-rank models~\cite{Qian2013,He2019,Zhuang2018,Cao2016}. Matrix-based sparse/low-rank models inevitably destroy the spectral-spatial structure of HSIs when converting 3D HSIs into a 2D matrix and require more parameters.  For this reason, tensor-based sparse/low-rank representations are introduced and successfully employed in HSI denoising~\cite{Qi2018,Chang2021,Liu2012,Wang2021,Peng2014,Chang2017}.  

Although these methods work well when the prior model assumptions match well with the HSIs to be processed, they normally require iterative numerical optimization which is computationally expensive. Exhaustive hyperparameter tuning also limits their  effectiveness in complex scenarios.

\subsection{Deep Learning Based Approaches}
Instead of subjective physical assumptions on the underlying HSIs, DL-based methods leverage their strong image representation capability and learn a direct mapping function between noisy and clean HSIs~\cite{Shi2021,Yuan2021}.  Chang~\emph{et al.}~\cite{Chang2019} learned a series of multiple channels of 2D filters for the consideration of both spatial and spectral structures of HSIs. Yuan~\emph{et al.}~\cite{Yuan2019} proposed multiscale spectral-spatial denoising through multi-scale feature extraction and multilevel feature representation using deep neural networks. To effectively extract the joint spectral-spatial information, Zhao~\emph{et al.}~\cite{Zhao2020} and Maffei~\emph{et al.}~\cite{Maffei2020} took the current band and its $K$ adjacent bands as input. Alternatively, structural spectral-spatial correlations in HSIs can also be extracted by 3D convolutions~\cite{Dong2019,Wei2020}. Instead of directly operating on the original HSIs, noisy HSIs can be converted into low-dimensional HSIs before applying CNNs on the reduced images~\cite{Lin2020,Nguyen2020}.

 Although these methods are able to recover the underlying image via training deep models on a large number of noisy-clean HSI pairs, they often ignore very important intrinsic prior structures of  HSIs such as sparsity and low-rankness as mentioned previously.  Instead, they perform denoising in a ``black box" mechanism whose architecture is empirically determined in a heuristic manner and exhausting trials, making further model improvement difficult. Therefore, the interpretability of the actual functionality is a matter of concern.

\subsection{Deep Unfolding}

Deep unfolding starts from the iterative algorithm deduced by the model-based methods and maps each iteration into a typical layer of a deep neural network. By stacking a predefined number of layers, a hierarchical deep network architecture can be obtained. Deep unfolding provides an attractive way to inherit the advantages of model-based methods, for example, high interpretability and good generalization capability and absorb the merits of learning-based methods such as strong learning ability and computational efficiency.  Attributing to this, deep unfolding networks have been successfully applied in many tasks, such as hyperspectral fusion~\cite{Xie2019,Dong2021}, rain removal~\cite{Wang2020}, image super-resolution~\cite{Liu2016} and  deblurring~\cite{Li2020a}. Our work falls in the line of deep unfolding and converts the subspace based multidimensional sparse model into an end-to-end network to achieve more  more effective  HSI denoising.

\section{SMDS HSI Denoising Model}\label{sec:model}
In this section, we introduce our subspace based multidimensional sparse model and also its associated optimization procedure. 
\subsection{Notations}
In this paper, we represent a scalar by a lowercase letter, e.g. $y$; a vector by a   lowercase letter, e.g. $\y \in \Rbb^{I}$; a matrix by  a  capital letter, e.g. $\Y \in \Rbb^{I_1 \times I_2}$; and a tensor by  a  calligraphic letter, e.g. $\Ycal \in \Rbb^{I_1 \times I_2 \times \ldots \times I_N}$.  The $L_0$ norm of a tensor $\Ycal$ is defined as the number of non-zero entries, i.e, $\|\Ycal\|_0=\#\{\Ycal_{i_1,i_2,\cdots, i_N}\neq 0\}$.  The $L_1$ norm and Frobenius norm of $\Ycal$ are respectively defined as $\|\Ycal\|_1=\sum_{i_1}\sum_{i_2}\cdots\sum_{i_N}|\Ycal_{i_1,i_2,\cdots, i_N}|$,  $\|\Ycal\|_F=\sqrt{\sum_{i_1}\sum_{i_2}\cdots\sum_{i_N}(\Ycal_{i_1,i_2,\cdots, i_N})^2}$.  The n-mode unfolding  is the process of  arranging all the n-mode vectors as columns of a matrix, denoted by $\Y_{(n)}$.  The n-mode product of a tensor $\Ycal$ and matrix $\U \in \Rbb^{J\times I_n}$ is written as $\Ycal\times_n\U$ and computed by $(\Ycal\times_n\U)_{i_1\cdots i_{n-1}j i_{n+1}\cdots i_N}=\sum_{i_n=1}^{I_n}y_{i_1i_2\cdots i_N} u_{ji_n}$.

\subsection{Model Formulation}
Given an HSI $\Ycal$ with $H \times W$ pixels and $B$ bands corrupted by additive zero-mean Gaussian noise $\Ncal$, the observation model can be represented as
\begin{equation}
	\Ycal=\Xcal+\Ncal \label{eq:model}
\end{equation}
 where $\Xcal$ is the clean HSI to be estimated. This  is an ill-posed inverse problem and the prior knowledge on $\Xcal$ is very important to solve the problem. Specifically, we consider spectral-spatial correlation,  spatial sparsity and spectral low-rankness priors to model $\Xcal$.

\noindent\textbf{Spectral low-rankness:}  As stated in~\cite{BioucasDias2008,Khan2015}, the spectral correlations among bands make acquired spectra in fact lie in a low-dimensional subspace, i.e.,
\begin{equation}
	\Xcal=\Gcal\times_3\A \label{eq:spectral}
\end{equation}
Here $\A=[\a_1,\cdots,\a_r] \in \Rbb^{B\times R}$ is the spectral subspace spanned by $R$  bases with $R$ far less than $B$ and $\Gcal \in H \times W \times R$ is the representation tensor with respect to $\A$. In general, $\A$ can be obtained by many approaches, for example, unmixing and singular value decomposition (SVD) on the unfolded $\Ycal$ along spectral models. As orthogonal subspace shares many merits such as disentangling the correlation among bands~\cite{Zhuang2018,Nguyen2020,He2019} and also simplifying the optimization, we choose a set of orthonormal bases, i.e.,  $\A^{\Trm}\A=\I$. \\

%
%

\noindent\textbf{Spatial sparsity:} As the low-rank projection is performed in the spectral domain, the projected image can be considers  as a linear combinations of the bands of $\Xcal$, meaning that the pixels nearby shares similarity.  On the other hand, the spectral low-rankness is a global  transformation for all the pixels, indicating that for pixels in a local region, there still exist dependencies.   Instead of separately encoding each mode (channel) of $\Gcal$, we  model the local 3D cubes within $\Gcal$ under the framework of tensor based multidimensional (MD) sparse  representation.  In this way,  all the spatial structures in each mode remain  and  can be shared  across channels with less parameters, facilitating better handling  of noises. This can be implemented  by dividing $\Gcal$ into a number of small  overlapping cubes, calculating their sparsity, and putting them back to the original position using their denoised estimation by averaging with their overlapped cubes. Specifically, let $\Rcal_i$ denote an operator extracting the $i$-th cube from $\Gcal$, under the framework of  MD sparse representation,  the sparse representation problem of each cube $\Gcal_i=\Rcal_i\Gcal$ is formulated as
   \begin{equation}
   	\arg \min_{\Bcal_i}\frac{1}{2}\|\Gcal_i-\Bcal_i\times_1\D_1 \times_2\D_2 \times_3\D_3\|_F^2, \text{s.t.}\: \|\Bcal_i\|_0 \leq K \label{eq:spatial}
   	\end{equation} where $\D_j \in \Rbb^{I_j \times M_j } (j=1,2,3)$ is the overcomplete dictionary matrix along $j-$th mode with $I_j \leq M_j$. Unlike Tucker decomposition which only models the space that the tensor itself belongs to, the model in Eq.~(\ref{eq:spatial}) can simultaneously and adaptively characterize redundant basis of the space where multiple tensors reside in a data-driven way.

Substituting Eq.~(\ref{eq:spectral}) and  Eq.~(\ref{eq:spatial}) into Eq.~(\ref{eq:model}), the proposed subspace based MD sparse representation denoising method is written as
   	\begin{equation}
   	\begin{split}
   		&\!\{\A^*, \Gcal^*, \Bcal_{i}^*\}=\arg\min_{\A,\Gcal, \Bcal_i} \frac{1}{2}\|\Gcal\times_3\A-\Ycal\|_F^2\\
   		&+\mu\sum_{i}\left(\frac{1}{2}\|\Gcal_i-\Bcal_i\times_1\D_1 \times_2\D_2 \times_3\D_3\|_F^2+\lambda\|\Bcal_i\|_0\right)\\
   		  & \text{s.t.}\:\A^{\Trm}{\A}=\I
   		  \end{split}\label{eq:obj}
   	\end{equation}
where $\D_j$ is the dictionary along each dimension. In Eq.~(\ref{eq:obj}), the first term is the data fidelity term that demands proximity between the observed noisy $\Ycal$ and estimated image $\Gcal\times_3\A$. The second term stands for the prior assumption on $\Gcal$ that every cube $\Gcal_i$ has an MD sparse representation and $\lambda>0$ controls sparse regularization level. $\mu$ balances the data fidelity term and sparse representation term and can be implicitly set to 1 as in~\cite{He2019,Chang2017}.
\subsection{Model Optimization}
\begin{algorithm}[h]
\caption{SMDS for HSI denoising.} \label{alg:ALS-LL1}
\begin{algorithmic}[1]
\REQUIRE
Noisy HSI $\Ycal$. \\
\ENSURE Denoised HSI $\Xcal$. \\
\STATE Estimate $R$ using Hysime~\cite{BioucasDias2008}. \\
\STATE Conduct SVD on  $\Y_{(3)}$, i.e., $\Y_{(3)}=\U\S\V^{\Trm}$ and set $\A=[\v_1,\cdots, \v_R]$.   \\
\STATE Project $\Ycal$ to obtain $\Gcal$ with Eq.~(\ref{eq:projG}).//\textit{spectral subspace projection}\\
\REPEAT
	\STATE Extract overlapping cubes from $\Gcal$. //\textit{cube extraction}
	\STATE Compute $\Bcal_i$ with Eq.~(\ref{eq:tista}). //\textit{spatial MD sparse coding}
\UNTIL{TISTA converges.}
\STATE Obtain $\widehat{\Gcal}$ with Eq.~(\ref{eq:gupdate}). //\textit{cube reconstruction and aggregation} \\
\STATE Output the estimated $\widehat{\Xcal}$ with Eq.~(\ref{eq:spectral}). //\textit{image reconstruction}
\end{algorithmic}
\end{algorithm}
Eq.~(\ref{eq:obj}) is difficult to solve because of the complicated constraint on $\A$ and $\Bcal_i$. Similar to~\cite{Zhuang2018, He2019}, we solve Eq.~(\ref{eq:obj}) in a greedy way, which includes sequentially learning  the subspace $\A$ from $\Ycal$ and  MD sparse representation $\Bcal_{i}^*$ from $\Gcal_i$.

\noindent\textbf{Learning $\A$:} The optimization of $\A$ can be considered as a Tucker-1 decomposition problem. Under the framework of higher-order singular value decomposition (HOSVD), we can obtain $\A$ by performing singular decomposition on the unfolded matrix $\Y_{(3)}$ along the spectral dimension, i.e., $\Y_{(3)}=\U\S\V^{\Trm}$. Then, $\A$ is assigned with $\A=[\v_1,\cdots, \v_R]$ where $R$ is estimated by the Hysime algorithm~\cite{BioucasDias2008}. In order to more accurately estimate $R$ and $\A$, some fast denoising methods such as NLM~\cite{Buades2005} can be applied. Once $\A$ is obtained, $\Gcal$ can be obtained by
\begin{equation}
	\Gcal=\Ycal\times_3\A^{\Trm} \label{eq:projG}
\end{equation}

\noindent\textbf{Learning $\Bcal_{i}$:} The $L_0$ norm of $\Bcal$ is intractable because of its NP-hard property. It can be relaxed to the $L_1$ norm to result in  a convex optimization problem. Along this line, the optimization problem of $\Bcal_i$ can be formulated by
\begin{equation}
	\arg \min_{\Bcal_i}\frac{1}{2}\|\Gcal_i-\Bcal_i\times_1\D_1 \times_2\D_2 \times_3\D_3\|_F^2+\lambda\|\Bcal_i\|_1
\end{equation}  This problem can be solved with Tensor-based Iterative Shrinkage Thresholding Algorithm (TISTA)~\cite{Qi2016} with the following solution:
\begin{equation}
\begin{split}
	\Bcal_i^{(k+1)}&\!\!=\!P_{\lambda/L}(\Bcal_i^{(k)}\!\!-\!\!\frac{1}{L}(\Bcal_i^{(k)}\!\times_1\D_1^{\Trm}\D_1\!\times_2\D_2^{\Trm}\D_2\!\times_3\D_3^{\Trm}\D_3\\
	&-\Gcal_i\times_1 \D_1^{\Trm}\times_2\D_2^{\Trm}\times_3\D_3^{\Trm}))
\end{split} \label{eq:tista}
\end{equation} where $L$ is a Lipschitz constant, $P_{\lambda/L}=\text{sgn}(x)(|x|-\lambda/L)_+$ performs soft-thresholding operation to achieve the sparsity of $\Bcal_i^{(k+1)}$ and $k$ indexes the iteration number.

After  obtaining $\Bcal_i$,  $\widehat{\Gcal}$ can be obtained by averaging $m$ estimates for each entry in the cube, i.e.,
\begin{equation}
	\widehat{\Gcal}=\frac{1}{m}\sum \limits_{i}\Rcal_{i}^{'}\Bcal_i\times_1\D_1 \times_2\D_2 \times_3\D_3 \label{eq:gupdate}
\end{equation}
where $\Rcal_i^{'}$ is a linear operator  placing back the cube  $\Bcal_i\times_1\D_1 \times_2\D_2 \times_3\D_3$ at the position centered on entry $i$.  Finally, the output denoised HSI $\widehat{\Xcal}$ can be obtained  by Eq.~(\ref{eq:spectral}).  Algorithm~\ref{alg:ALS-LL1} summarizes the denoising procedure of  the SMDS method.  SMDS simultaneously considers the spectral-spatial correlation, spectral low-rankness and spatial sparsity structures of HSIs and has strong denoising ability. However, SMDS formulates denoising as an optimization problem, which normally requires computationally inefficient iterations and exhausting hyperparameter tuning, making the denoising efficiency and effectiveness still limited. In the next section, we will introduce an unfolded SMDS network to solve this problem.
\section{Unfolded SMDS Network}\label{sec:net}
In this section, we unfold all the steps of SMDS in Algorithm~\ref{alg:ALS-LL1} as network layers to yield an end-to-end neural network for discriminative training.  
\subsection{Network Architecture}
Algorithm~\ref{alg:ALS-LL1} can be decomposed into five stages, which corresponds to five parts in the network shown in Fig.~\ref{fig:framework}.  Here, we describe the details of key parts as follows:  

\noindent\textbf{Spectral subspace projection:} The first stage of the network is mapping $\Ycal$ into a low-dimensional subspace held by $\A$ following Eq.~(\ref{eq:projG}).  Since different HSIs may reside in different subspaces, it is unrealistic to encode $\A$ as a network parameter to learn the unique subspace for all the HSIs.    For this reason,  $\A$ is precomputed and taken as network input along with  $\Ycal$. Consequently, our network can not only  employ the internal data, i.e., the noisy HSI itself for adaptive learning of spectral subspace but also  the extern data, i.e., the training set to learn the shared dictionaries  for all the HSIs.  Moreover, with the spectral subspace as input, our network can be used for input HSI with an arbitrary number of bands, offering much flexibility.

\noindent\textbf{MD sparse coding:} After cube extraction, MD sparse coding stage learns the sparse coding of each cube with respect to multidimensional dictionaries, corresponding to Eq.~(\ref{eq:tista}). Eq.~(\ref{eq:tista}) can be  decomposed into the following three steps:
\begin{align}
	&\Ecal_i^{(k+1)}=\Gcal_i-\Bcal_i^{(k)}\times_1\D_1 \times_2\D_2 \times_3\D_3\label{eq:objE}\\
	&\Hcal_i^{(k+1)}=\Bcal_i^{(k)}+\Ecal_i^{(k+1)}\times_1\C_1^{\Trm}\times_2\C_2^{\Trm}\times_3\C_3^{\Trm}\label{eq:objH}\\
	&\Bcal_i^{(k+1)}=P_{\LAmbda^{(k)}}(\Hcal_i^{(k+1)})\label{eq:objB}
\end{align} where $\LAmbda^{(k)} \in \Rbb^{M_1\times M_2\times M_3}$ denotes the thresholding parameter tensor  for $\Hcal_i^{(k)}$ in the $k$-th layer of MD sparse coding module. $1/L$ is absorbed into $\LAmbda^{(k)}$ and $\D_j$. Following~\cite{Liu2019}, we  decouple $\C$ from  $\D$  and set different $\LAmbda$ in each layer for accelerated convergence.

 Eq.~(\ref{eq:objE}) and  Eq.~(\ref{eq:objH})  are similar in function, mapping the processed tensor to a new dimension along three modes. They can  be regarded as three linear transformation layers.  $P_{\LAmbda_{(k)}}$ in Eq.~(\ref{eq:objB}) corresponds to the nonlinear function allowing for the sparsity of $\Hcal^{(k+1)}$, which can be implemented with $\text{Relu}(x-\theta)-\text{Relu}(-x-\theta)$. Eqs.~(\ref{eq:objE}),~(\ref{eq:objH}) and~(\ref{eq:objB}) make up  a typical MD sparse coding block in  Fig.~\ref{fig:framework}. Stacking this block  for $K$ times  is equivalent to running Eqs.~(\ref{eq:objE}),~(\ref{eq:objH}) and~(\ref{eq:objB}) for $K$ times, resulting in a subnetwork for MD sparse coding. \\ 
 
 \noindent\textbf{Cube reconstruction and aggregation:} Similarly, Eq.~(\ref{eq:gupdate}) can  be interpreted as three linear transformation layers followed by an averaging layer.  As in~\cite{Simon2019,Lecouat2019}, we decouple $\W$ from $\D$ for more effective end-to-end training, i.e., $\widehat{\Gcal_i}=\Bcal_i^{(K)}\times_1\W_1 \times_2\W_2 \times_3\W_3$. This corresponds to the cube reconstruction block in Fig.~\ref{fig:framework}. After that the overlapped cubes are aggregated, i.e., $\widehat{\Gcal}=\frac{1}{m}\sum \Rcal_i^{'}\widehat{\Gcal_i}$. 
 

\noindent\textbf{Image reconstruction:}  The last image reconstruction stage uses Eq.~(\ref{eq:spectral}) to produce the denoised HSIs $\widehat{\Xcal}$. 

\begin{table*}[thbp]
\caption{Comparison of different methods on 50 testing HSIs of ICVL  Dataset. The top two values are marked \textcolor{red}{\textbf{Red}} and \textcolor{blue}{\textbf{Blue}}. }\label{tab:ICVL}
\centering
	\tablesizetwo{	
\begin{tabular}{c|c|c|c|c|c|c|c|c|c|c|c|c|c|c|c|c}
\hline
&&&\multicolumn{4}{c|}{\textbf{Sparse methods}}&\multicolumn{5}{c|}{\textbf{Low-rank  methods}}&\multicolumn{5}{c}{\textbf{DL  methods}}\\
\hline
\multirow{2}*{$\sigma$}&\multirow{2}*{Index}&Noisy&BM3D&BM4D&TDL&MTSNMF&PARAFAC&LLRT&NGMeet&LRMR&LRTDTV&Dn-CNN&HSI-SDe&HSID-&QRNN3D&SMDS-Net\\
&&&~\cite{Dabov2007}&~\cite{Maggioni2013}&~\cite{Peng2014}&~\cite{Ye2015}&~\cite{Liu2012}&~\cite{Chang2017}&~\cite{He2019}&~\cite{Zhang2014}&~\cite{Wang2018}&~\cite{Zhang2017}&CNN~\cite{Maffei2020}&CNN~\cite{Yuan2019}&~\cite{Wei2020}&\textbf{(Ours)}\\
\hline  
\multirow{3}*{\textbf{[0-15]}}&PSNR&33.47&45.68& 45.12& 38.84&45.62&39.09& \textcolor{blue}{\textbf{46.17}}& 39.42& 40.48&42.17   &41.92&41.06&38.08&43.36&   \textcolor{red}{\textbf{47.81}}       		  	
		    		\\	
&SSIM&0.6295&0.9786&0.9753& 0.8481& 0.9562&0.9468&0.9649&0.8647&0.9601&0.9717   &0.9596&0.9565&0.9669&\textcolor{blue}{\textbf{0.9884}}&	\textcolor{red}{\textbf{0.9899}}	
		\\
&SAM&0.1622&0.0252& 0.0254& 0.0829& 0.0409& 0.0299&0.0319&0.0891&0.0177&\textcolor{blue}{\textbf{0.0229}}   &0.0274&0.0329&0.0511&0.0336&\textcolor{red}{\textbf{0.0139}} 
\\ 
\hline 
\multirow{3}*{\textbf{[0-55]}}&PSNR&21.67&38.67&38.21&29.45&37.57&35.71&37.19&30.48&36.20&39.69&37.90&35.66&32.55&\textcolor{blue}{\textbf{40.15}}&\textcolor{red}{\textbf{42.00}}\\	 	
&SSIM&0.2361&0.9363&0.9217&0.5238&0.8569&0.8808&0.8389&0.6748&0.9295&0.9625&0.9293&0.8736&0.8421&\textcolor{blue}{\textbf{0.9729}}&\textcolor{red}{\textbf{0.9733}}		
		\\
&SAM&0.5014&0.0541&0.0552&0.2409&0.1372&0.0542&0.1020&0.2830&\textcolor{blue}{\textbf{0.0334}}&0.0335&0.0504&0.0602&0.0918
&0.0382&\textcolor{red}{\textbf{0.0242}}
\\
\hline       
\multirow{3}*{\textbf{[0-95]}}&PSNR&16.97&35.99& 35.27& 25.40&34.20&32.80& 32.49 &27.20&34.27& \textcolor{blue}{\textbf{38.12}}& 34.65&32.45&29.12&37.66&\textcolor{red}{\textbf{39.34}}\\
&SSIM& 0.1442&0.9080&0.8764&  0.3735& 0.7998&0.7810&0.7191& 0.5506&0.9153&\textcolor{blue}{\textbf{0.9539}}&0.8442&0.7875&0.7049&	0.9479&			\textcolor{red}{\textbf{0.9593}}		\\
&SAM&0.7199& 0.0738&0.0799&0.3641&0.2128&0.0997&0.1687& 0.4275&0.0484&\textcolor{blue}{\textbf{0.0402}}&	0.1094&0.0876&	0.1316&	0.0468&\textcolor{red}{\textbf{0.0303}}\\
\hline 		
\end{tabular}}
\end{table*}

\subsection{Network Training}
\noindent\textbf{Training loss:}
The training loss for a given training set of noisy-clean pairs is defined as the Euclidean distance  between the output of SMDS-Net  and the ground truth $\Xcal$, i.e.,
\begin{equation}
	L=\|\text{SMDS-Net}(\Ycal, \Theta)-\Xcal\|_F^2
\end{equation}
where $\Theta=\{(\C_j,\D_j,\W_j)_{j=1,2,3}, (\LAmbda^{(k)})_{k=1,\cdots,K}\}$ represents all parameters to be learned.
 
\noindent\textbf{Implement details:} SMDS-Net is implemented on the Pytorch platform and trained on  two  NVIDIA GeForce RTX 3090 GPUs with 300 epochs. We adopt the Adam optimizer with the batch size of 2 and  patch size of $56\times 56$.  The initial learning rate is $5\times10^{-3}$ and multiplied by 0.35 for every 80 epochs.  The cube size $[I_1, I_2, I_3]$ is set as $[9,9,9]$.   The dictionaries were initialized with discrete cosine transform (DCT) basis whose number  $[M_1, M_2, M_3] $ are given by [9, 9, 9], leading to dictionaries size of $9\times 9$ along three modes. The number of unfolding $K$ is set to 6. \\
\noindent\textbf{Training dataset:}  The same as~\cite{Wei2020}, 100 images containing $1392\times 1300$ pixels and 31 spectral bands are selected from ICVL hyperspectral dataset~\footnote{http://icvl.cs.bgu.ac.il/hyperspectral/} to construct the training set.   Data augmentation is performed  to enlarge the training  set including random flipping, cropping and resizing. After that  images with a size of $56\times 56\times 31 $ are used to train the network.

 \begin{figure*}[!htbp]
\makeatletter 
\subfigure[Clean]{\includegraphics[scale=0.175,clip=true]{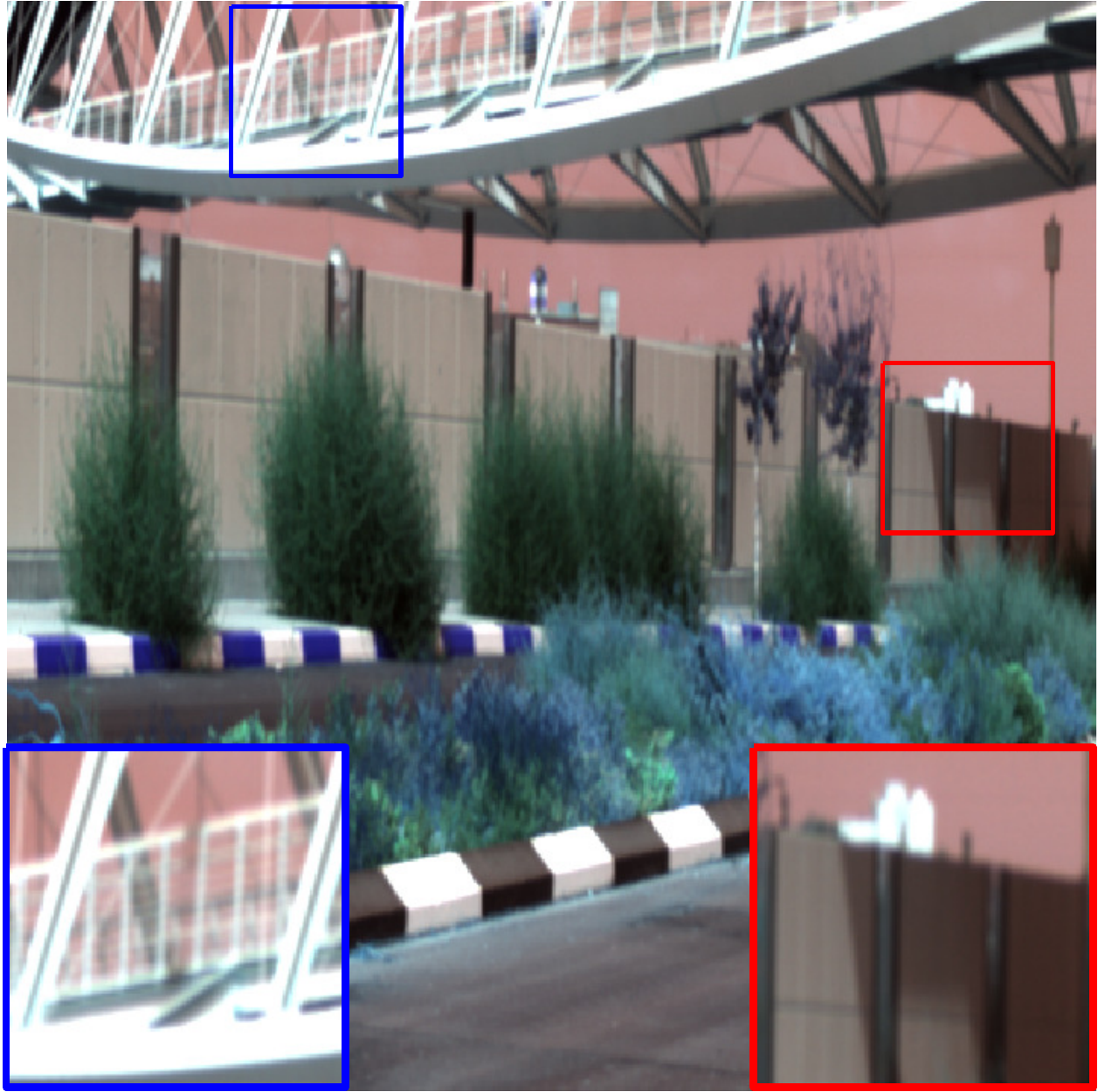}}\hspace{-0.8mm}
\subfigure[Noisy]{\includegraphics[scale =0.175,clip=true]{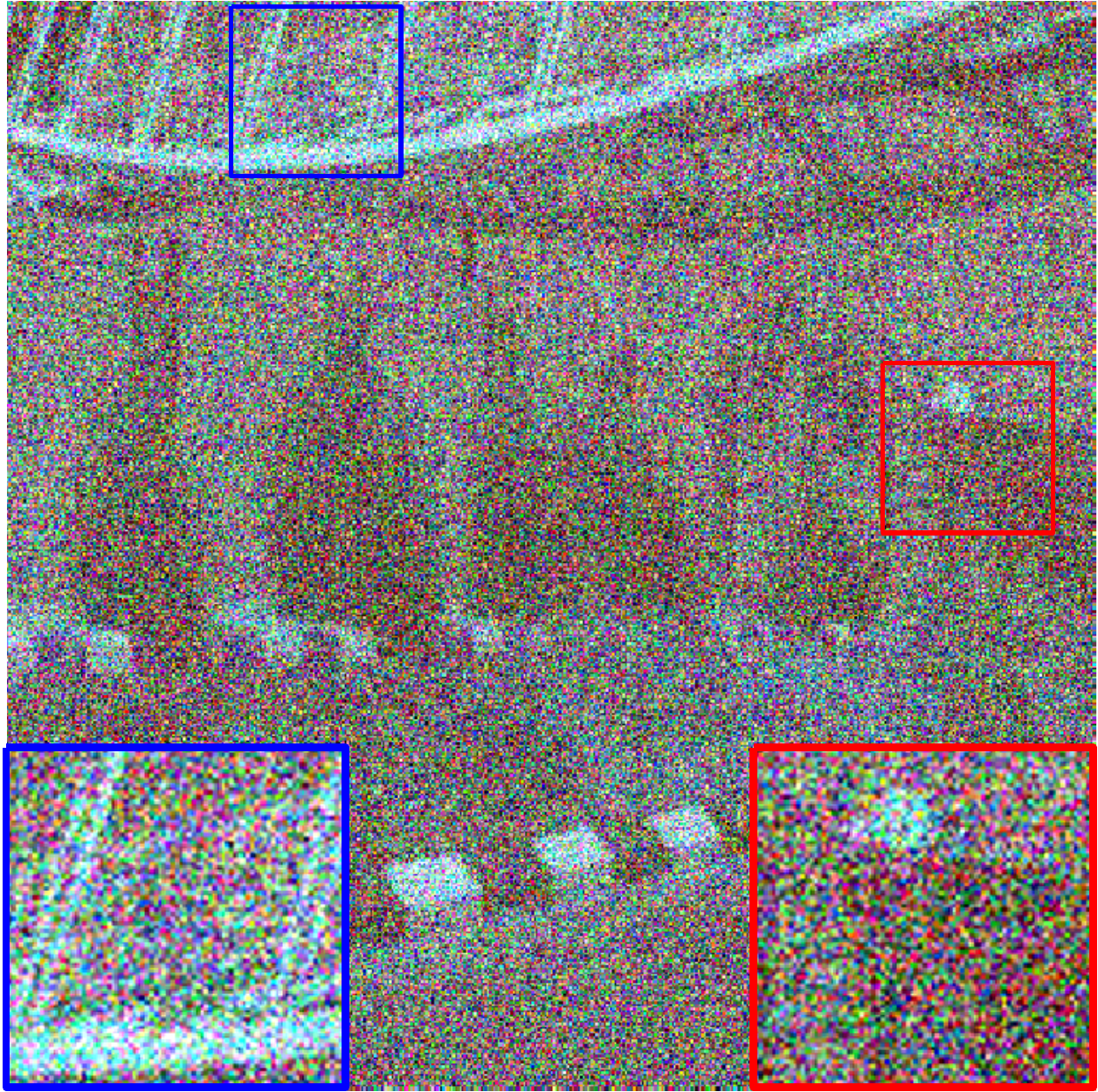}}\hspace{-0.8mm}
\subfigure[BM3D]{\includegraphics[scale =0.175,clip=true]{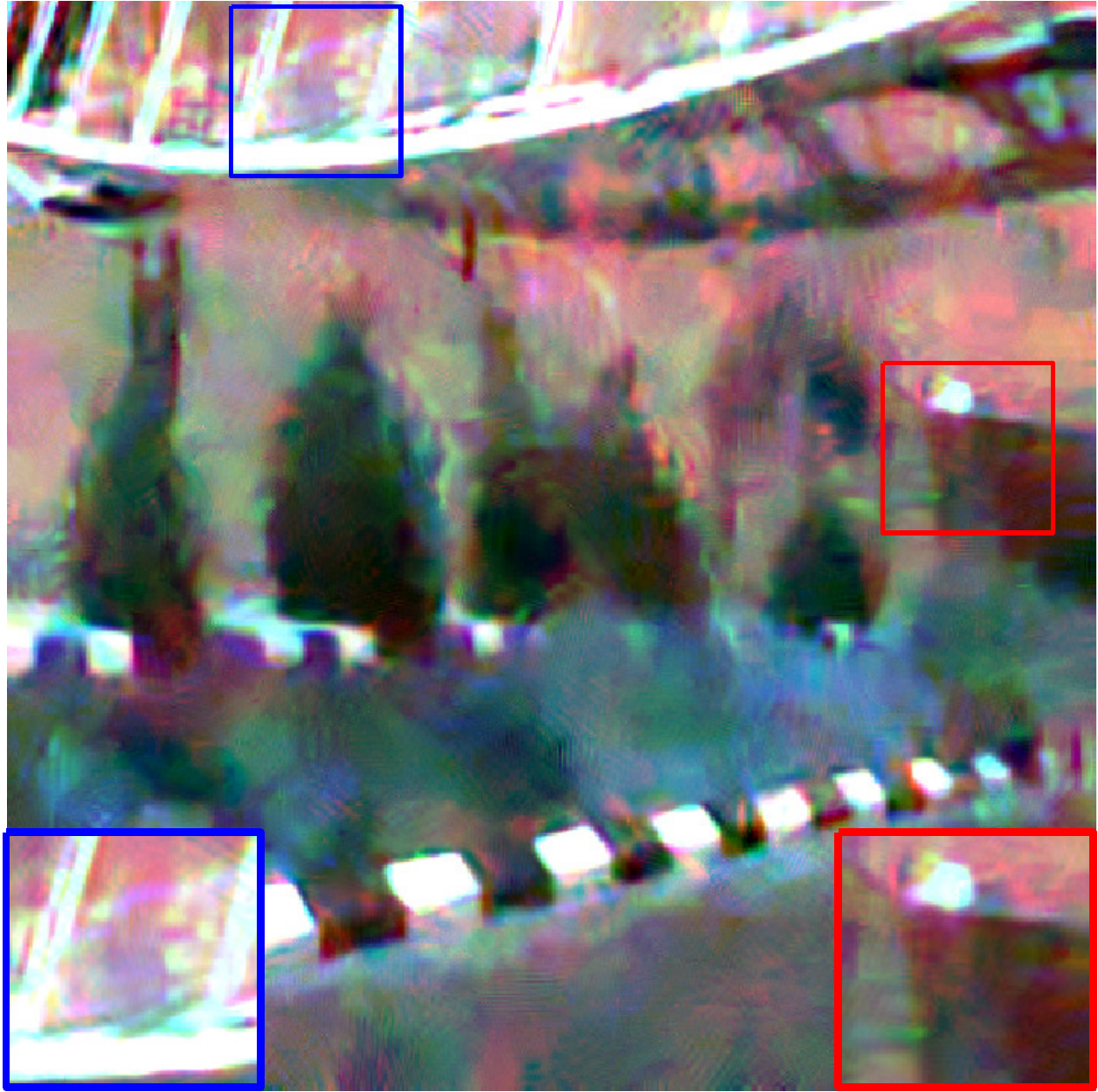}}\hspace{-0.8mm}
\subfigure[BM4D]{\includegraphics[scale =0.175,clip=true]{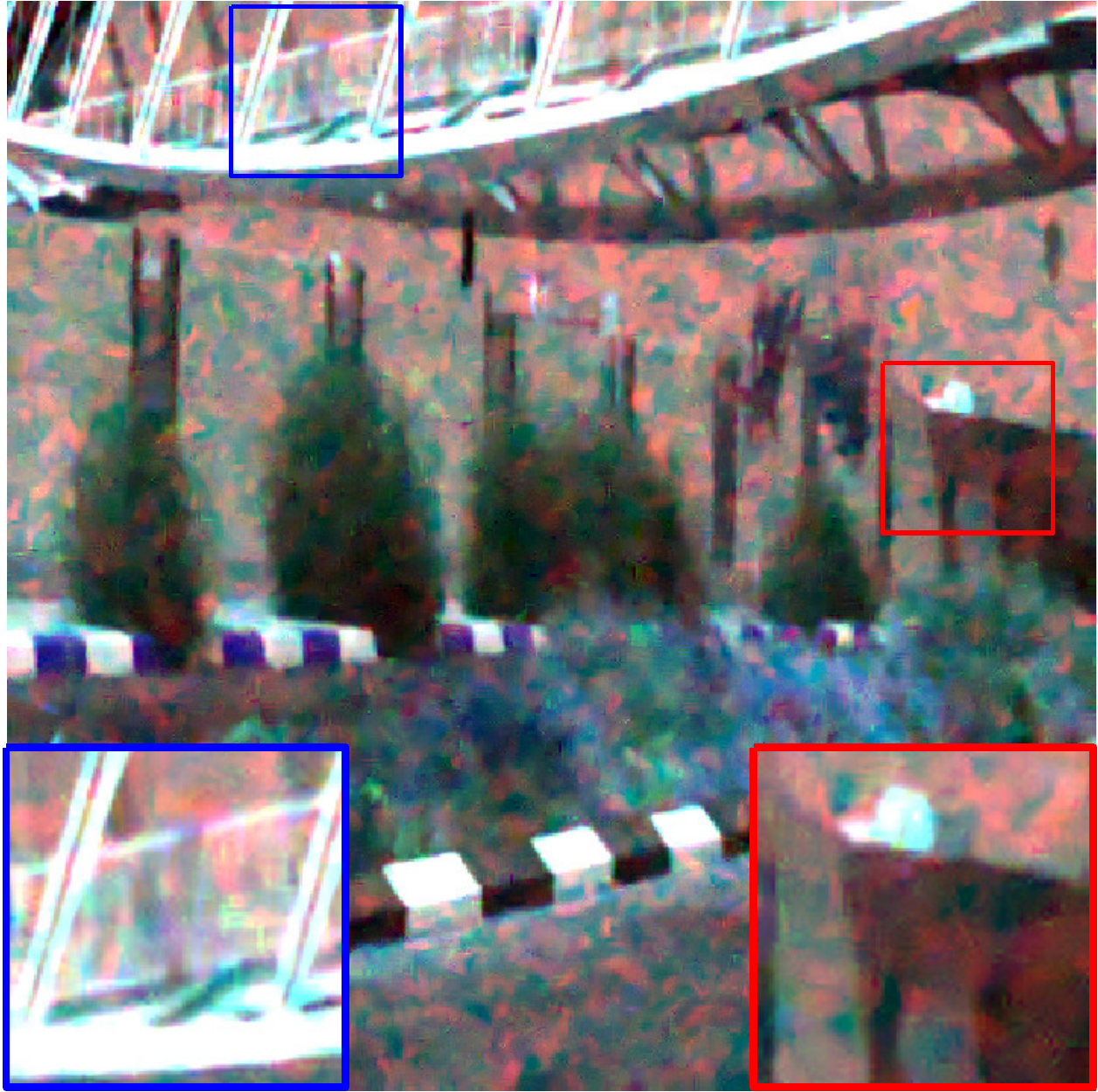}}\hspace{-0.8mm}
\subfigure[TDL]{\includegraphics[scale =0.175,clip=true]{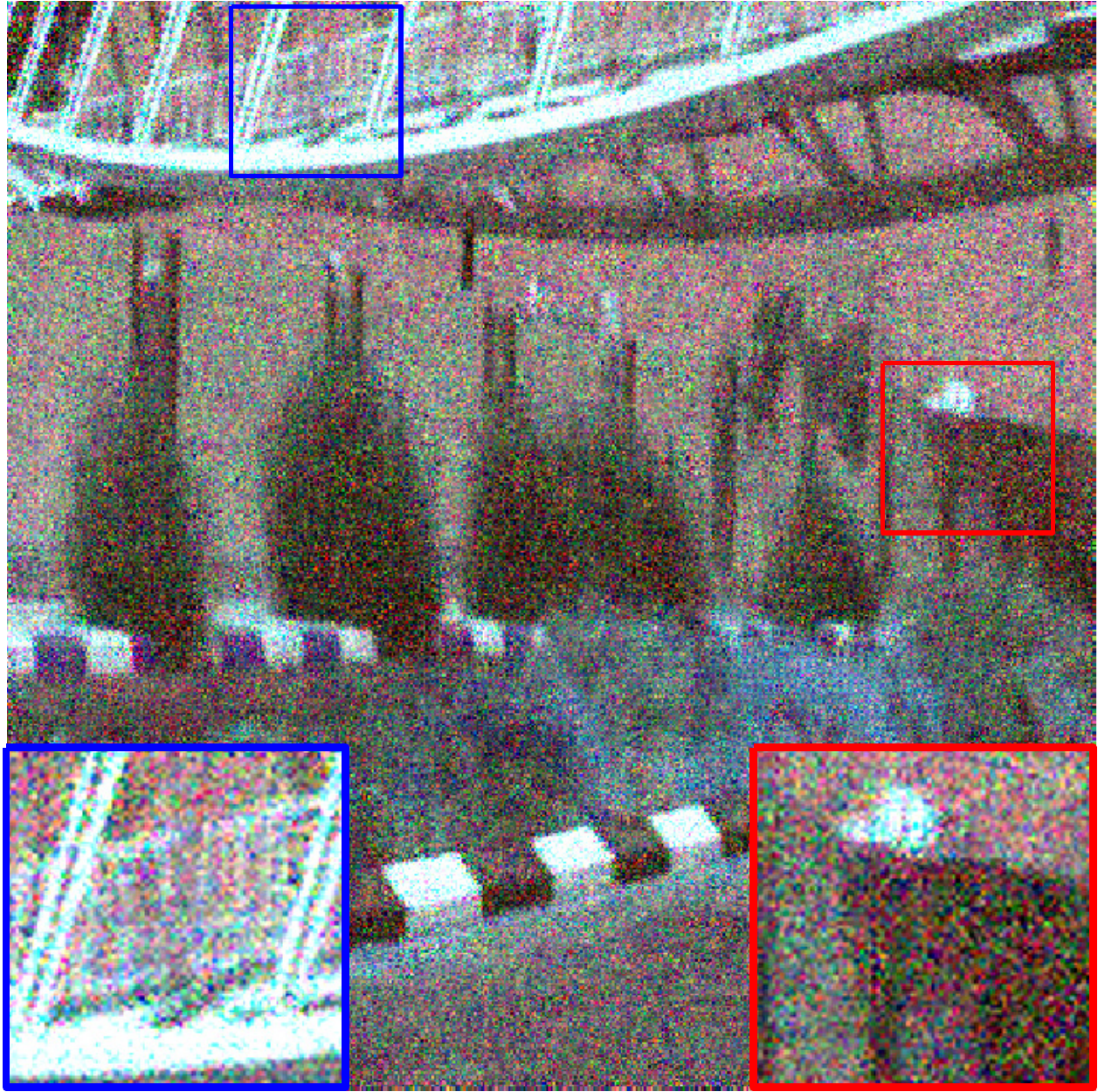}}\hspace{-0.8mm}
\subfigure[MTSNMF]{\includegraphics[scale =0.175,clip=true]{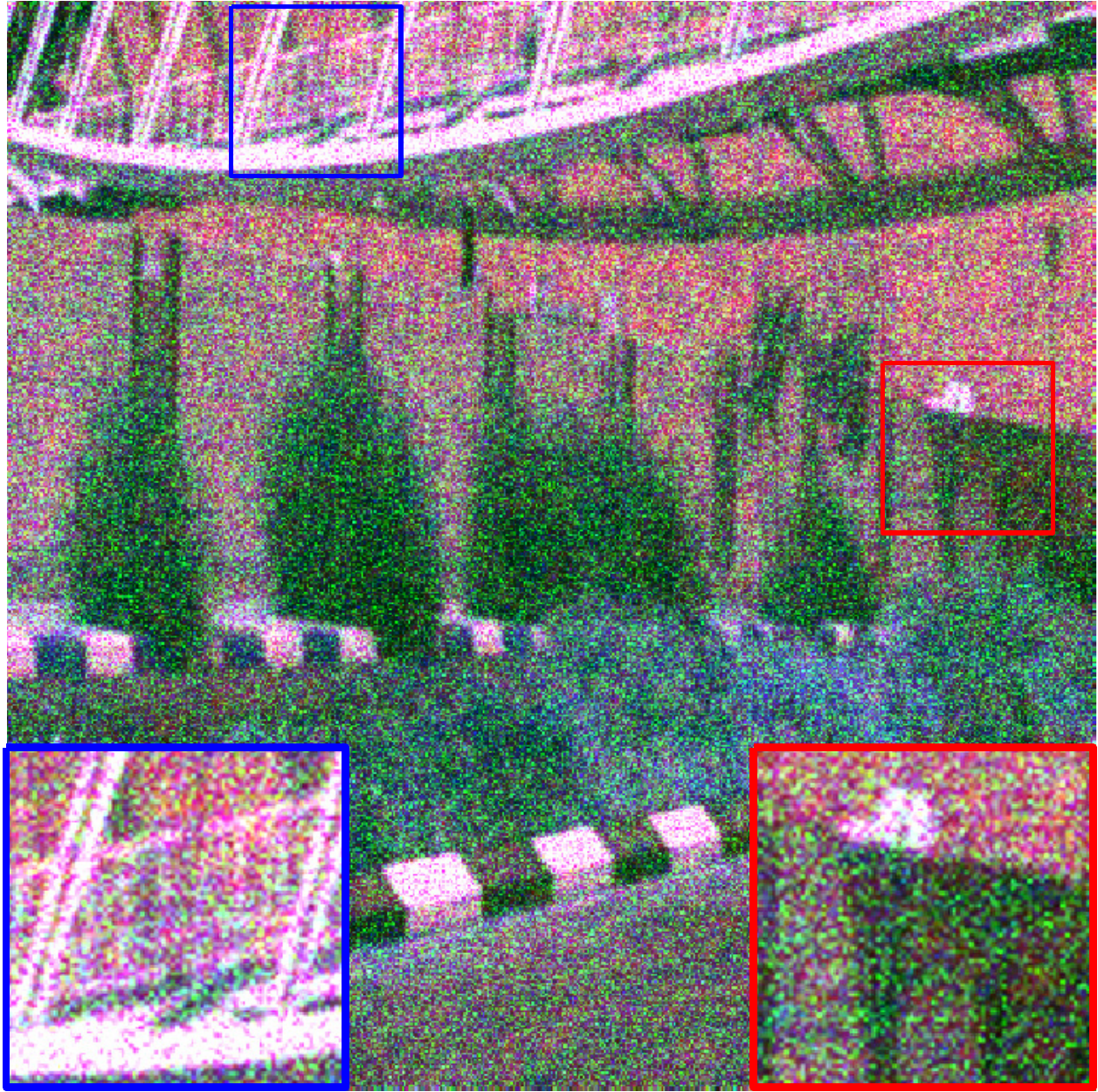}}\hspace{-0.8mm}
\subfigure[PARAFAC]{\includegraphics[scale =0.175,clip=true]{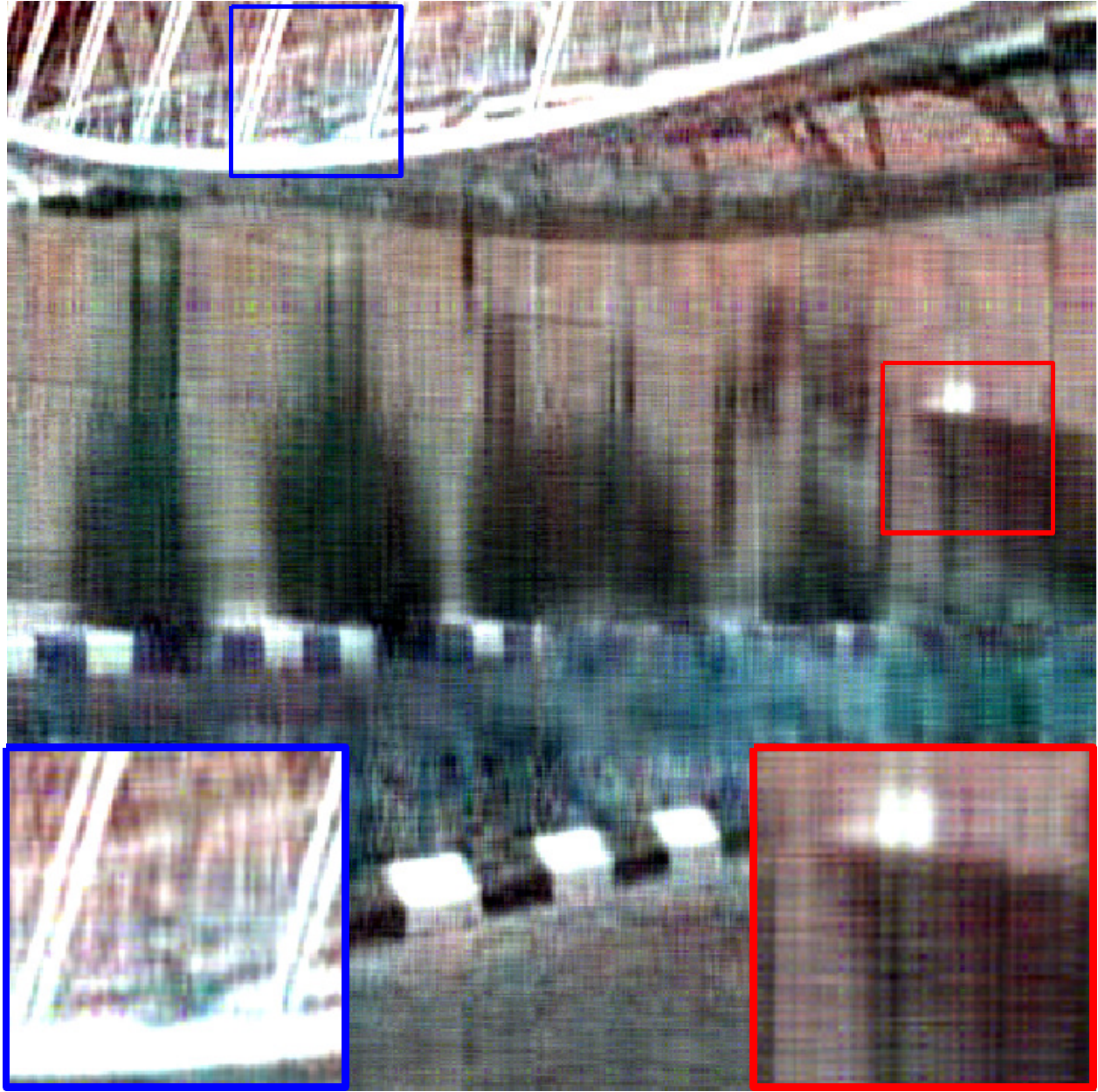}}\hspace{-0.8mm}
\subfigure[LLRT]{\includegraphics[scale =0.175,clip=true]{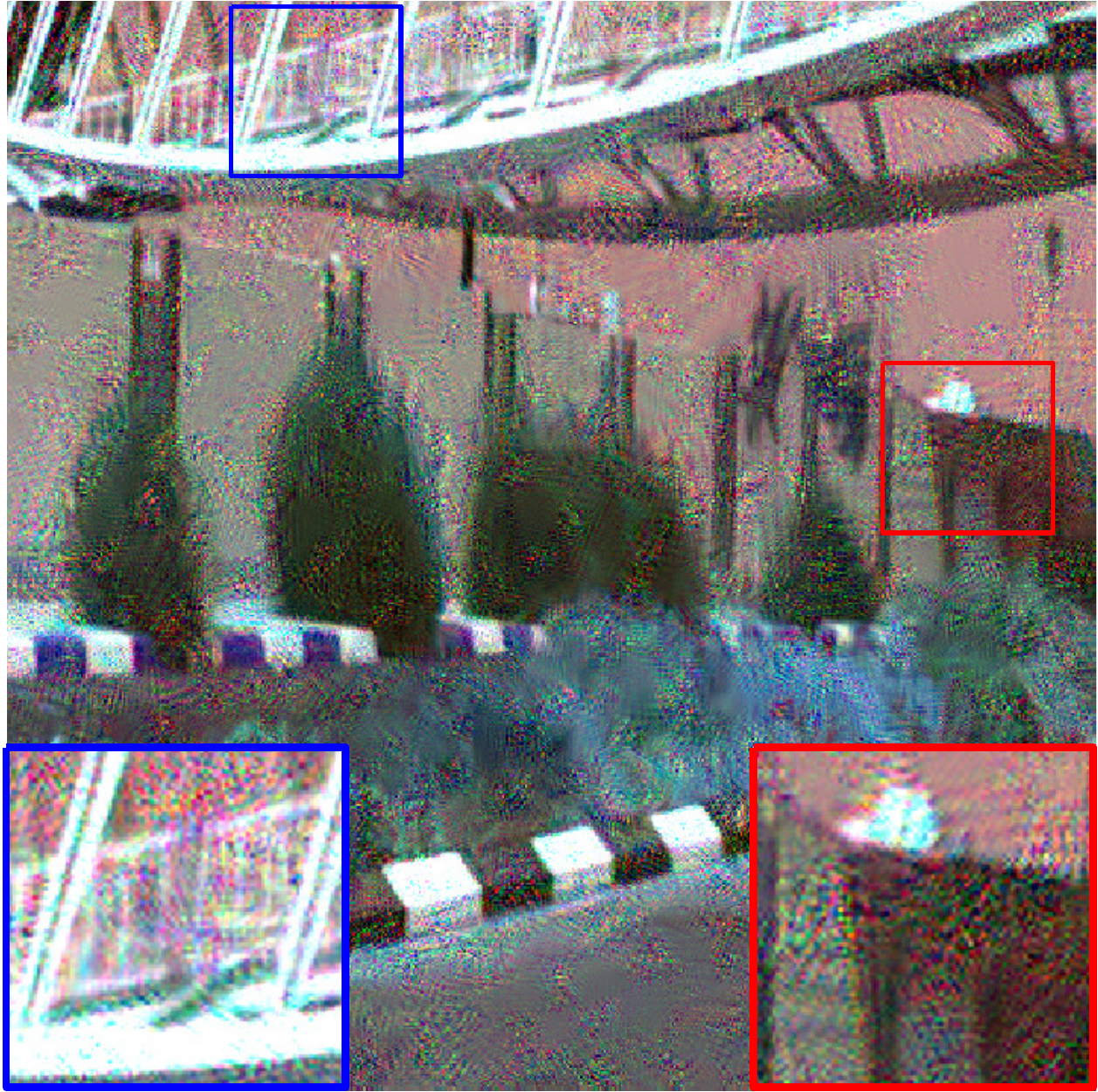}}\\
\subfigure[NGMeet]{\includegraphics[scale =0.175,clip=true]{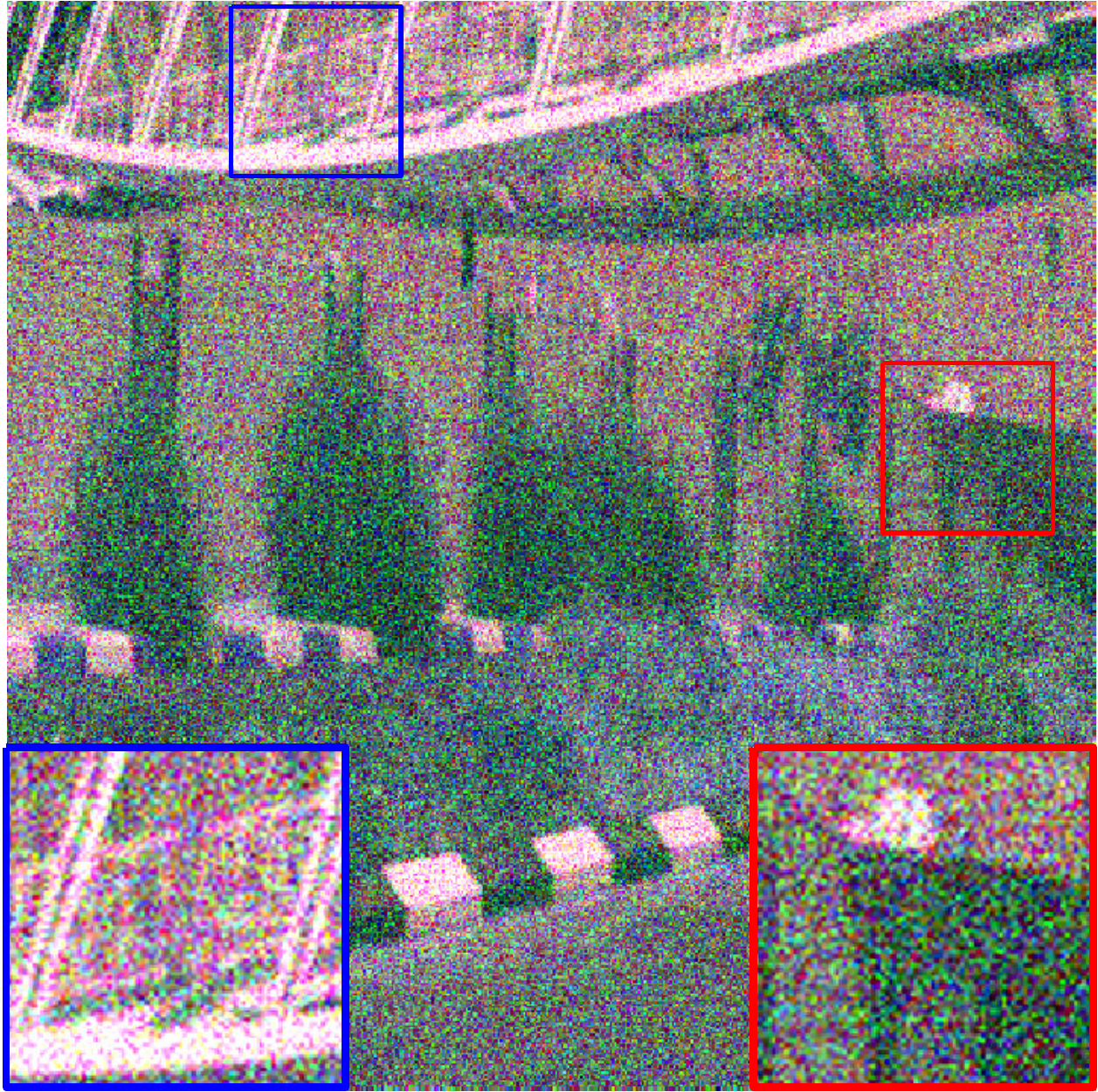}}\hspace{-0.8mm}
\subfigure[LRMR]{\includegraphics[scale =0.175,clip=true]{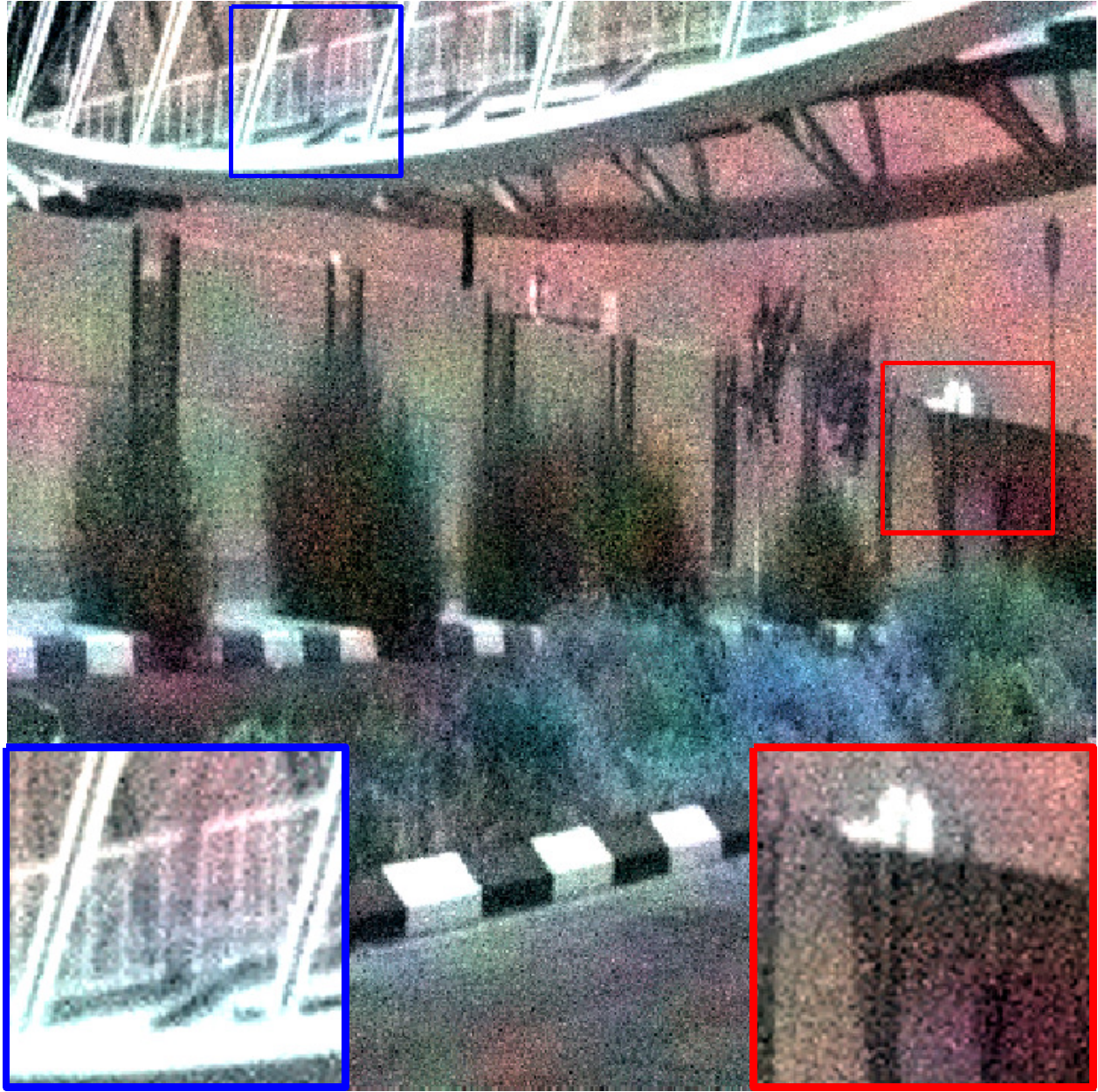}}\hspace{-0.8mm}
\subfigure[LRTDTV]{\includegraphics[scale =0.175,clip=true]{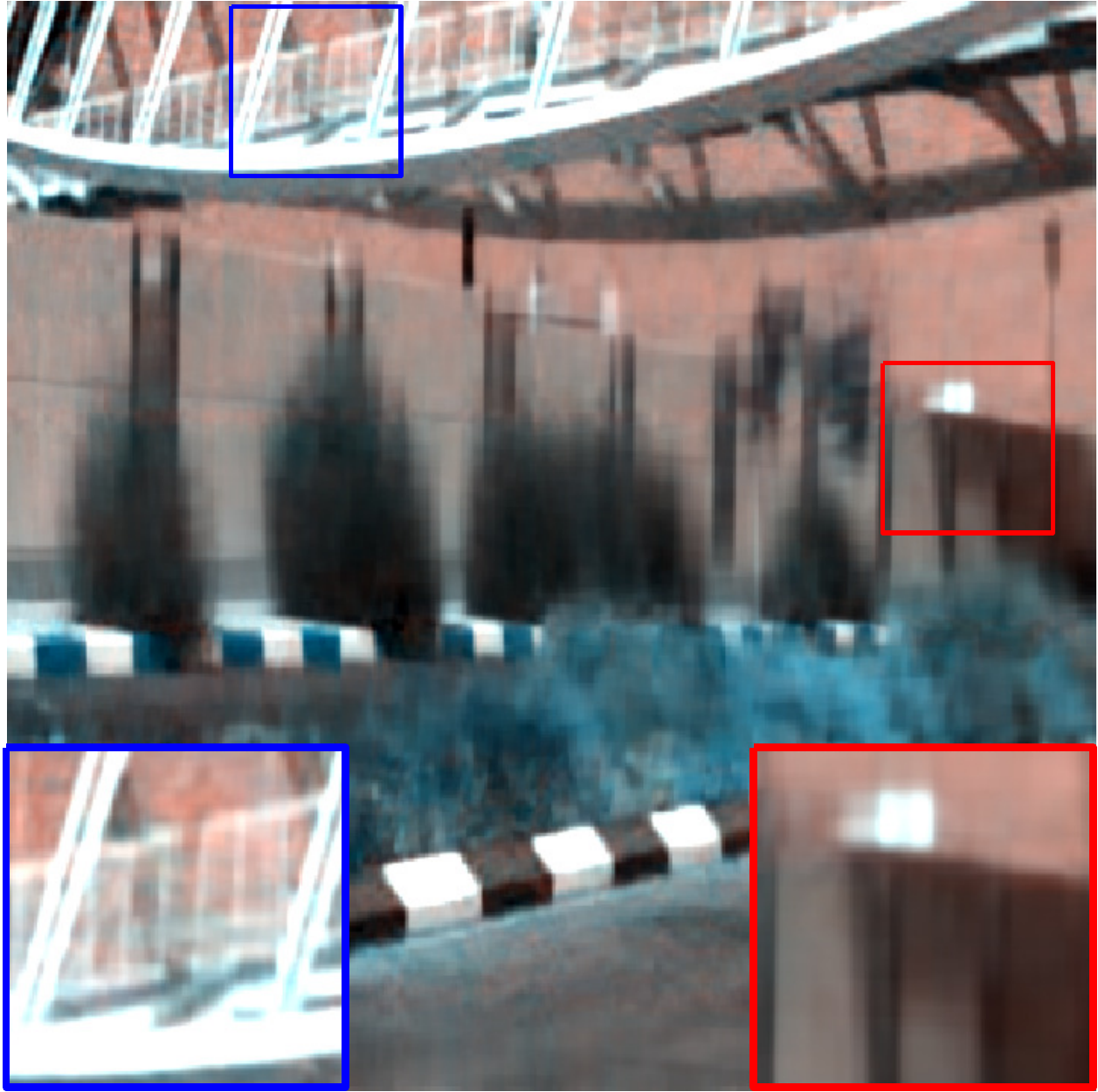}}\hspace{-0.8mm}
\subfigure[DnCNN]{\includegraphics[scale =0.175,clip=true]{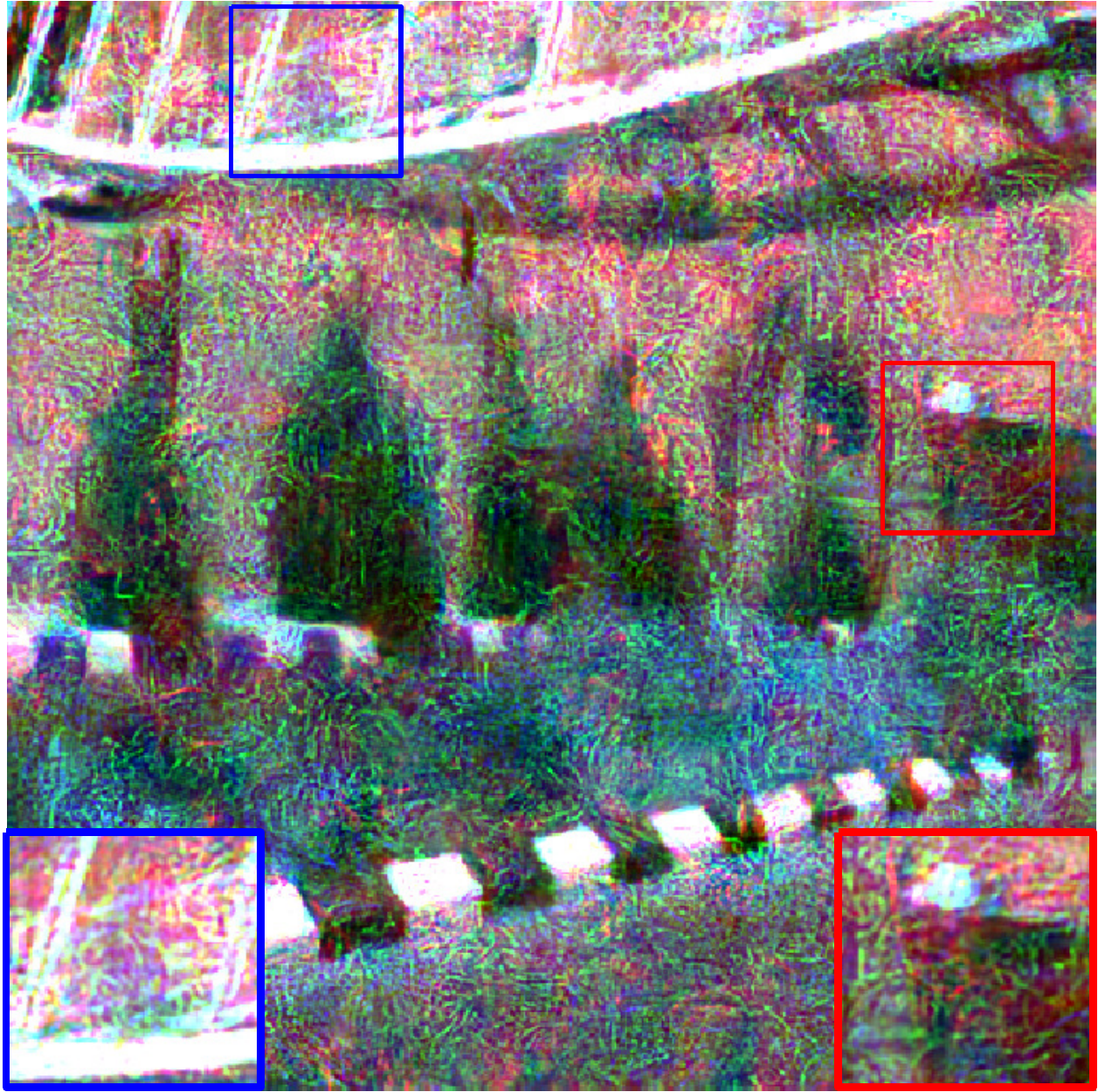}}\hspace{-0.8mm}
\subfigure[HSI-SDeCNN]{\includegraphics[scale =0.175,clip=true]{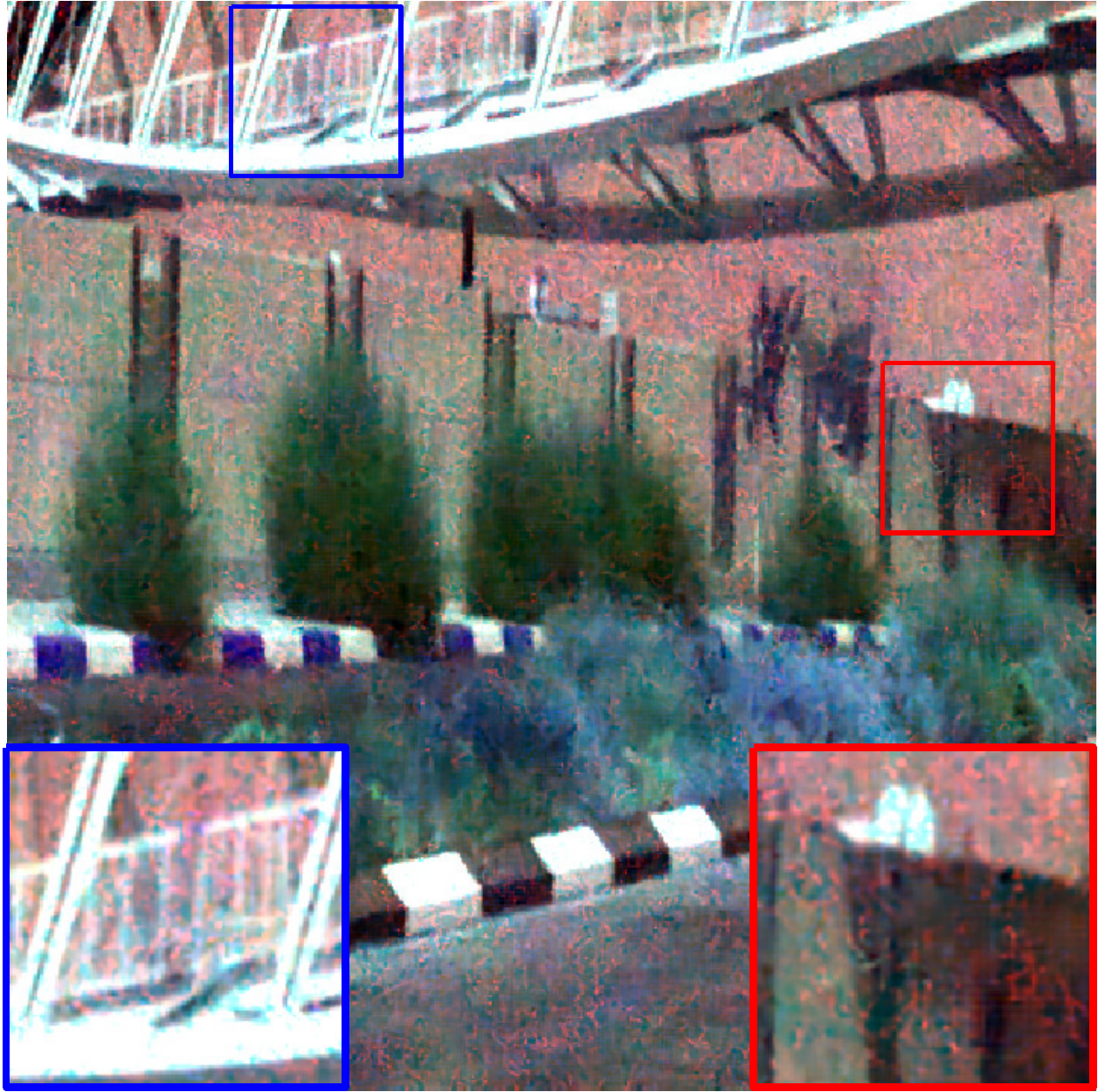}}\hspace{-0.8mm}
\subfigure[HSID-CNN]{\includegraphics[scale =0.175,clip=true]{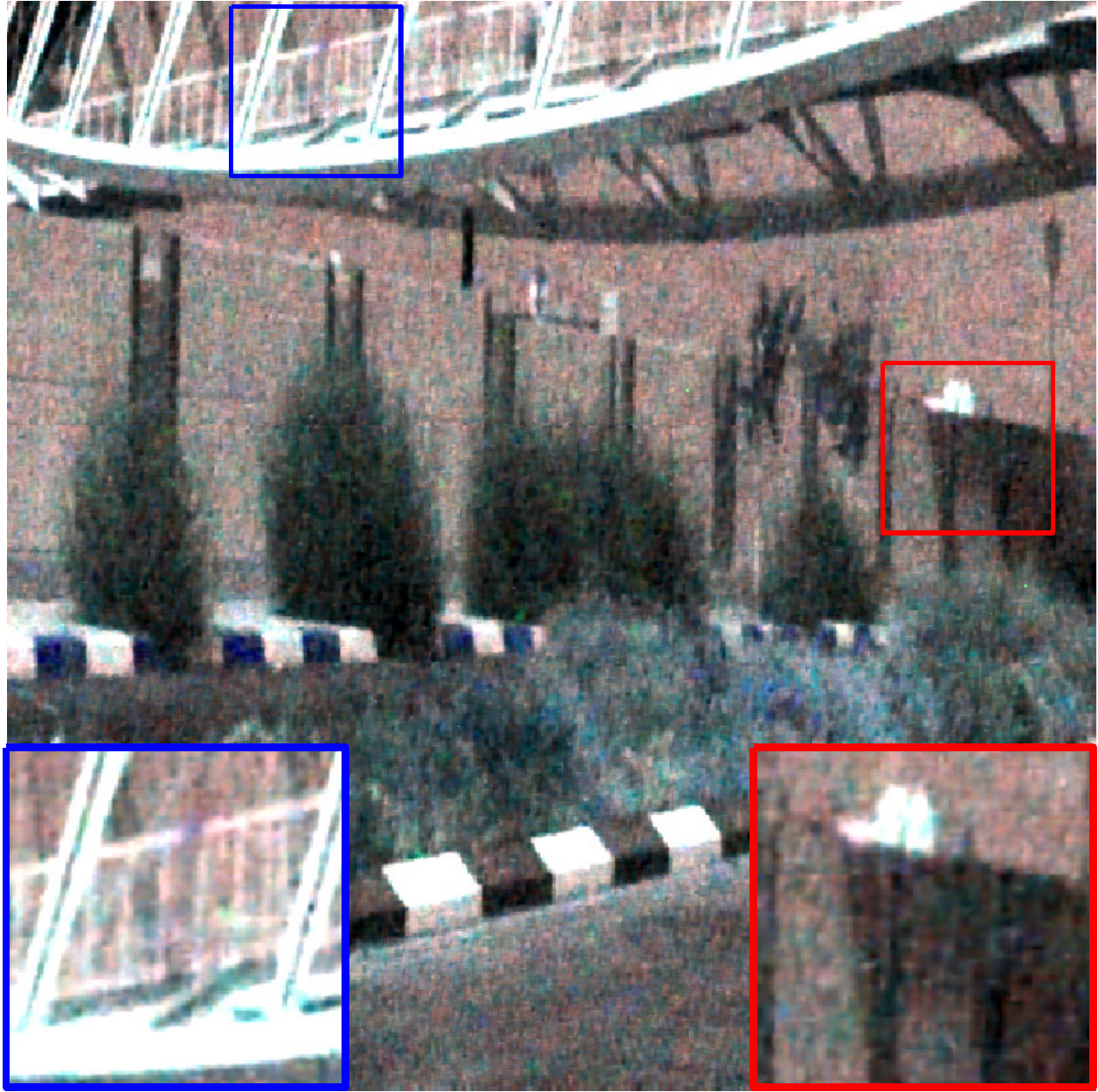}}\hspace{-0.8mm}
\subfigure[QRNN3D]{\includegraphics[scale =0.175,clip=true]{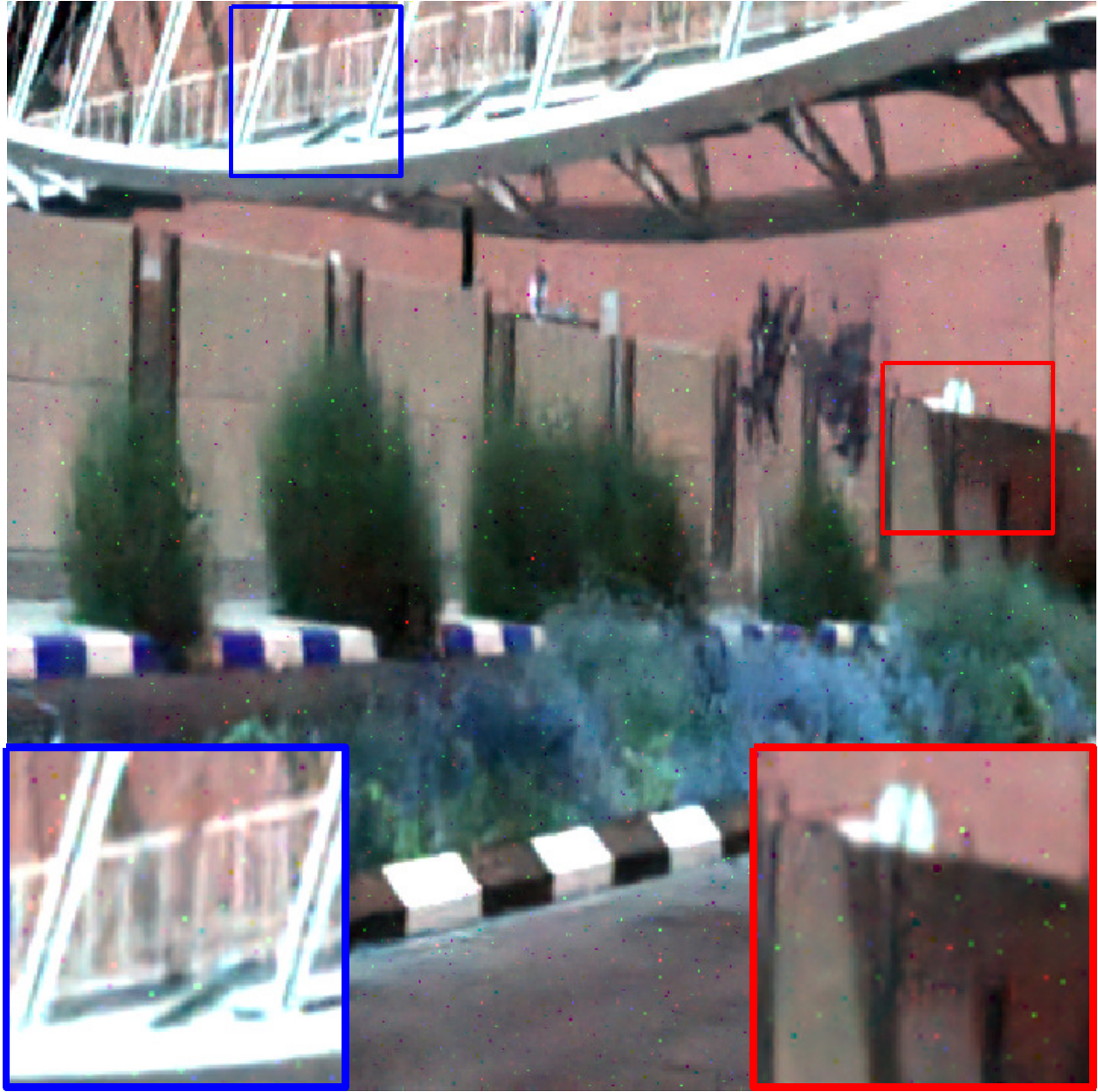}}\hspace{-0.8mm}
\subfigure[SMDS-Net]{\includegraphics[scale =0.175,clip=true]{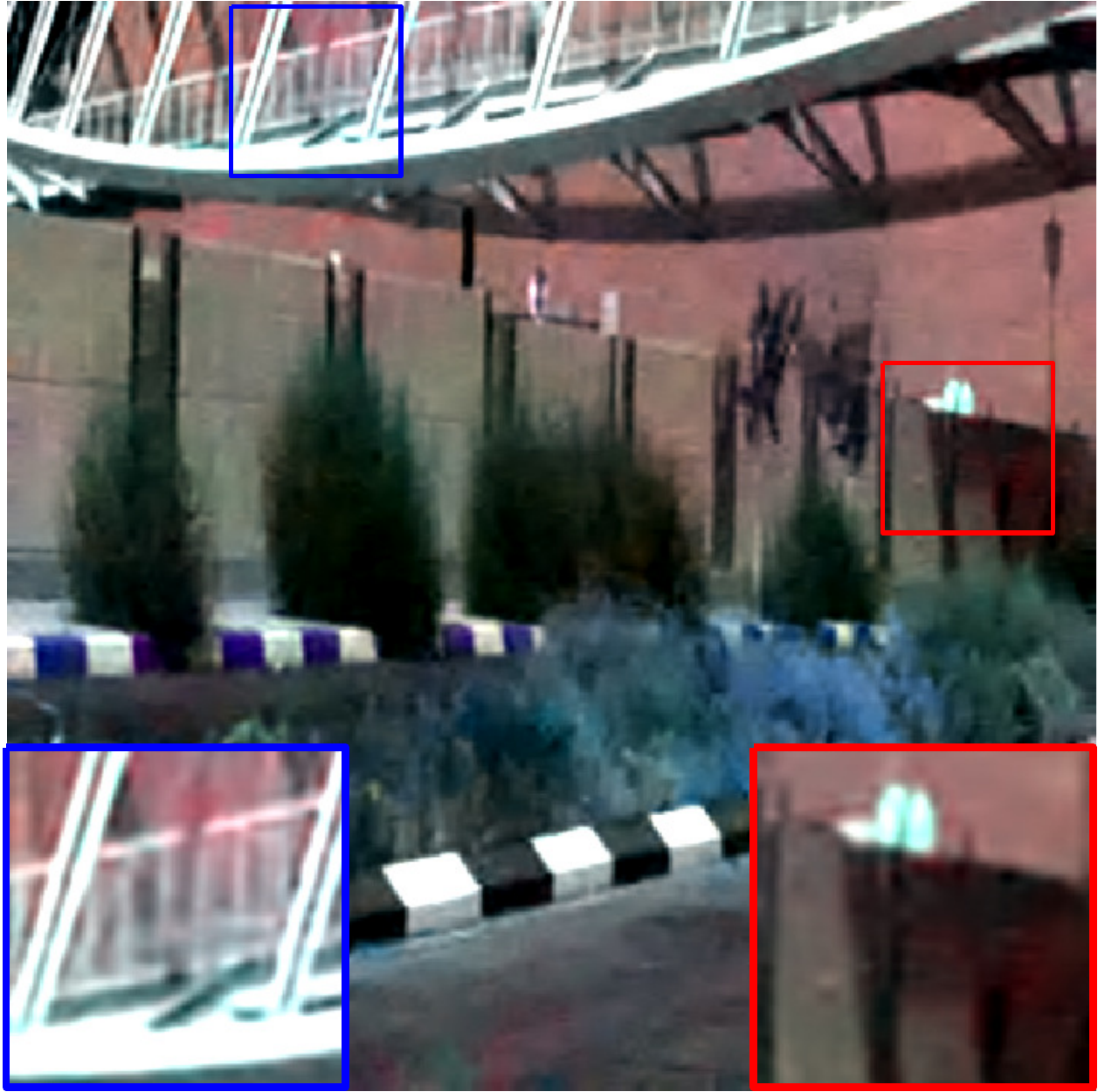}}
\caption{Denoising results on the  \emph{gavyam\_0823-0933} image with the noise variance in [0-95]. The false-color images were generated by combining bands 5, 18, 25. SMDS-Net achieves the best visual results with less artifacts. (\textbf{Best view on screen with zoom})} \label{fig:icvl2}
 \end{figure*}
 
 	\begin{figure*}[!htbp]
  \centering
\subfigure[Noisy]{\includegraphics[scale =0.125,clip=true]{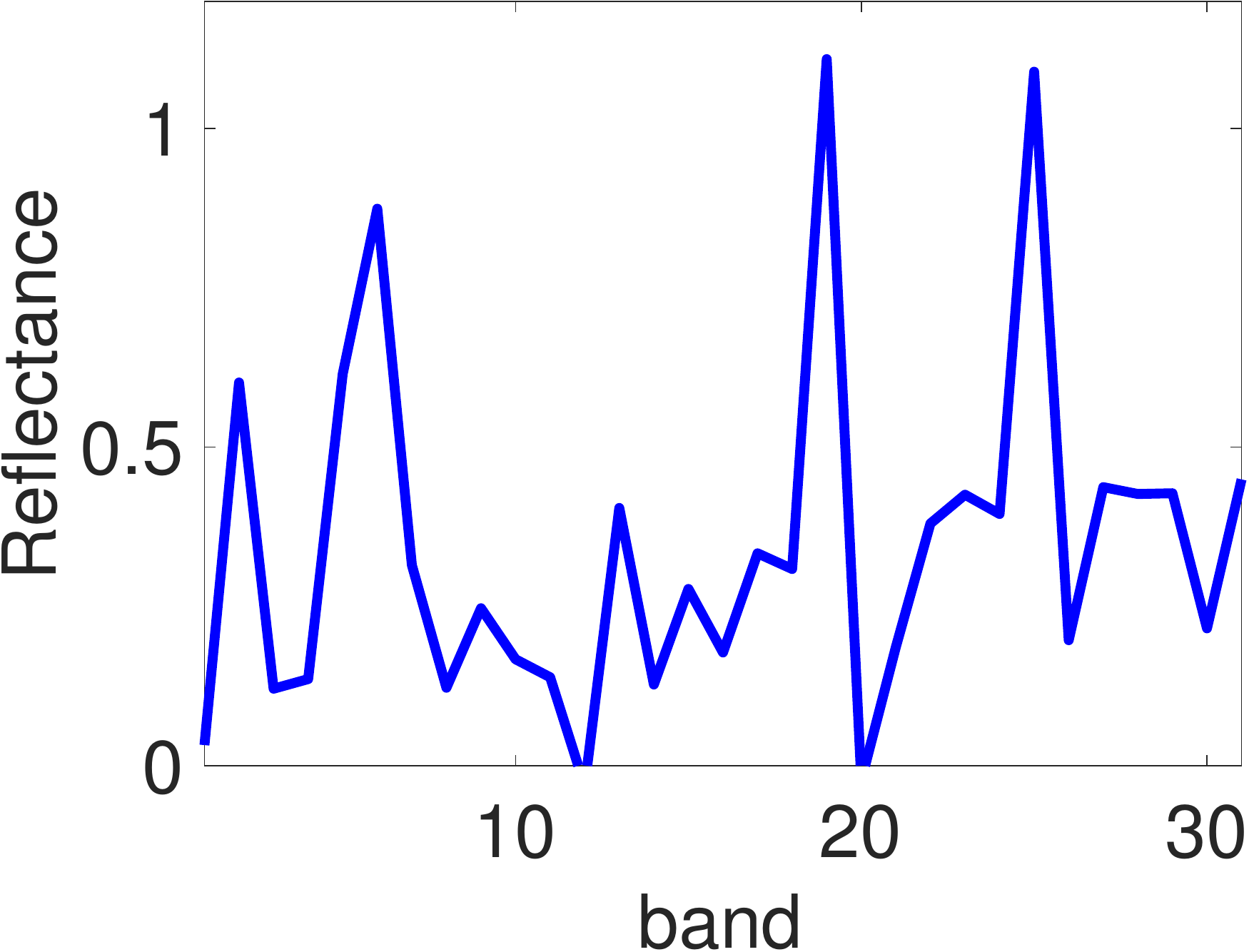}}\hspace{-0.8mm}
\subfigure[Clean]{\includegraphics[scale =0.125,clip=true]{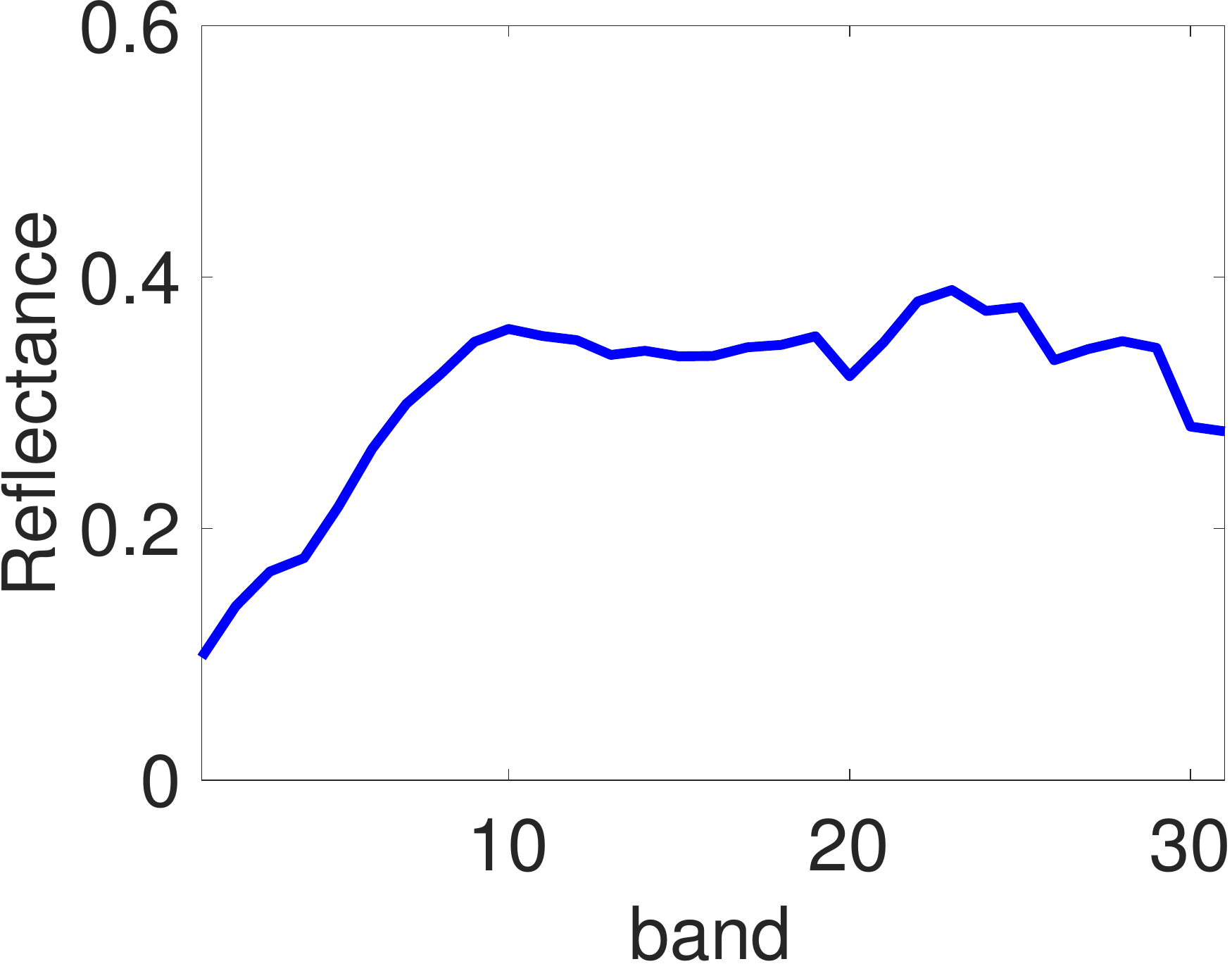}}\hspace{-0.8mm}
\subfigure[BM3D]{\includegraphics[scale =0.125,clip=true]{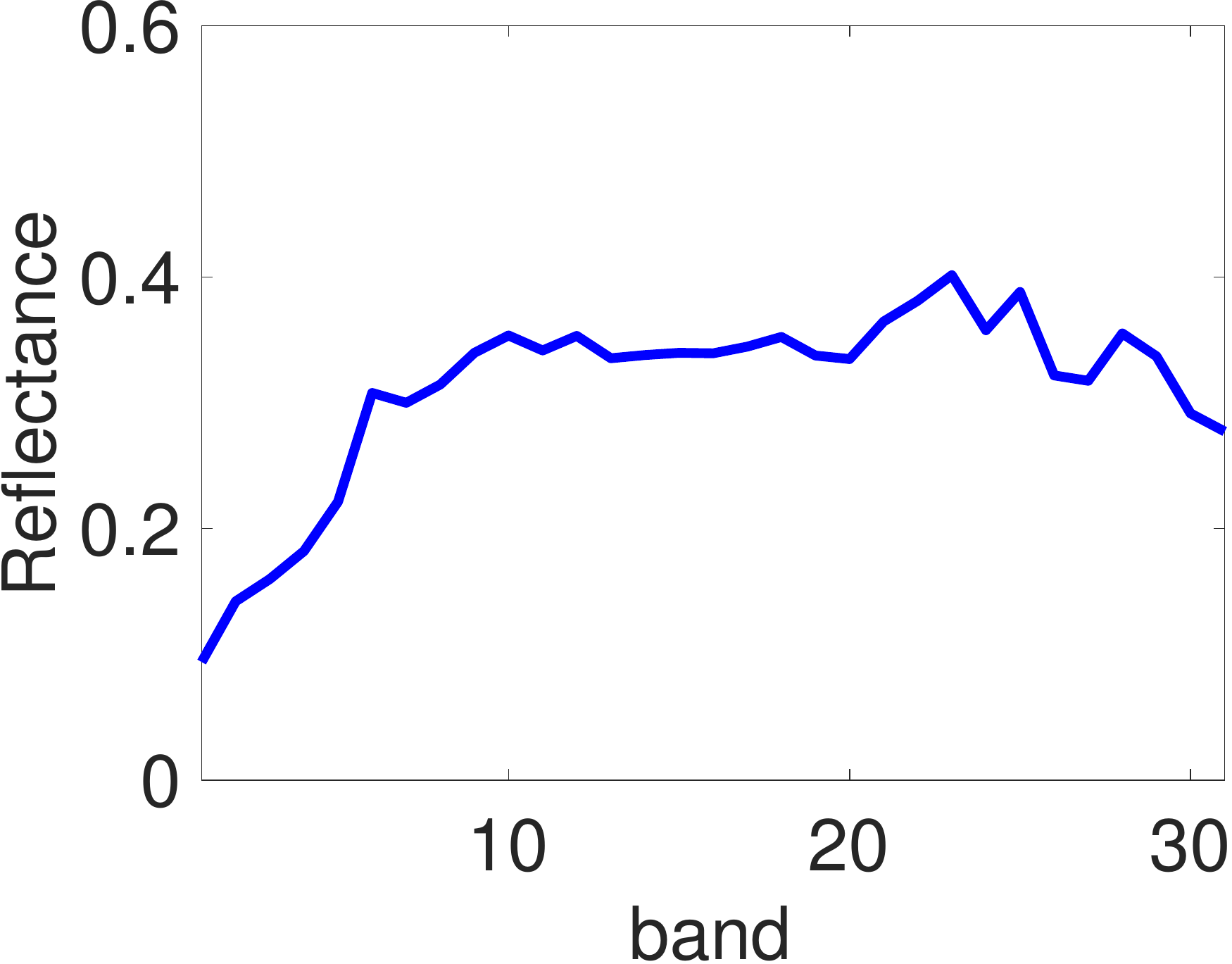}}\hspace{-0.8mm}
\subfigure[BM4D]{\includegraphics[scale =0.125,clip=true]{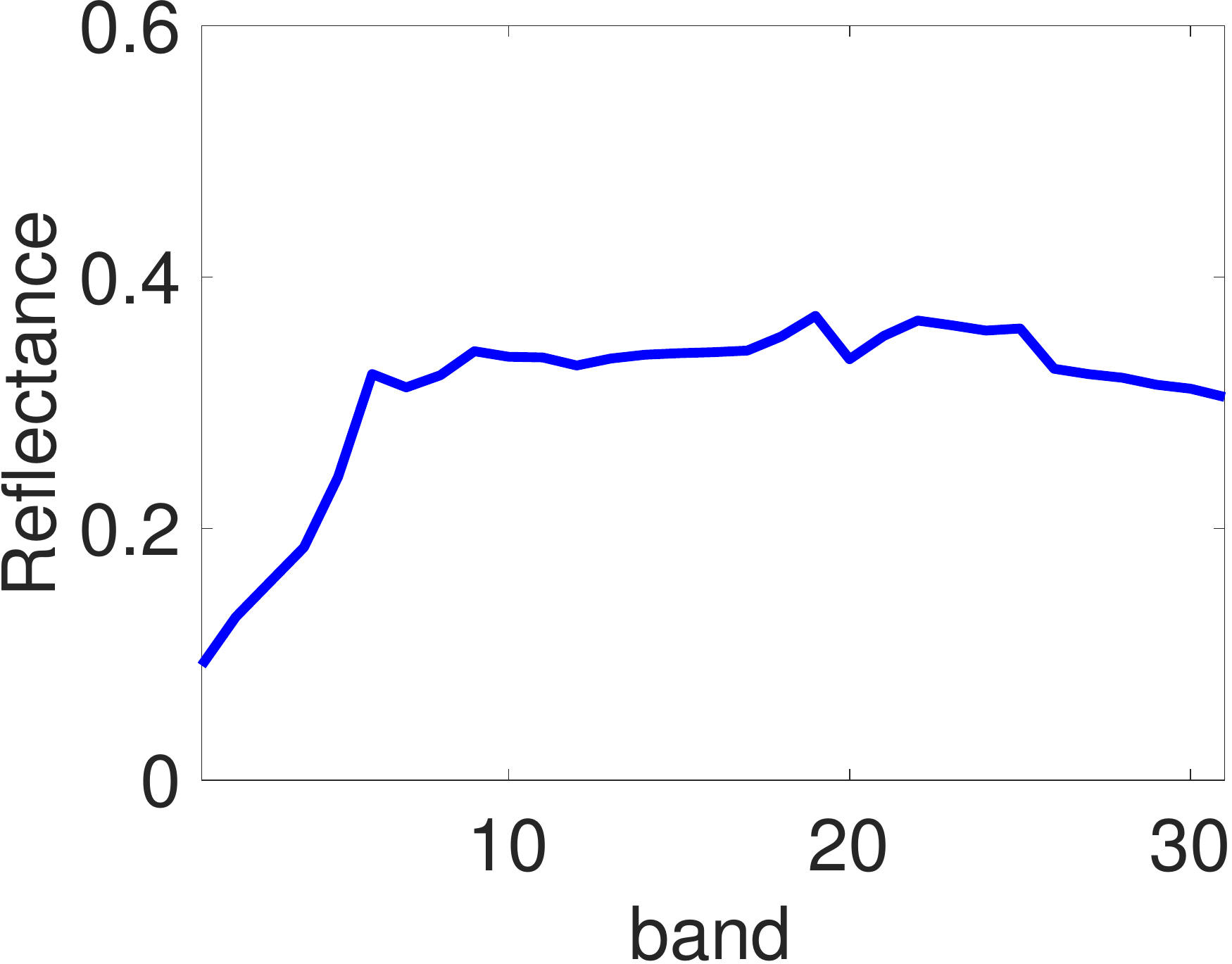}}\hspace{-0.8mm}
\subfigure[TDL]{\includegraphics[scale =0.125,clip=true]{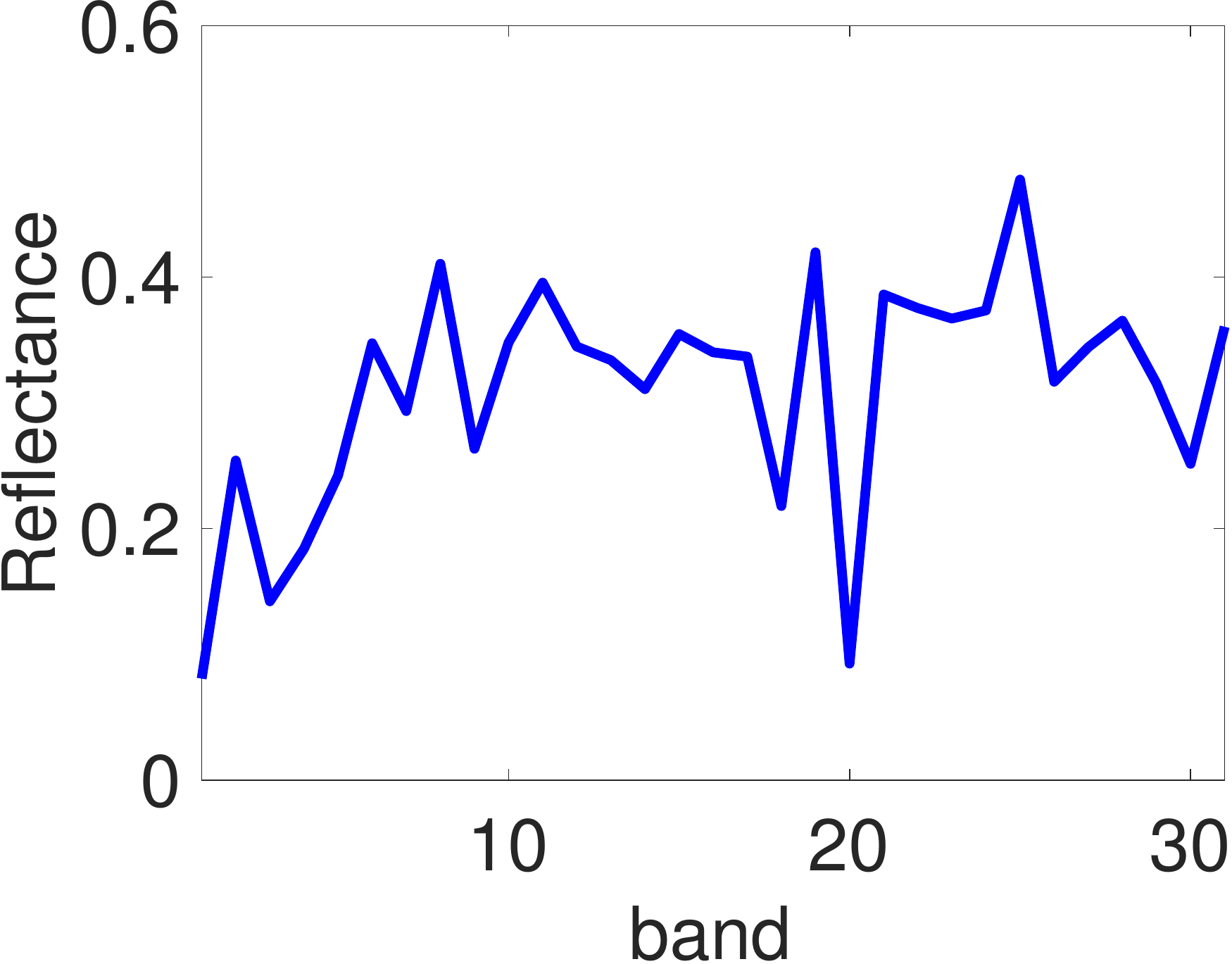}}\hspace{-0.8mm}
\subfigure[PARAFAC]{\includegraphics[scale =0.125,clip=true]{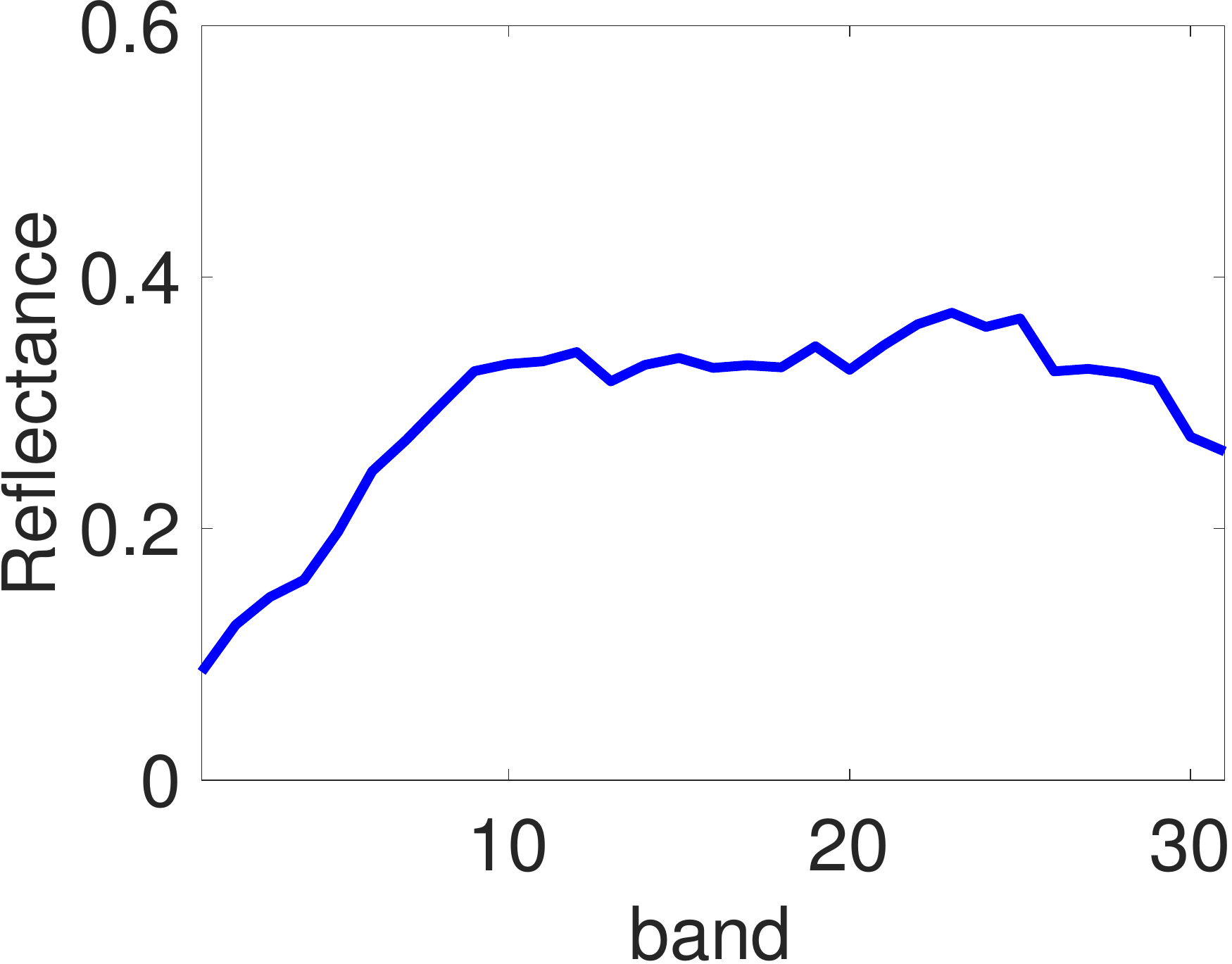}}\hspace{-0.8mm}
\subfigure[MTSNMF]{\includegraphics[scale =0.125,clip=true]{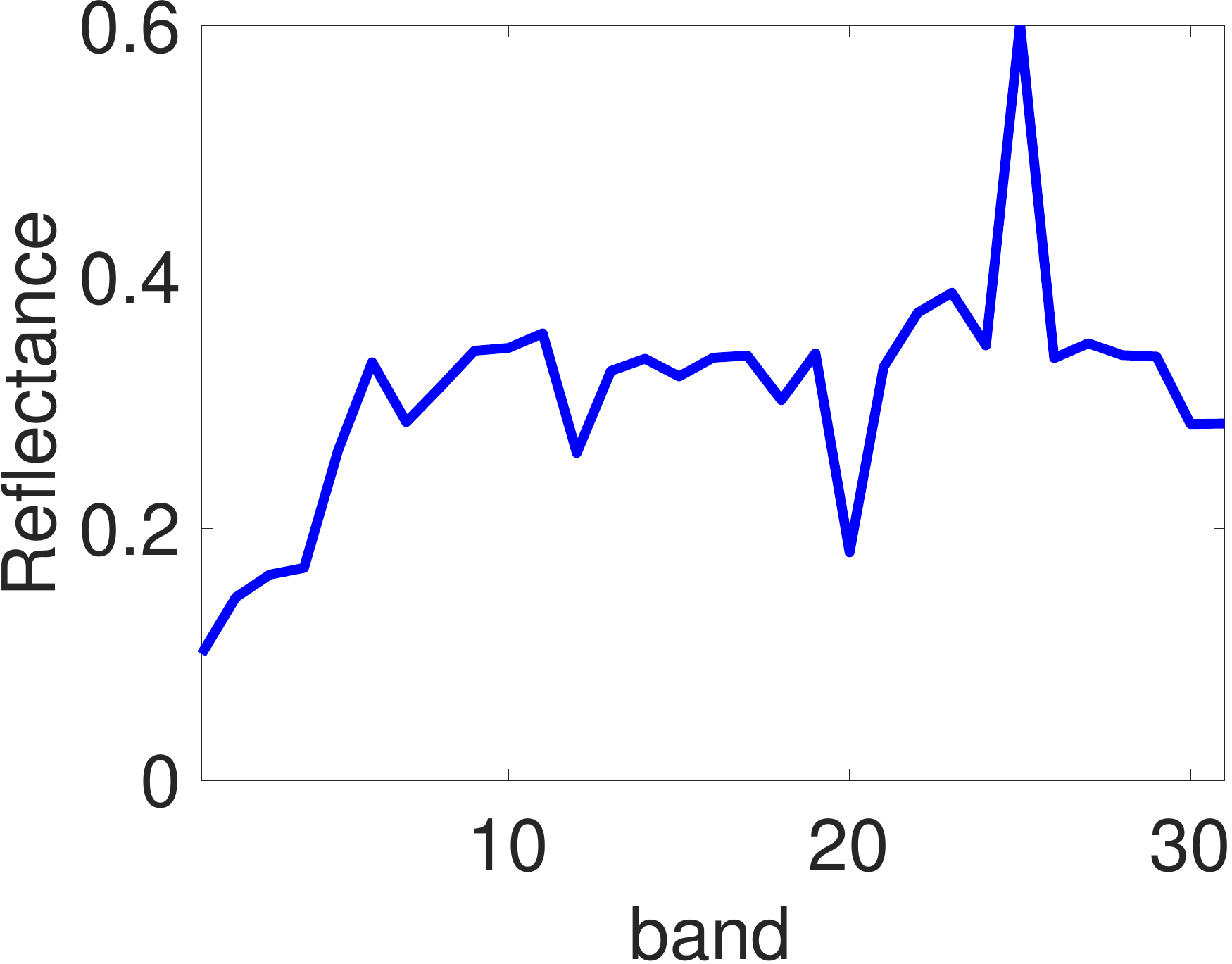}}\hspace{-0.8mm}
\subfigure[LLRT]{\includegraphics[scale =0.125,clip=true]{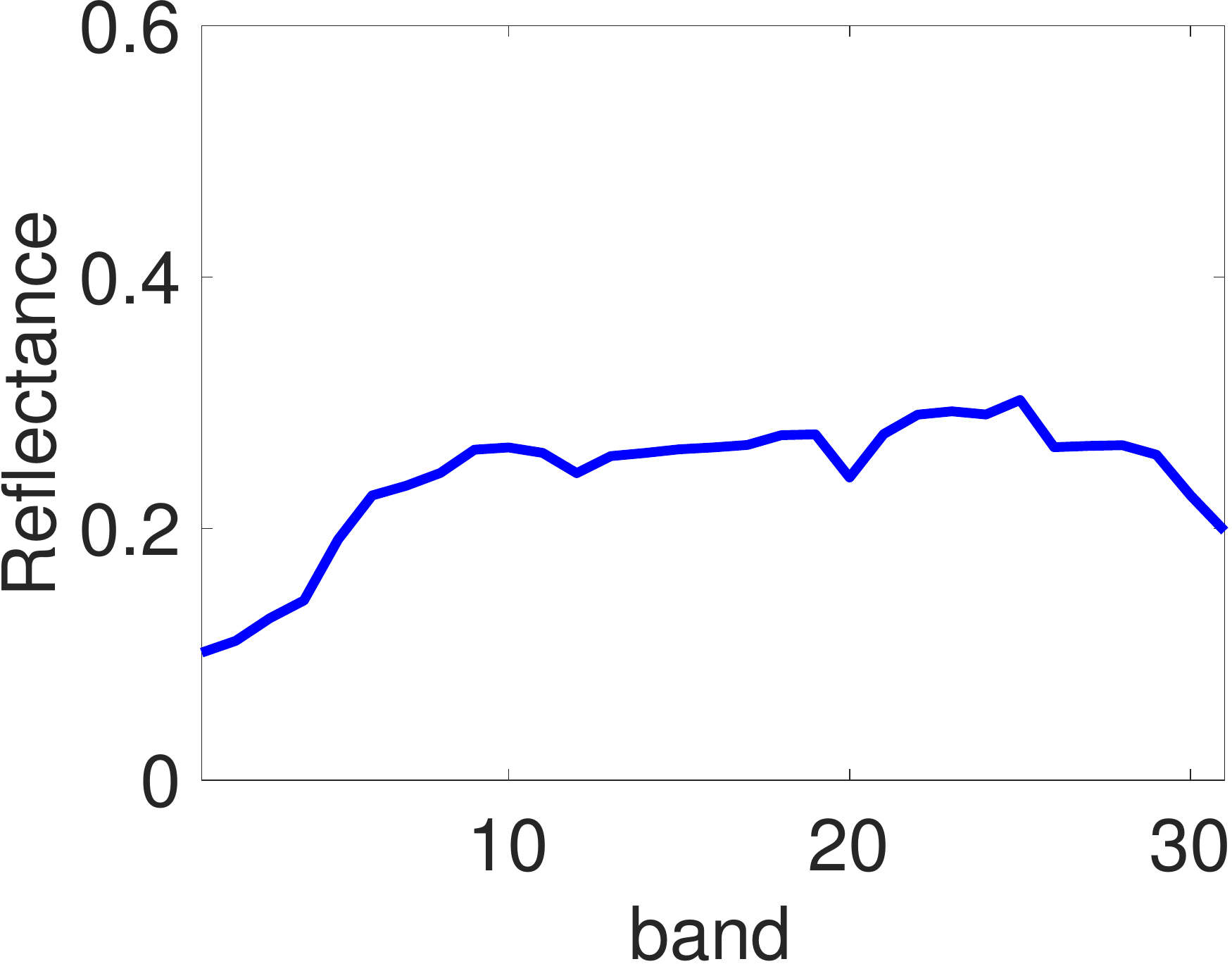}}\hspace{-0.8mm}\\
\subfigure[NGMeet]{\includegraphics[scale =0.125,clip=true]{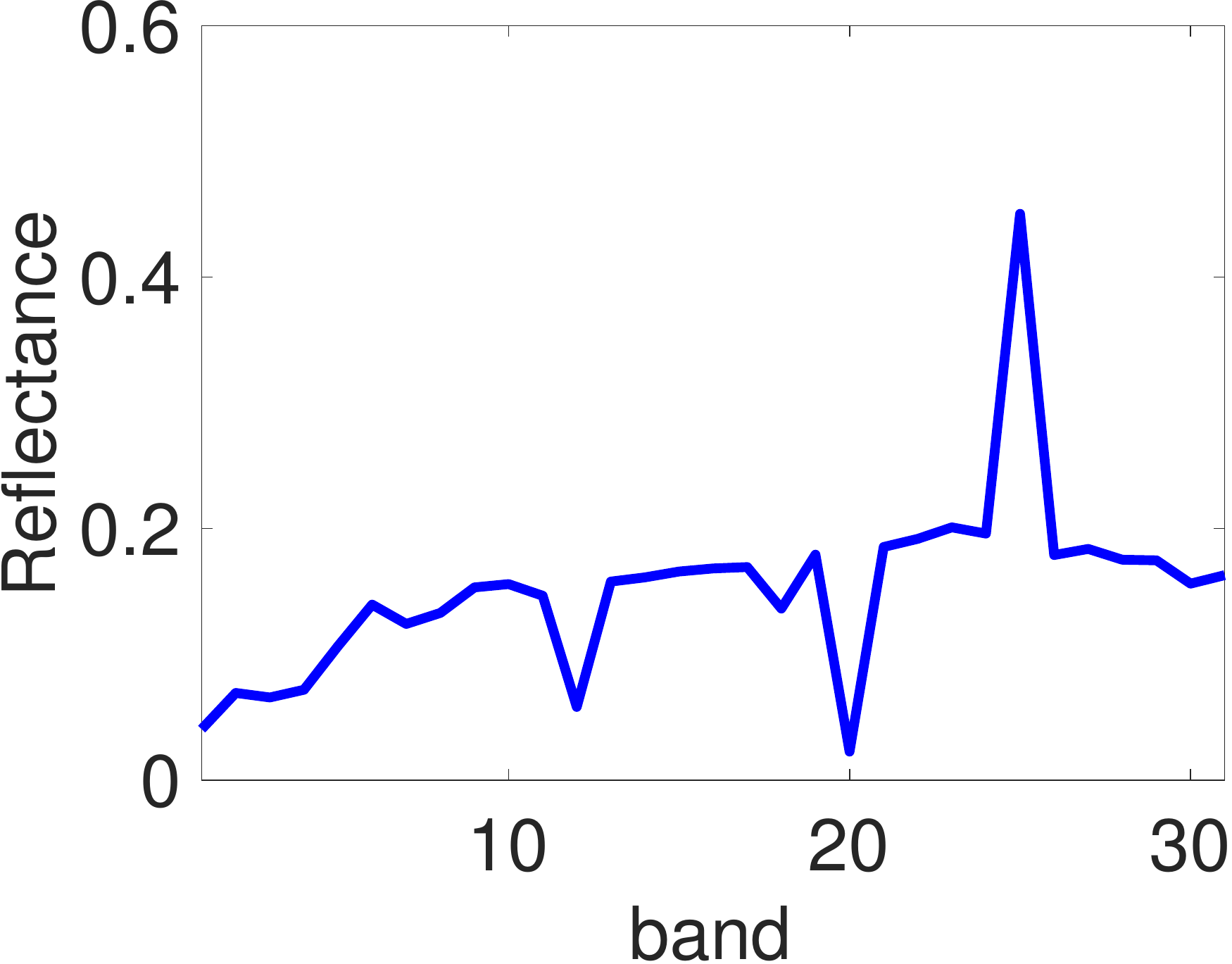}}\hspace{-0.8mm}
\subfigure[LRMR]{\includegraphics[scale =0.125,clip=true]{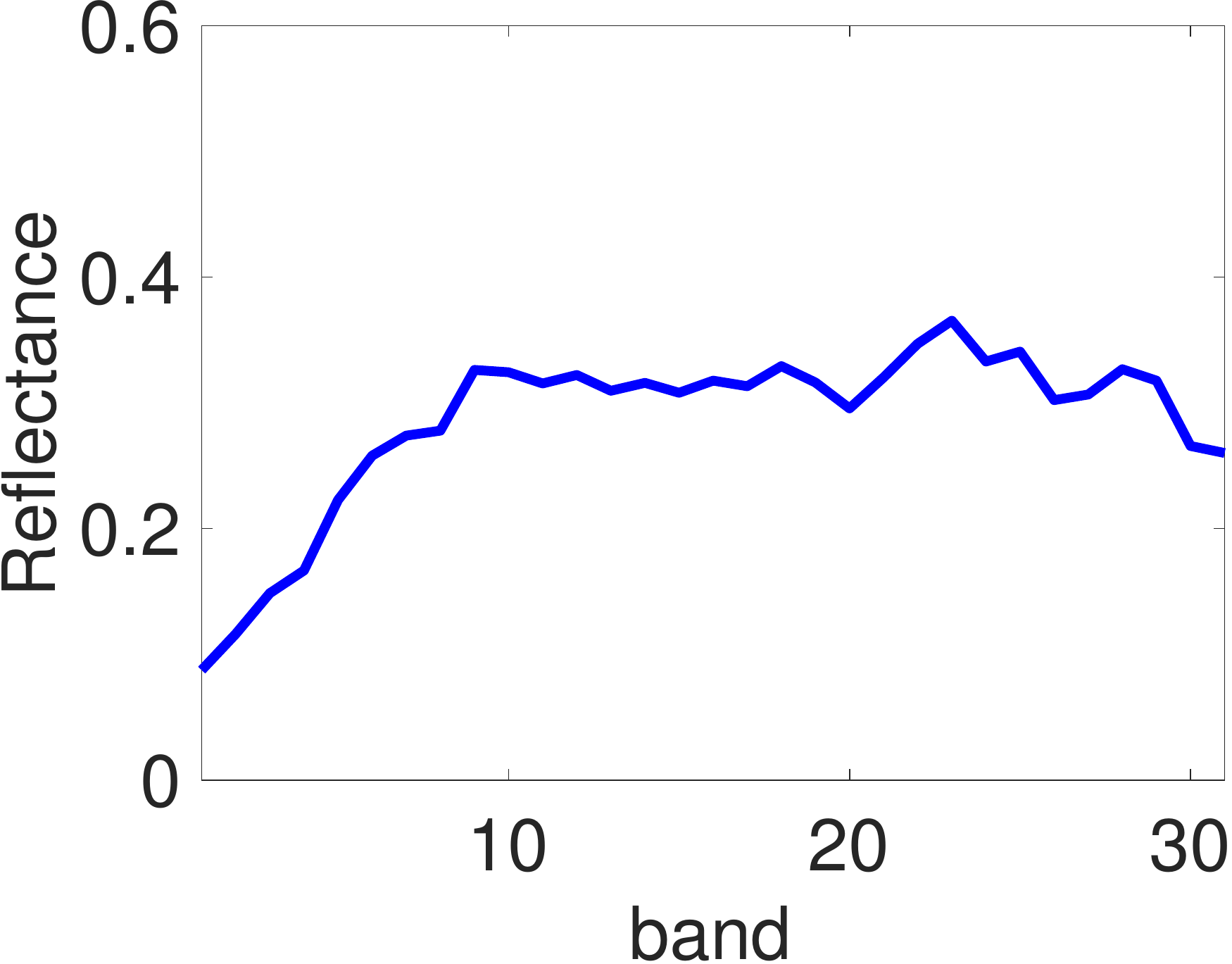}}\hspace{-0.8mm}
\subfigure[LRTDTV]{\includegraphics[scale =0.125,clip=true]{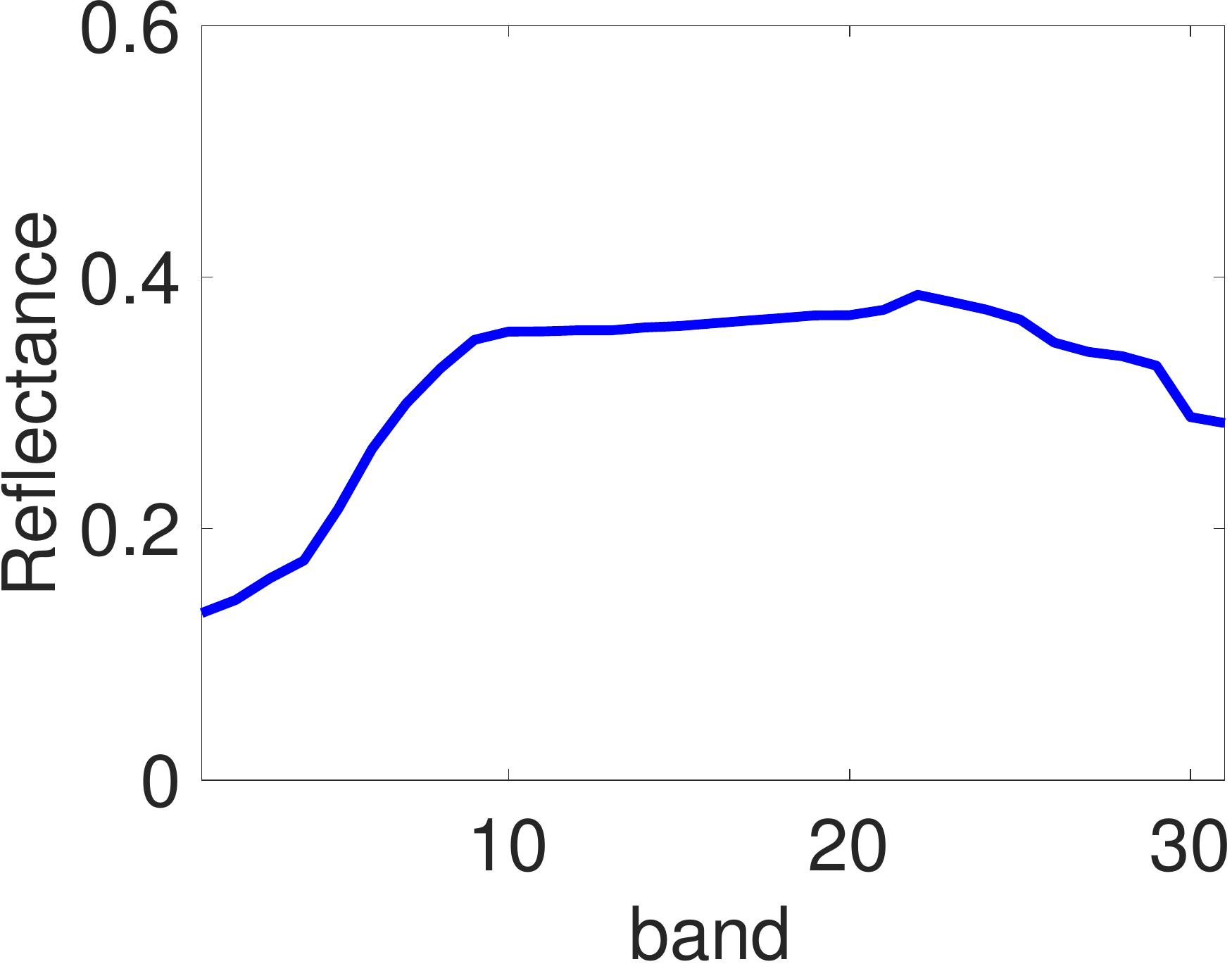}}\hspace{-0.8mm}
\subfigure[DnCNN]{\includegraphics[scale =0.125,clip=true]{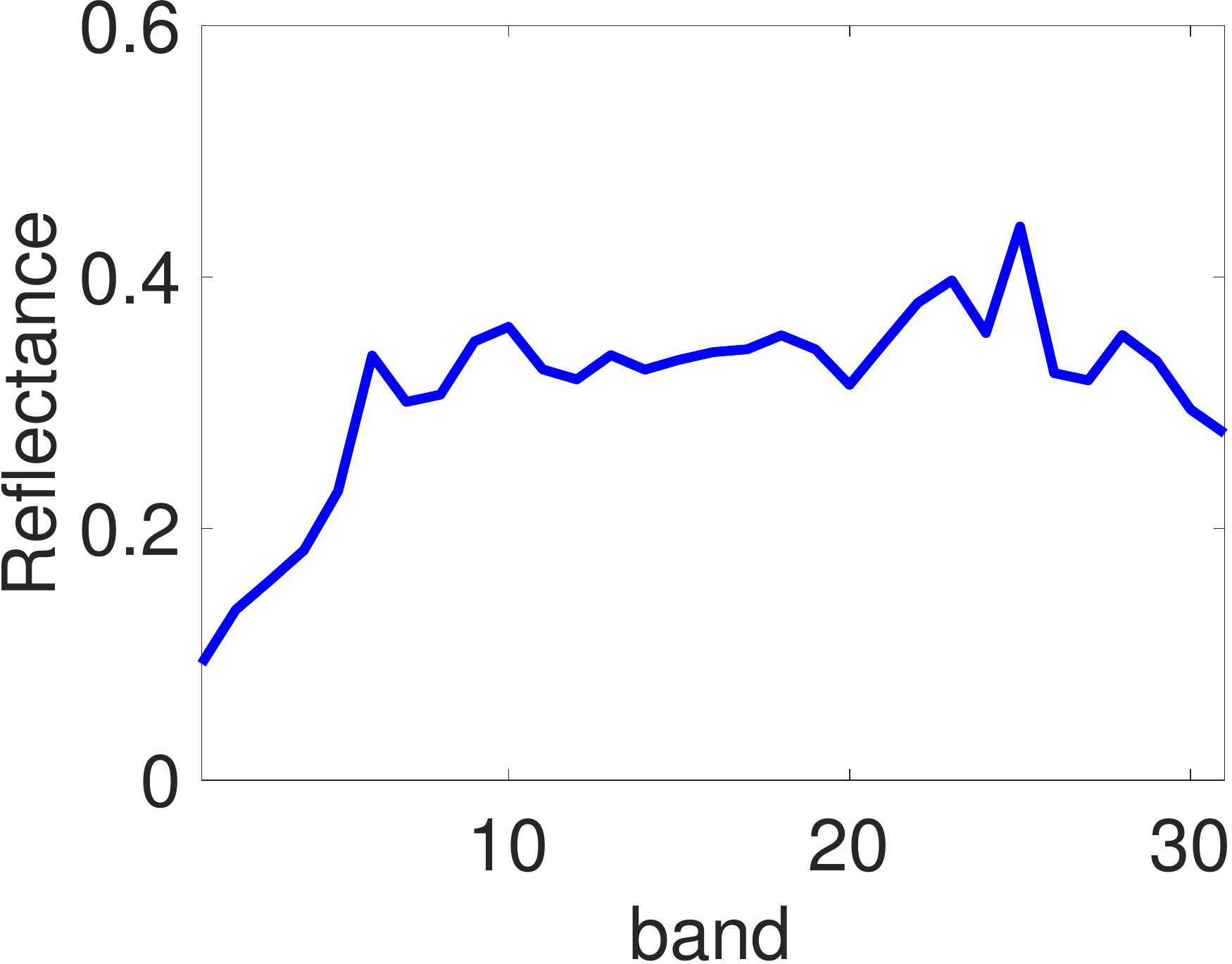}}\hspace{-0.8mm}
\subfigure[HSI-SDeCNN]{\includegraphics[scale =0.125,clip=true]{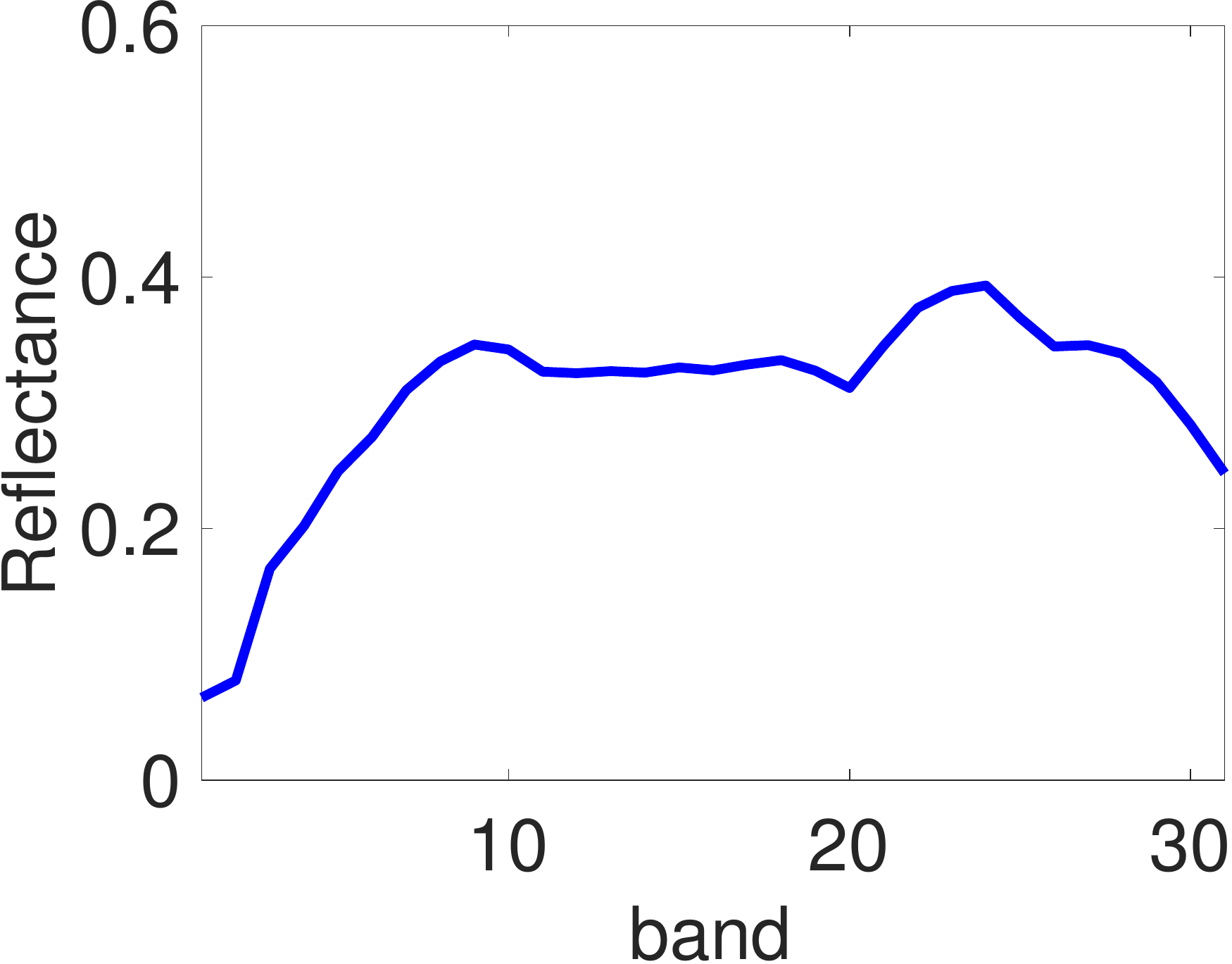}}\hspace{-0.8mm}
\subfigure[HSID-CNN]{\includegraphics[scale =0.125,clip=true]{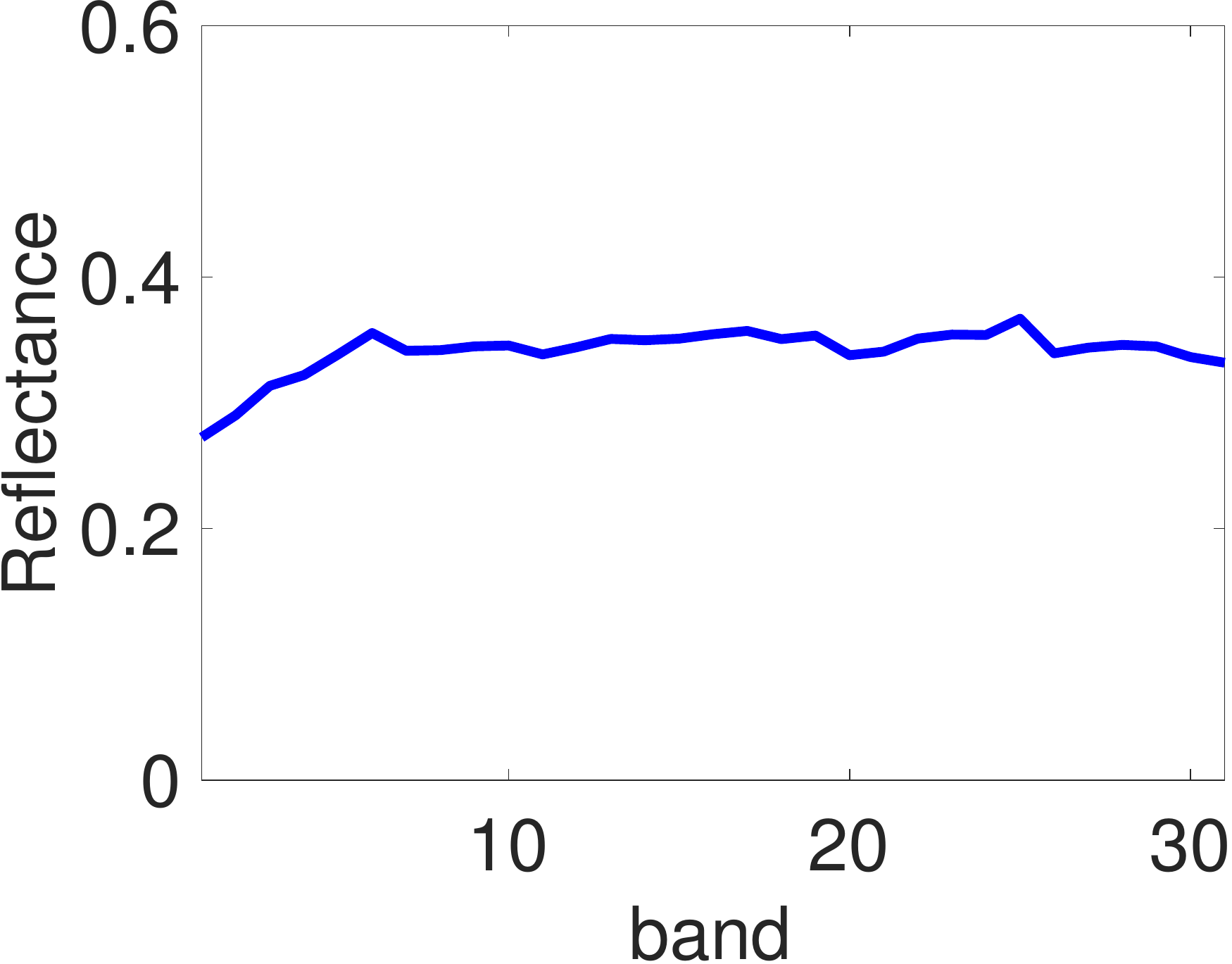}}\hspace{-0.8mm}
\subfigure[QRNN3D]{\includegraphics[scale =0.125,clip=true]{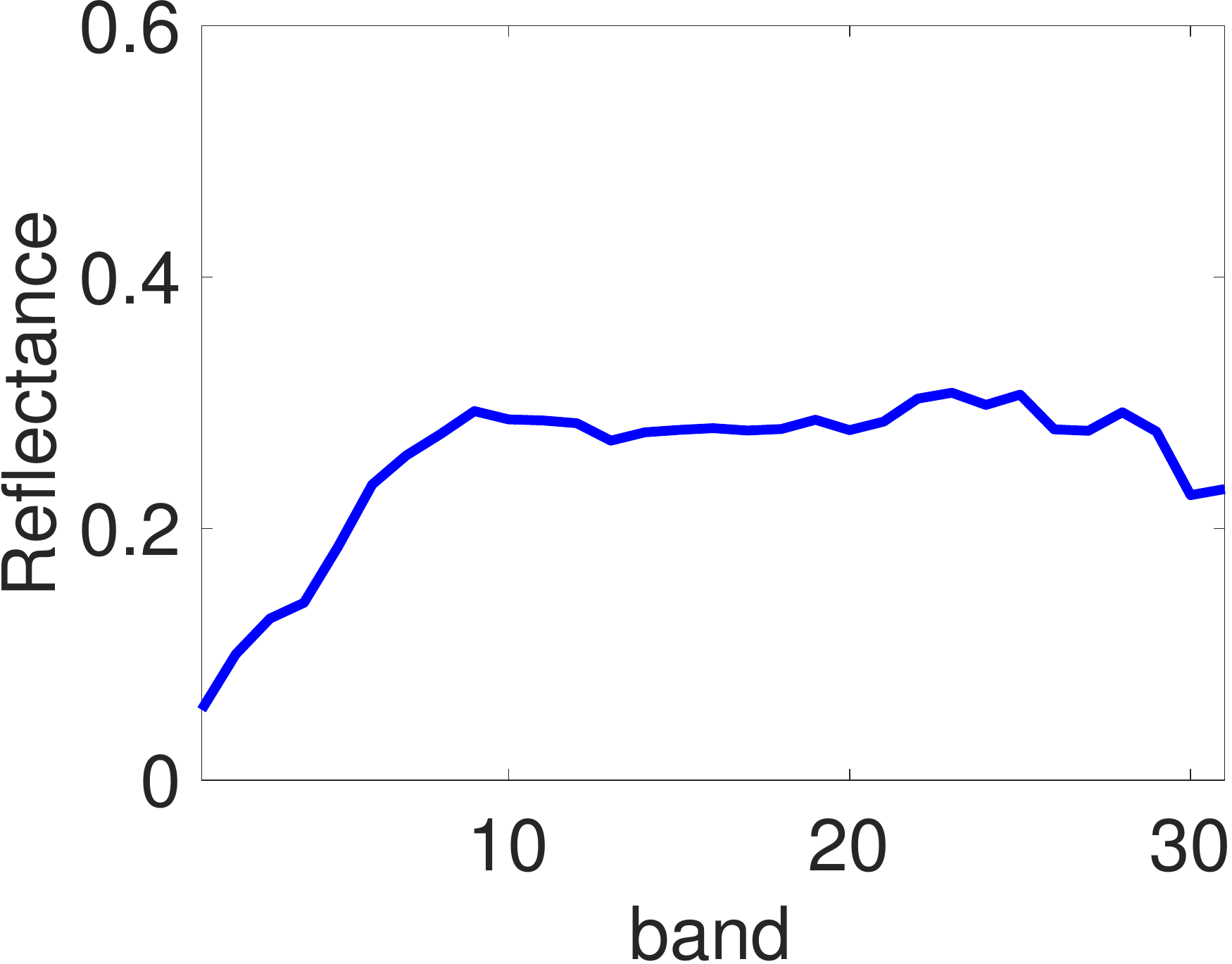}}\hspace{-0.8mm}
\subfigure[SMDSNet]{\includegraphics[scale =0.125,clip=true]{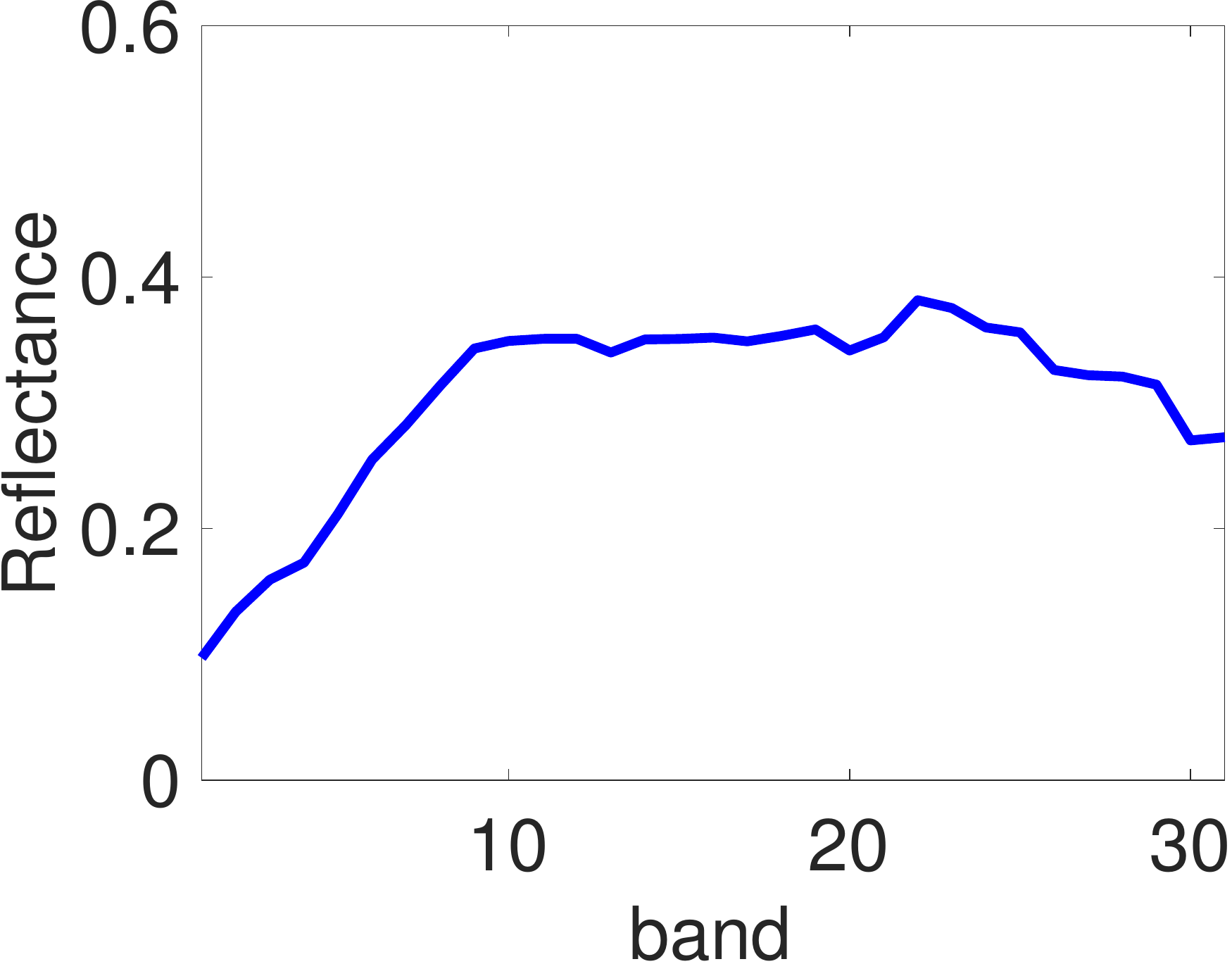}}
\caption{Denoising results of pixel (451, 135) in  \emph{gavyam\_0823-0933} HSI with the noise variance in [0-95].} \label{fig:icvlspec}
 \end{figure*}

\section{Experimental Results} \label{sec:exp}
In this section, we conduct experiments on both close-range HSIs and also remote sensing HSIs to show the denoising ability of our methods.  Moreover, comprehensive studies are conducted to illustrate its high learning ability. 

\subsection{Experimental Settings }

\noindent\textbf{Methods of Comparison:}  The competing denoising methods include: four sparse representation based methods, i.e., BM3D~\cite{Dabov2007}~\footnote{https://webpages.tuni.fi/foi/GCF-BM3D/}, BM4D~\cite{Maggioni2013}~\footnote{http://www.cs.tut.fi/foi/}, TDL~\cite{Peng2014}~\footnote{http://gr.xjtu.edu.cn/web/dymeng/}, MTSNMF~\cite{Ye2015}, five low-rank representation  based methods, PARAFAC~\cite{Liu2012}, LLRT~\cite{Chang2017}~\footnote{https://owuchangyuo.github.io/}, NGMeet~\cite{He2019}~\footnote{github.com/quanmingyao/NGMeet}, LRMR~\cite{Zhang2014}~\footnote{https://sites.google.com/site/rshewei/home}, LRTDTV~\cite{Wang2018}~\footnote{http://gr.xjtu.edu.cn/en/web/dymeng/3} and four deep learning based method, i.e., DnCNN~\cite{Zhang2017}~\footnote{https://github.com/cszn/DnCNN}, HSI-SDeCNN~\cite{Maffei2020}~\footnote{https://github.com/mhaut/HSI-SDeCNN}, HSID-CNN~\cite{Yuan2019}~\footnote{https://github.com/qzhang95/HSID-CNN}, QRNN3D~\cite{Wei2020}~\footnote{https://github.com/Vandermode/QRNN3D}. BM3D and DnCNN are executed in a band by band  manner.

\noindent\textbf{Noise Setting:} Due to the sensitivity of the sensor at different wavelengths, the illumination condition, and other factors, the noise levels in different bands are usually different~\cite{Ye2015,Wei2020,He2016}. That is to say, there exist nonindependent and identically distributed (non-i.i.d.) noises in the spectral domain,  see Fig.~\ref{fig:icvl}(a) for an example on the Urban HSI. The noise distribution in bands 1, 109, and 208 show obvious differences. For this reason, following~\cite{Wang2018, Ye2015}, the synthetic data are accordingly generated by manually adding different noise variances ($\sigma^2$) of additive white Gaussian noise (AWGN)  to each band of clean HSI.   Specifically $\sigma$ is chosen in three ranges, i.e, [0-15], [0-55] and  [0-95].  In these three cases, the accurate noise variances for all the bands  are provided for testing.  Except for BM3D and DnCNN, all other competing methods which require the noise level as input are provided with the averaged noise level.


\noindent\textbf{Evaluation Measures:} PSNR, SSIM, and SAM are used for quantitative assessment. SAM depicts the spectral differences between clean and denoised HSIs. Smaller SAM and larger PSNR and SSIM imply better denoising. Moreover, we compare the restored HSI with clean HSI and noisy HSI to illustrate their differences for qualitative analysis.

As in the training, the input HSIs are cropped into multiple overlapping full-band cubes containing $56\times 56$ pixels, with a stride of 12, to test our SMDS-Net. After that, the denoising results of all the cubes are summarized to produce estimated HSI.

\subsection{Experiments with ICVL Dataset}
\subsubsection{Quantitative Comparison} As in~\cite{Wei2020}, we use 50 images in ICVL  whose main region size of  $512\times 512\times 31$ are cropped to quantitatively evaluate the denoising performance of all the methods.     The denoising results are shown in Table~\ref{tab:ICVL}, in which the best two results are highlighted in red and blue.  As can be seen, the proposed SMDS-Net achieves the state-of-the-art results and provides the best denoising performance in all the cases, confirming its high denoising capacity.   Among model-based methods,  TDL, PARAFAC, LLRT, and NGMeet are less effective, especially in the case where the noise levels largely differ between bands, for example, $\sigma \in [0-95]$.  The main reason is that all of them inherently assume the noise distributions are identical over all bands, which is far from the requirements of real-world HSI denoising that the noises are spectrally non-i.i.d. Additionally, the spectral non-i.i.d noises make it more difficult to select parameters that are suitable for all the bands, even after intensive parameter tuning.  This however is not the case with the proposed SMDS-Net as it learns all the parameters from a large scale of data and treats the denoising as a feedforward process without parameter tuning, making it more capable of removing complex noises.  

Compared with MTSNMF and LRMR,  LRTDTV and SMDS-Net achieve more satisfactory results as the former two methods treat HSIs as a matrix in the process of denoising, unavoidably losing the spatial structural information contained in HSIs.  QRNN3D and SMDS-Net dominate alternative DL-based methods, i.e., HSI-SDeCNN, HSID-CNN, and band-wise Dn-CNN because of better incorporation of spectral-spatial structural correlation and global spectral correlation.  The hybrid advantages of the low-rank multidimensional sparse model and the strong learning ability of DL help our SMDS-Net outperform QRNN3D, especially concerning PSNR and SAM index. This experiment evidently shows the benefit of developing DL-based denoising methods considering the physical model of HSIs. 

 \begin{table*}[!htbp]
\caption{Comparison of different methods on  32 testing HSIs of CAVE dataset. The top two values are marked \textcolor{red}{\textbf{Red}} and \textcolor{blue}{\textbf{Blue}}. }\label{tab:cave}
\centering
	\tablesizetwo{	
\begin{tabular}{c|c|c|c|c|c|c|c|c|c|c|c|c|c|c|c|c}
\hline
&&&\multicolumn{4}{c|}{\textbf{Sparse methods}}&\multicolumn{5}{c|}{\textbf{Low-rank  methods}}&\multicolumn{5}{c}{\textbf{DL  methods}}\\
\hline
\multirow{2}*{$\sigma$}&\multirow{2}*{Index}&Noisy&BM3D&BM4D&TDL&MTSNMF&PARAFAC&LLRT&NGMeet&LRMR&LRTDTV&Dn-CNN&HSI-SDe&HSID-&QRNN3D&SMDS-Net\\
&&&~\cite{Dabov2007}&~\cite{Maggioni2013}&~\cite{Peng2014}&~\cite{Ye2015}&~\cite{Liu2012}&~\cite{Chang2017}&~\cite{He2019}&~\cite{Zhang2014}&~\cite{Wang2018}&~\cite{Zhang2017}&CNN~\cite{Maffei2020}&CNN~\cite{Yuan2019}&~\cite{Wei2020}&\textbf{(Ours)}\\
\hline
\multirow{3}*{\textbf{[0-15]}}&PSNR&33.42&45.01&44.78&37.84& 42.79&34.91& \textcolor{blue}{\textbf{45.02}}&37.81& 35.95&36.88& 40.10& 41.23&35.97&  40.10& \textcolor{red}{\textbf{45.40}}\\	
&SSIM&0.6256& \textcolor{blue}{\textbf{0.9753}}&0.9744&0.8248&	0.9688&	0.8678&	0.9634&0.8619&0.9425&0.9268& 0.9692&	0.9593&0.9186&	0.9692&\textcolor{red}{\textbf{0.9826}} 	
		\\
&SAM&	0.6113& \textcolor{blue}{\textbf{0.1143}}&0.1325&0.4348&0.1388&0.2174&0.1773&0.4259&0.1286&0.1835&	0.1767&0.1476&0.2646&0.1767&\textcolor{red}{\textbf{0.1023}}
\\
\hline
\multirow{3}*{\textbf{[0-55]}}&PSNR&21.51& 37.87& \textcolor{blue}{\textbf{38.04}}&28.52&35.63&32.87&     36.34&28.22&32.74& 34.49& 37.11& 36.10& 33.06&  37.95& \textcolor{red}{\textbf{39.28}}\\	 	
&SSIM&0.2389&0.9149&0.9102&0.4994&0.8413&0.8124&0.8244&0.5788&0.9058&0.9015&0.8510&0.8830&0.8499&\textcolor{blue}{\textbf{0.9481}}&\textcolor{red}{\textbf{0.9510}}
		\\
&SAM&1.0008&0.2224&0.2936&0.7630&0.3764&0.3324&0.4446&0.8359&\textcolor{blue}{\textbf{0.1975}}&0.2917&0.2374&0.2234&0.3951&0.2242&\textcolor{red}{\textbf{0.1671}}
\\
\hline
\multirow{3}*{\textbf{[0-95]}}&PSNR&   17.55&35.39&34.93&24.46&32.57&30.61&31.59&24.47&31.17&33.92&34.23&33.30&30.16&\textcolor{blue}{\textbf{35.94}}& \textcolor{red}{\textbf{36.22}}
\\
&SSIM& 0.1632&0.8707&0.8330&0.3546&0.7677&0.7124&0.6868&0.4367&0.8780&0.8939&0.8213&0.8061&0.7374&\textcolor{blue}{\textbf{0.9153}}&\textcolor{red}{\textbf{0.9247}}	\\
&SAM&1.1417&0.2855&0.4305&0.9139&0.5191&0.4814&0.6103&0.9619&\textcolor{blue}{\textbf{0.2724}}&0.3051&0.3097&0.2694&0.5437&0.2873&\textcolor{red}{\textbf{0.1977}}  	\\
\hline 		
\end{tabular}}
\end{table*}

   \begin{figure*}[!htbp]
\makeatletter 
\subfigure[Clean]{\includegraphics[scale=0.175,clip=true]{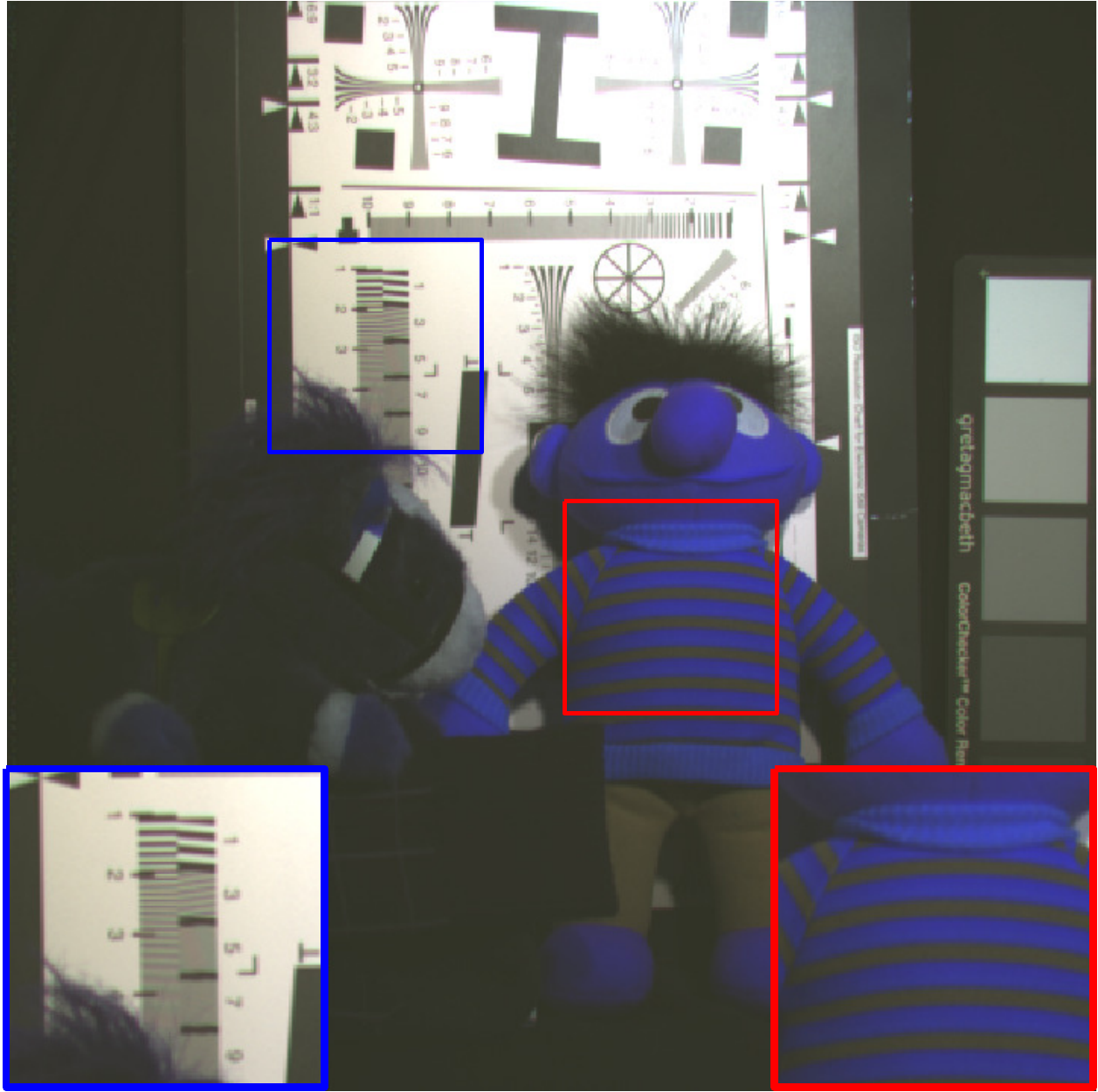}}\hspace{-0.8mm}
\subfigure[Noisy]{\includegraphics[scale =0.175,clip=true]{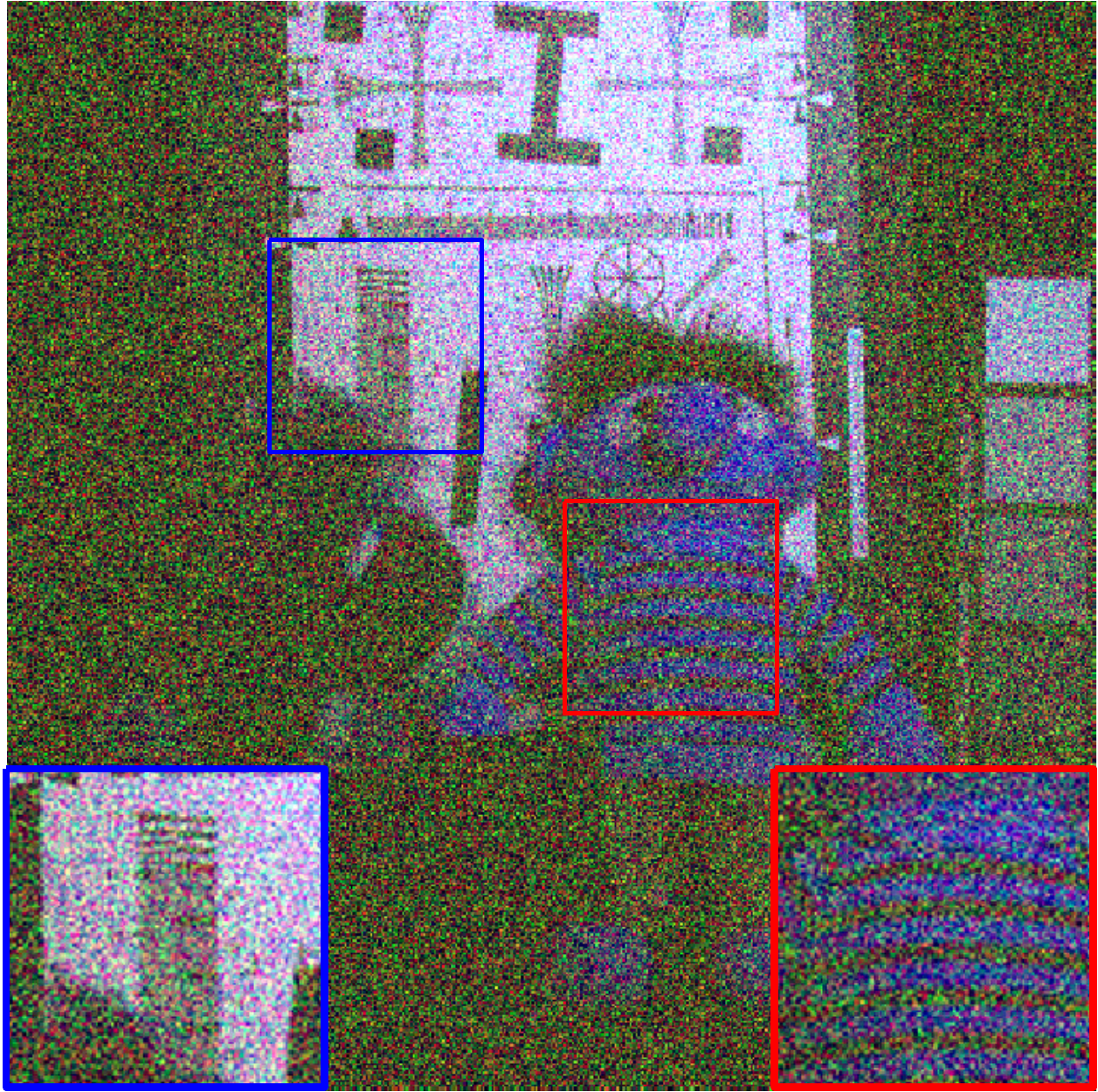}}\hspace{-0.8mm}
\subfigure[BM3D]{\includegraphics[scale =0.175,clip=true]{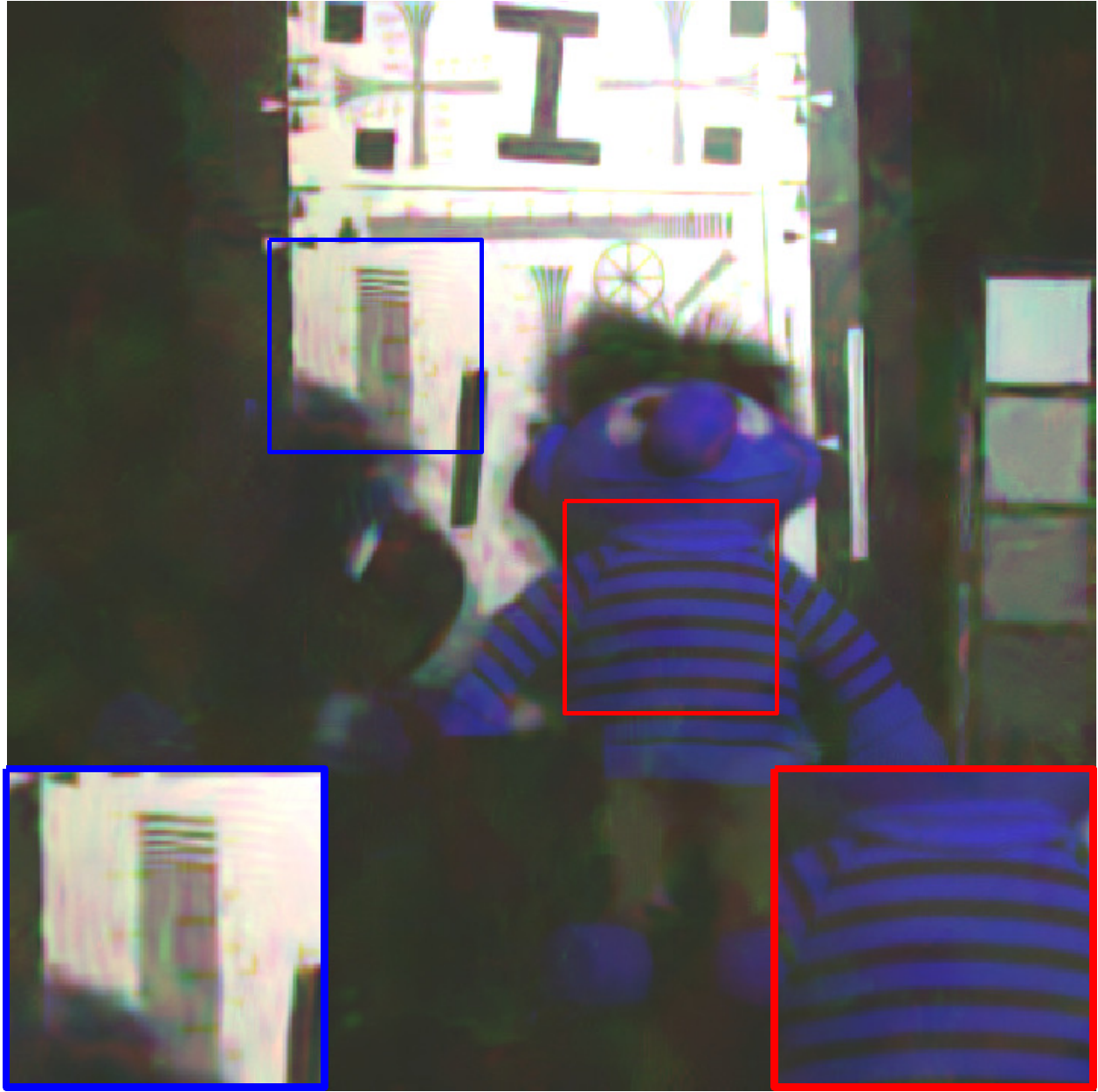}}\hspace{-0.8mm}
\subfigure[BM4D]{\includegraphics[scale =0.175,clip=true]{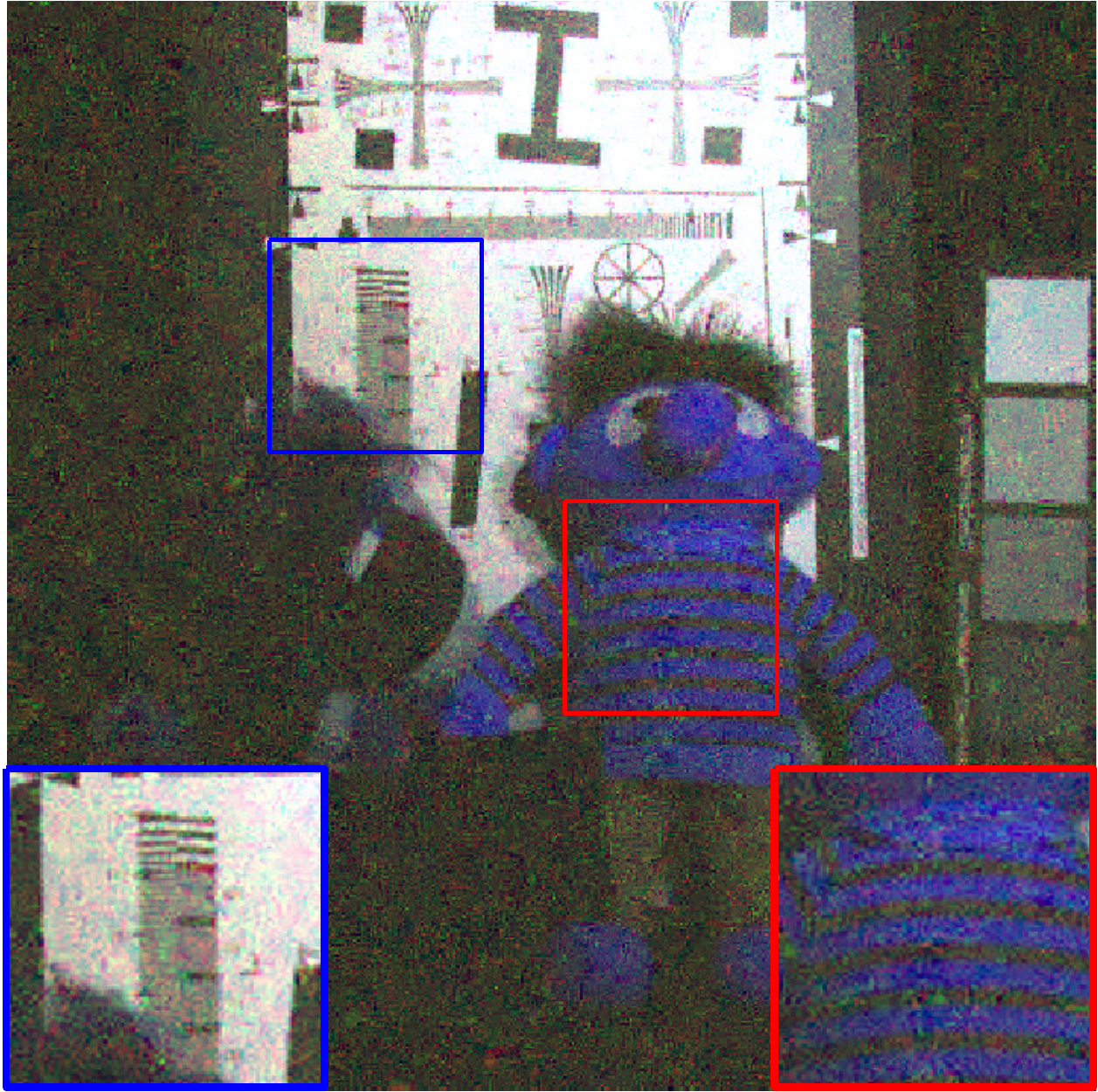}}\hspace{-0.8mm}
\subfigure[TDL]{\includegraphics[scale =0.175,clip=true]{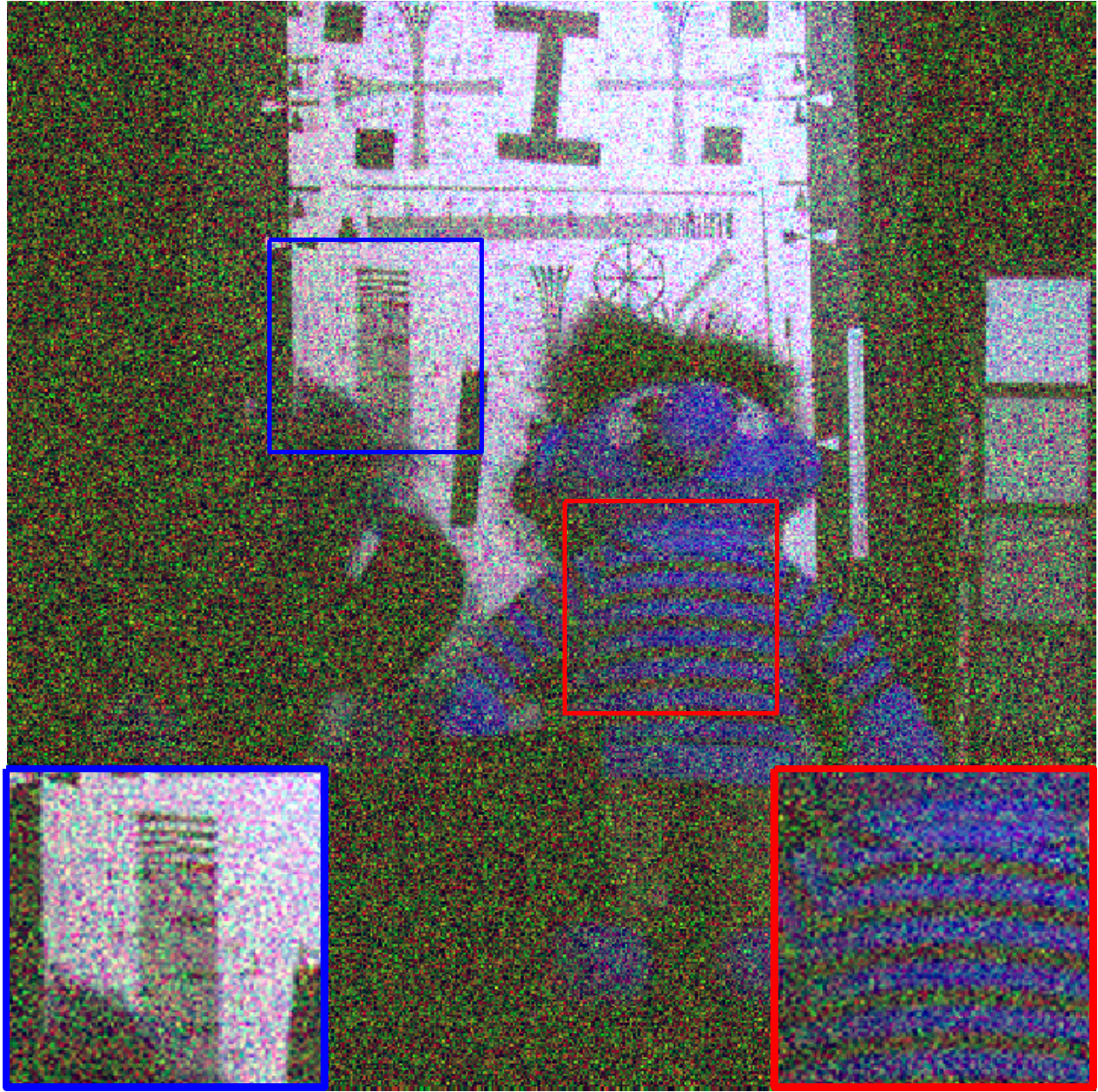}}\hspace{-0.8mm}
\subfigure[MTSNMF]{\includegraphics[scale =0.175,clip=true]{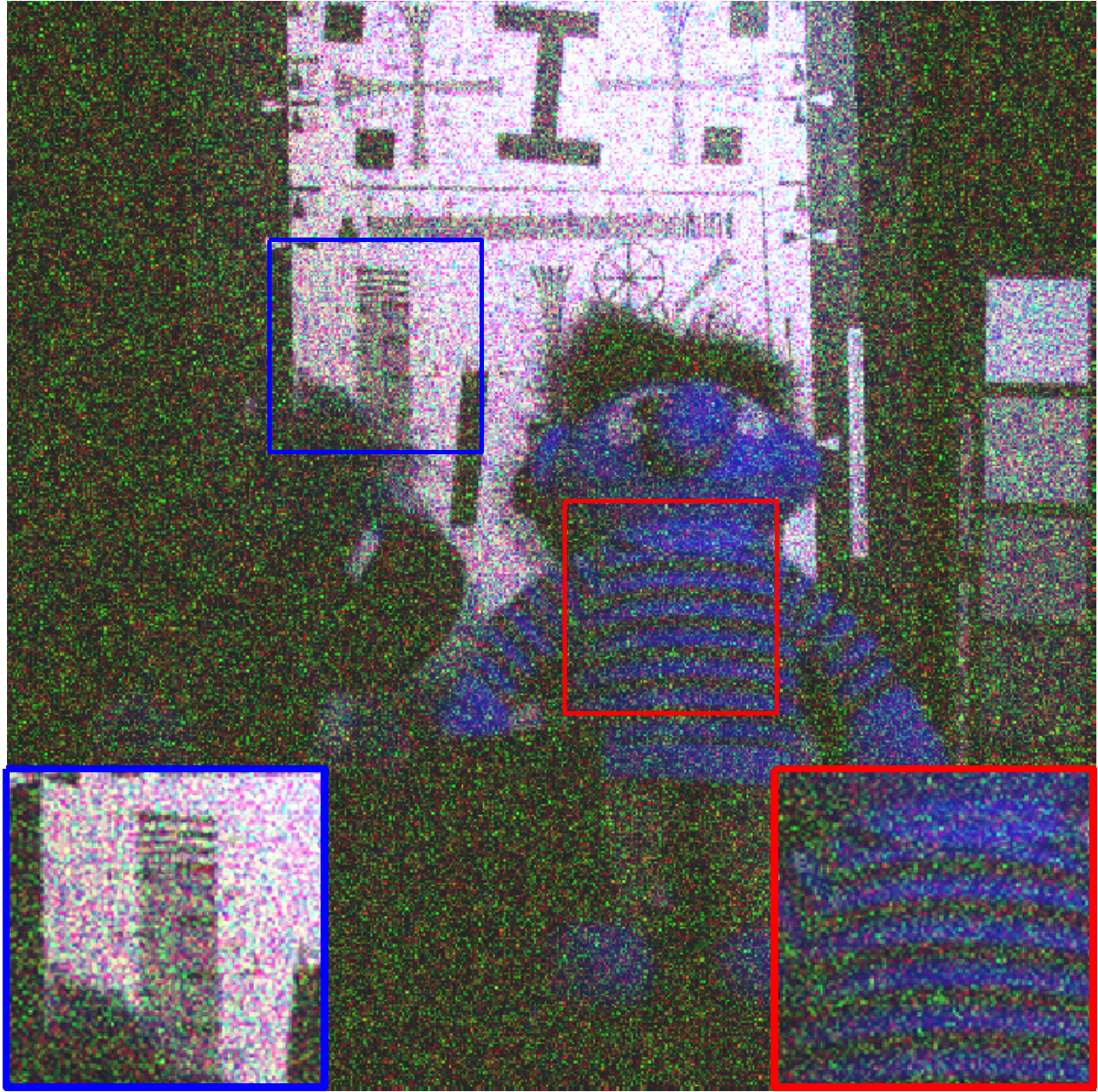}}\hspace{-0.8mm}
\subfigure[PARAFAC]{\includegraphics[scale =0.175,clip=true]{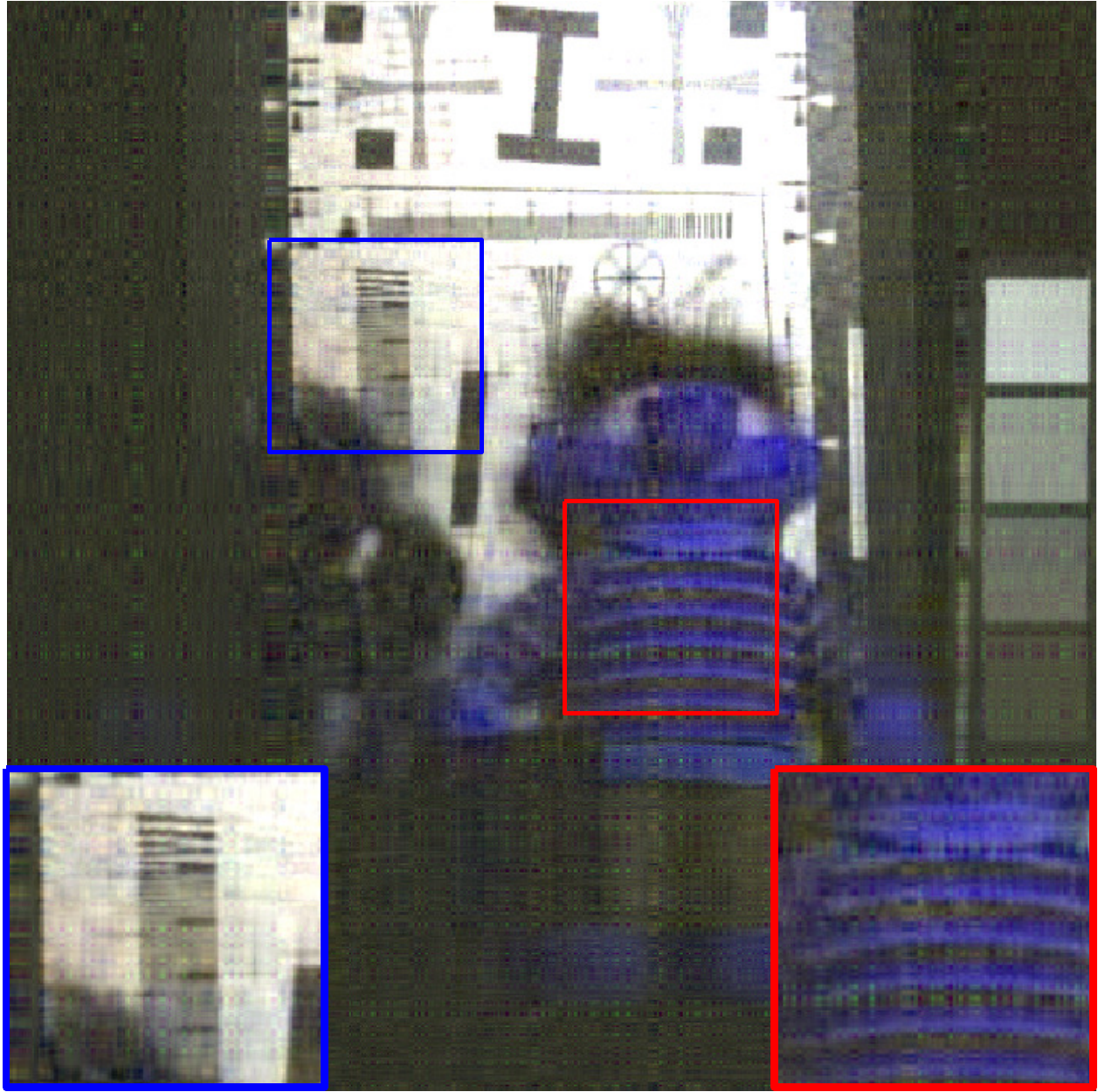}}\hspace{-0.8mm}
\subfigure[LLRT]{\includegraphics[scale =0.175,clip=true]{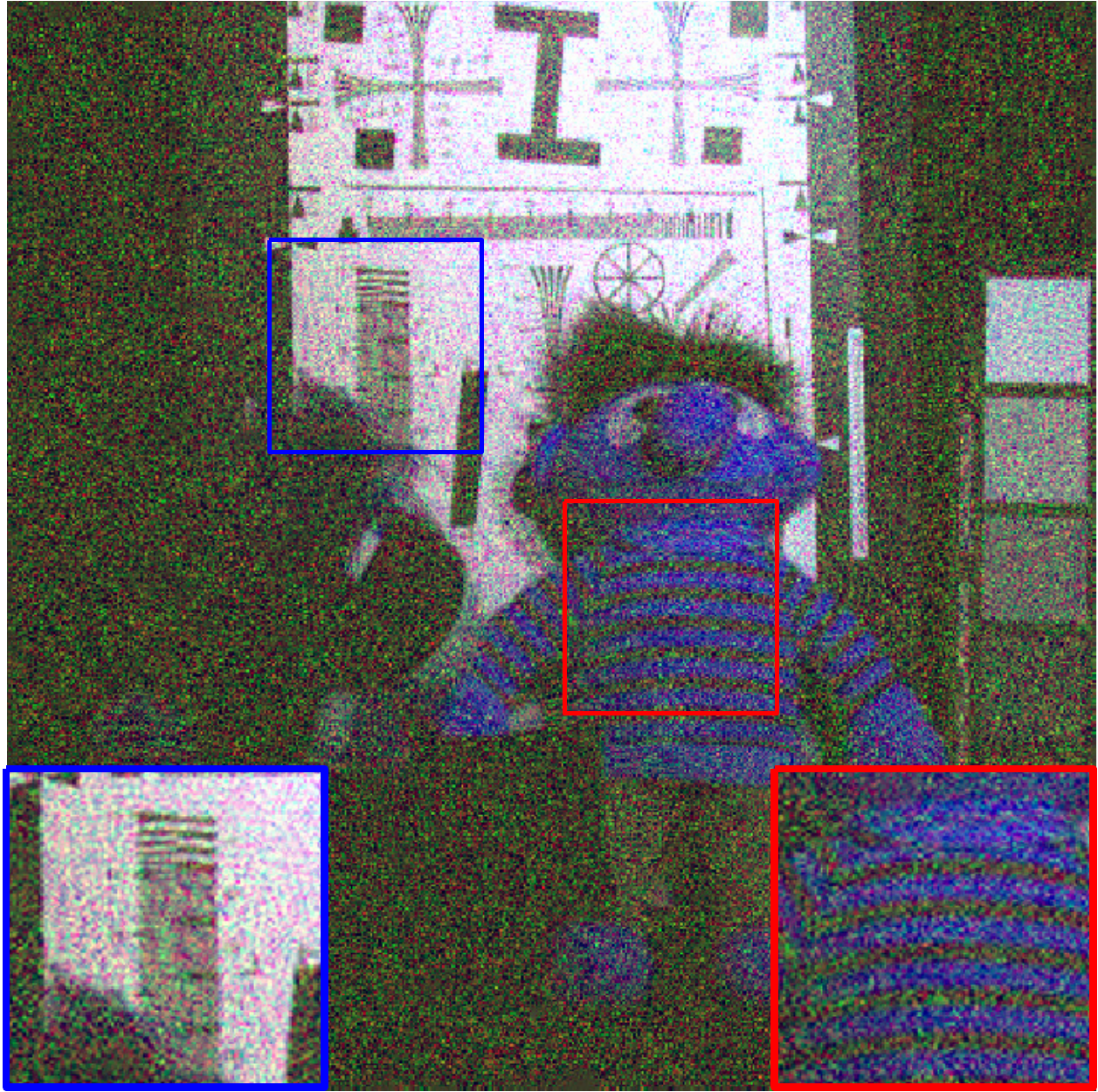}}\\
\subfigure[NGMeet]{\includegraphics[scale =0.175,clip=true]{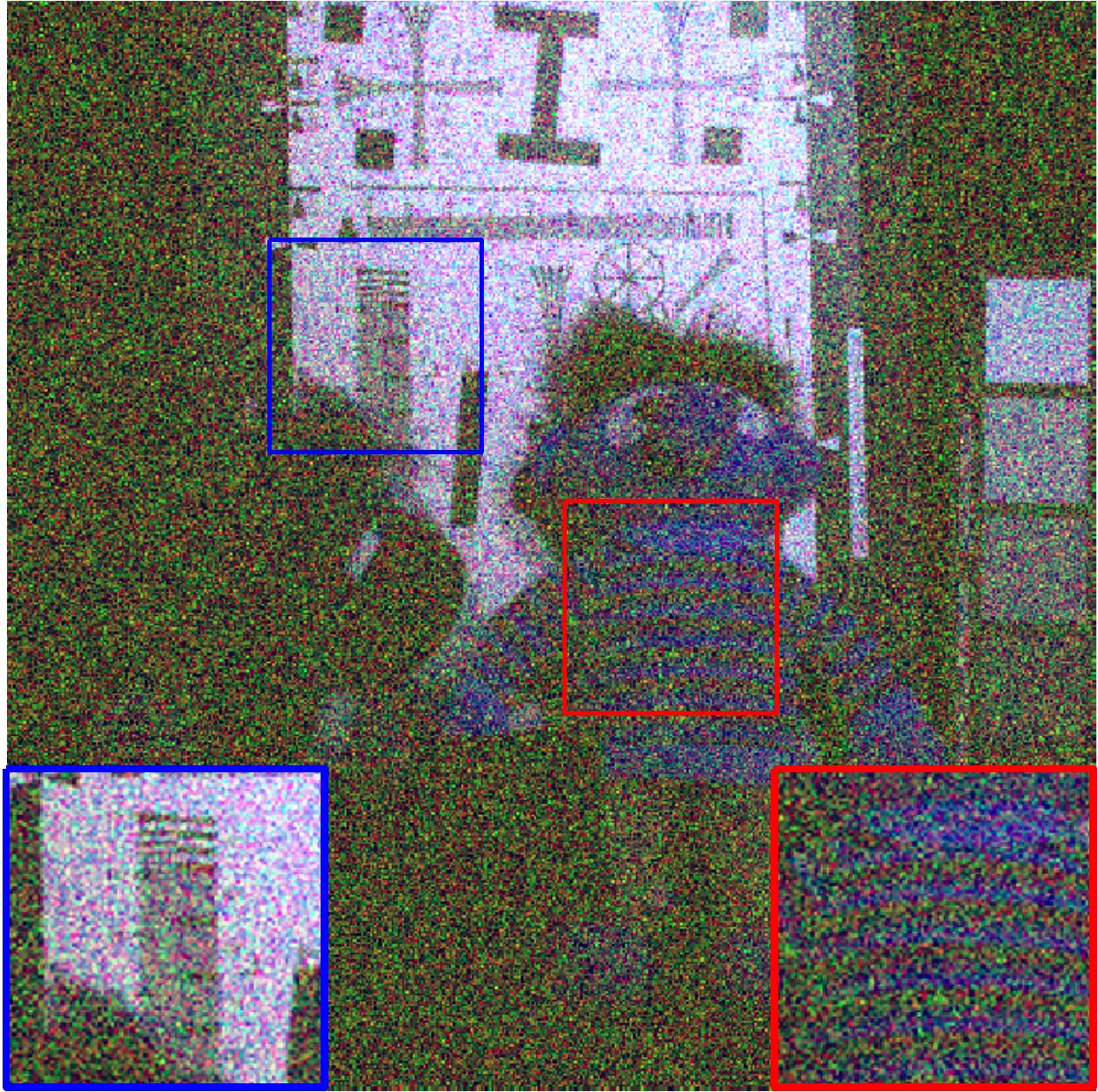}}\hspace{-0.8mm}
\subfigure[LRMR]{\includegraphics[scale =0.175,clip=true]{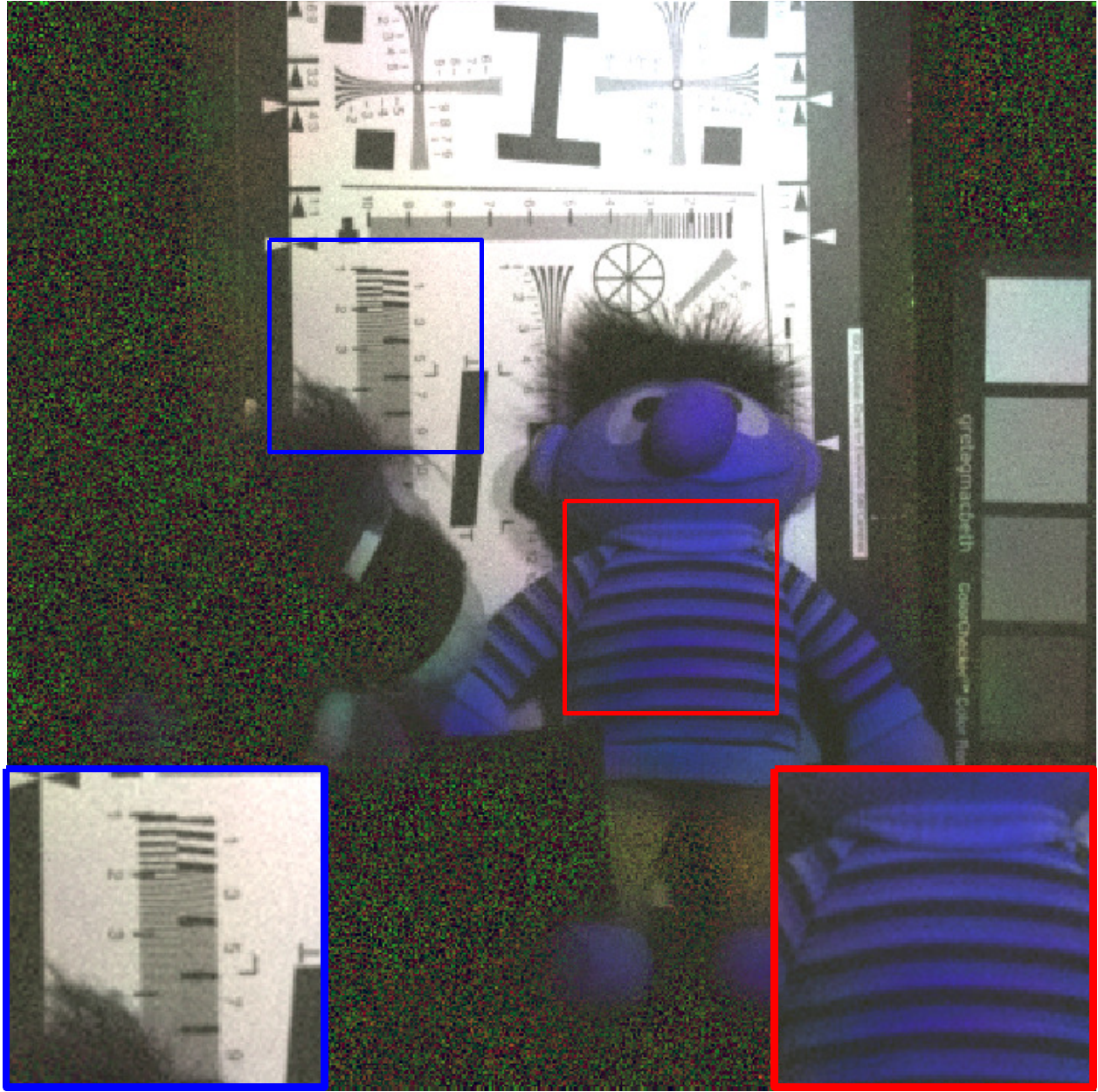}}\hspace{-0.8mm}
\subfigure[LRTDTV]{\includegraphics[scale =0.175,clip=true]{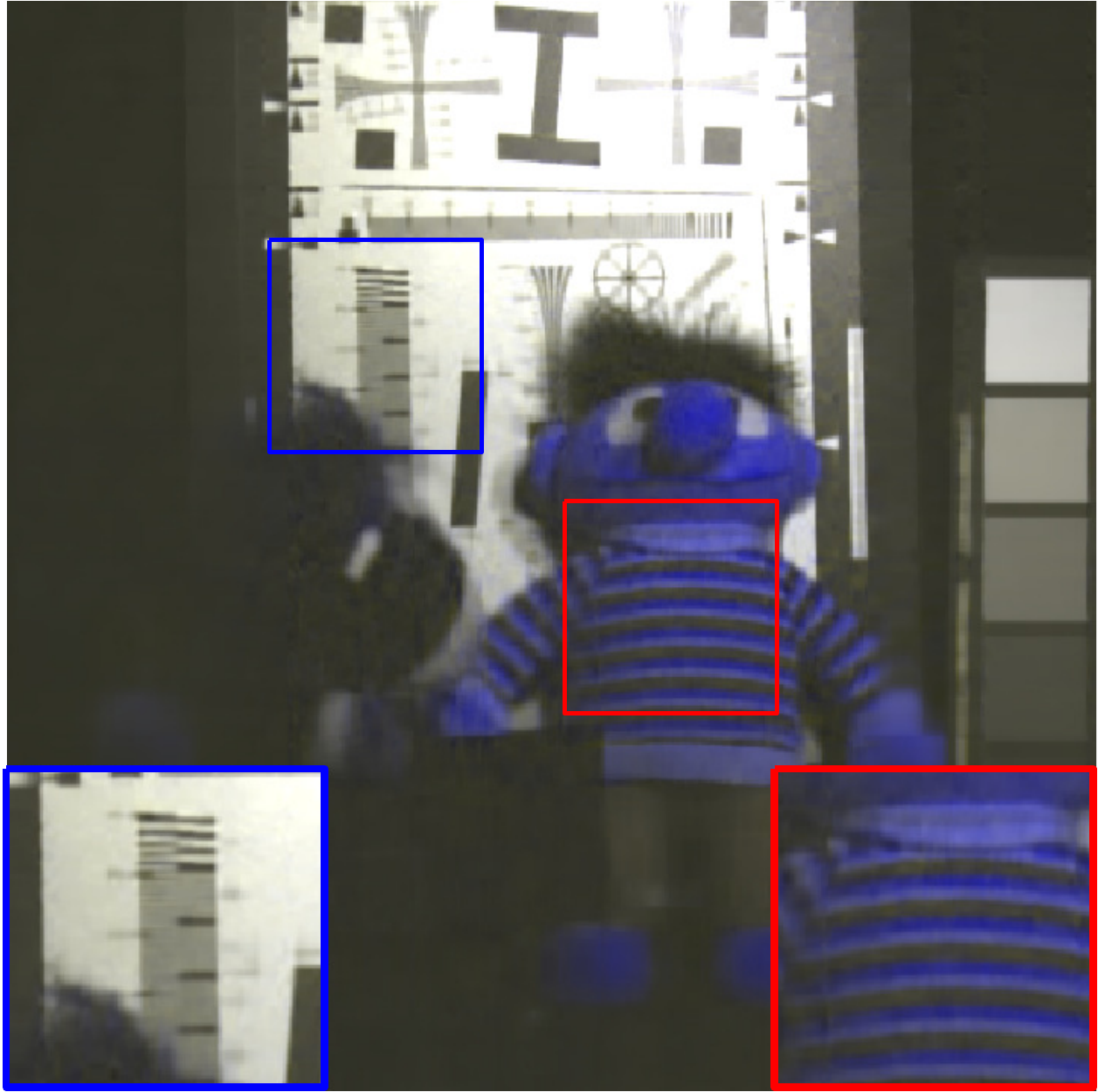}}\hspace{-0.8mm}
\subfigure[DnCNN]{\includegraphics[scale =0.175,clip=true]{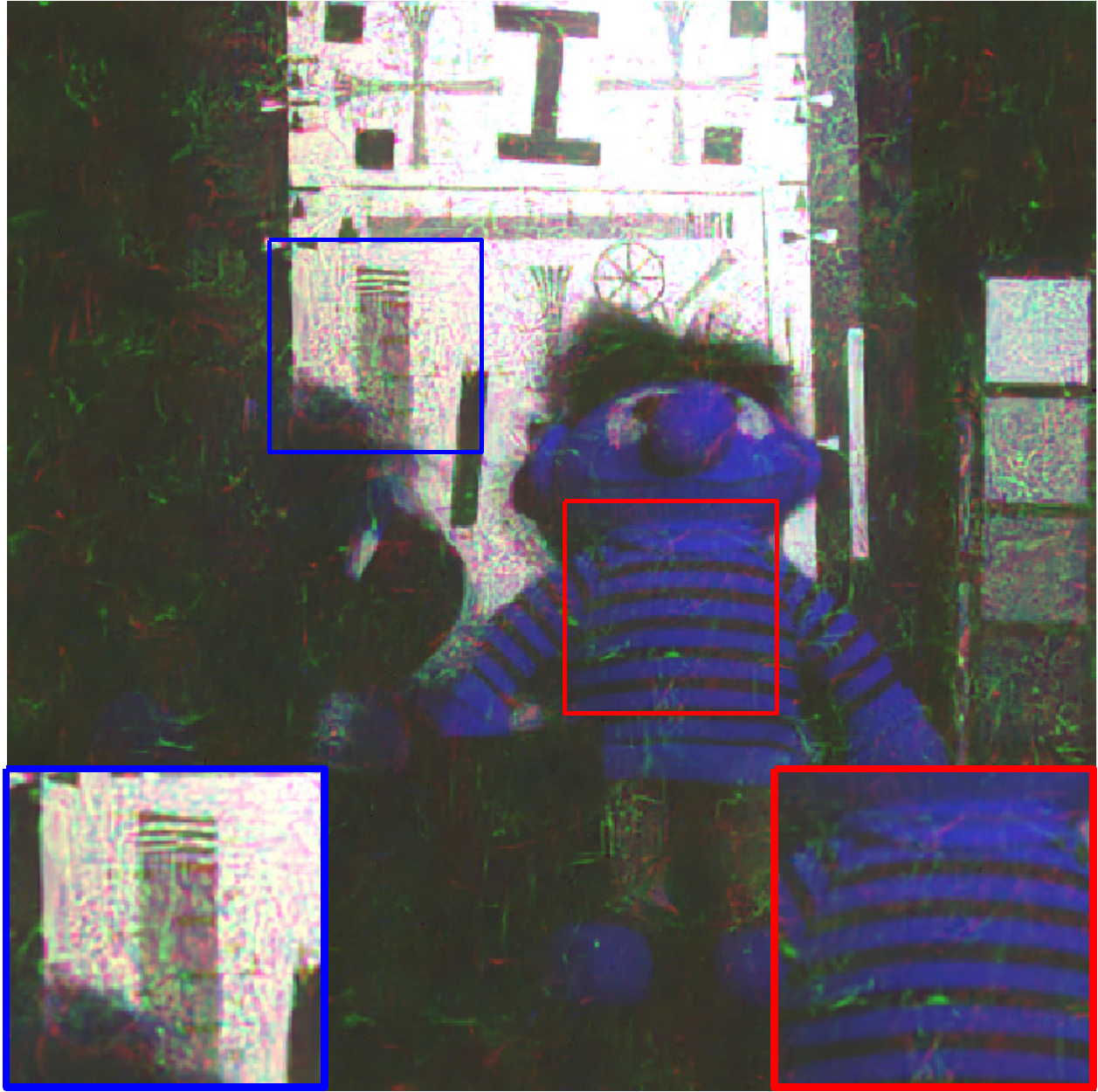}}\hspace{-0.8mm}
\subfigure[HSI-SDeCNN]{\includegraphics[scale =0.175,clip=true]{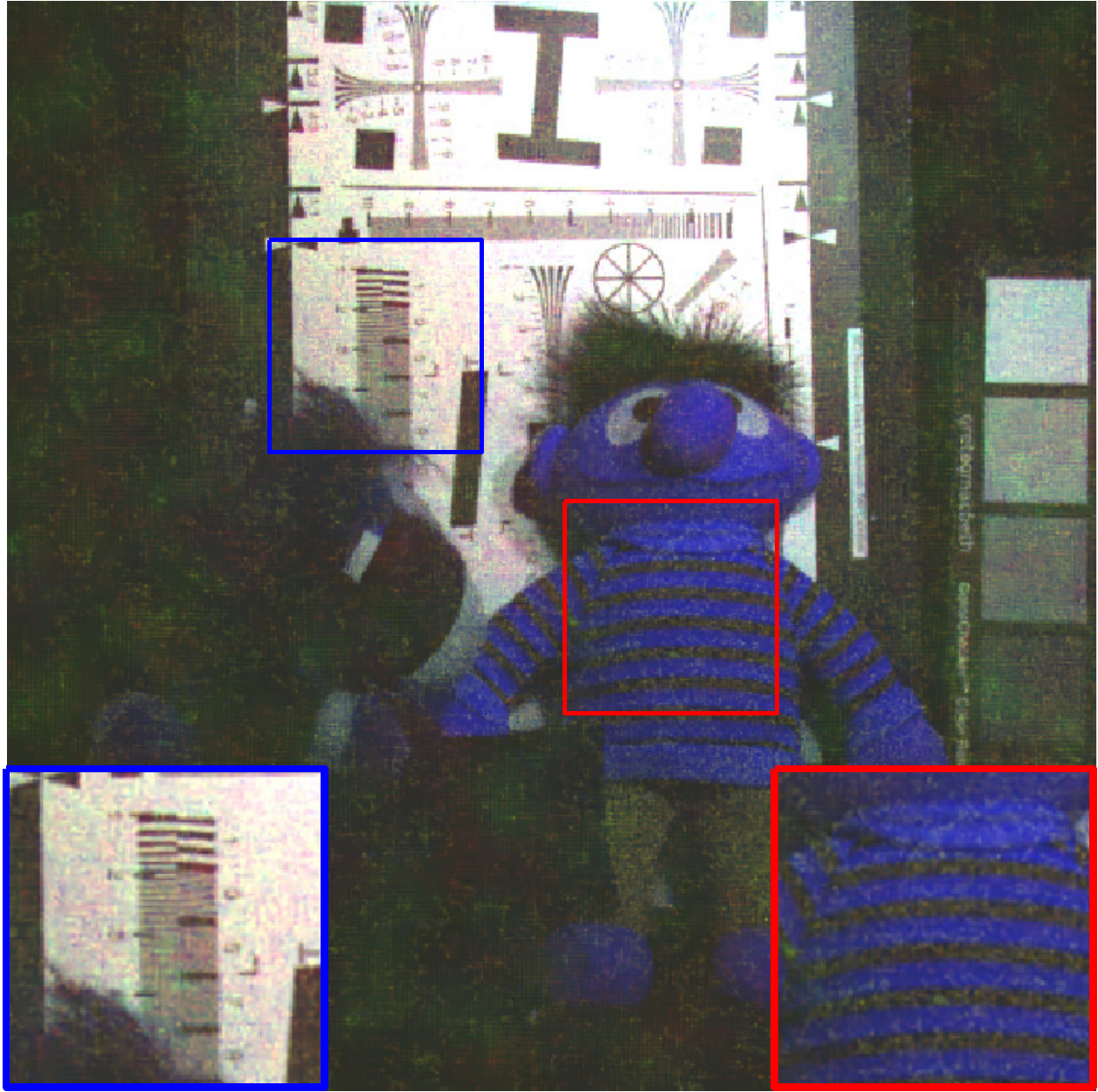}}\hspace{-0.8mm}
\subfigure[HSID-CNN]{\includegraphics[scale =0.175,clip=true]{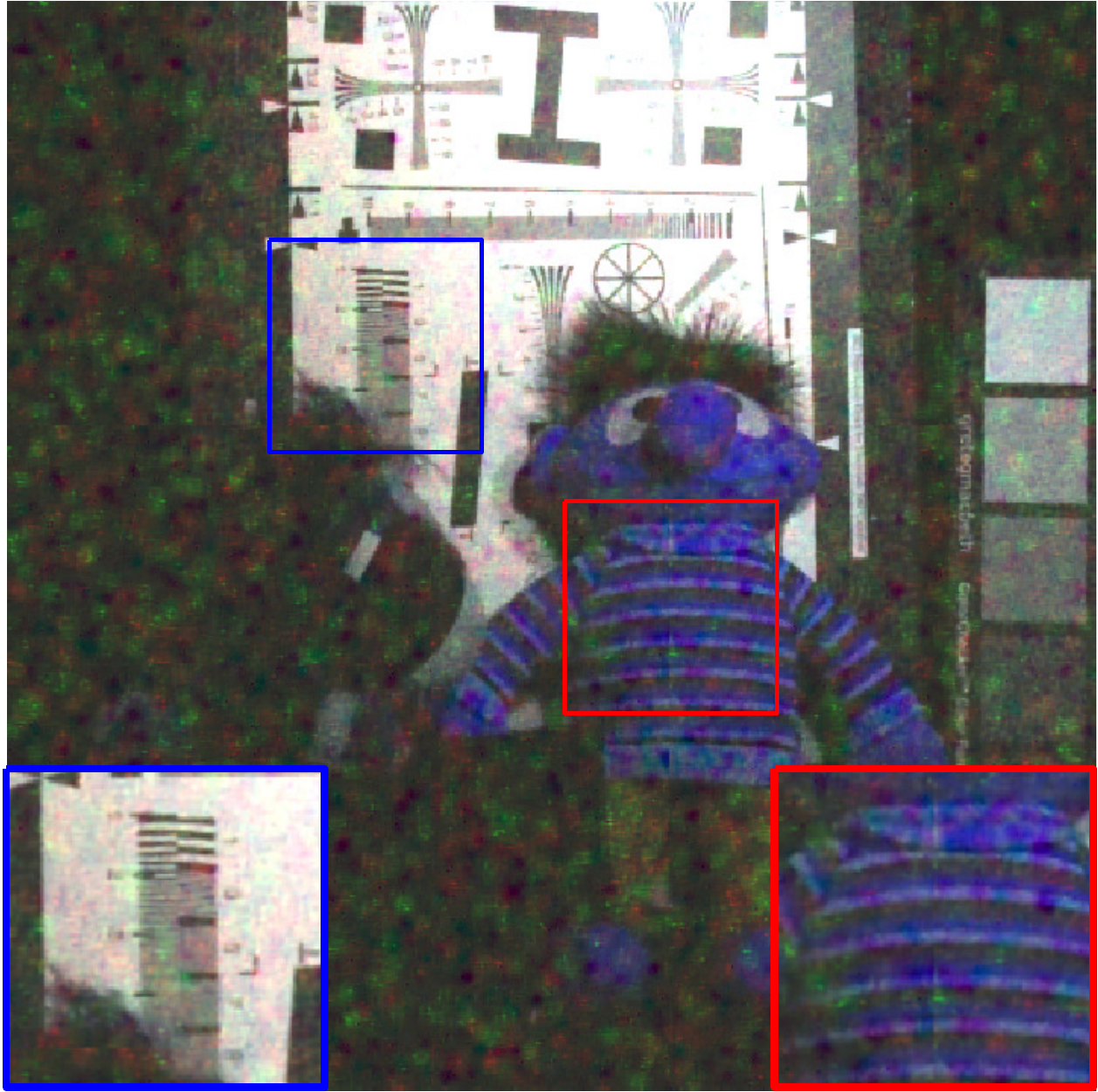}}\hspace{-0.8mm}
\subfigure[QRNN3D]{\includegraphics[scale =0.175,clip=true]{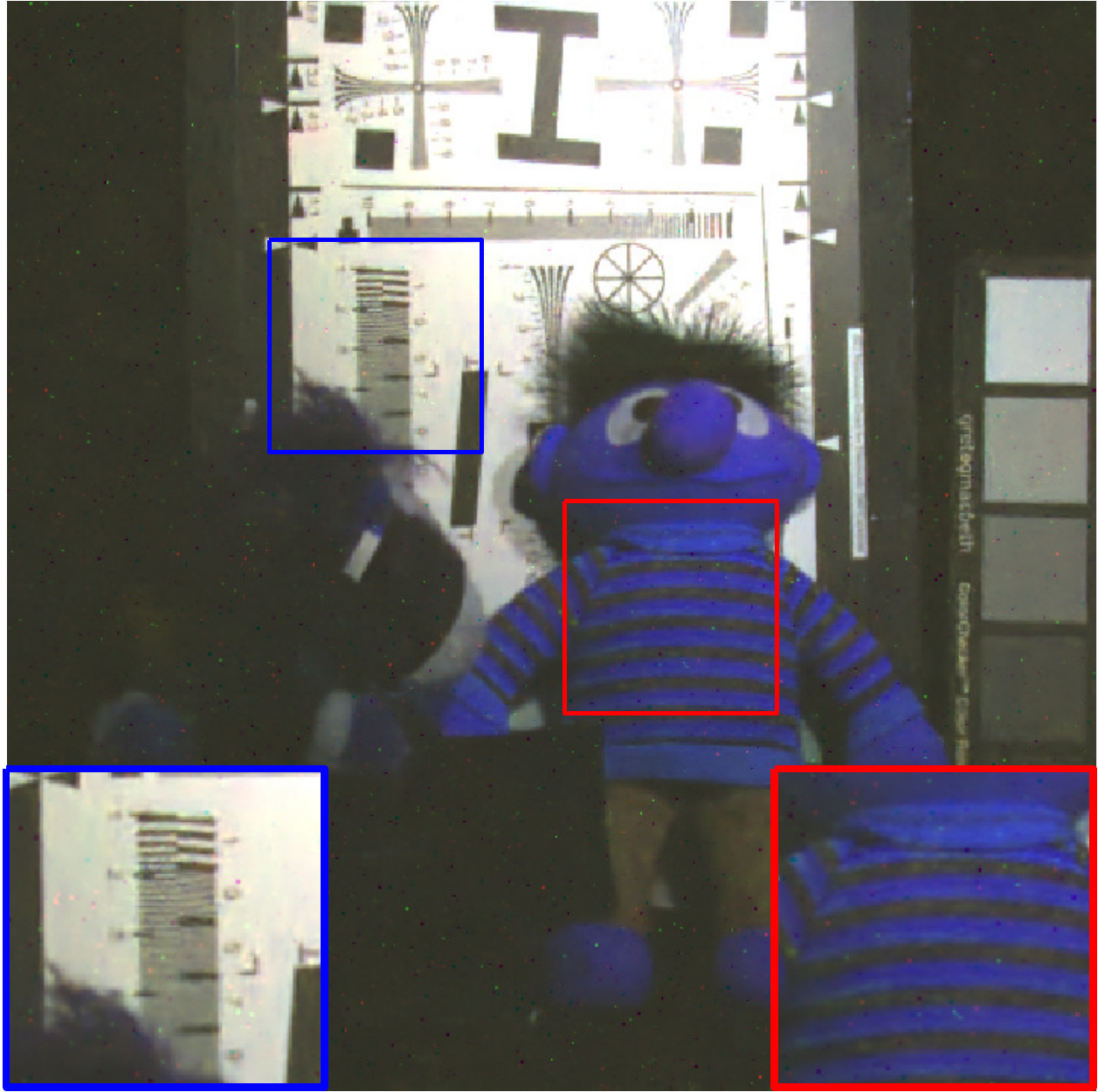}}\hspace{-0.8mm}
\subfigure[SMDS-Net]{\includegraphics[scale =0.175,clip=true]{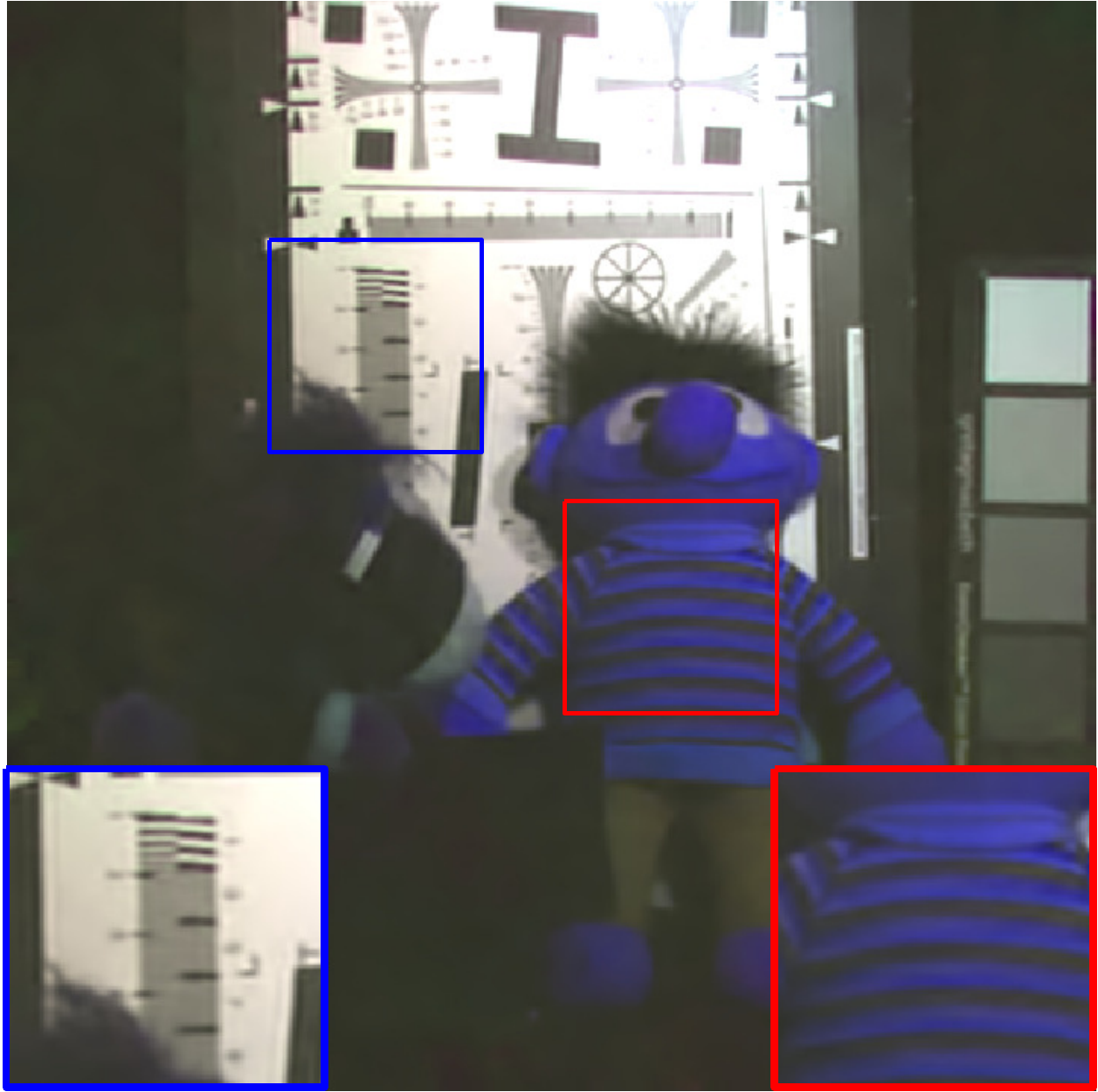}}
\caption{Denoising results on the  \emph{chart and stuffed toy} HSI with the noise variance in [0-95]. The false-color images were generated by combining bands 5, 18, 25. SMDS-Net achieves the best visual results with less artifacts. (\textbf{Best view on screen with zoom}) } \label{fig:cave}
 \end{figure*}

  	\begin{figure*}[!htbp]
  \centering
\subfigure[Noisy]{\includegraphics[scale =0.125,clip=true]{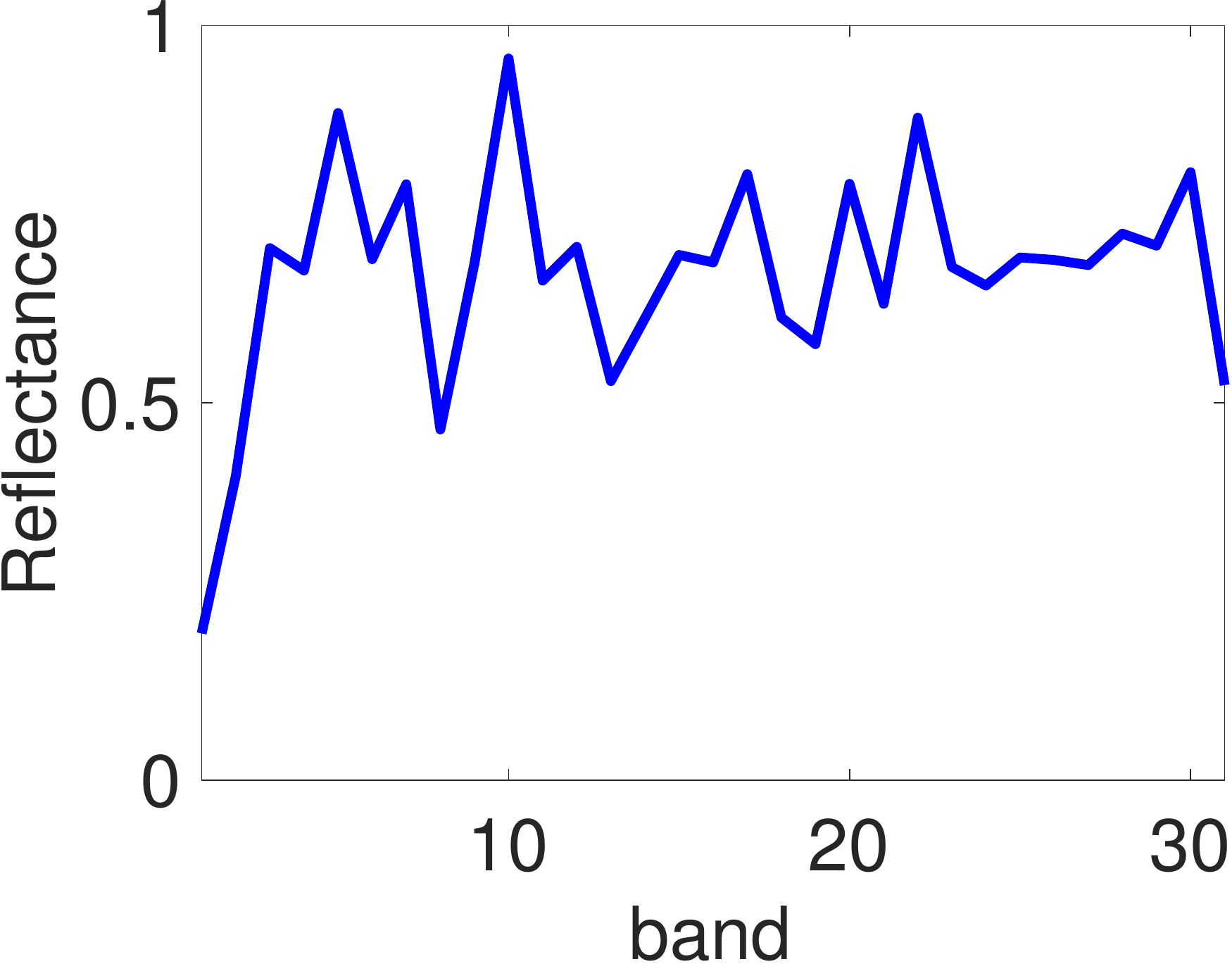}}\hspace{-0.8mm}
\subfigure[Clean]{\includegraphics[scale =0.125,clip=true]{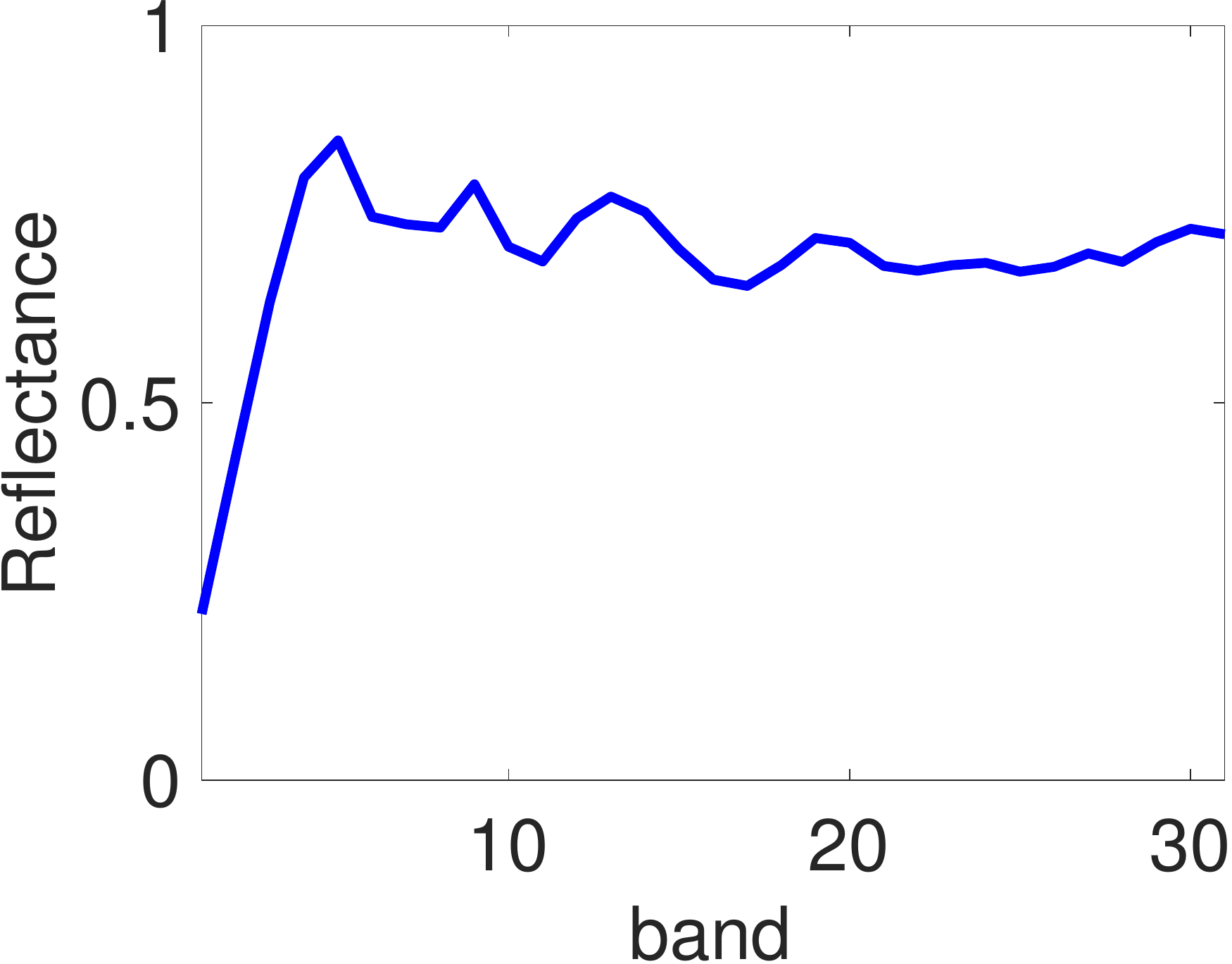}}\hspace{-0.8mm}
\subfigure[BM3D]{\includegraphics[scale =0.125,clip=true]{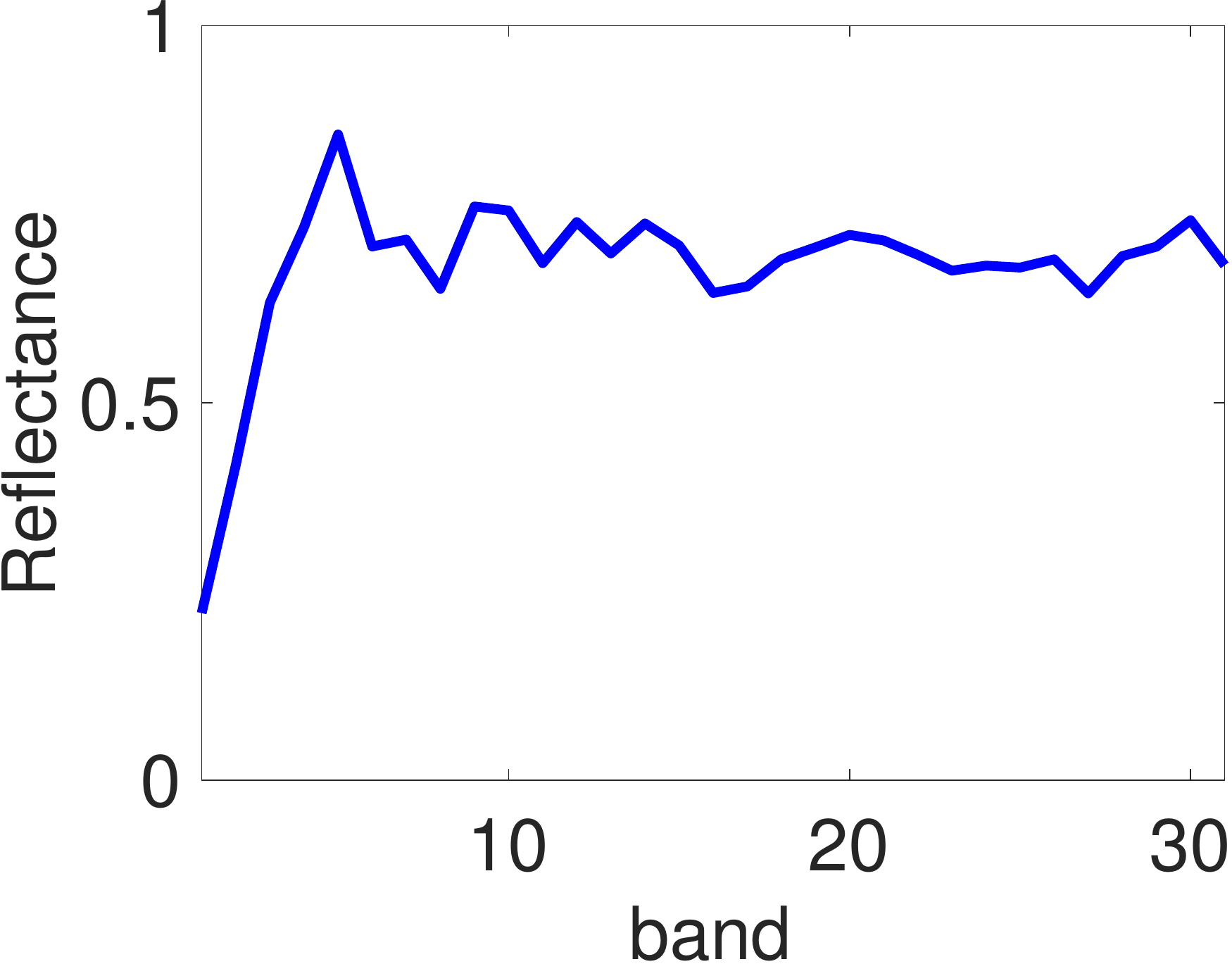}}\hspace{-0.8mm}
\subfigure[BM4D]{\includegraphics[scale =0.125,clip=true]{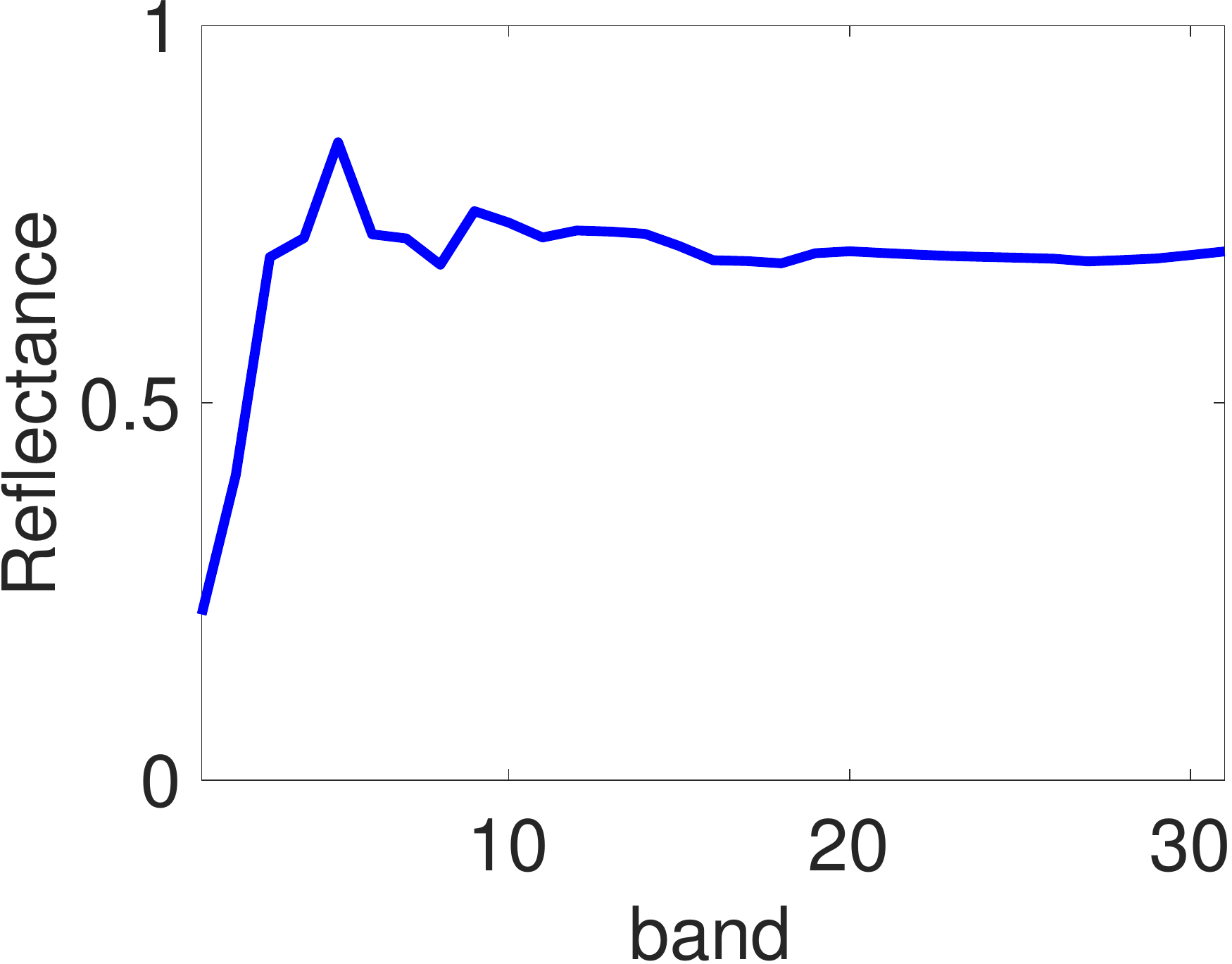}}\hspace{-0.8mm}
\subfigure[TDL]{\includegraphics[scale =0.125,clip=true]{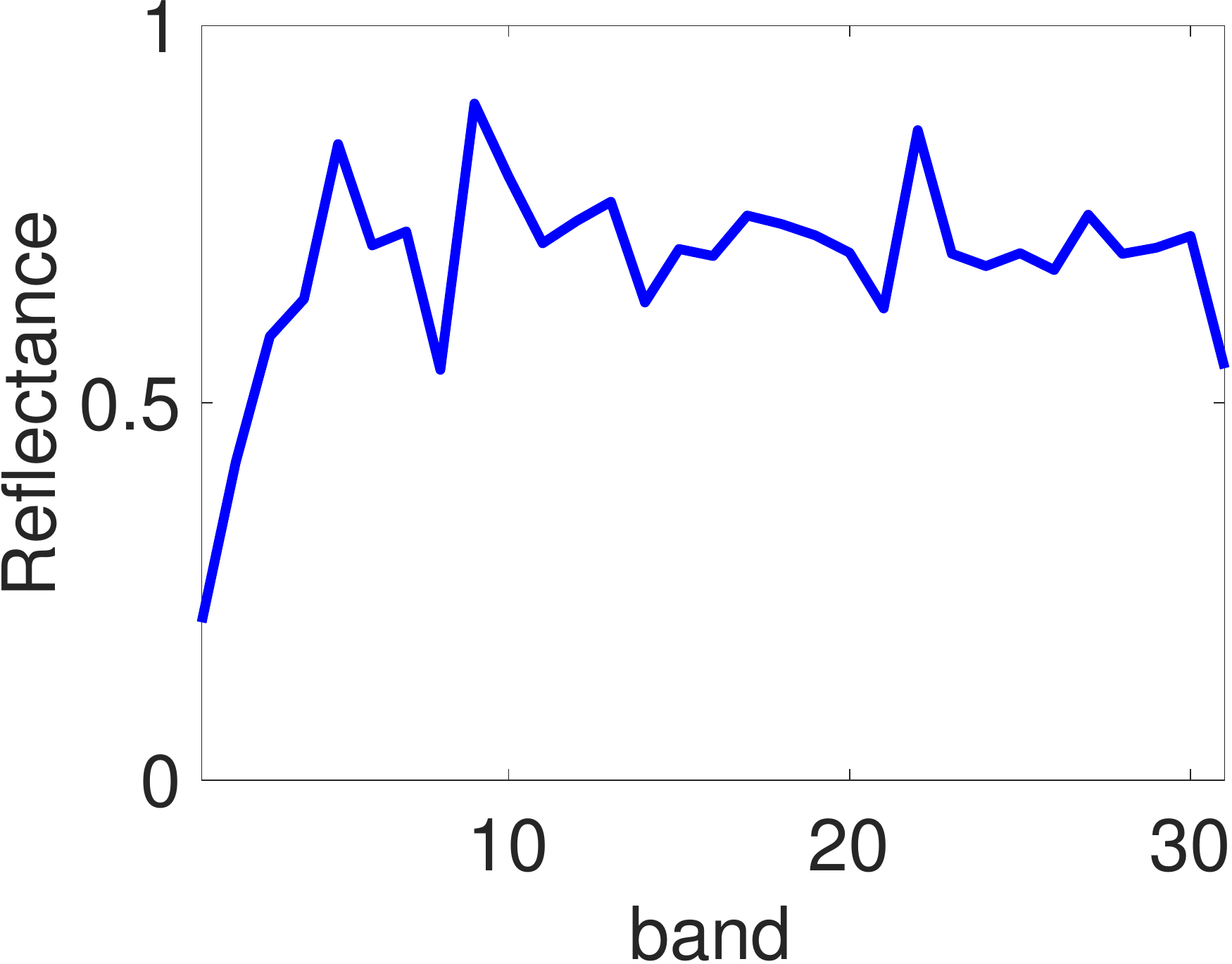}}\hspace{-0.8mm}
\subfigure[PARAFAC]{\includegraphics[scale =0.125,clip=true]{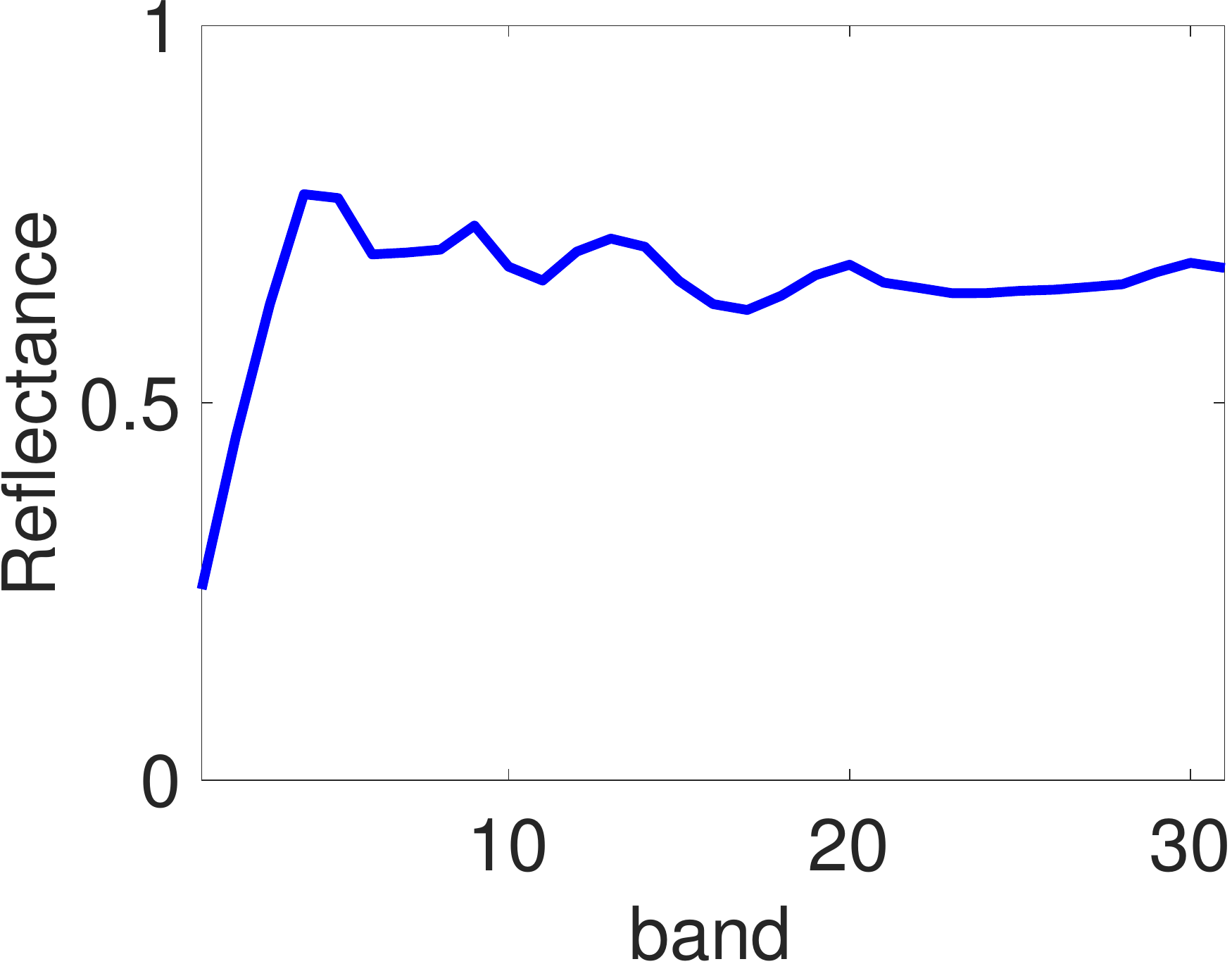}}\hspace{-0.8mm}
\subfigure[MTSNMF]{\includegraphics[scale =0.125,clip=true]{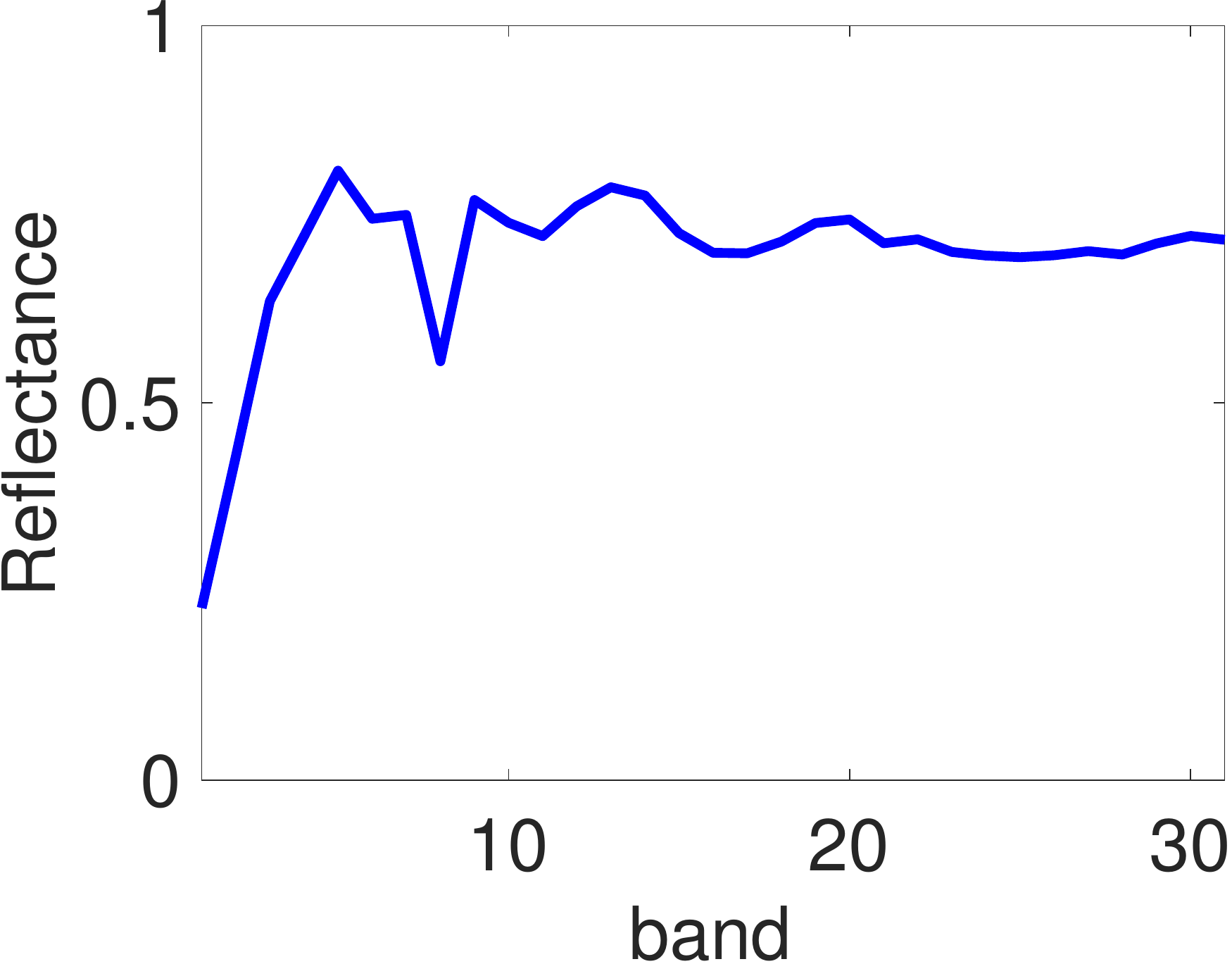}}\hspace{-0.8mm}
\subfigure[LLRT]{\includegraphics[scale =0.125,clip=true]{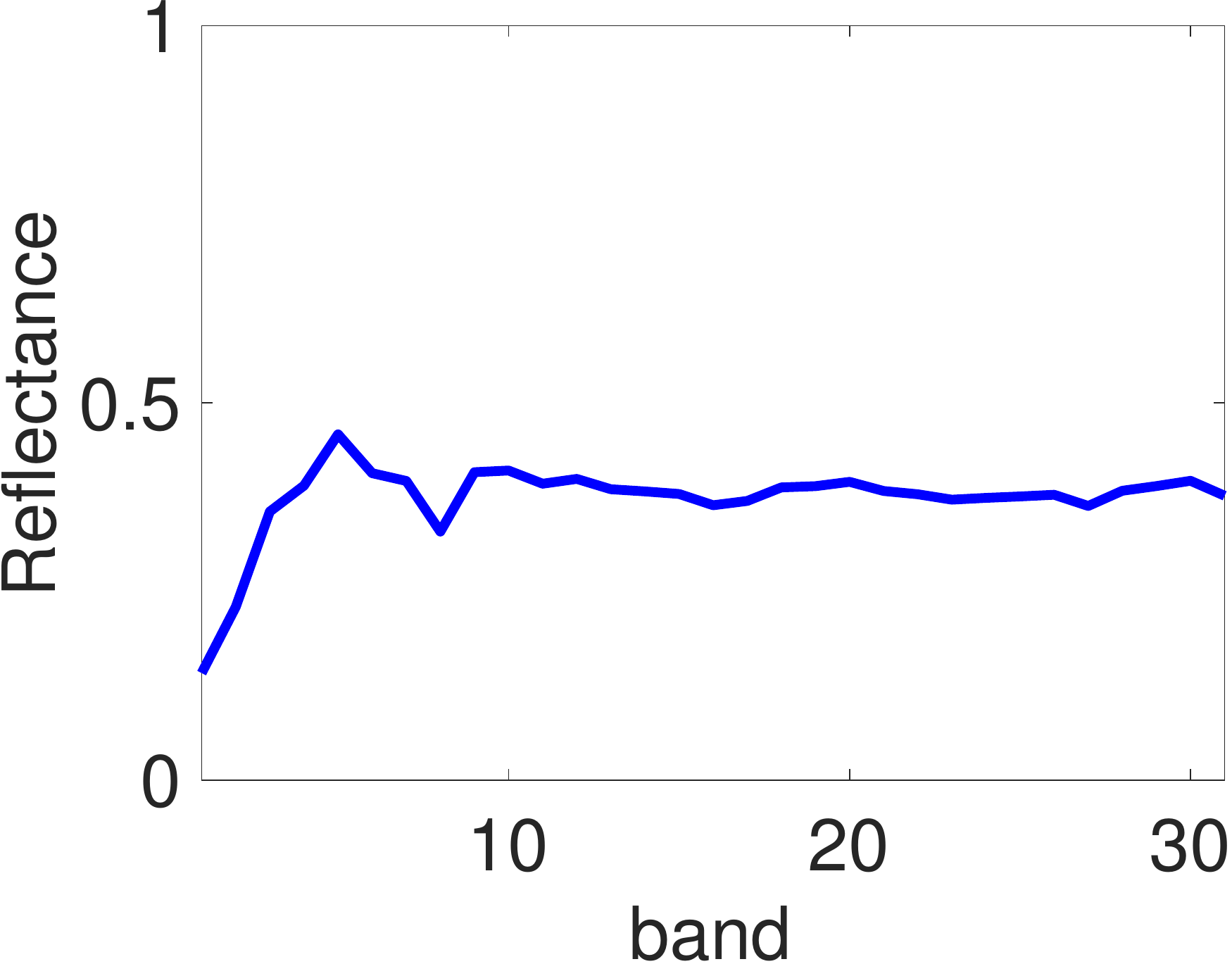}}\hspace{-0.8mm}\\
\subfigure[NGMeet]{\includegraphics[scale =0.125,clip=true]{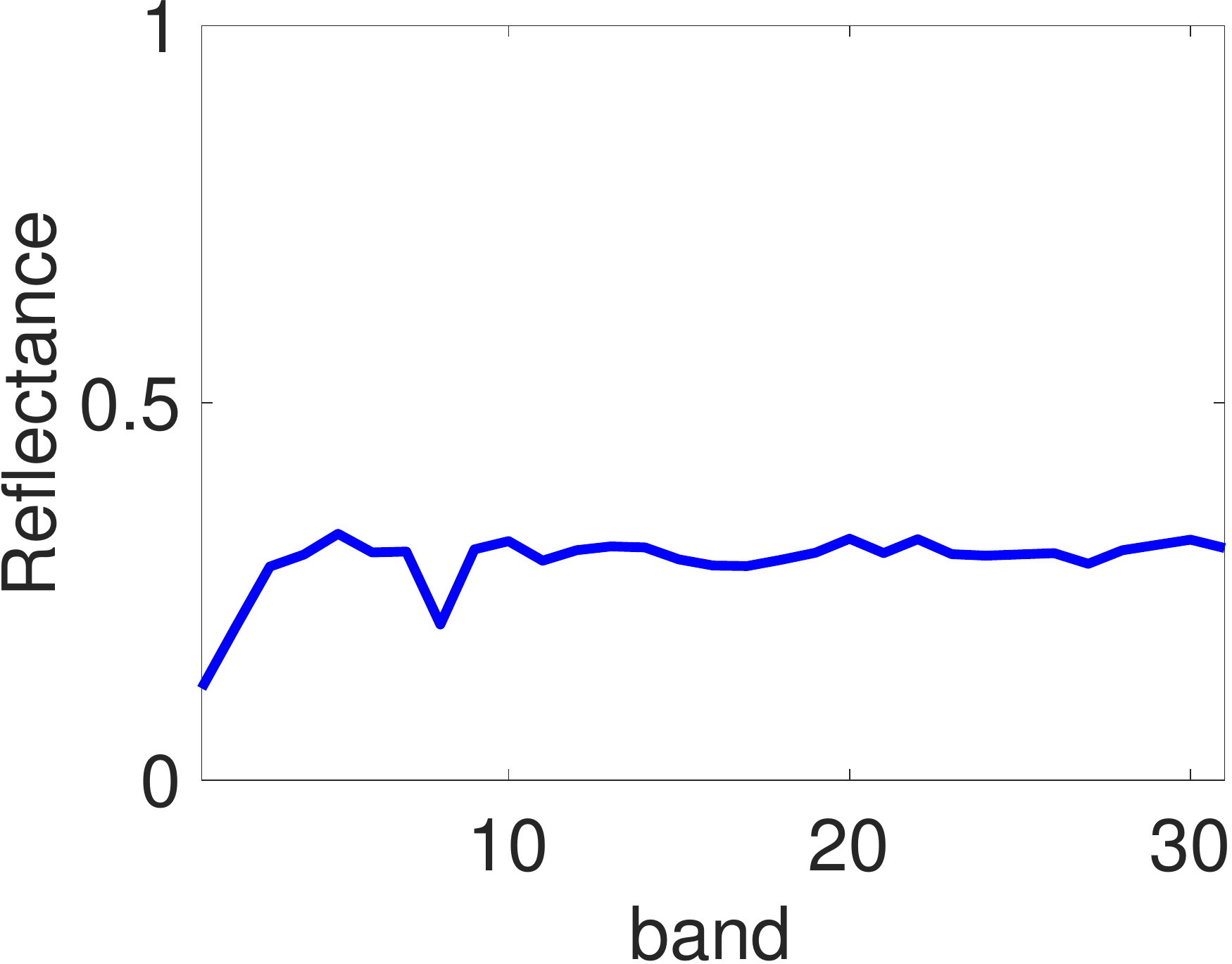}}\hspace{-0.8mm}
\subfigure[LRMR]{\includegraphics[scale =0.125,clip=true]{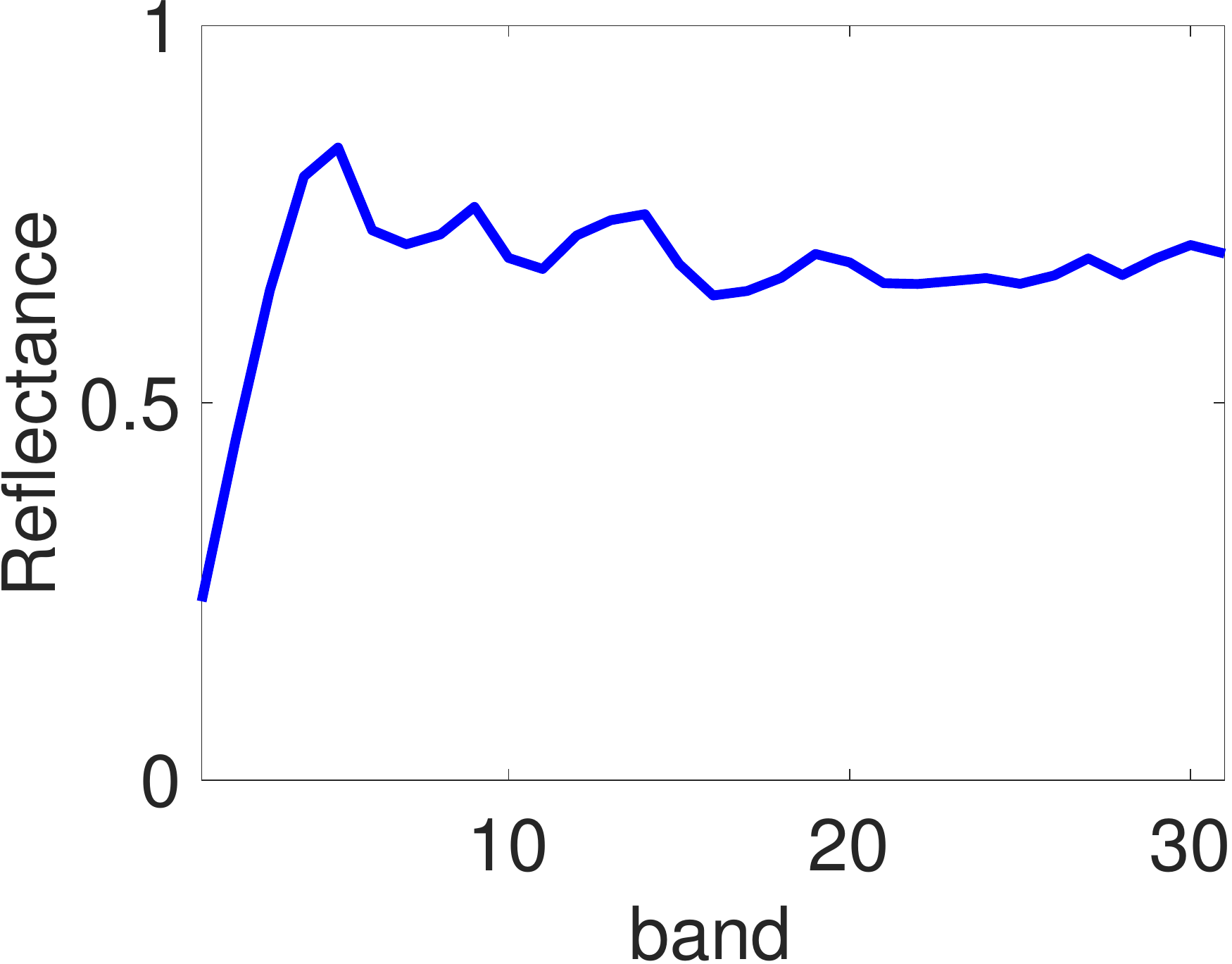}}\hspace{-0.8mm}
\subfigure[LRTDTV]{\includegraphics[scale =0.125,clip=true]{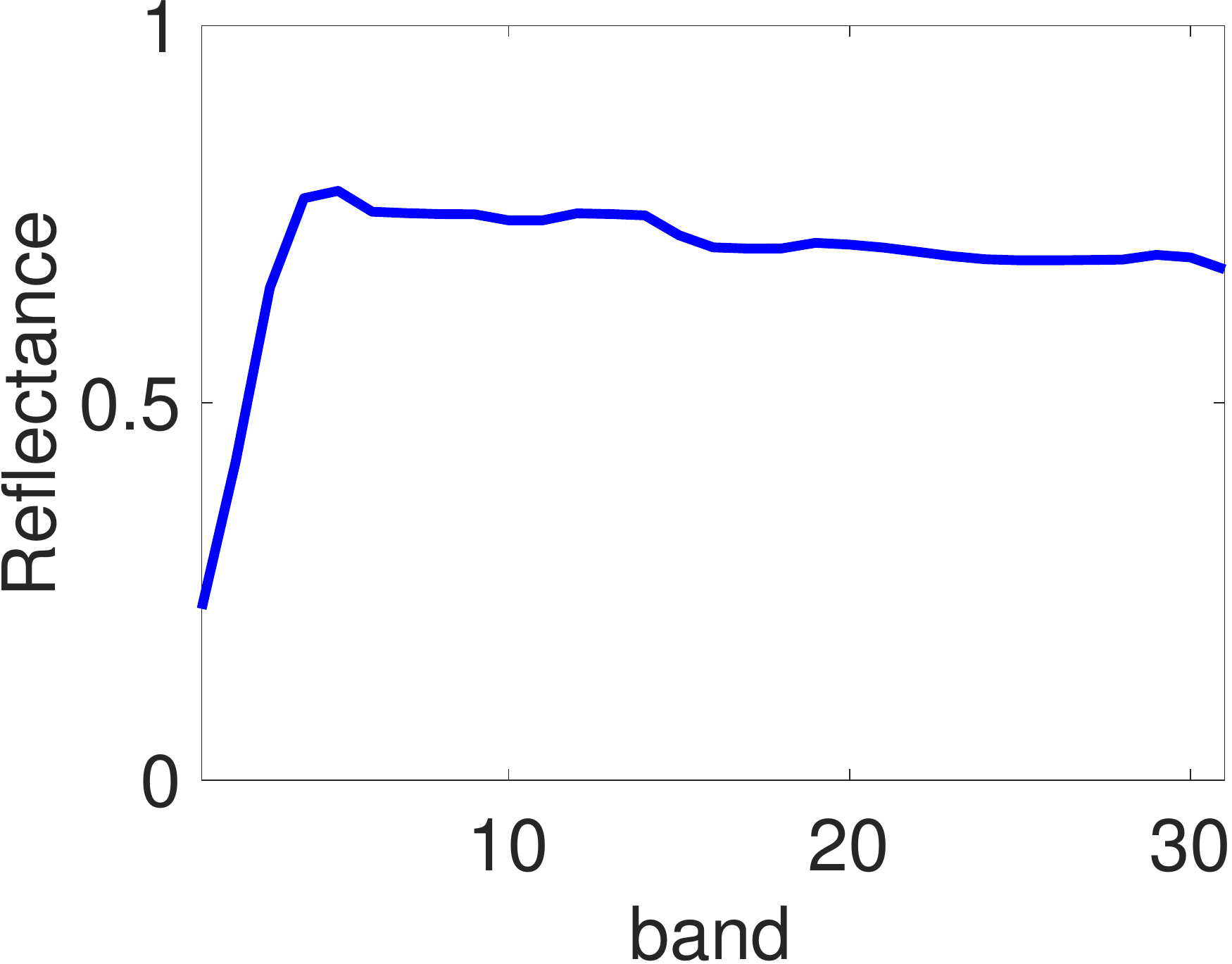}}\hspace{-0.8mm}
\subfigure[DnCNN]{\includegraphics[scale =0.125,clip=true]{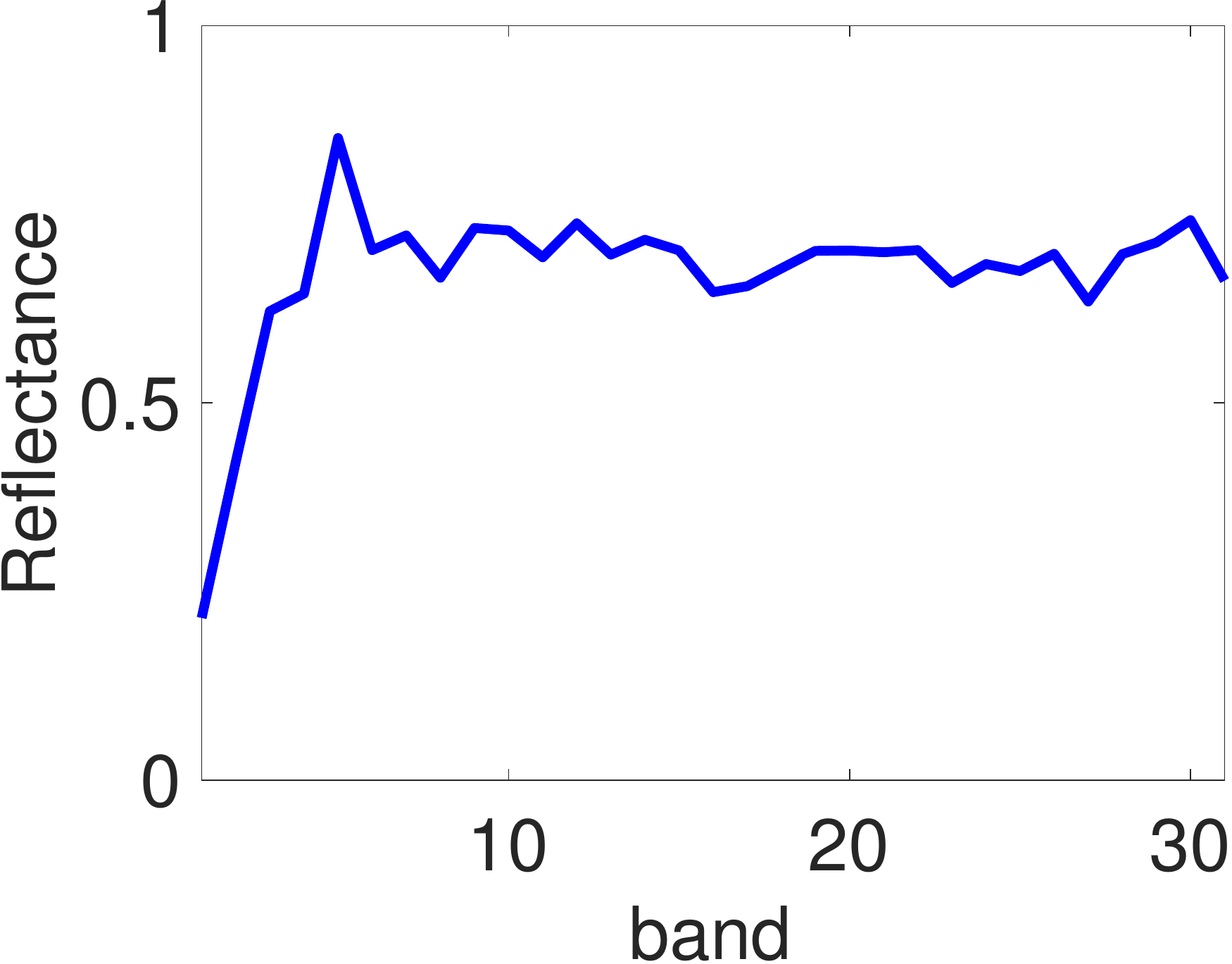}}\hspace{-0.8mm}
\subfigure[HSI-SDeCNN]{\includegraphics[scale =0.125,clip=true]{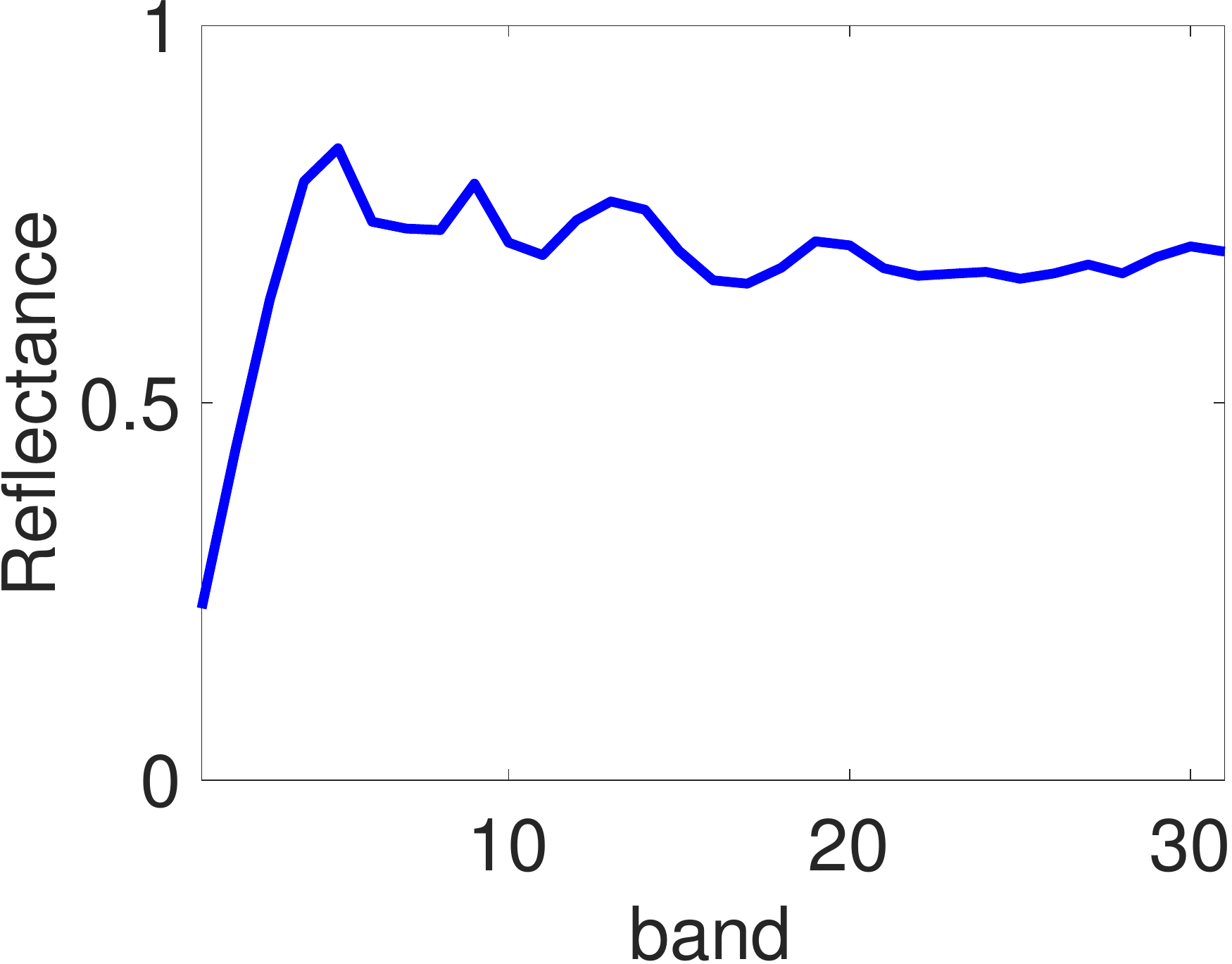}}\hspace{-0.8mm}
\subfigure[HSID-CNN]{\includegraphics[scale =0.125,clip=true]{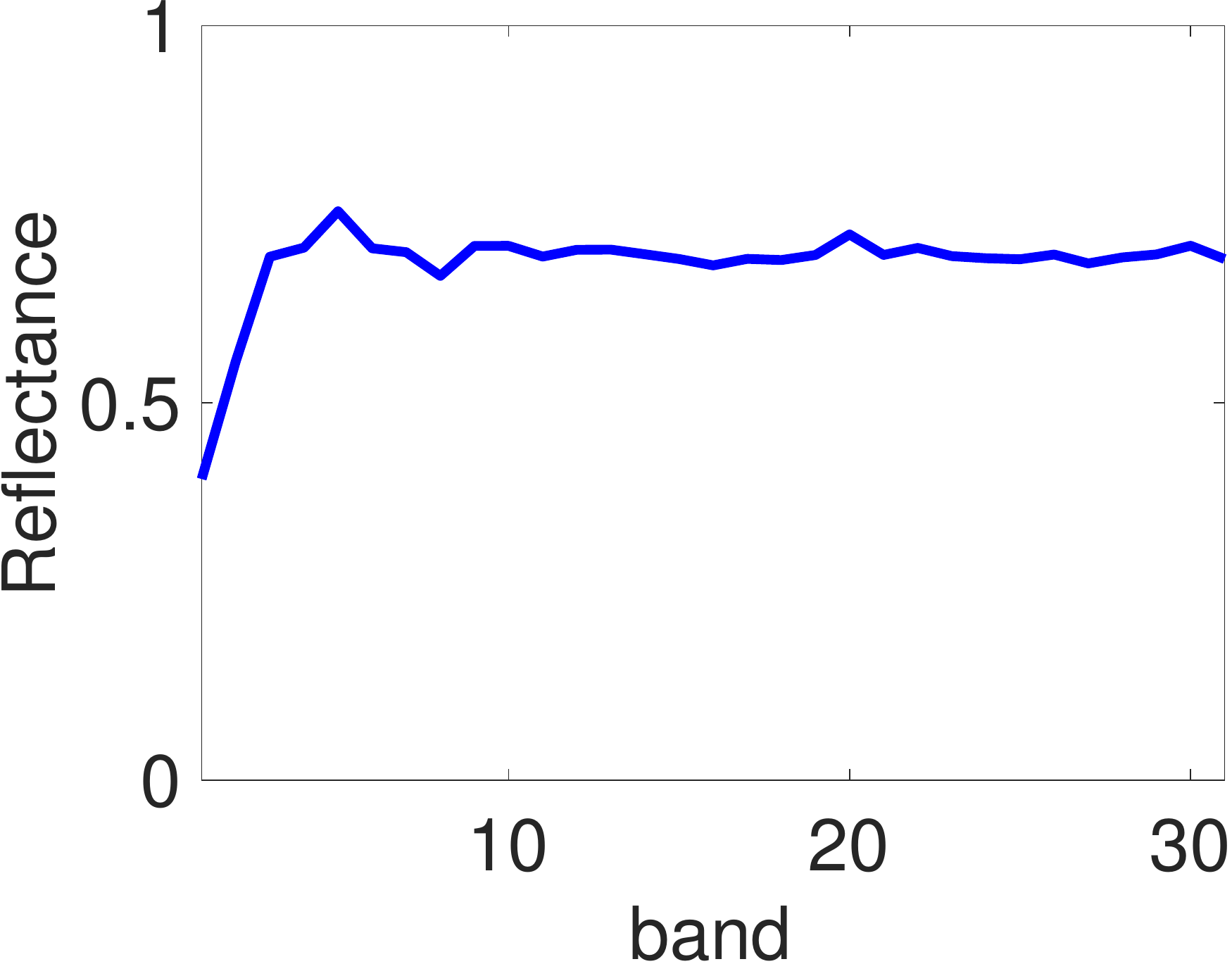}}\hspace{-0.8mm}
\subfigure[QRNN3D]{\includegraphics[scale =0.125,clip=true]{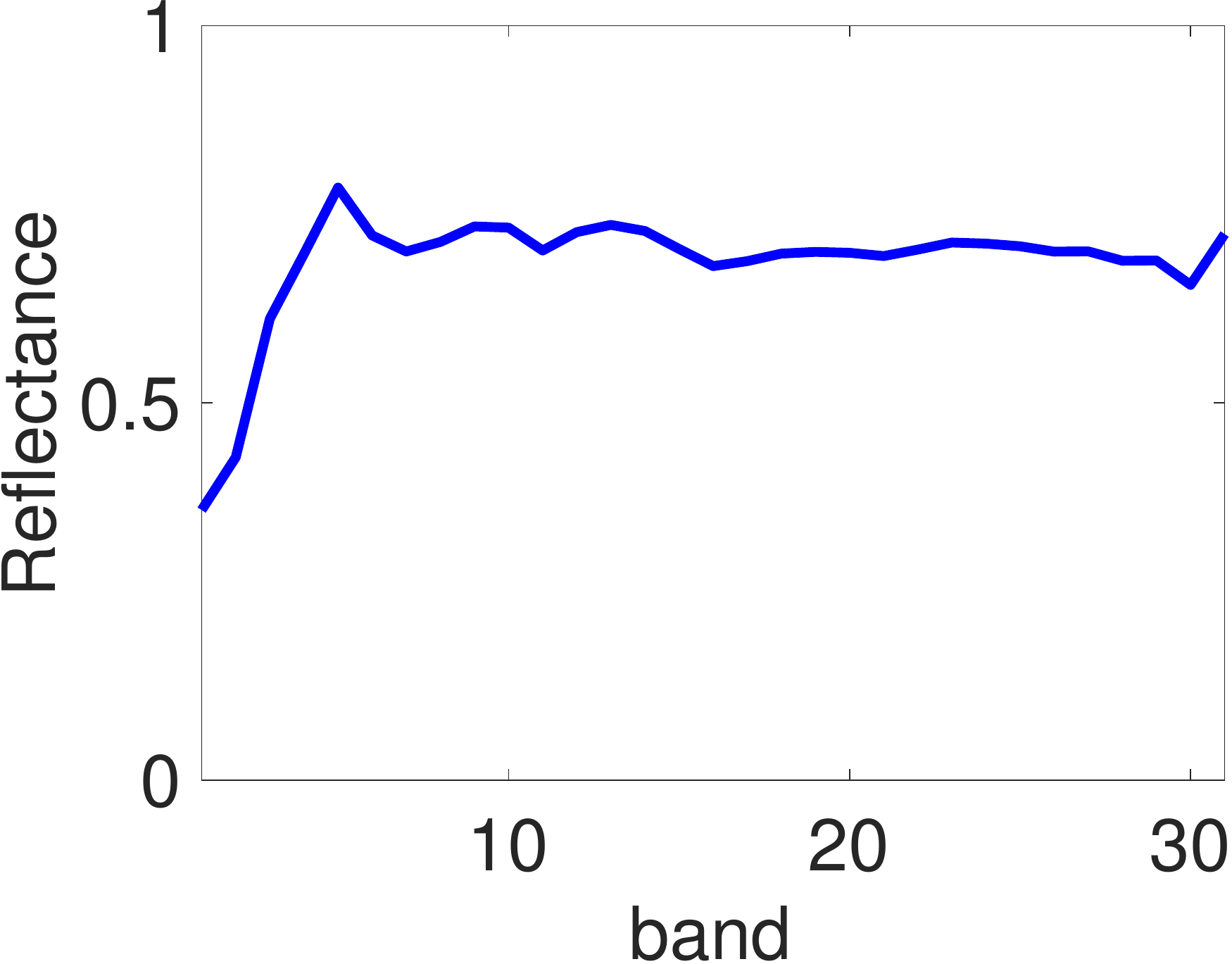}}\hspace{-0.8mm}
\subfigure[SMDSNet]{\includegraphics[scale =0.125,clip=true]{figure/CAVE_95_SMDSSpectral}}
\caption{Denoising results of pixel (200, 100) in  \emph{chart and stuffed toy} HSI.} \label{fig:cavespec}
 \end{figure*}

\subsubsection{Visual Comparison}
To thoroughly show the denoising effectiveness of our method, we present the visual denoising results of all the methods on \emph{gavyam\_0823-0933} image in Fig.~\ref{fig:icvl2}, where the false-color images are generated using bands 5, 18, 25.   Compared with other methods, band-wise BM3D and DnCNN lose some structures because of the ignorance of the spectral-spatial correlation among bands. The advantages of tensor in spectral-spatial modeling and the consideration of spectral low-rankness help PARAFAC and LRTDTV   provide more appealing visual results.  From highlighted areas, we can see LRMR, MTSNMF, BM4D, HSIDCNN, and HSI-SDeCNN can not remove all the noises. QRNN3D removes noises but introduces impulse noises in the denoised image.  Thanks to the strong physical model that considers the spatial and spectral properties of HSIs and the learning ability of the unfolded network,  SMDS-Net produces the best visual quality. 

Fig.~\ref{fig:icvlspec} illustrates the spectral signature of pixel (453,135) in   \emph{gavyam\_0823-0933}  HSIs before and after denoising. As a intuitive comparison shown in  Fig.~\ref{fig:icvlspec} (a) and (b), there are obvious spectral distortion because of the 
the contamination of heavy noises.  Among all the compared methods, our SMDS-Net can more capable of recovering the spectrum by best matching with the reference one.   Overall, the qualitive comparisons in Fig.~\ref{fig:icvl2} and Fig.~\ref{fig:icvlspec} further verifies strong denoising capacity of proposed SMDS-Net.

\subsection{Experiments with CAVE Dataset}
Besides ICVL dataset,  we ran all the methods on the CAVE dataset. The CAVE dataset contains 32 HSIs with size of $512\times 512\times 31$. we directly used the trained models from ICVL dataset to test deep learning based methods.  As can be seen from Table~\ref{tab:cave}, our method produces the best denoising results in all three cases and surpasses state-of-the-art  deep learning-based QRNN3D and model-based BM3D as well as BM4D.   Fig.~\ref{fig:cave} and Fig.~\ref{fig:cavespec} respective depicts false-color image and spectrum of \emph{chart and stuffed toy} HSI before and after denoising.  It can be observed that our SMDS-Net produces more appealing denoising quality than competing methods which either produce oversmooth image/spectrum, see Fig.~\ref{fig:cave} (c) and Fig.~\ref{fig:cavespec} (k), fail to remove all the noises, see Fig.~\ref{fig:cave} (d)-(i) or introduce annoying artefacts, see Fig.~\ref{fig:cave} (o). This experiment further shows the generalization and strong denoising capability of our SMDS-Net.

 \begin{figure*}[!htbp]
  \centering
\subfigure[Noisy]{\includegraphics[scale =0.17,clip=true]{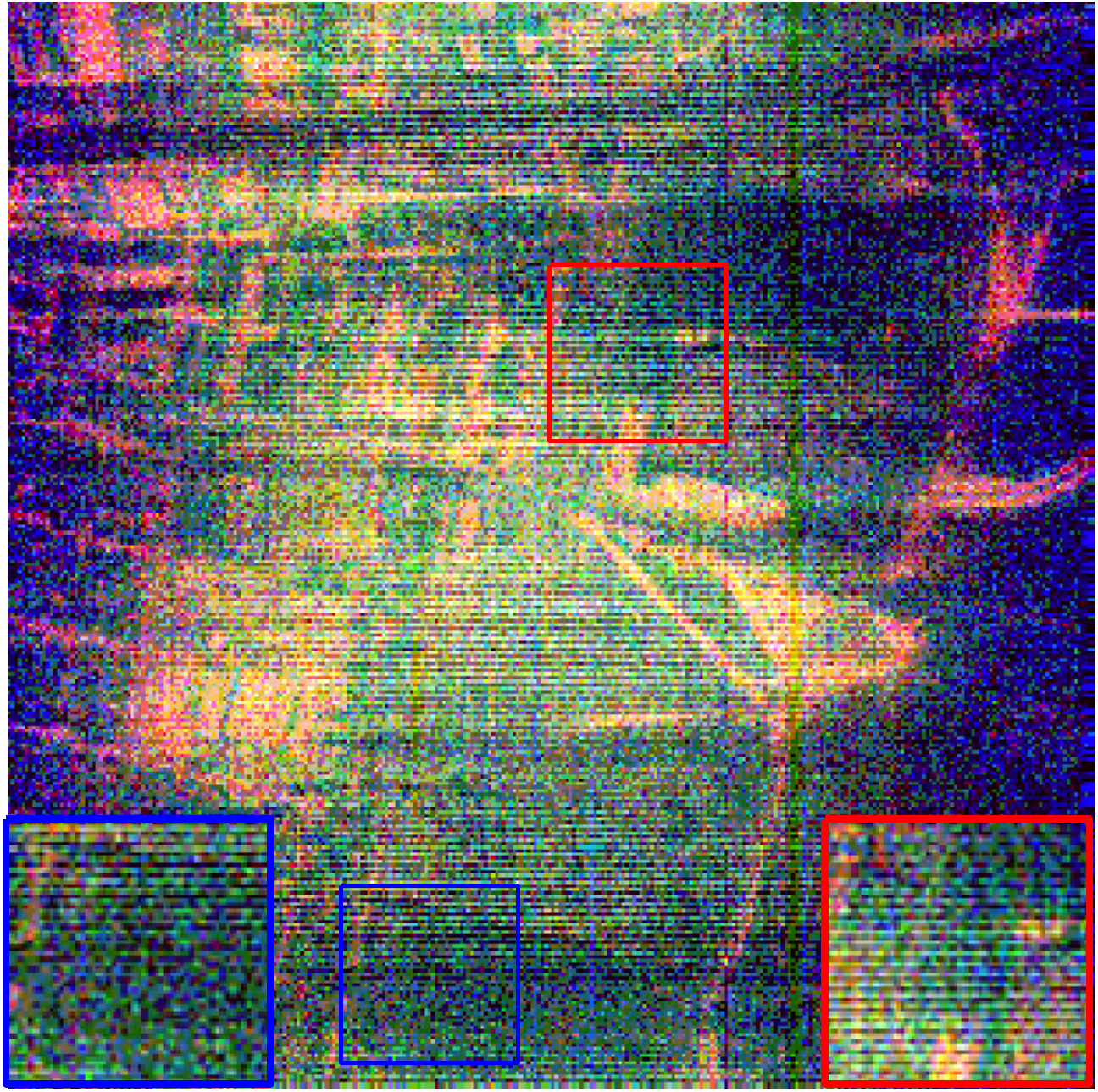}}\hspace{-0.8mm}
\subfigure[BM3D]{\includegraphics[scale =0.17,clip=true]{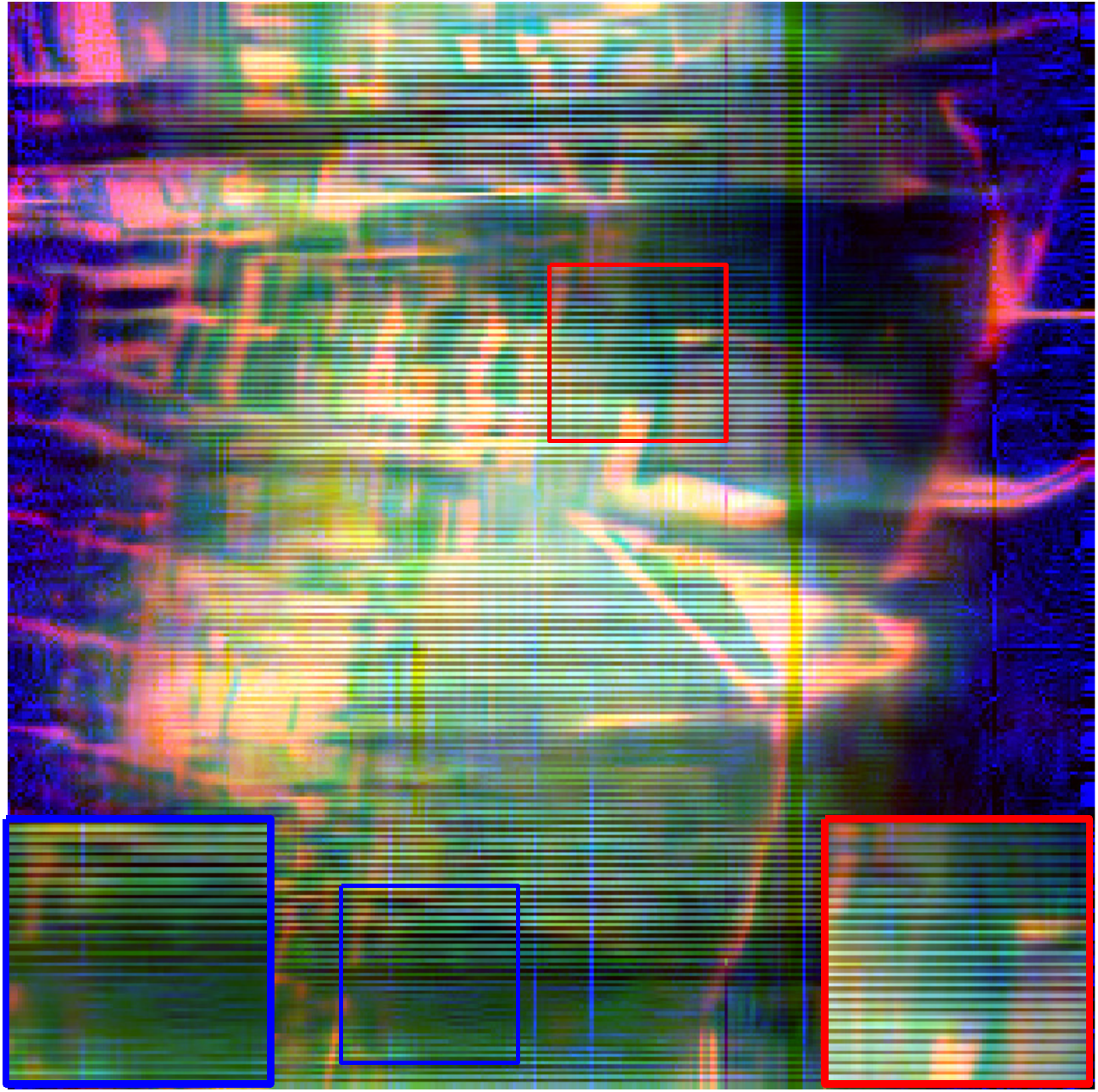}}\hspace{-0.8mm}
\subfigure[BM4D]{\includegraphics[scale =0.17,clip=true]{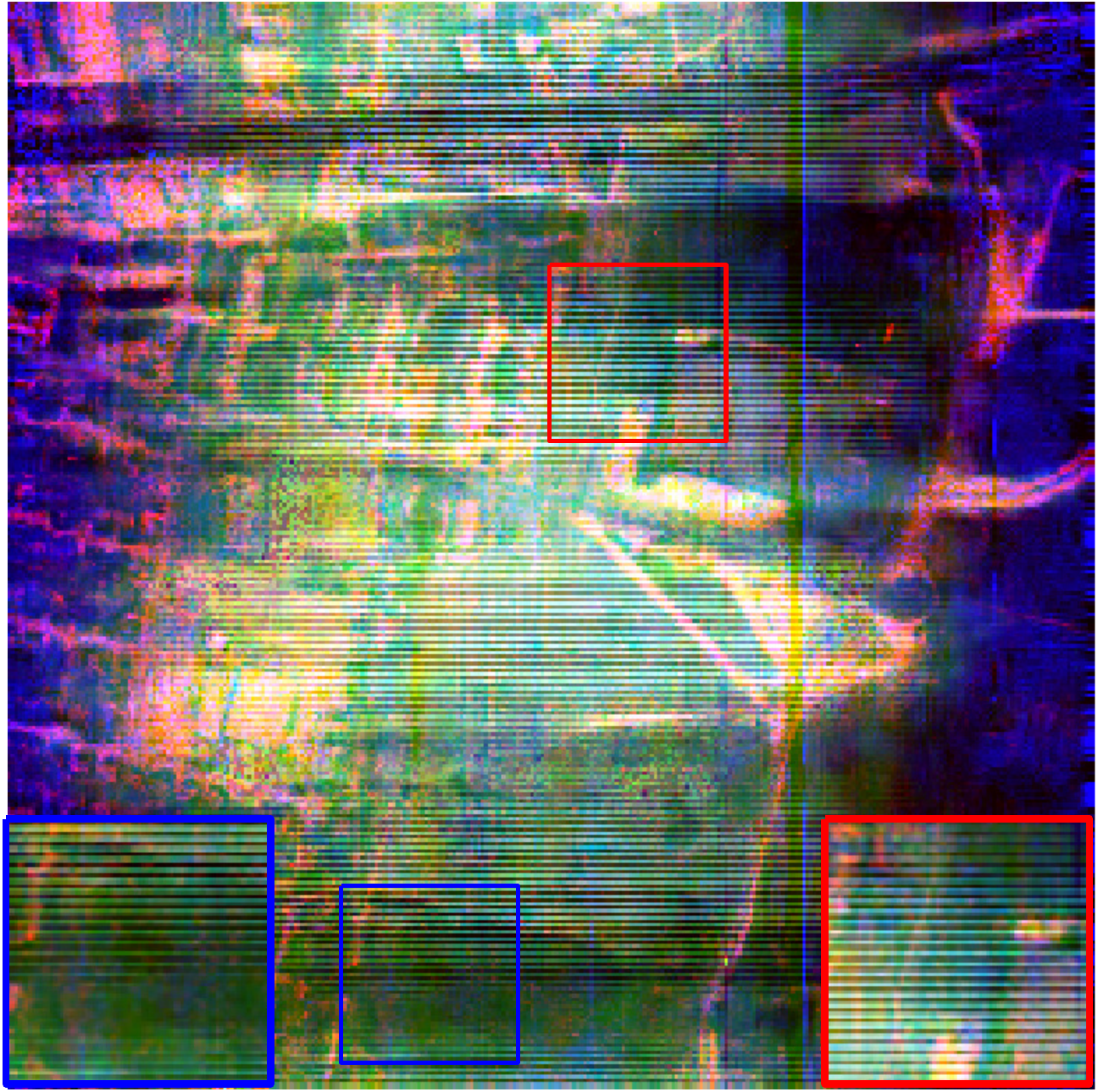}}\hspace{-0.8mm}
\subfigure[TDL]{\includegraphics[scale =0.17,clip=true]{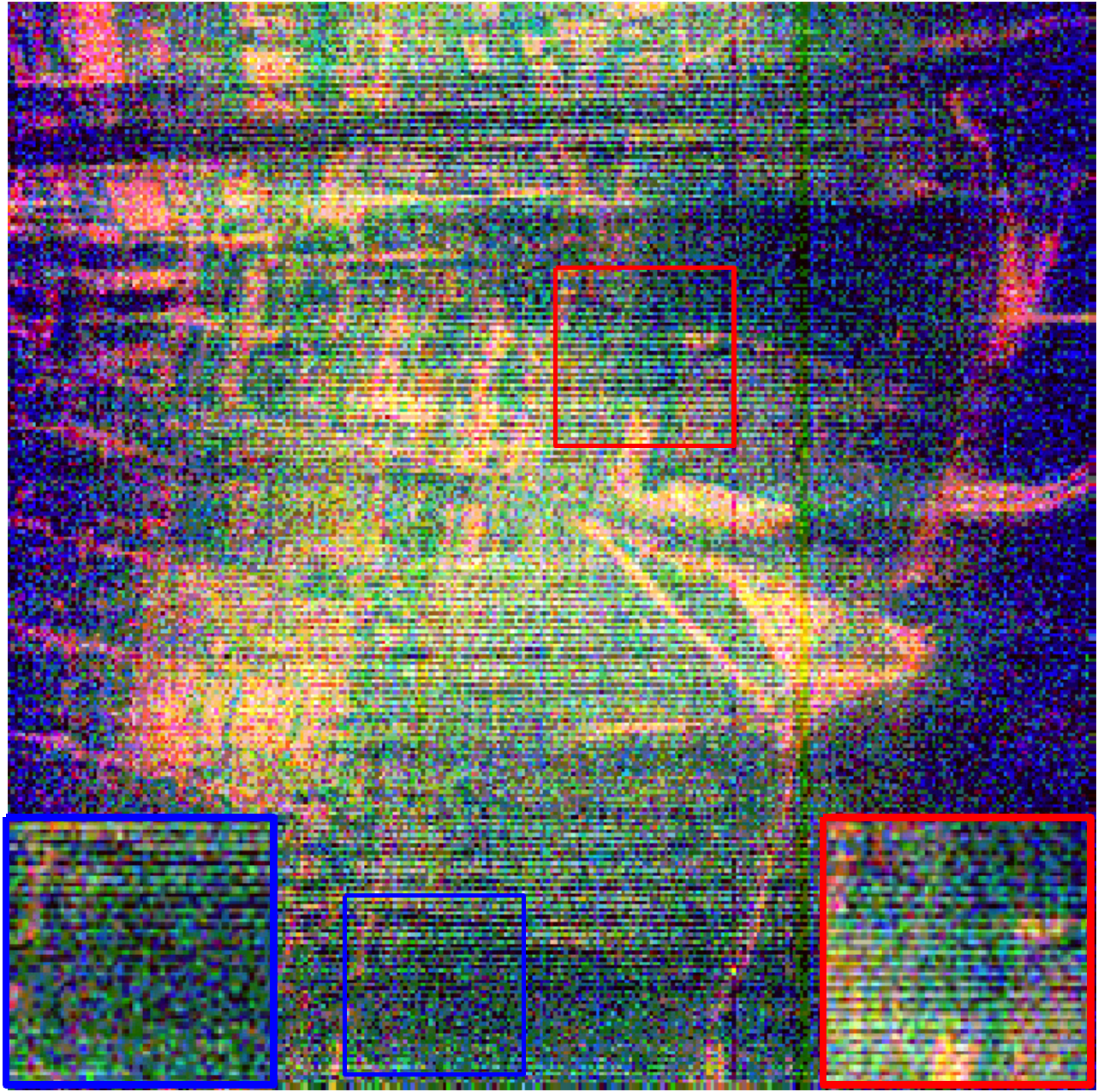}}\hspace{-0.8mm}
\subfigure[MTSNMF]{\includegraphics[scale =0.17,clip=true]{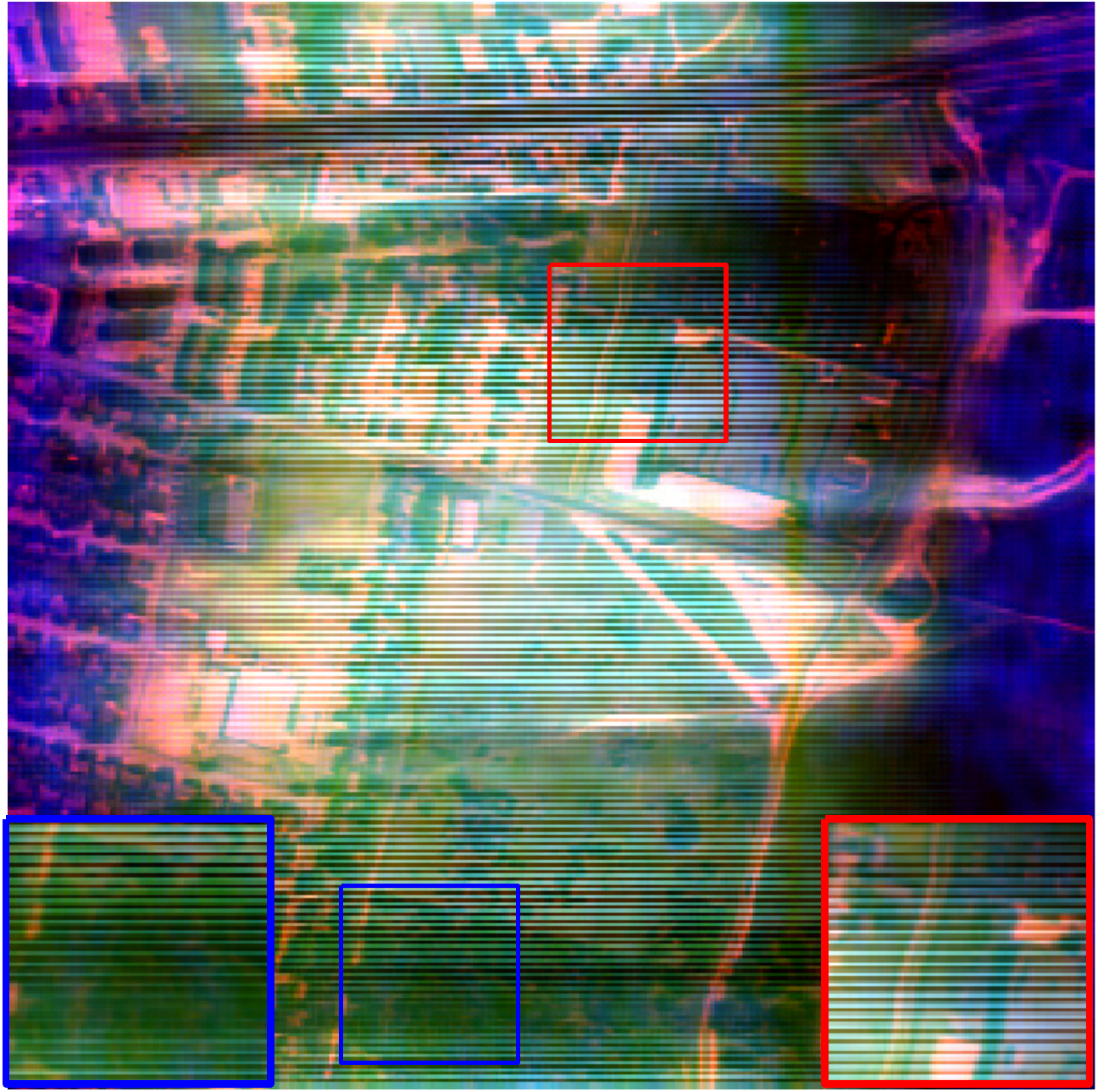}}\hspace{-0.8mm}
\subfigure[PARAFAC]{\includegraphics[scale =0.17,clip=true]{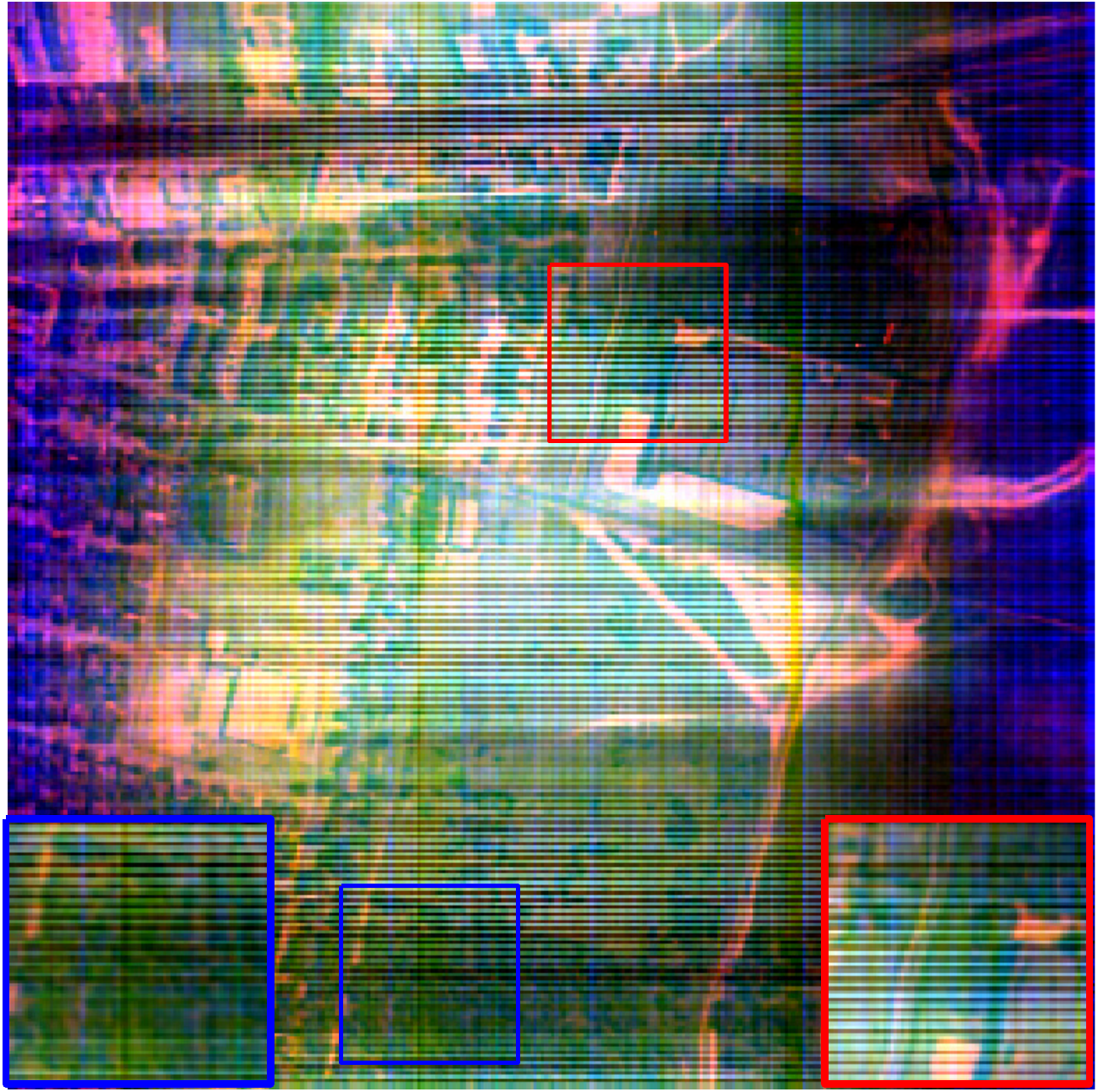}}\hspace{-0.8mm}
\subfigure[LLRT]{\includegraphics[scale =0.17,clip=true]{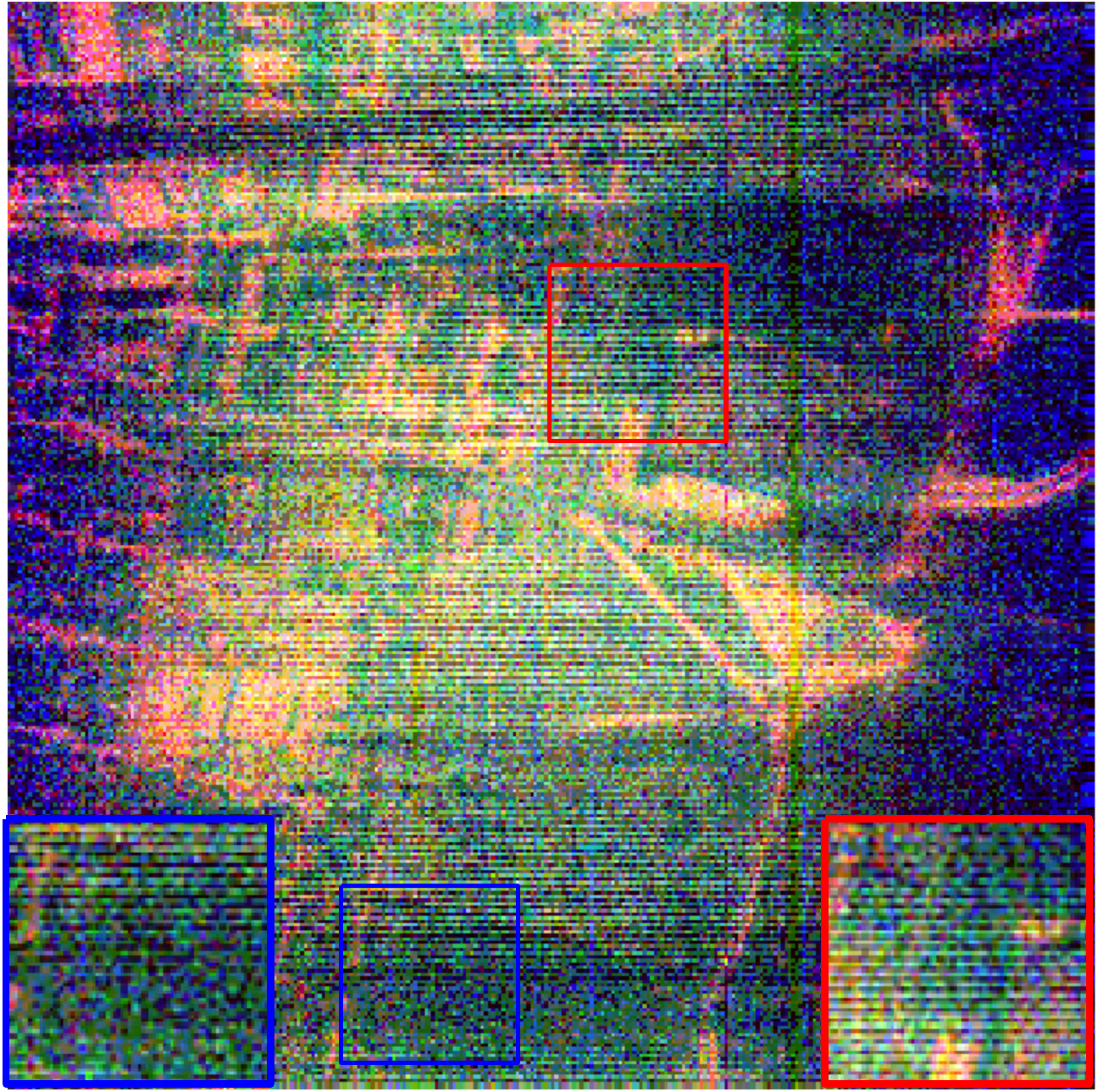}}\hspace{-0.8mm}
\subfigure[NGMeet]{\includegraphics[scale =0.17,clip=true]{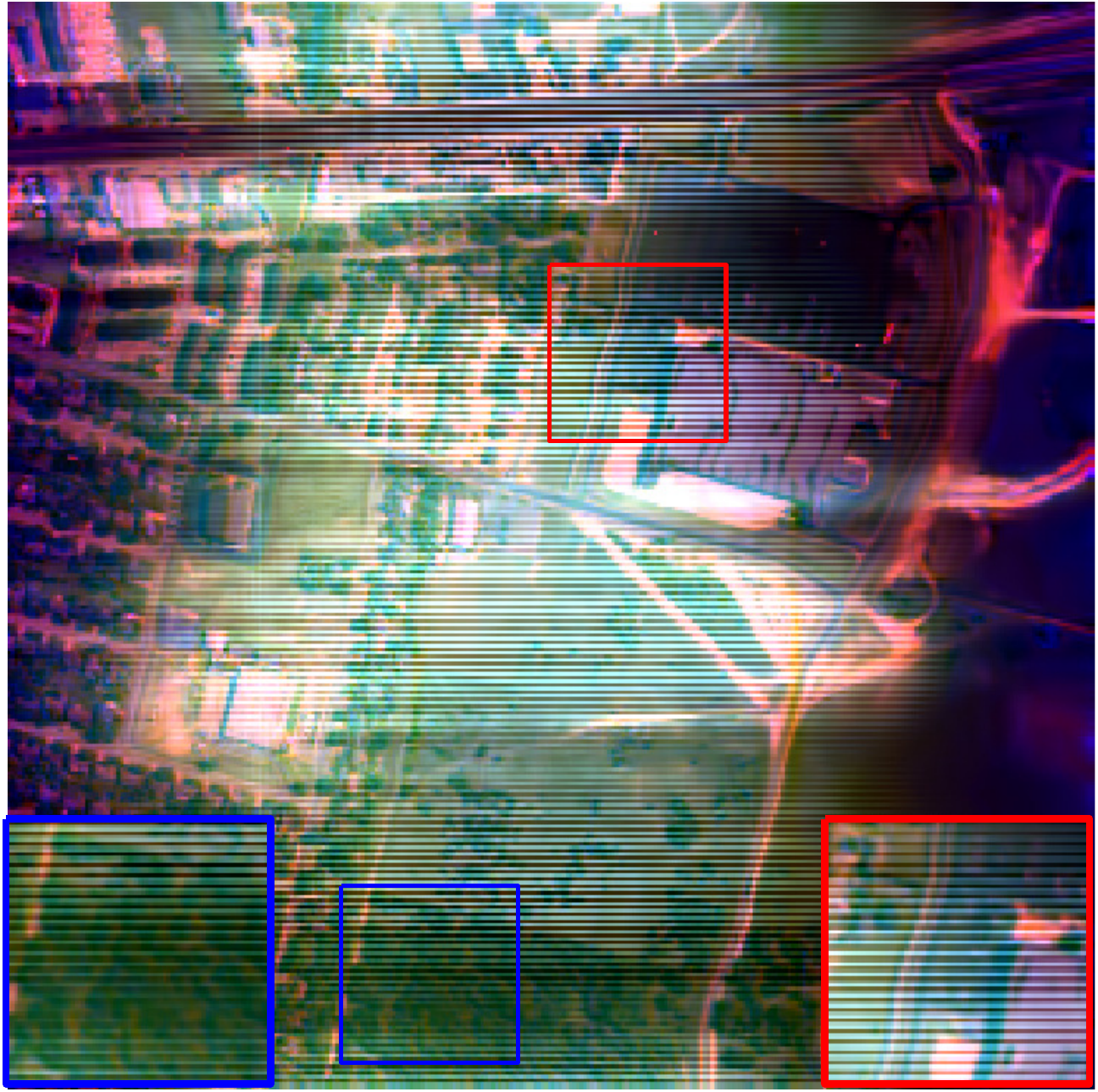}}\hspace{-0.8mm}\\
\subfigure[LRMR]{\includegraphics[scale =0.195,clip=true]{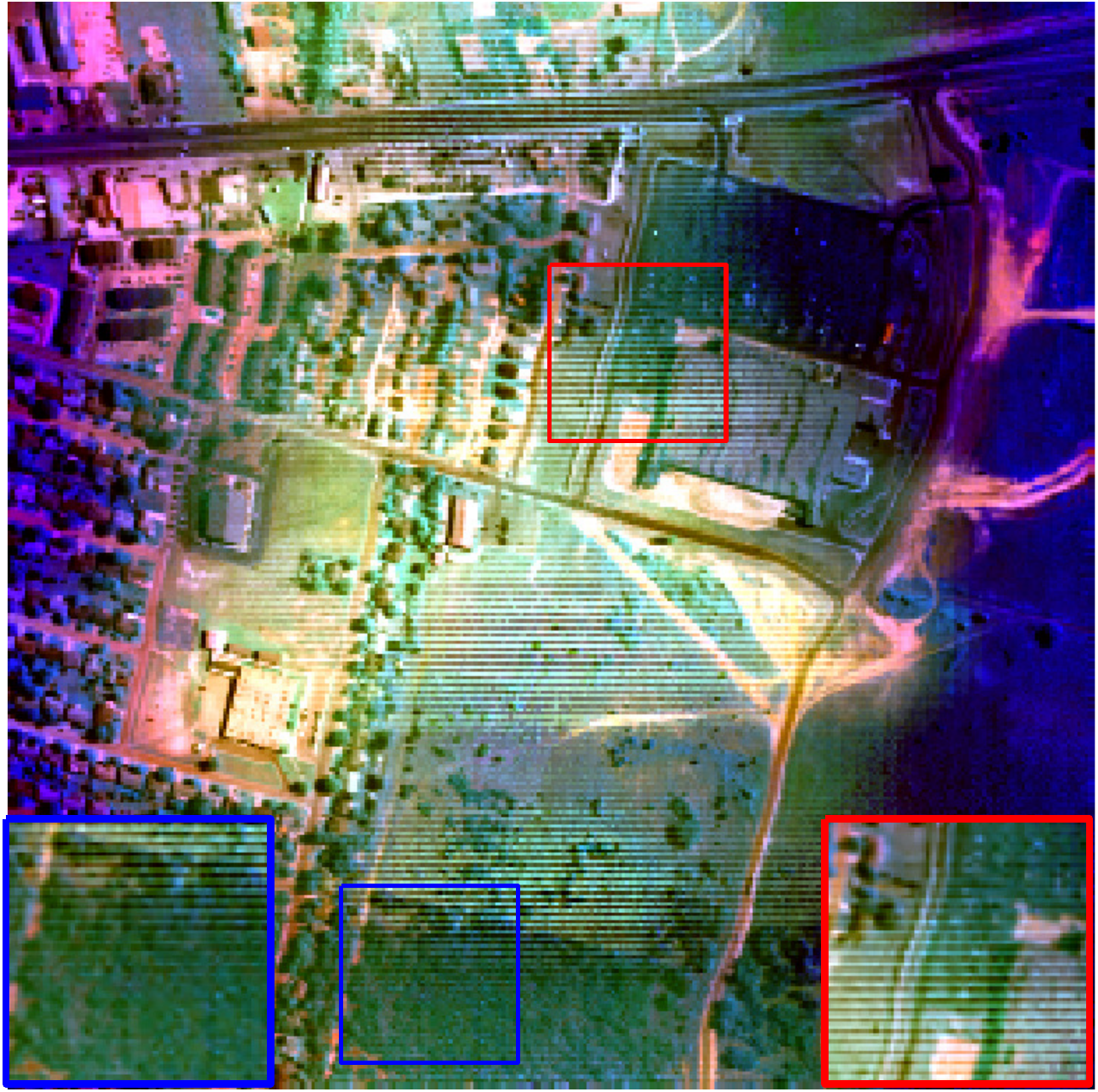}}\hspace{-0.8mm}
\subfigure[LRTDTV]{\includegraphics[scale =0.195,clip=true]{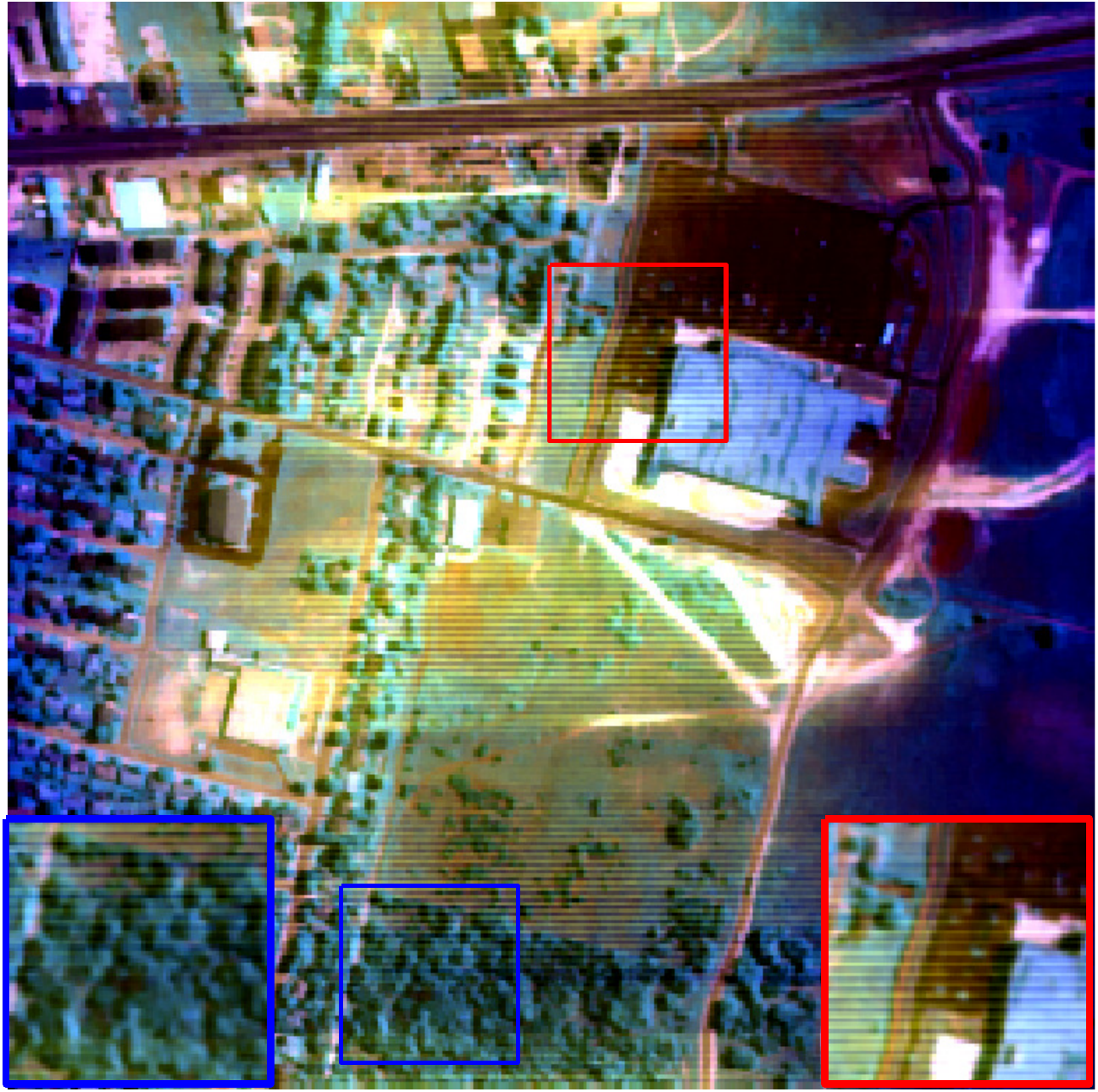}}\hspace{-0.8mm}
\subfigure[DnCNN]{\includegraphics[scale =0.195,clip=true]{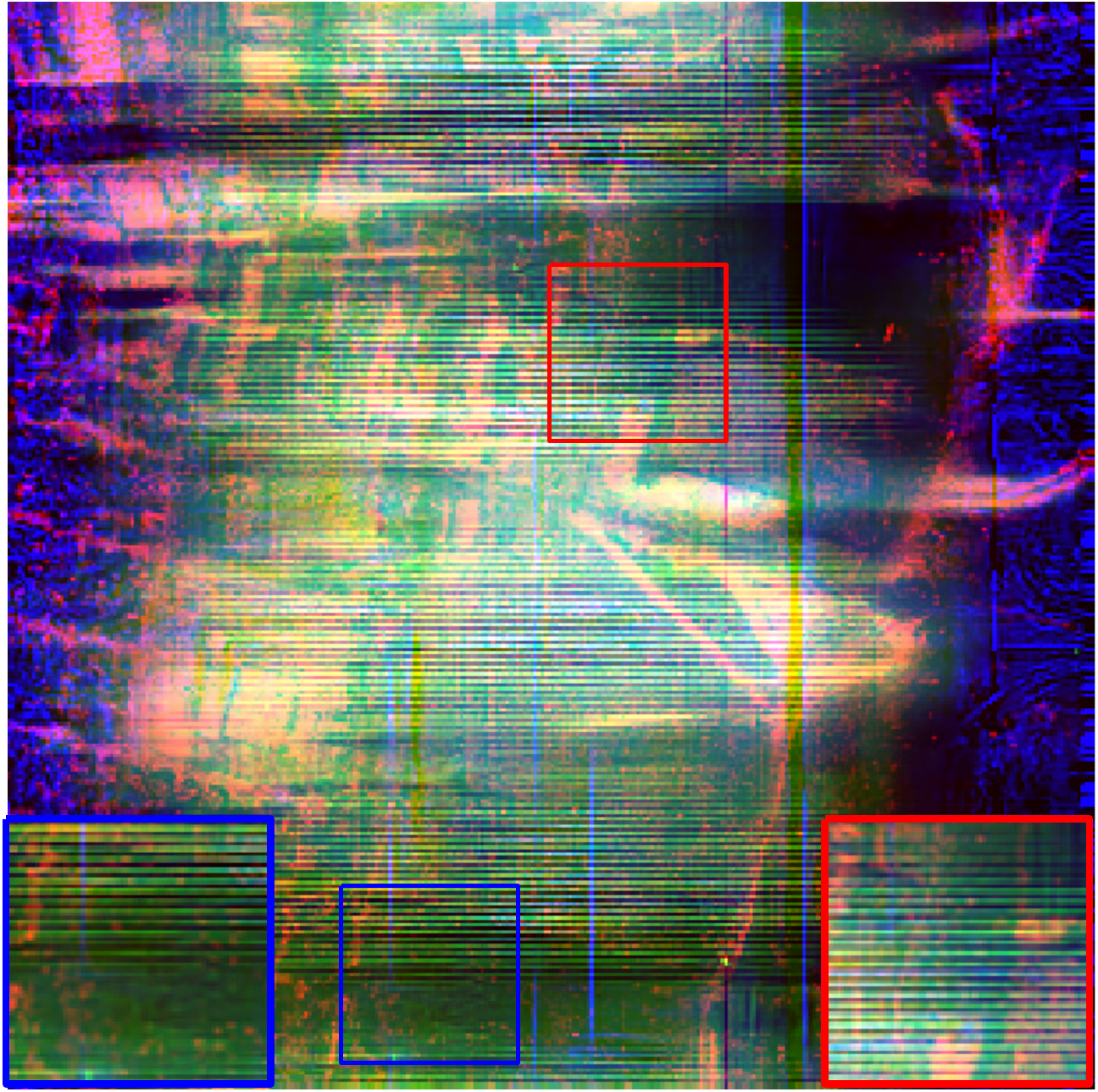}}\hspace{-0.8mm}
\subfigure[HSI-SDeCNN]{\includegraphics[scale =0.195,clip=true]{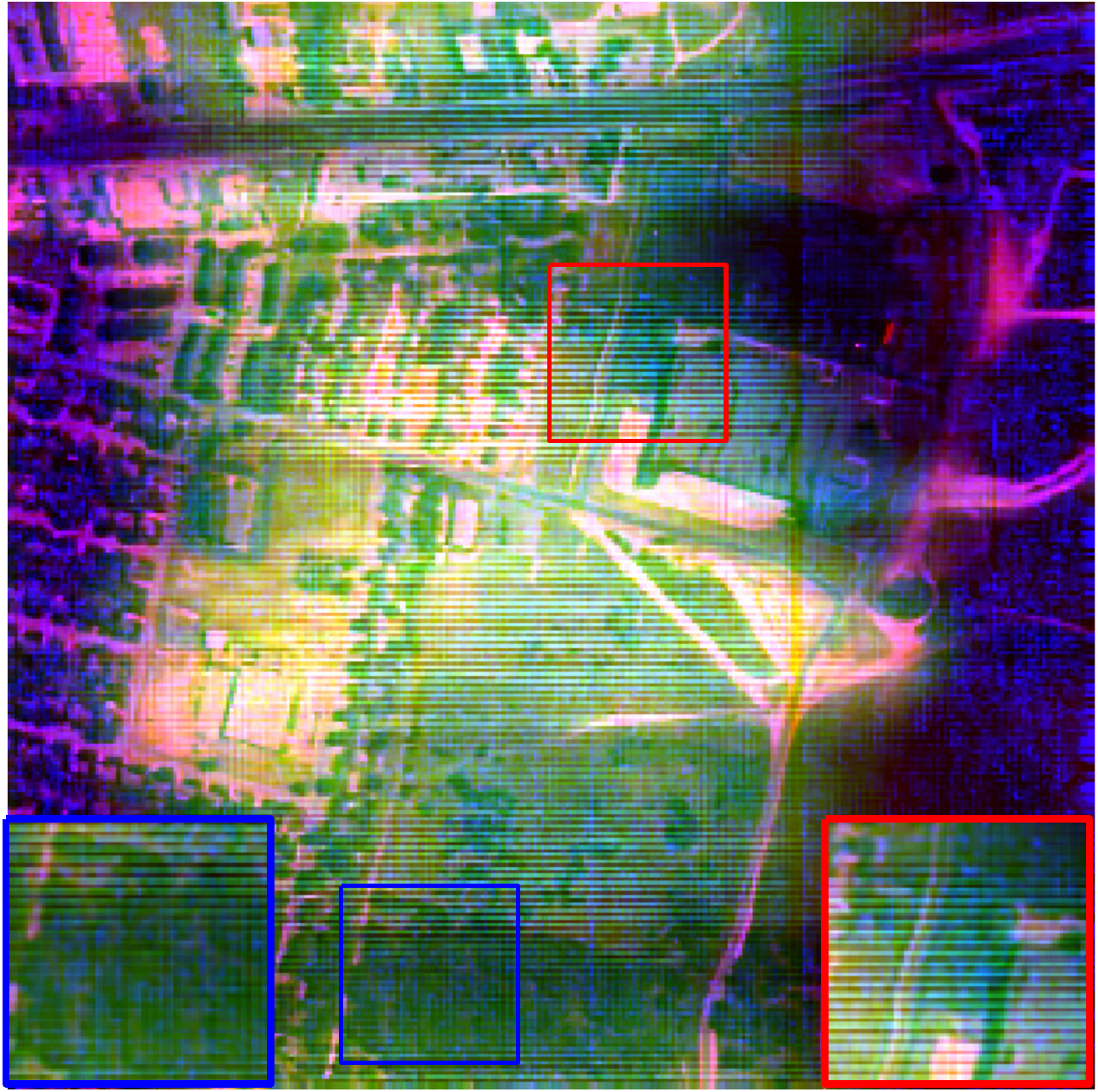}}\hspace{-0.8mm}
\subfigure[HSID-CNN]{\includegraphics[scale =0.195,clip=true]{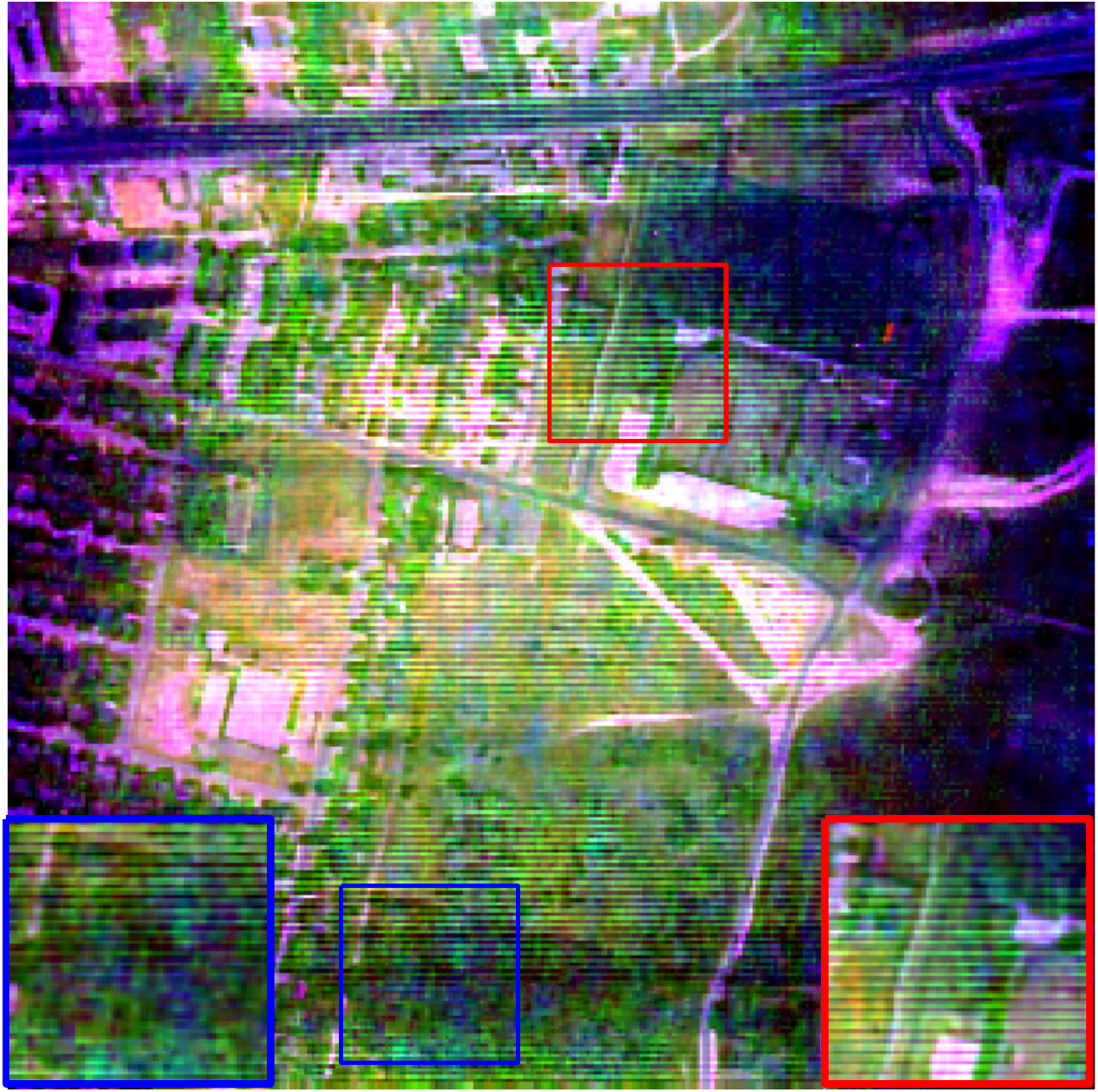}}\hspace{-0.8mm}
\subfigure[QRNN3D]{\includegraphics[scale =0.195,clip=true]{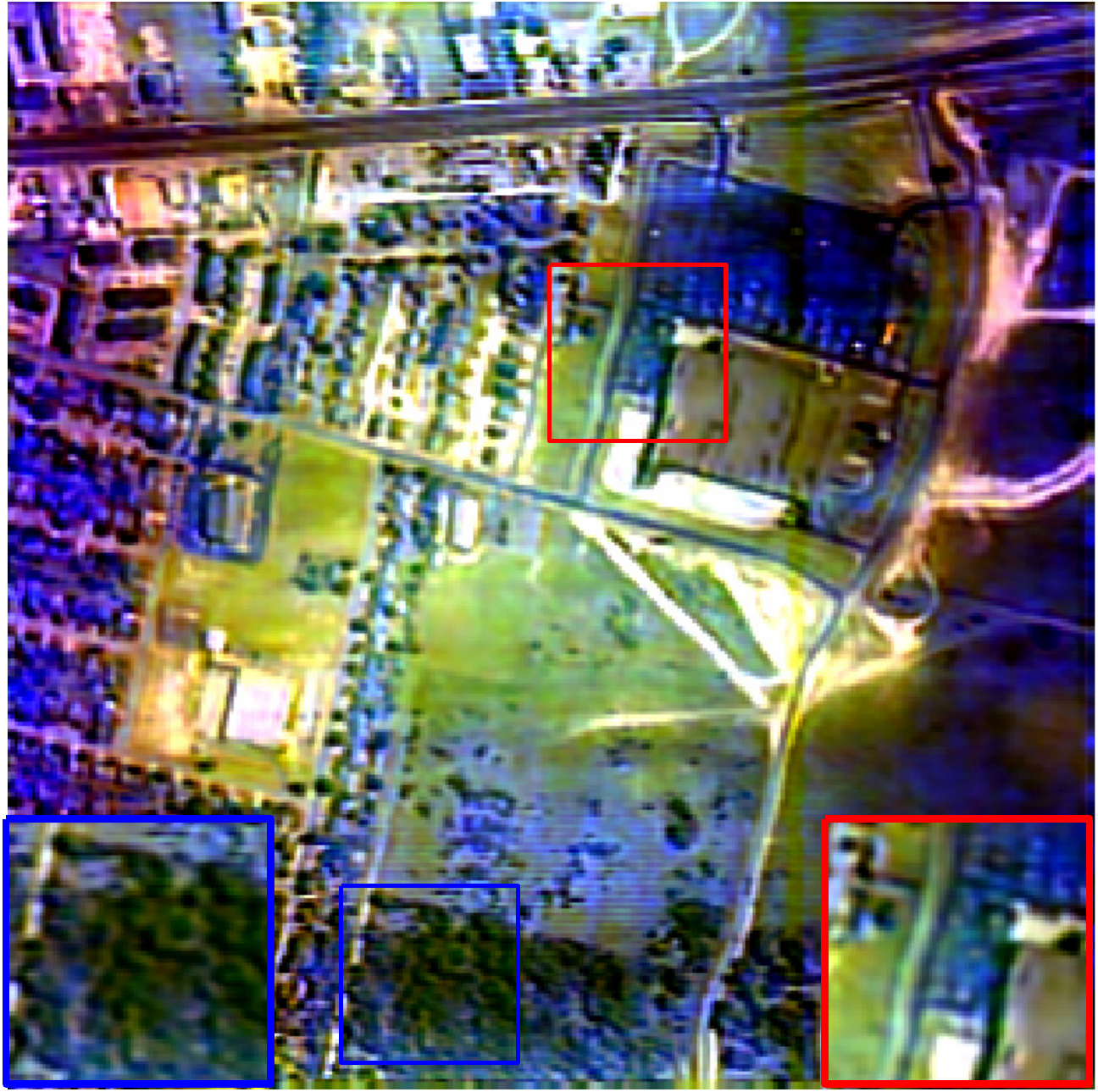}}\hspace{-0.8mm}
\subfigure[SMDS-Net]{\includegraphics[scale =0.195,clip=true]{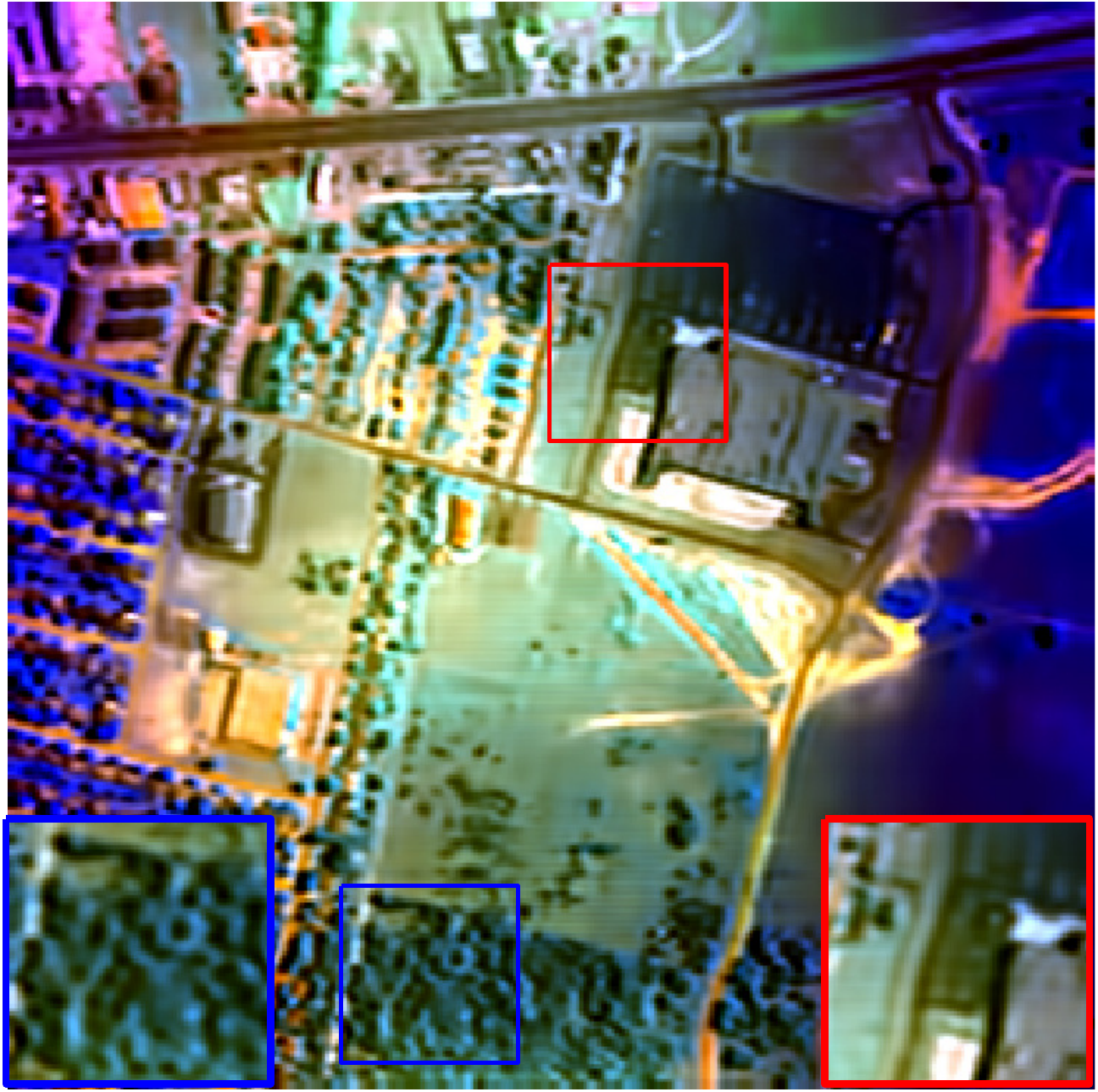}}
\caption{Denoising results on real-world Urban dataset. The false-color images were generated by combining bands 109, 141, 209.  The proposed SMDS-Net provides the best visual results. (\textbf{Best view on screen with zoom})} \label{fig:urban}
 \end{figure*} 
 
 	\begin{figure*}[!htbp]
  \centering
\subfigure[Noisy]{\includegraphics[scale =0.125,clip=true]{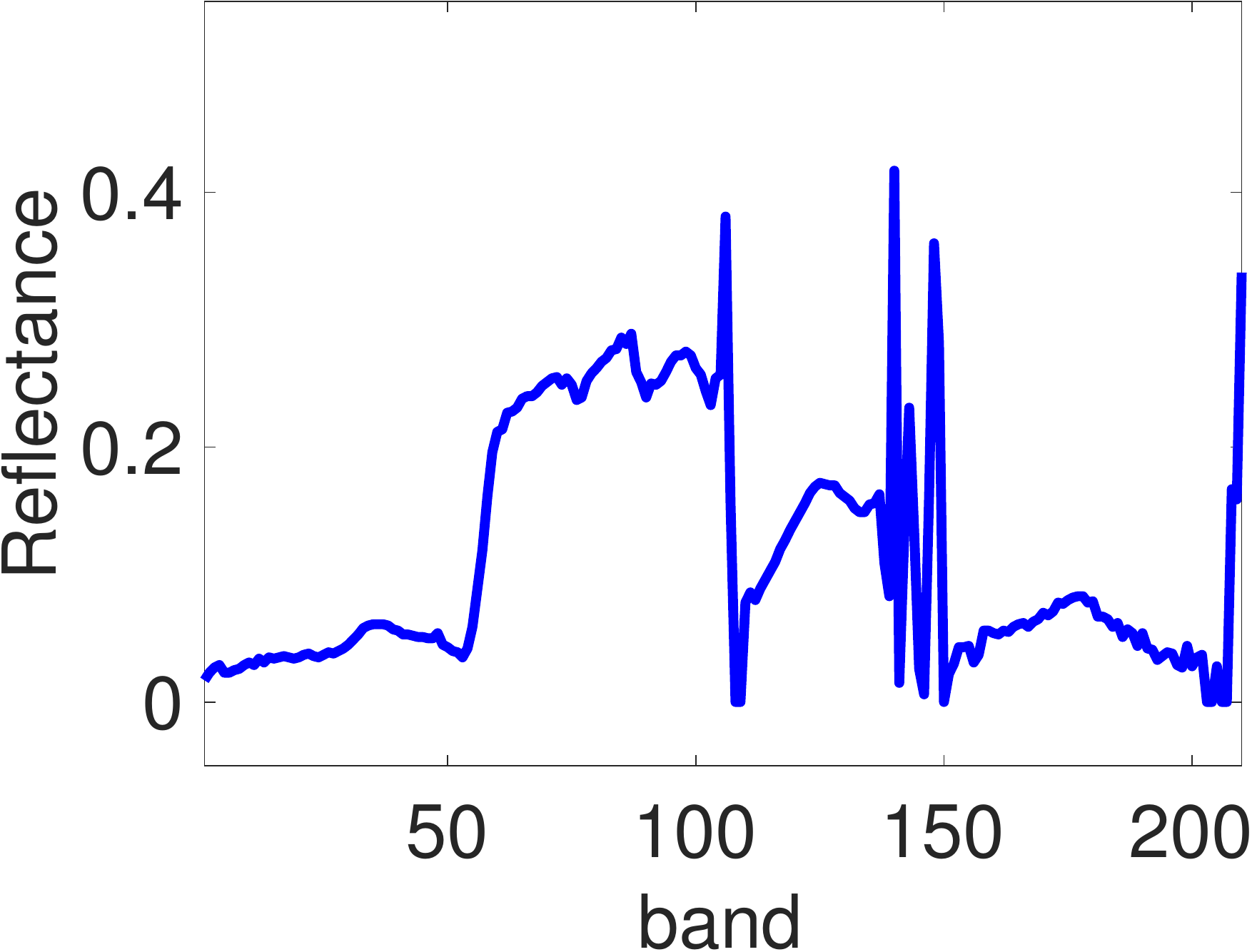}}\hspace{-0.8mm}
\subfigure[BM3D]{\includegraphics[scale =0.125,clip=true]{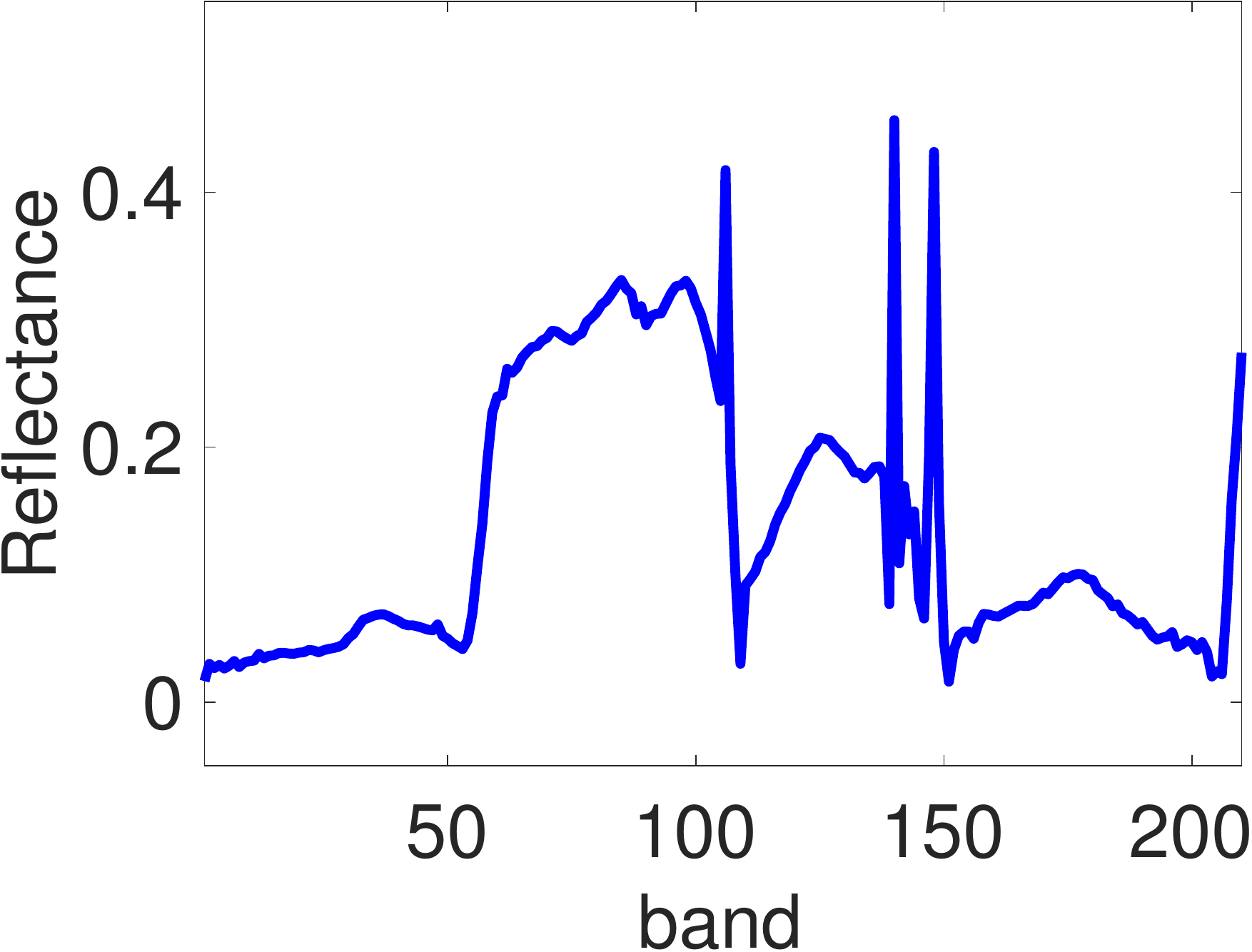}}\hspace{-0.8mm}
\subfigure[BM4D]{\includegraphics[scale =0.125,clip=true]{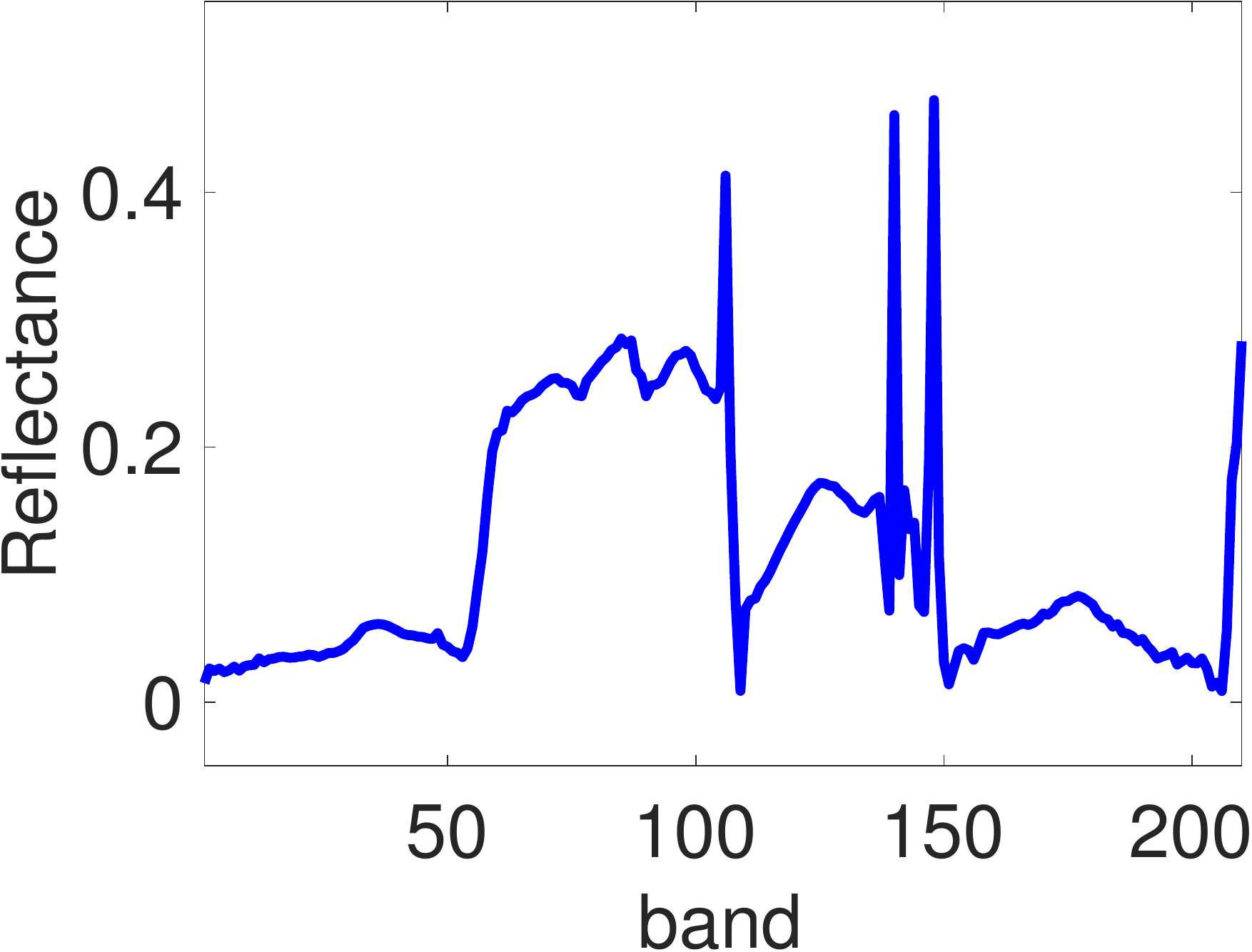}}\hspace{-0.8mm}
\subfigure[TDL]{\includegraphics[scale =0.125,clip=true]{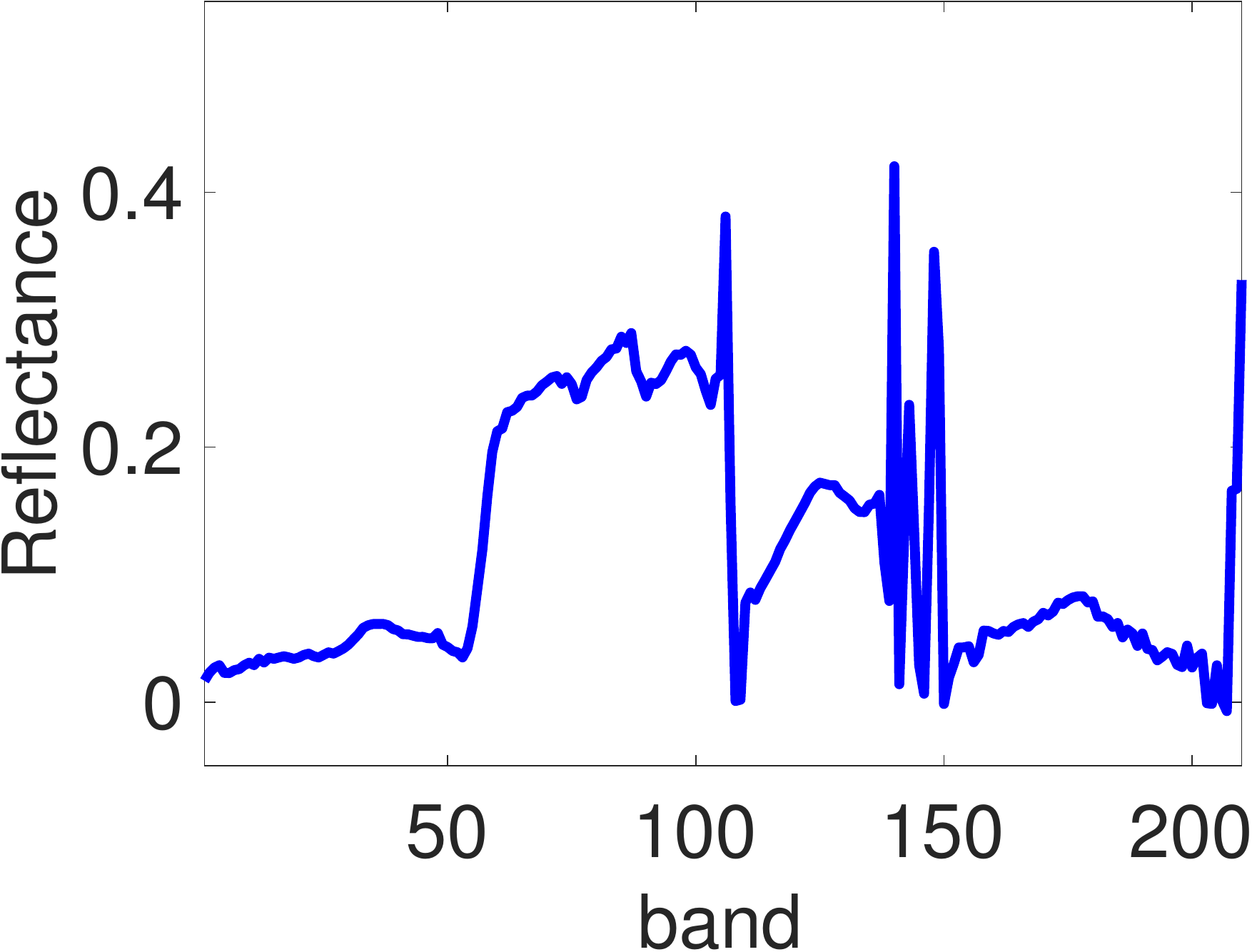}}\hspace{-0.8mm}
\subfigure[MTSNMF]{\includegraphics[scale =0.125,clip=true]{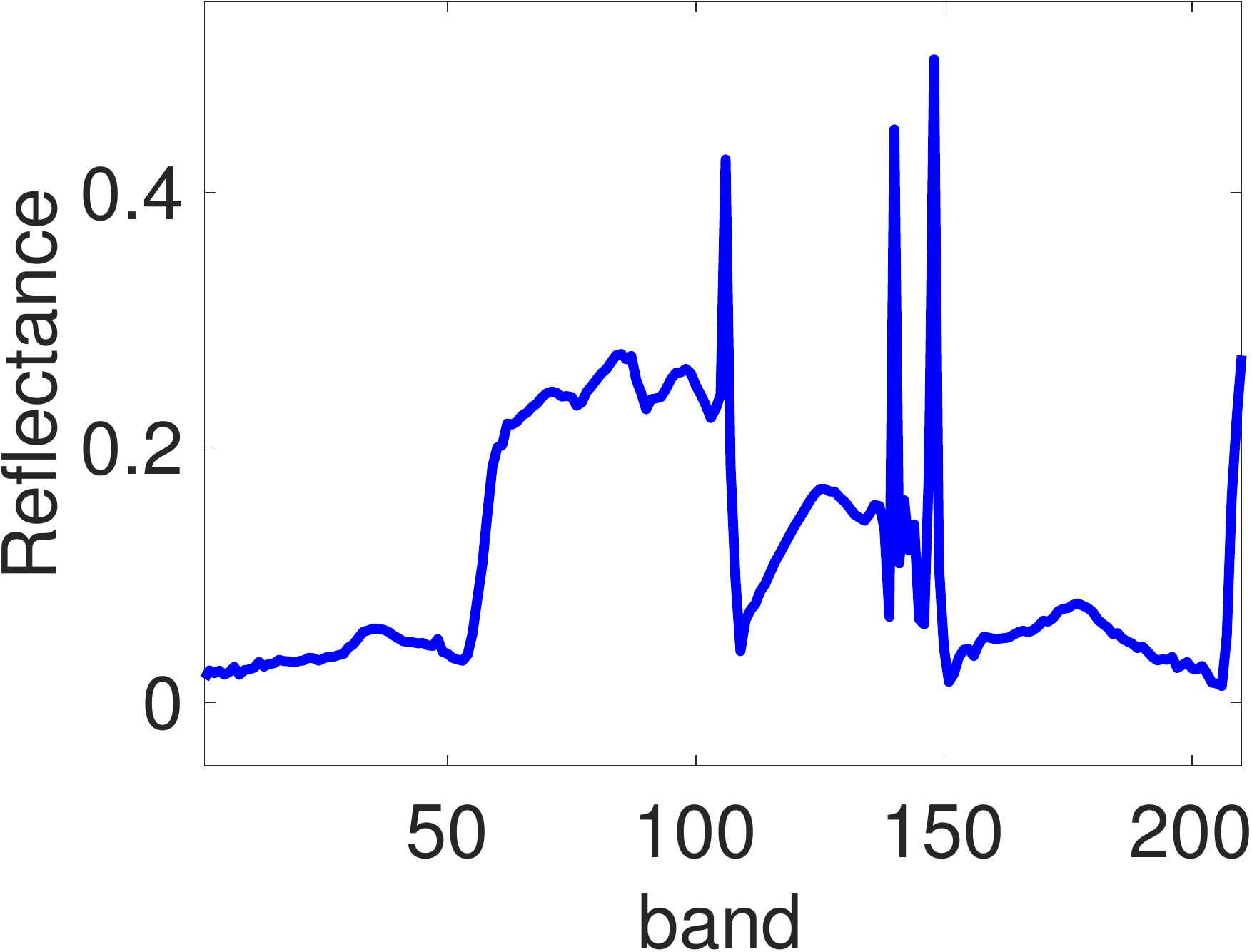}}\hspace{-0.8mm}
\subfigure[PARAFAC]{\includegraphics[scale =0.125,clip=true]{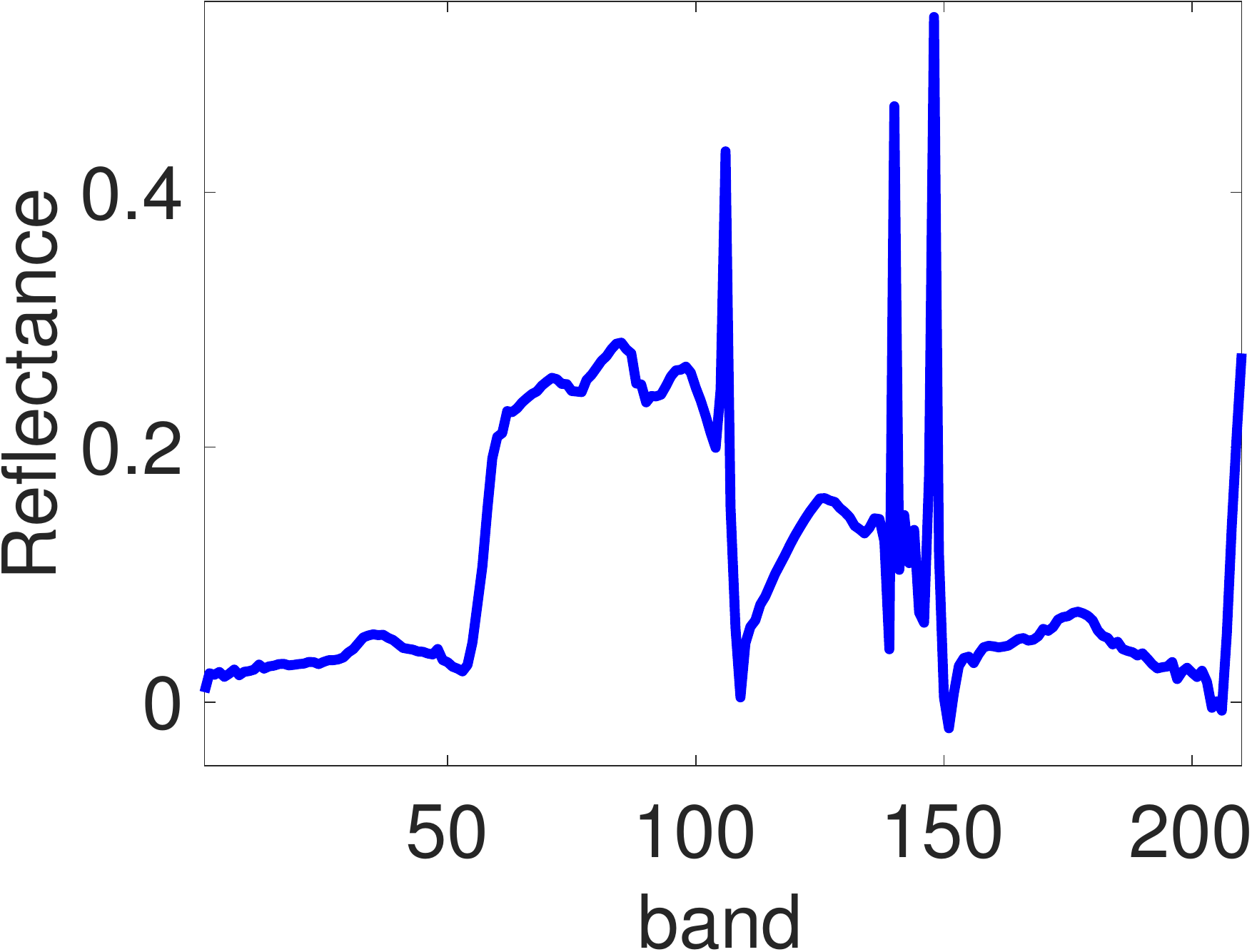}}\hspace{-0.8mm}
\subfigure[LLRT]{\includegraphics[scale =0.125,clip=true]{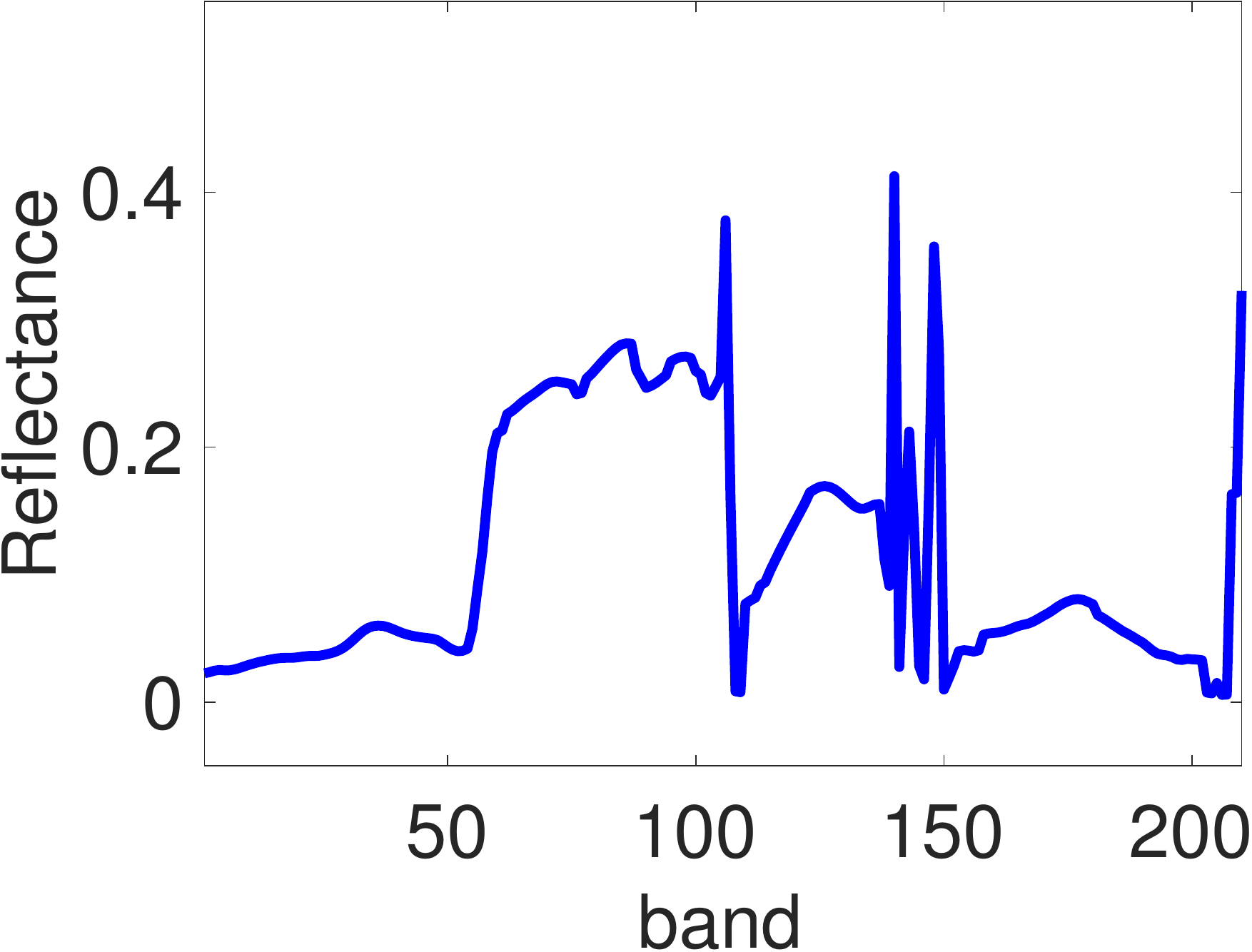}}\hspace{-0.8mm}
\subfigure[NGMeet]{\includegraphics[scale =0.125,clip=true]{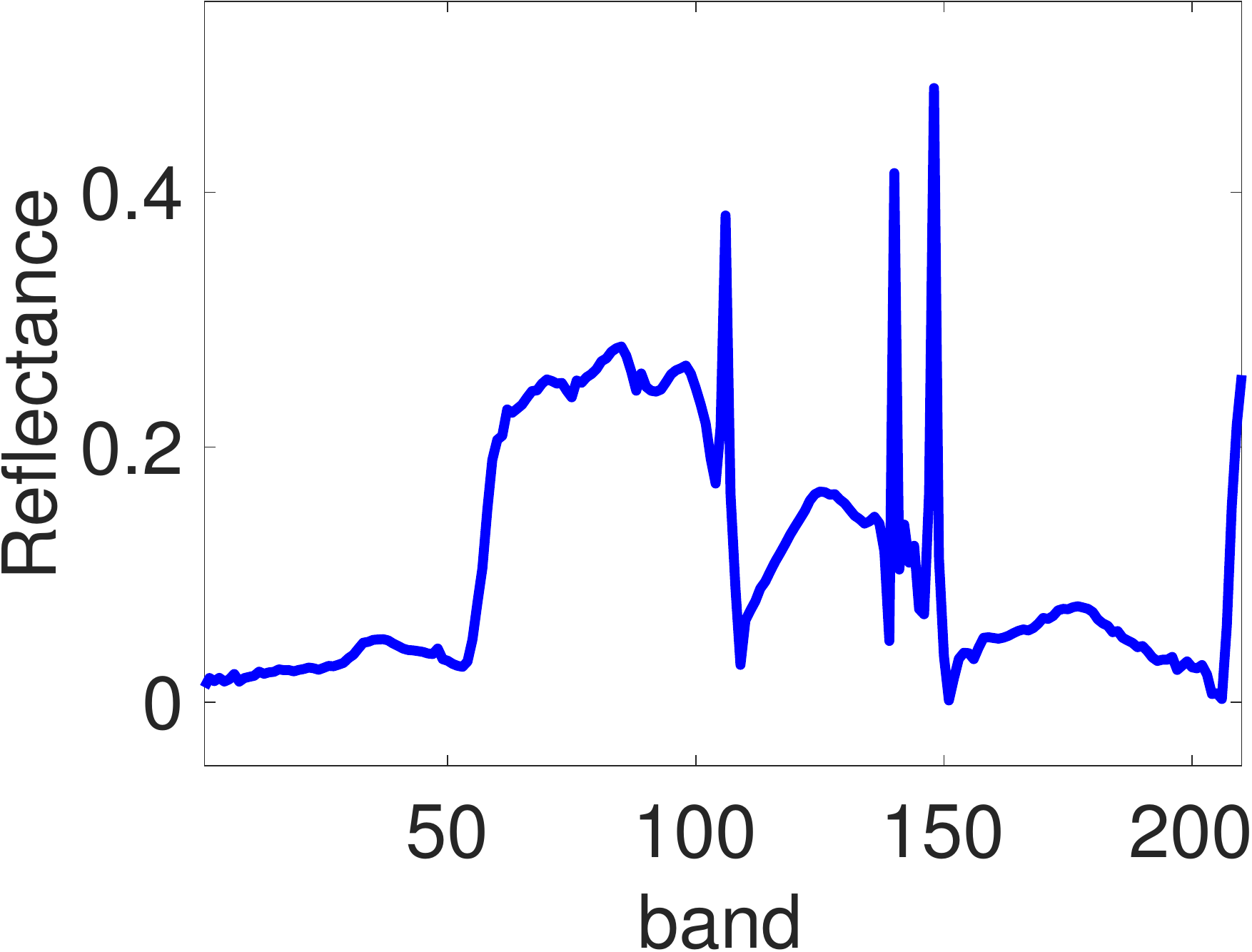}}\\\hspace{-0.8mm}
\subfigure[LRMR]{\includegraphics[scale =0.14,clip=true]{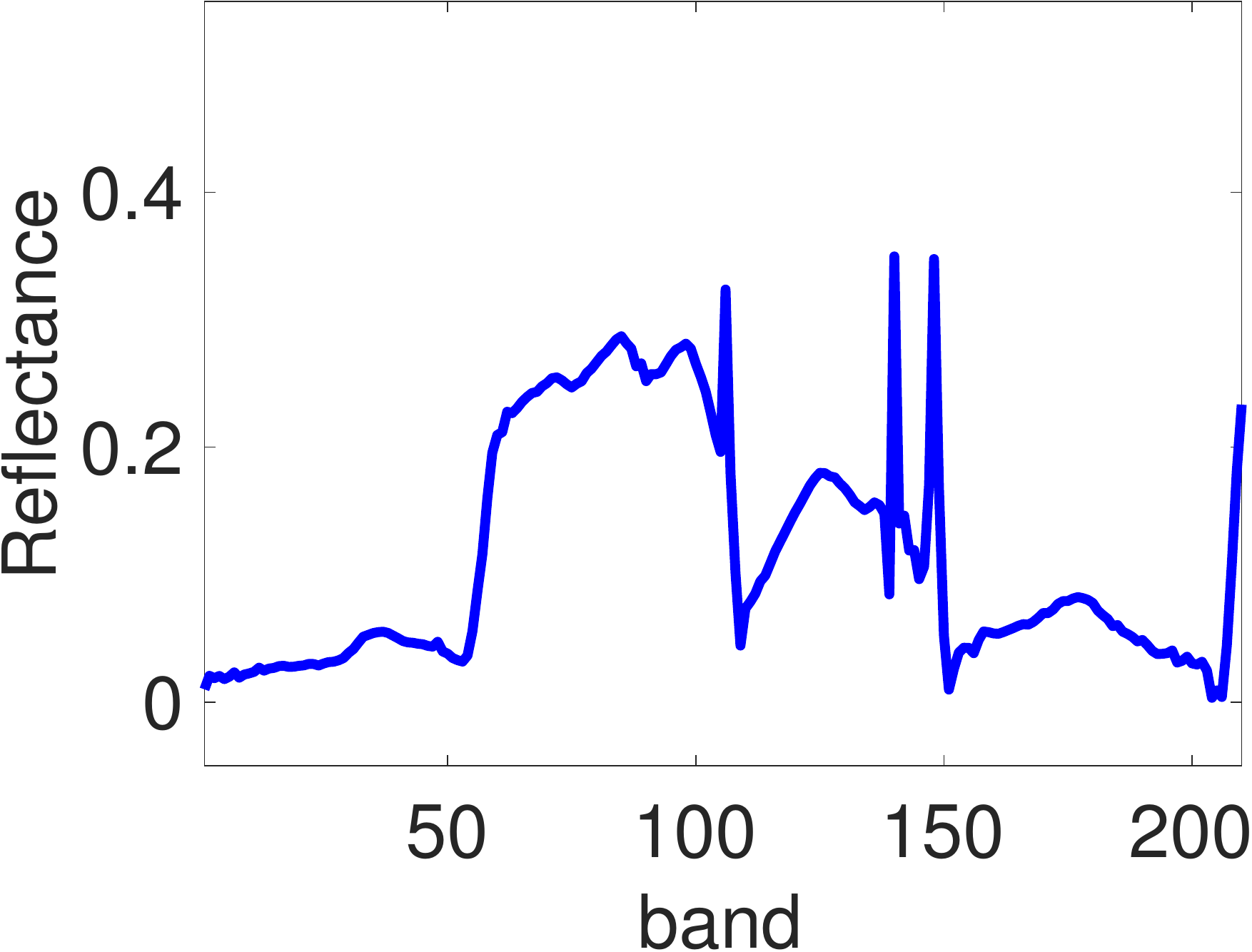}}\hspace{-0.8mm}
\subfigure[LRTDTV]{\includegraphics[scale =0.14,clip=true]{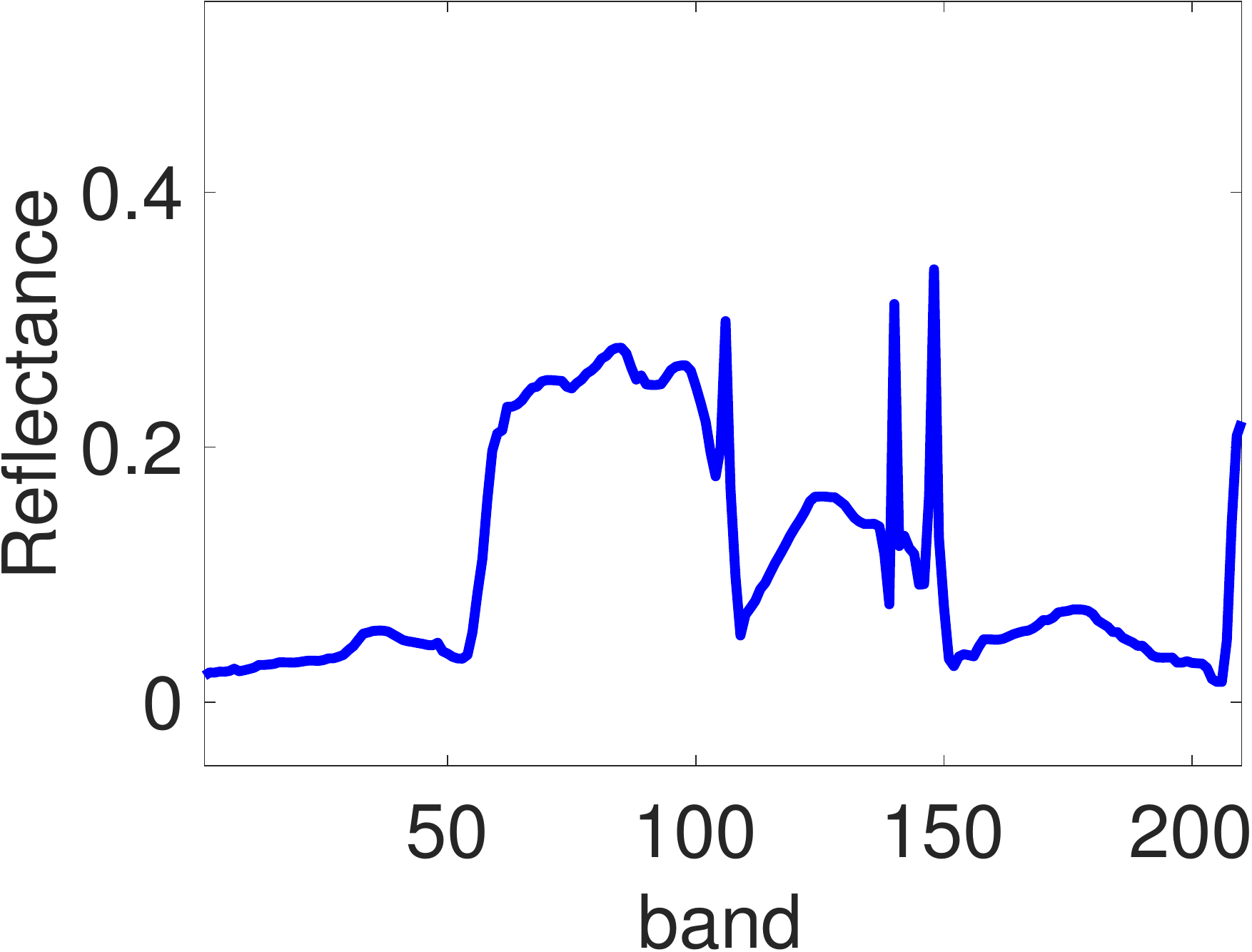}}\hspace{-0.8mm}
\subfigure[DnCNN]{\includegraphics[scale =0.14,clip=true]{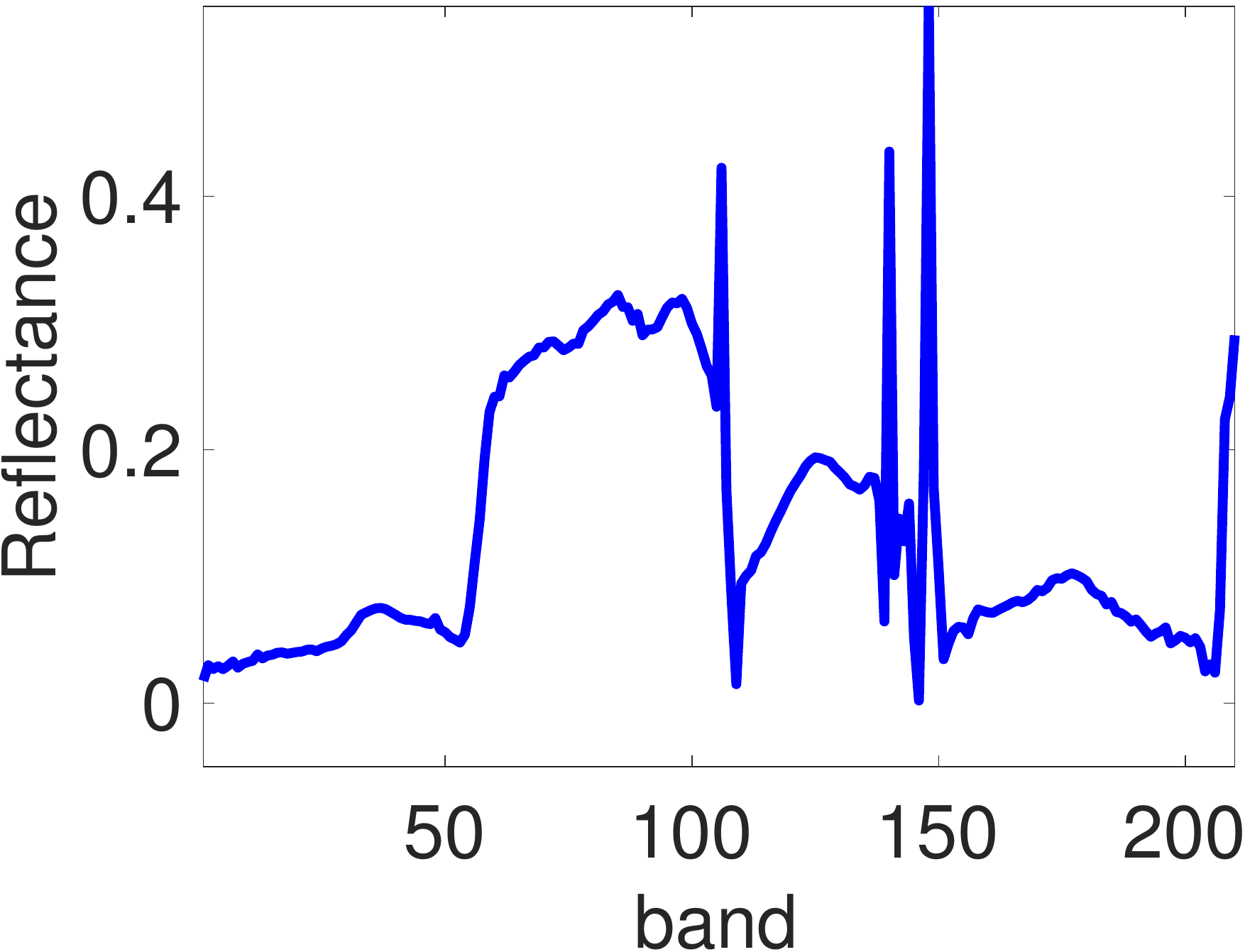}}\hspace{-0.8mm}
\subfigure[HSI-SDeCNN]{\includegraphics[scale =0.14,clip=true]{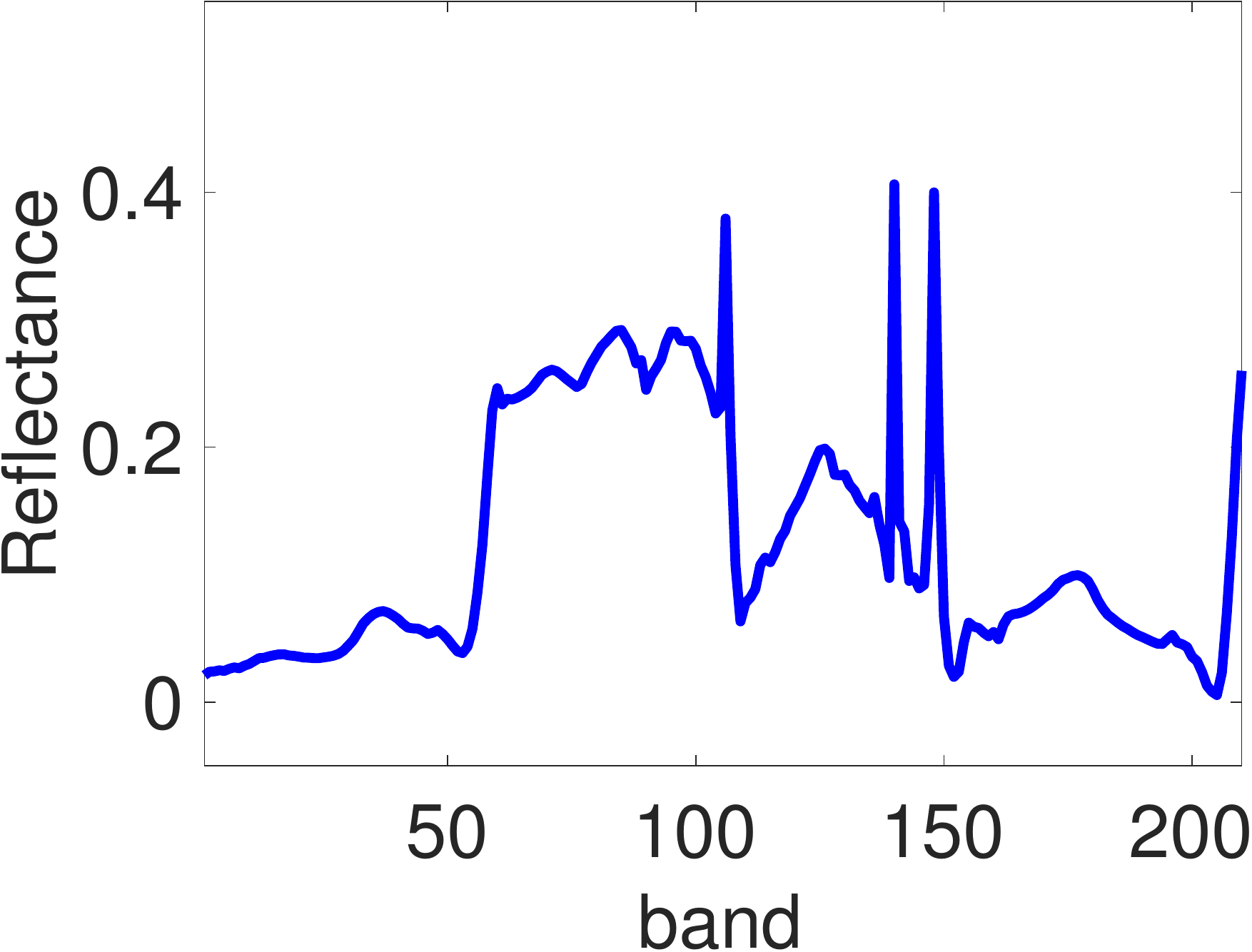}}\hspace{-0.8mm}
\subfigure[HSID-CNN]{\includegraphics[scale =0.14,clip=true]{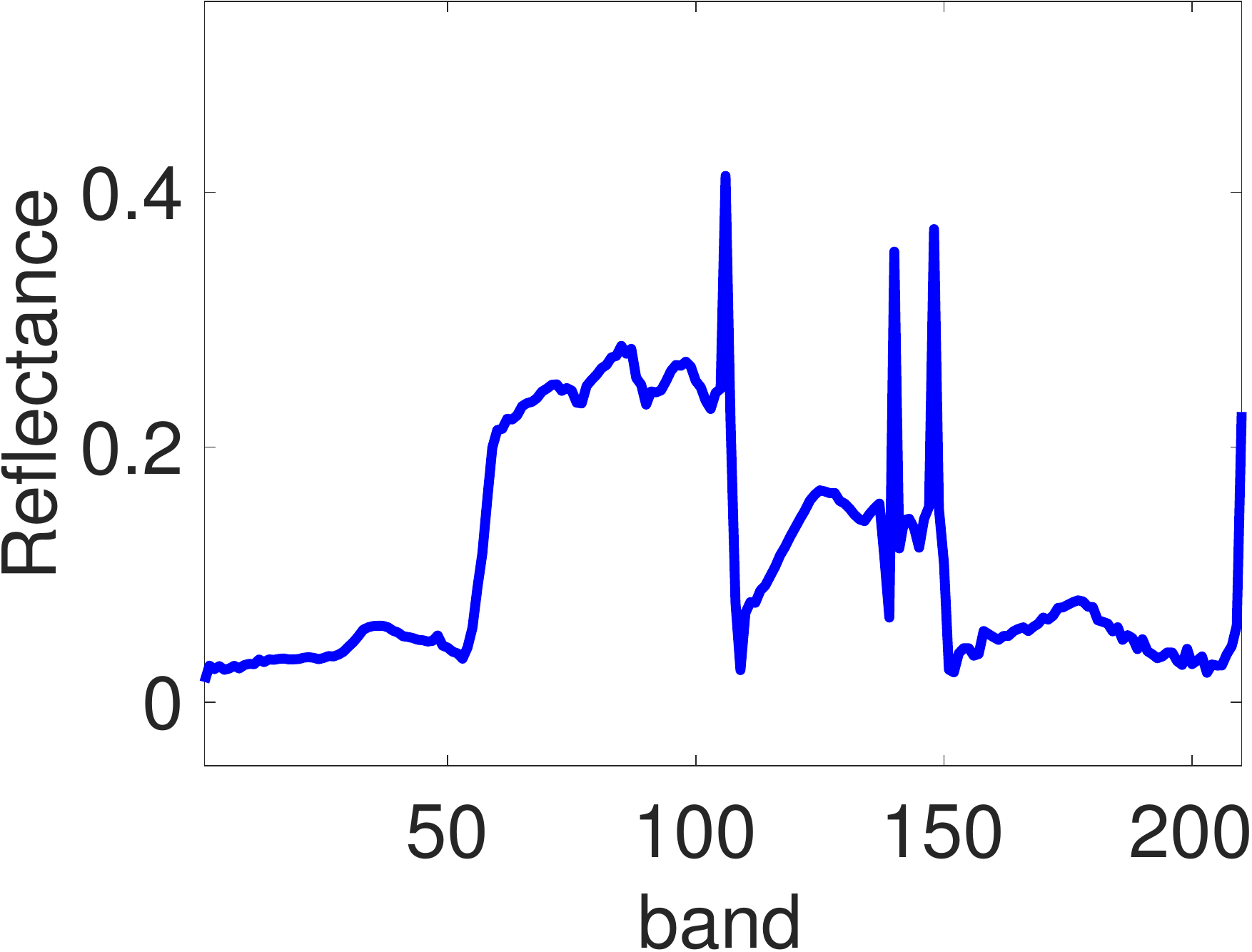}}\hspace{-0.8mm}
\subfigure[QRNN3D]{\includegraphics[scale =0.14,clip=true]{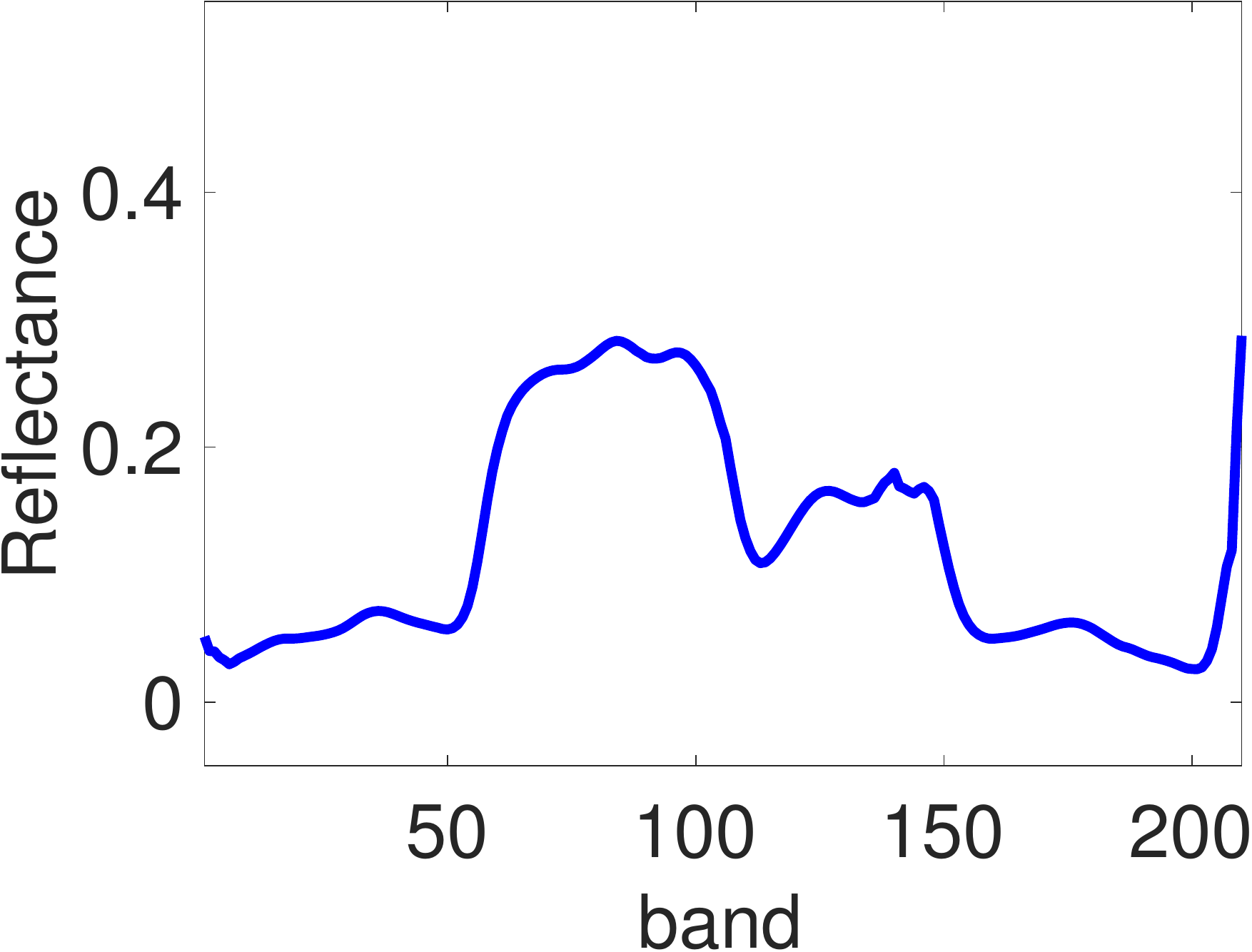}}\hspace{-0.8mm}
\subfigure[SMDS-Net]{\includegraphics[scale =0.14,clip=true]{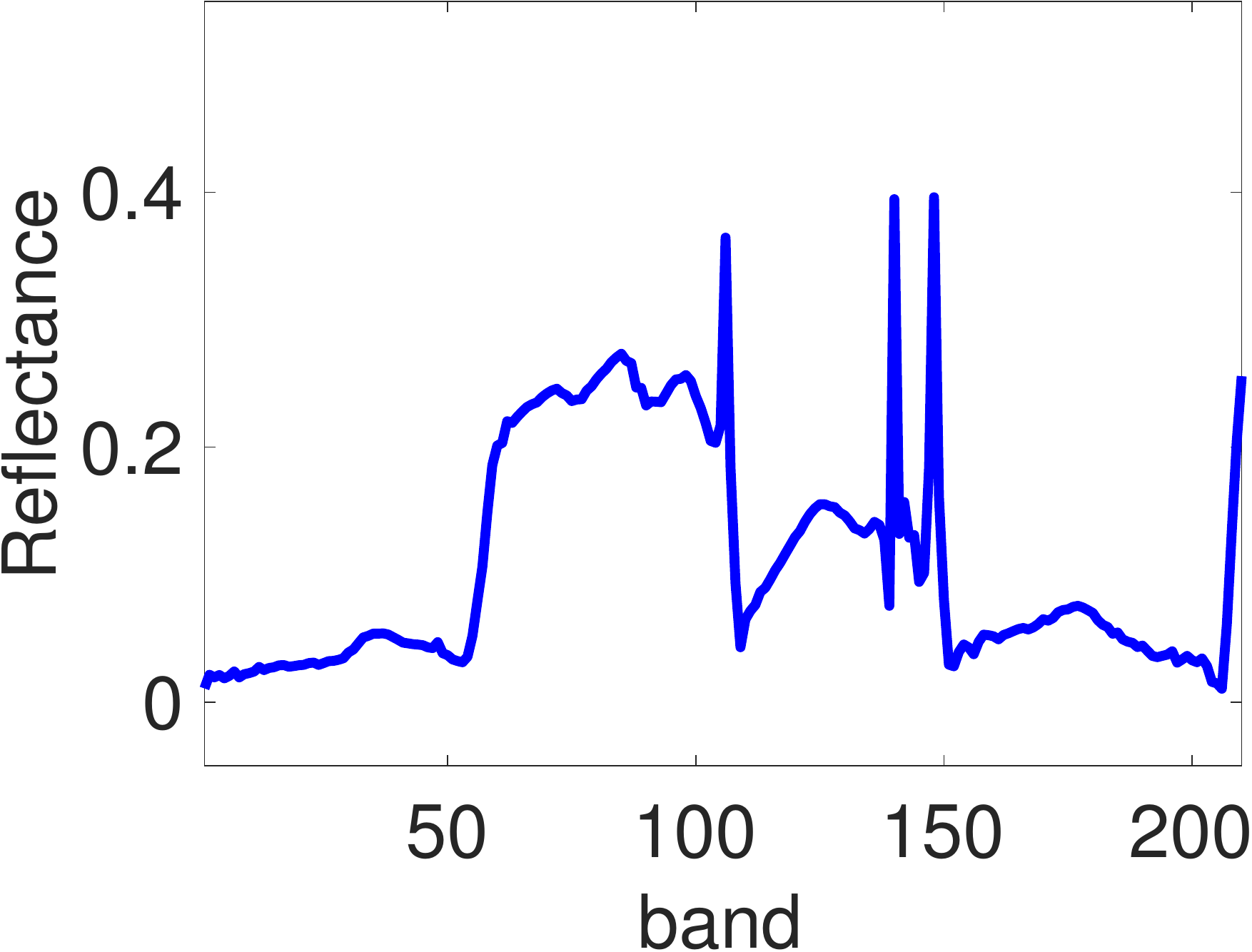}}
\caption{Denoising results of pixel (249, 216) in real-world  Urban data set.} \label{fig:urbanspec}
 \end{figure*}

\subsection{Experiments with Real-world Remote Sensing HSI}
 
Further, we ran all the competing methods on the widely used Urban dataset~\footnote{https://www.agc.army.mil/} to evaluate their real-world denoising ability. Urban HSI is a remote sensing HSI containing $307\times 307$ pixels and 210 bands. It is contaminated with very complex noises, especially in bands 1, 109, 141, 208, 209, making it very suitable to test the denoising ability of methods. Some typical bands are shown in Fig.~\ref{fig:icvl}. Though there are large differences between close-range HSI and remote sensing HSIs in terms of spatial and spectral resolution, we directly use the model trained on ICVL HSIs rather than fine-tuning the model on remote sensing HSIs.  Fig.~\ref{fig:icvl} shows the band-wise denoising results with respect to bands 1, 109 and 208 and Fig.~\ref{fig:urban} presents the false-color images  by compositing bands 109, 141, 209.   As can be seen, alternative methods either can not remove all the noises, for example, LLRT, MTSNMF, LRTDTV, HSI-SDeCNN, and HSID-CNN, or lose some important textures such as QRNN3D.  In contrast, our method achieves the best denoising results by removing most of the noises while retaining important structures. Fig.~\ref{fig:urbanspec} presents the spectral signature of pixel (249, 216) before and after denoising.  Our SMDS-Net also gives very competitive performance.  This experiment further verifies the strong denoising capability and generalization ability of the proposed SMDS-Net.


\subsection{Network Analysis}

Here, we conduct a throughout study on our SMDS-Net to show its virtue of strong learning ability,  attractive interpretability,  less requirement of training samples and also less number of parameters. All the experiments were conducted on ICVL HSI dataset.

\begin{figure*}[!htbp]
  \centering
\subfigure[PSNR]{\includegraphics[scale =0.32,clip=true]{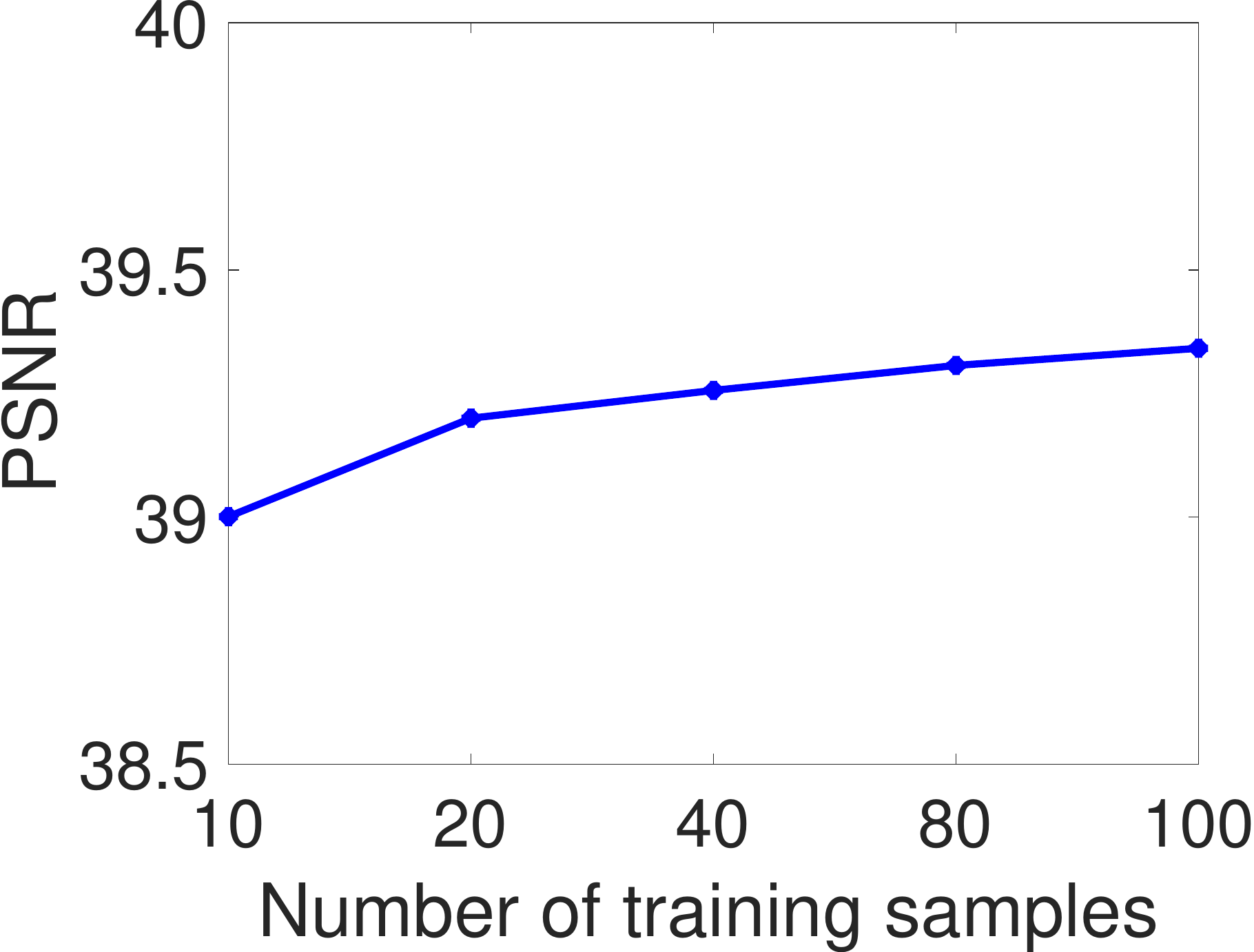}}\hspace{-0.8mm}
\subfigure[SSIM]{\includegraphics[scale =0.32,clip=true]{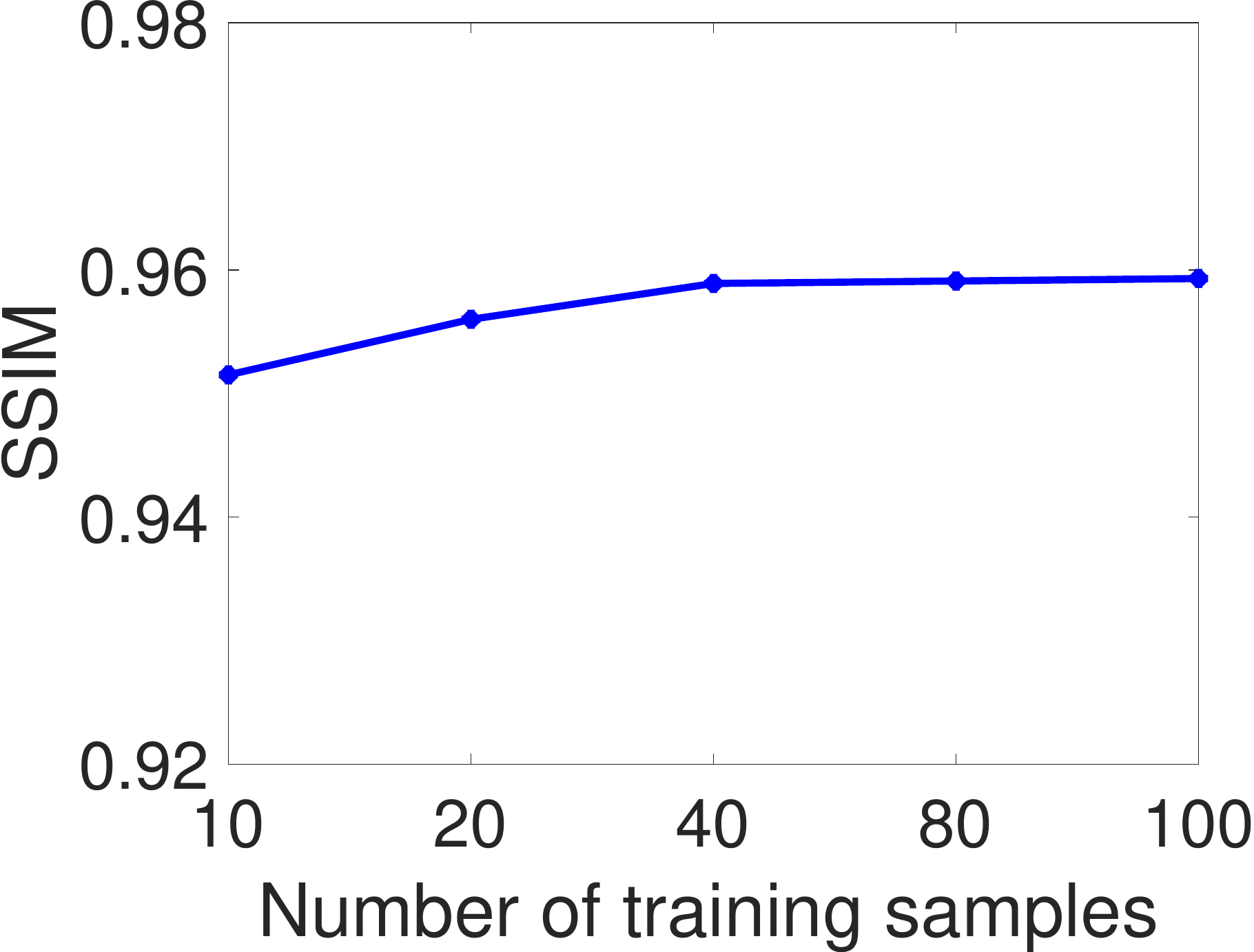}}\hspace{-0.8mm}
\subfigure[SAM]{\includegraphics[scale =0.32,clip=true]{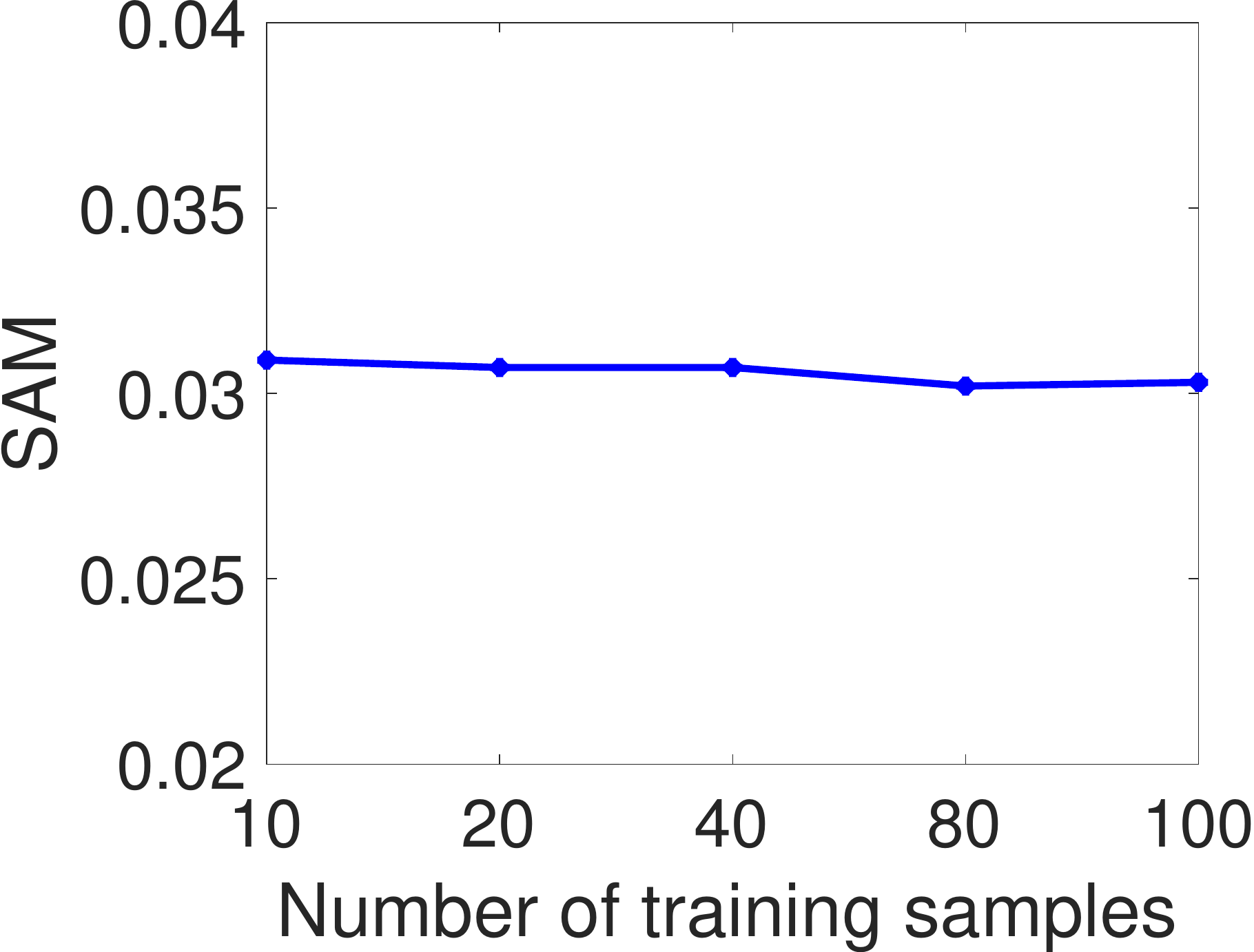}}\hspace{-0.8mm}
\caption{The impact of the number of training samples.} \label{fig:sample}
 \end{figure*}

\subsubsection{Study of the Number of Training Samples}  To further examine the learning capacity of our proposed network, we change the number of training samples from 10 to 100.  As shown in Fig.~\ref{fig:sample}, more training samples contribute to network learning, leading to better denoising performance. However, the denoising performance changes are not very significant when the number of training samples increases from 20 to 100, indicating the strong learning ability of SMDS-Net. This makes our method very practical for real-world HSI denoising where training samples are very rare and and expensive to acquire.

\begin{figure*}[!htbp]
  \centering
\subfigure[PSNR]{\includegraphics[scale =0.32,clip=true]{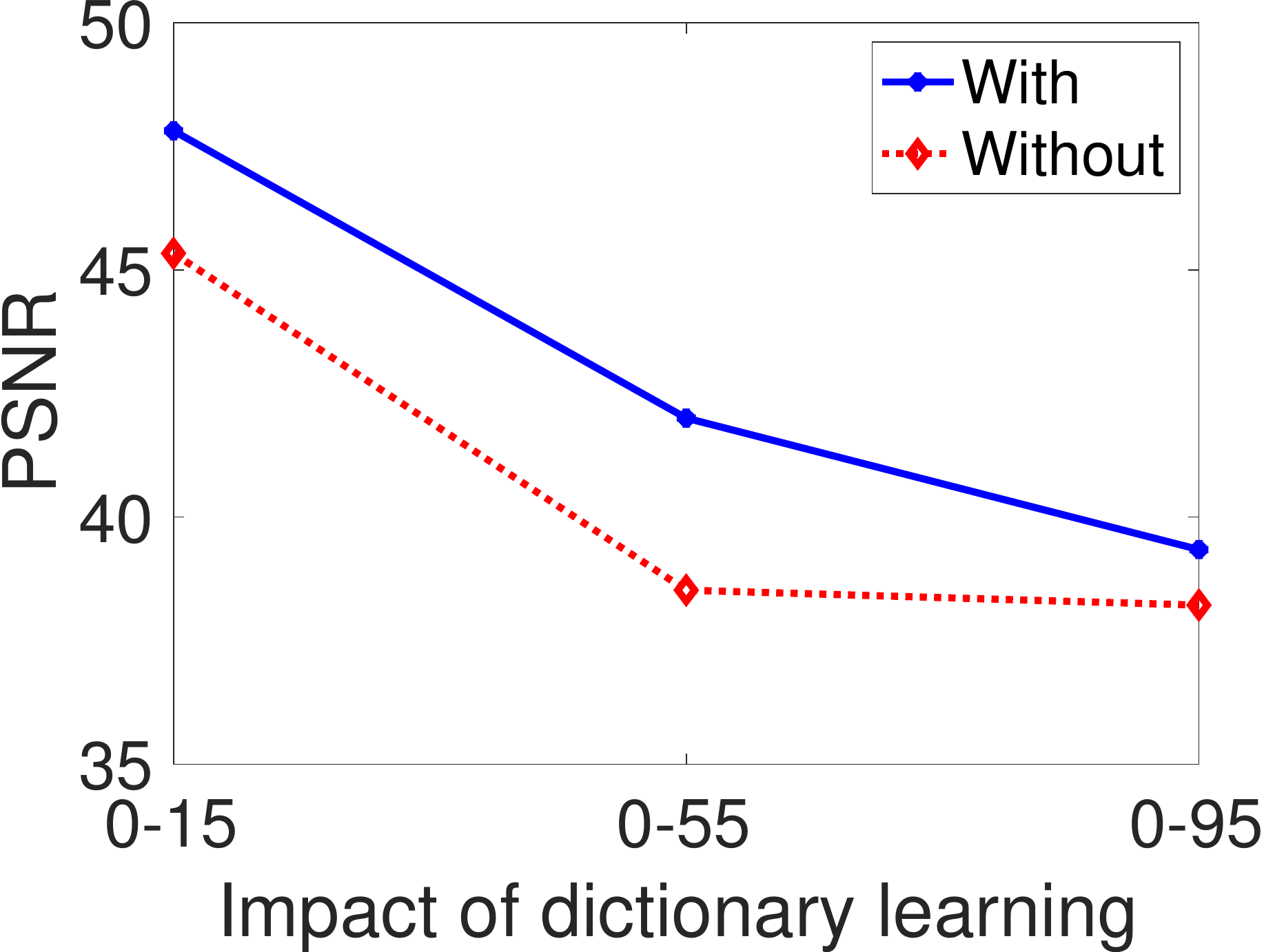}}\hspace{-0.8mm}
\subfigure[SSIM]{\includegraphics[scale =0.32,clip=true]{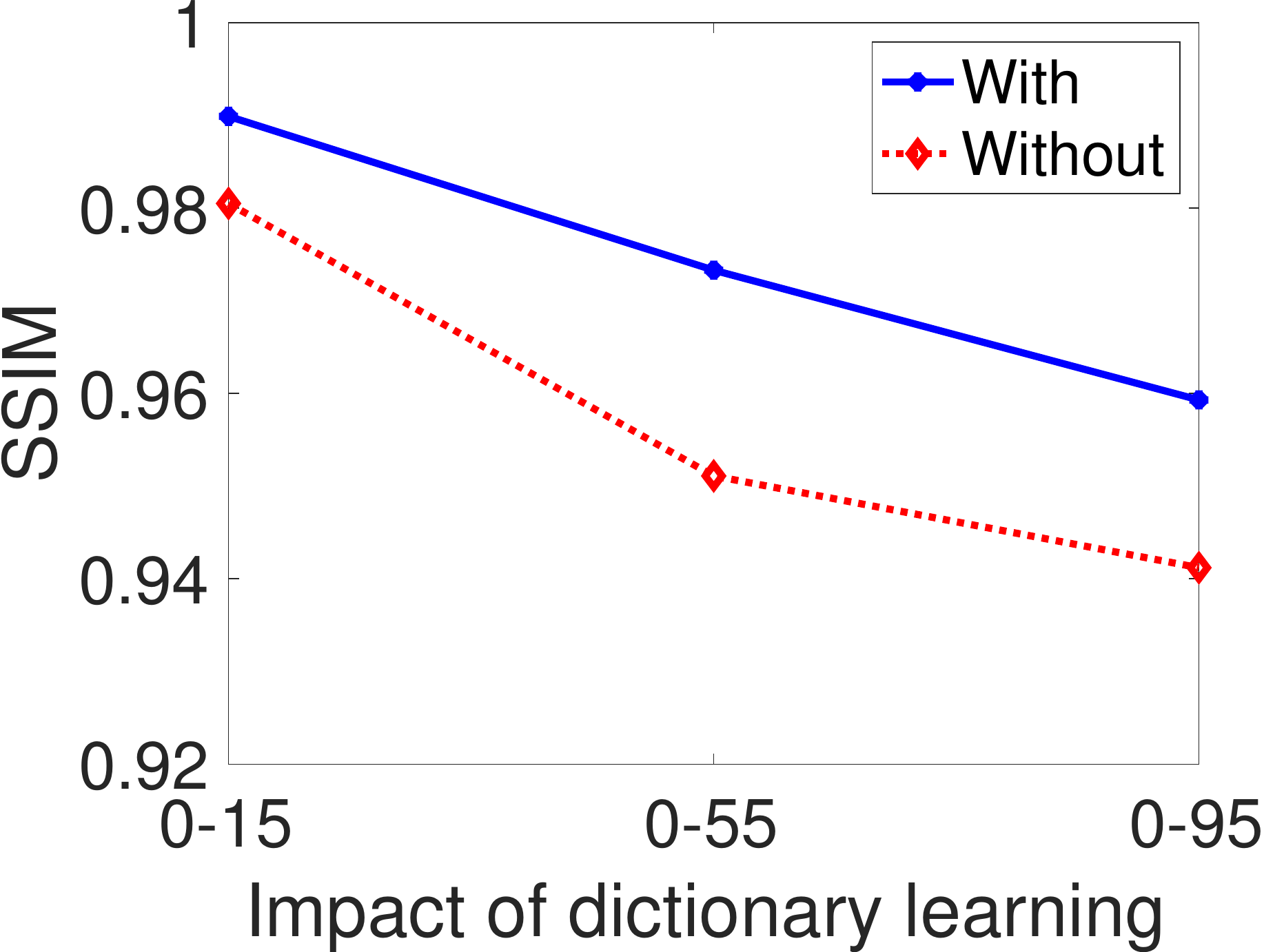}}\hspace{-0.8mm}
\subfigure[SAM]{\includegraphics[scale =0.32,clip=true]{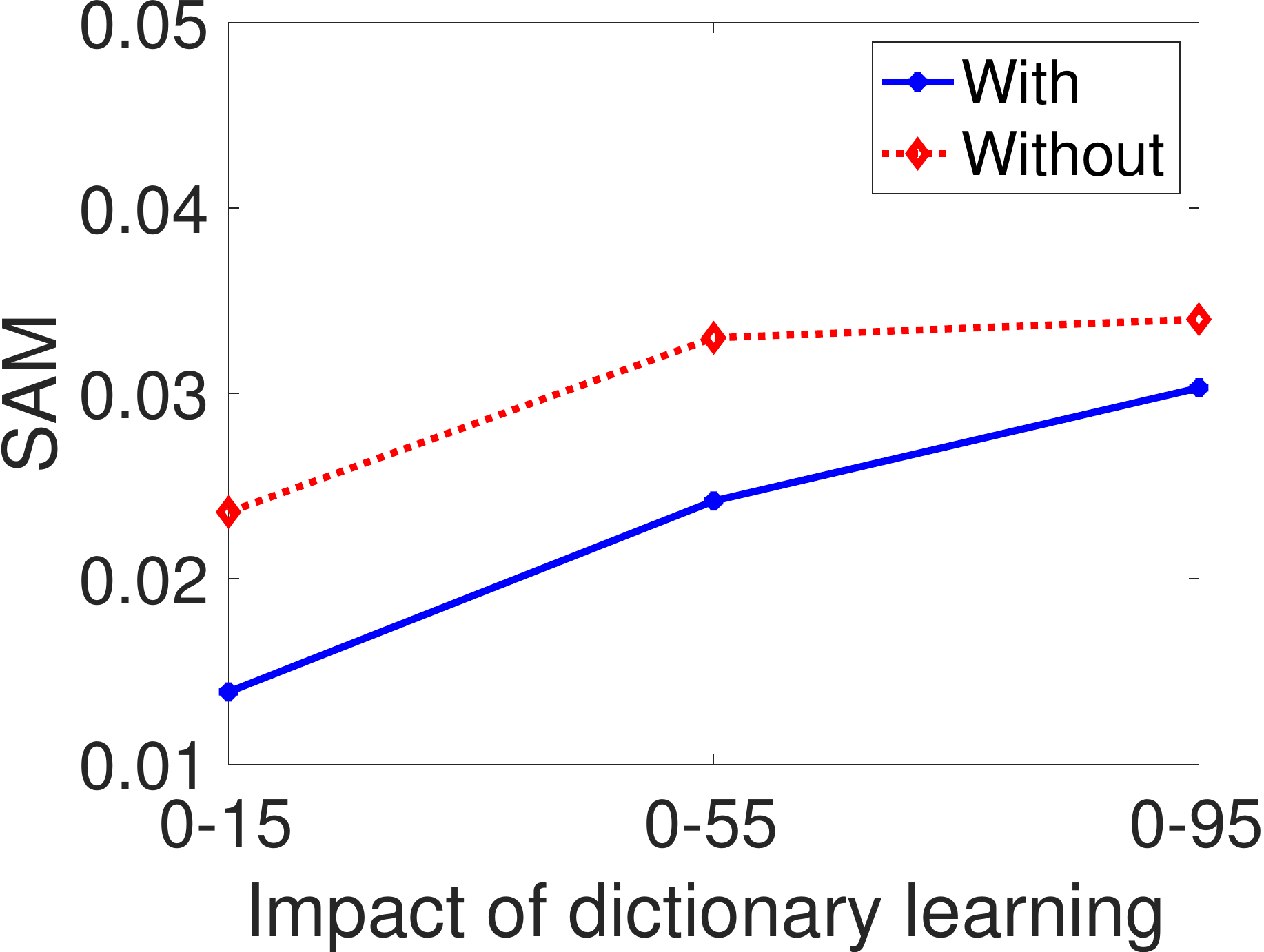}}\hspace{-0.8mm}
\caption{The impact of dictionary learning(with/without). } \label{fig:dict}
 \end{figure*} 

  \begin{figure*}[!htbp]
  \centering
\subfigure[Initial dictionary]{\includegraphics[scale=0.23,clip=true]{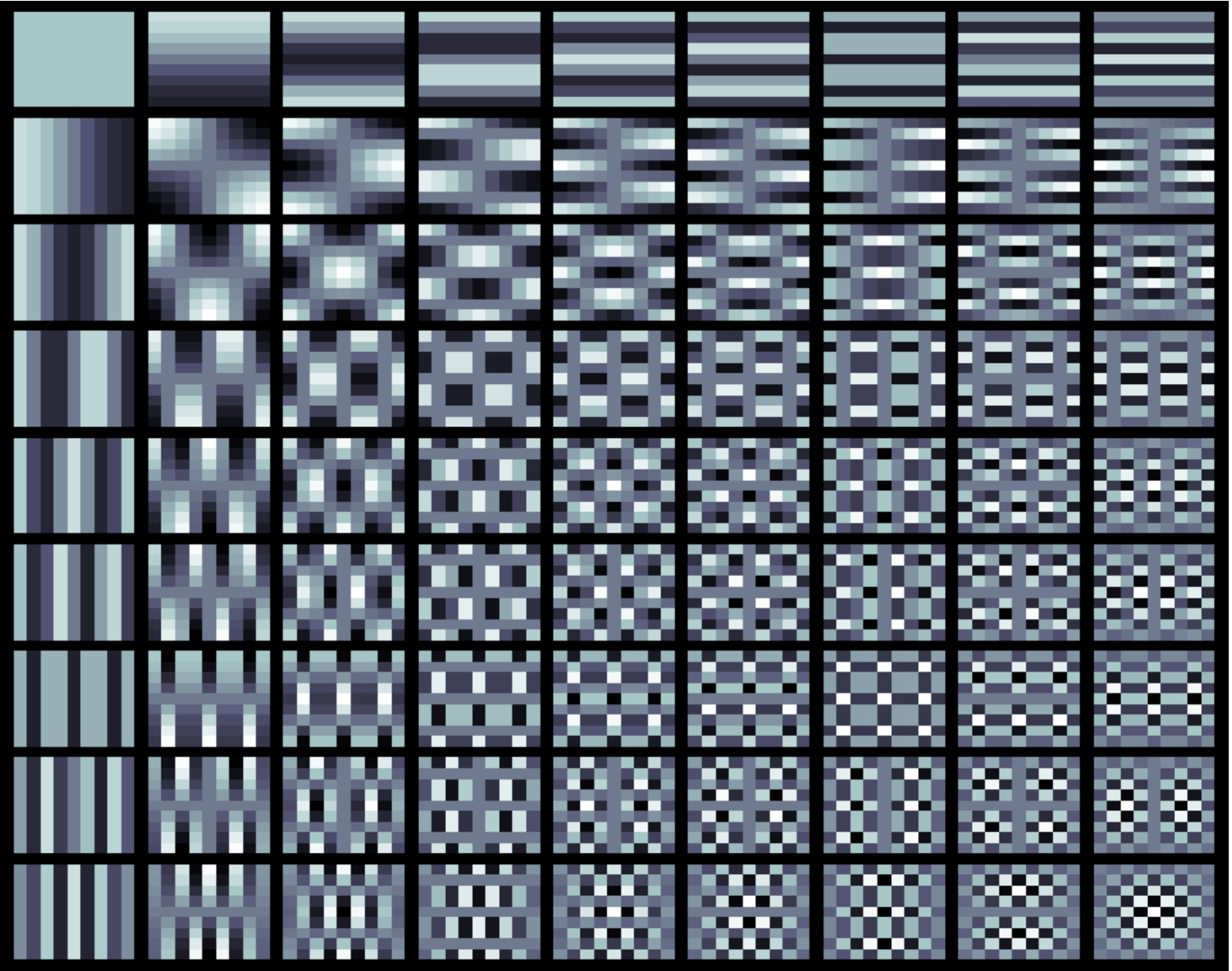}} 
\subfigure[$\D_1\otimes\D_2$]{\includegraphics[scale =0.23,clip=true]{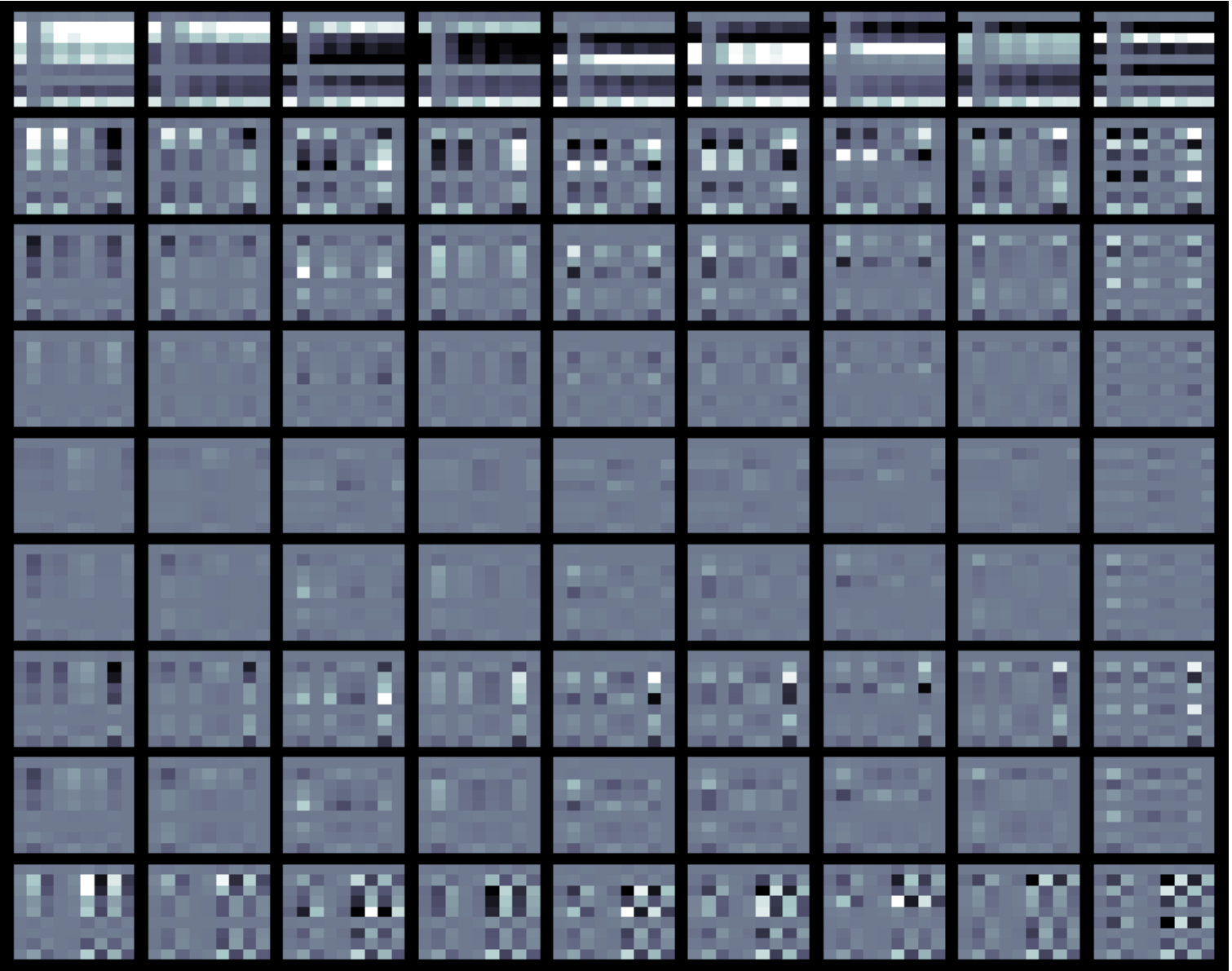}} 
\subfigure[$\D_2\otimes\D_3$]{\includegraphics[scale =0.23,clip=true]{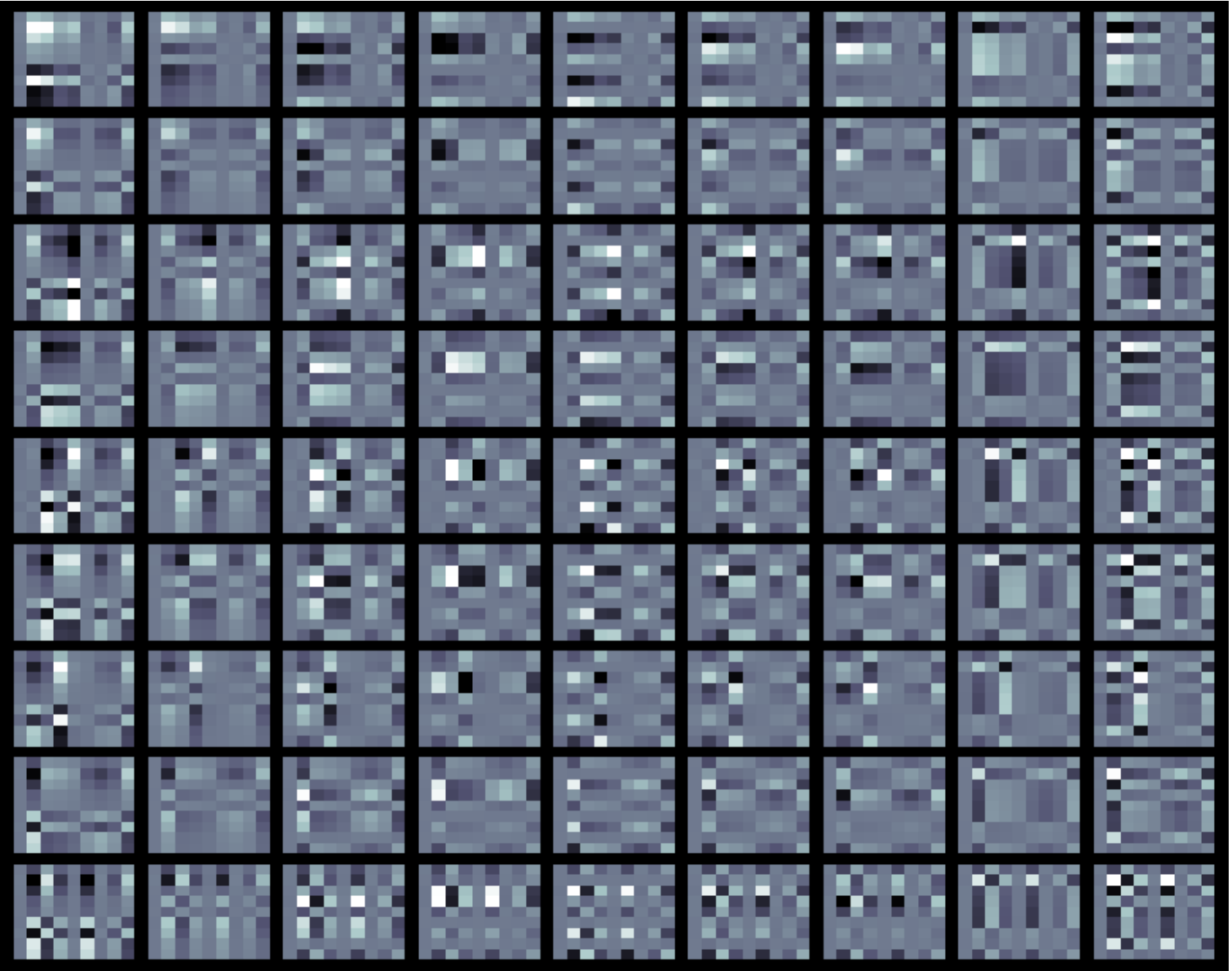}} 
\subfigure[$\D_1\otimes\D_3$]{\includegraphics[scale =0.23,clip=true]{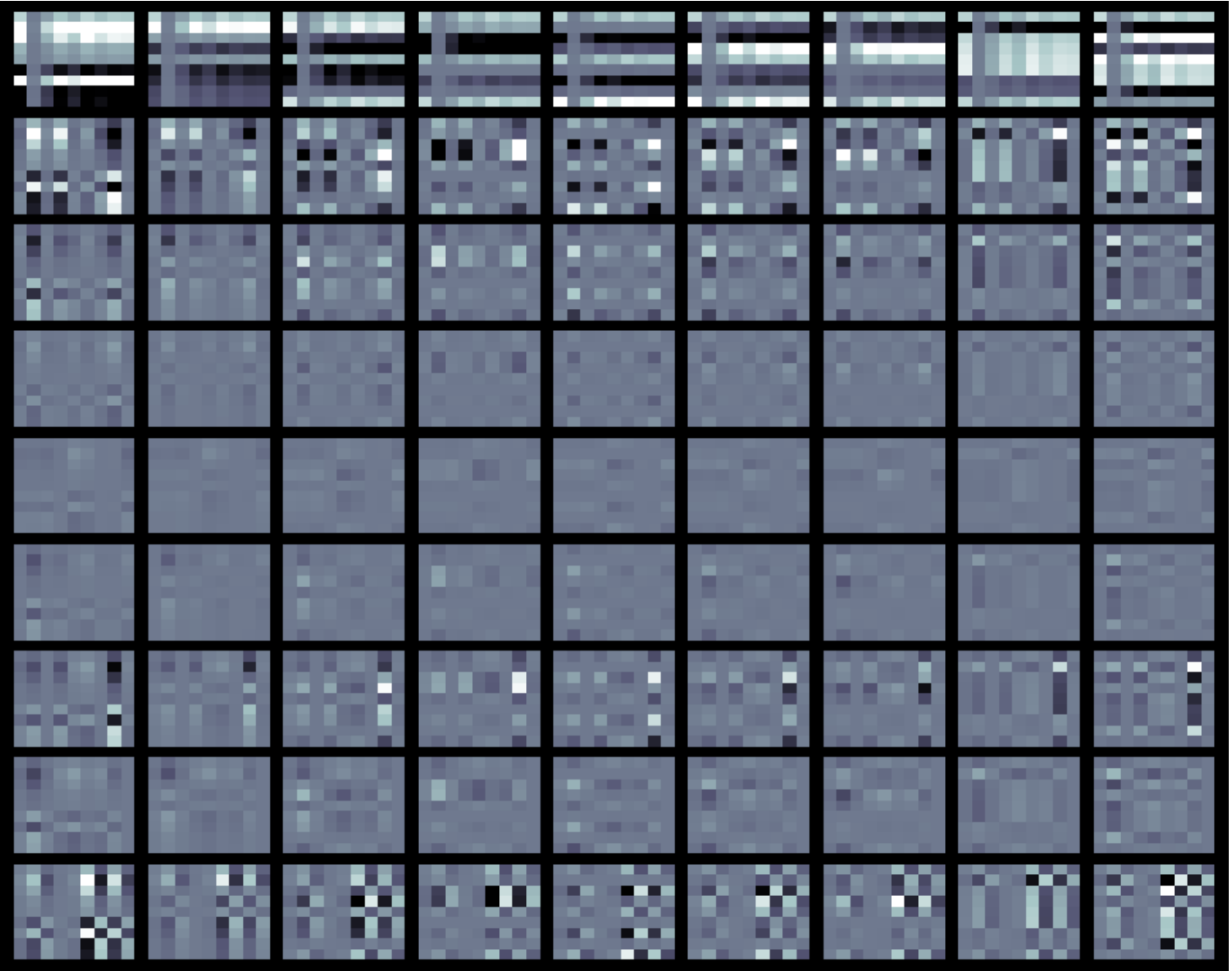}} 
\subfigure[$\C_1\otimes\C_2$]{\includegraphics[scale =0.23,clip=true]{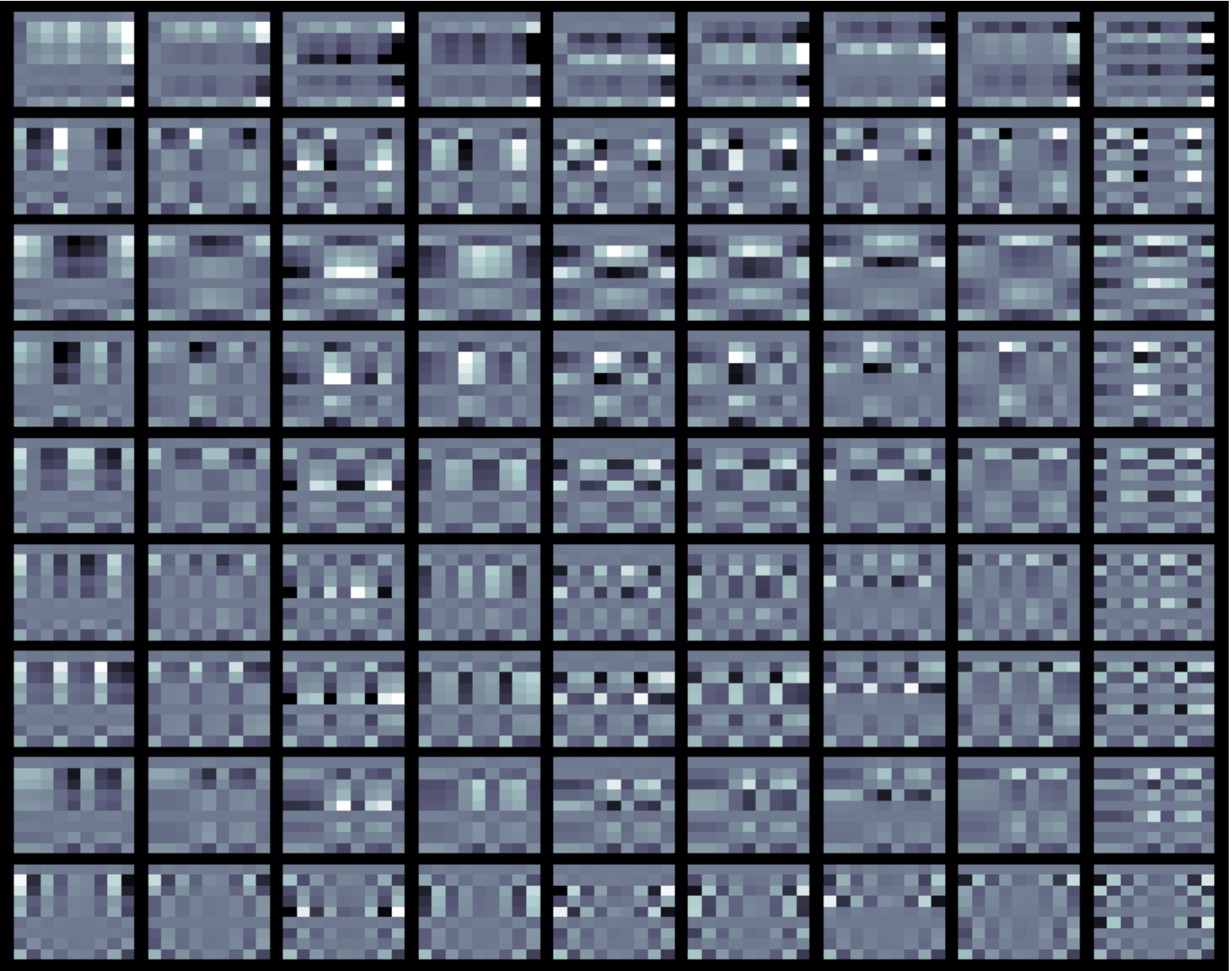}}\\
\subfigure[$\C_2\otimes\C_3$]{\includegraphics[scale =0.23,clip=true]{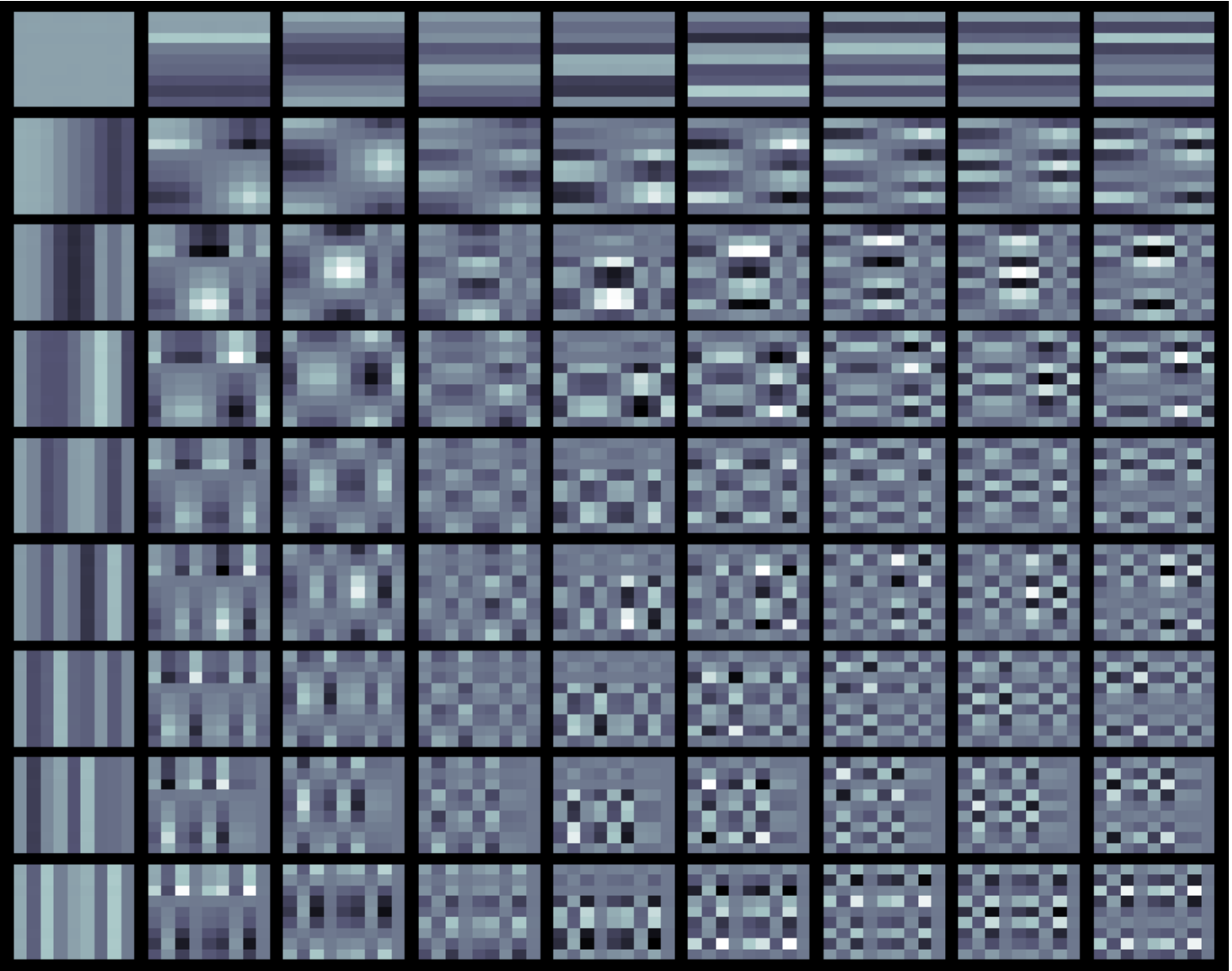}}
\subfigure[$\C_1\otimes\C_3$]{\includegraphics[scale =0.23,clip=true]{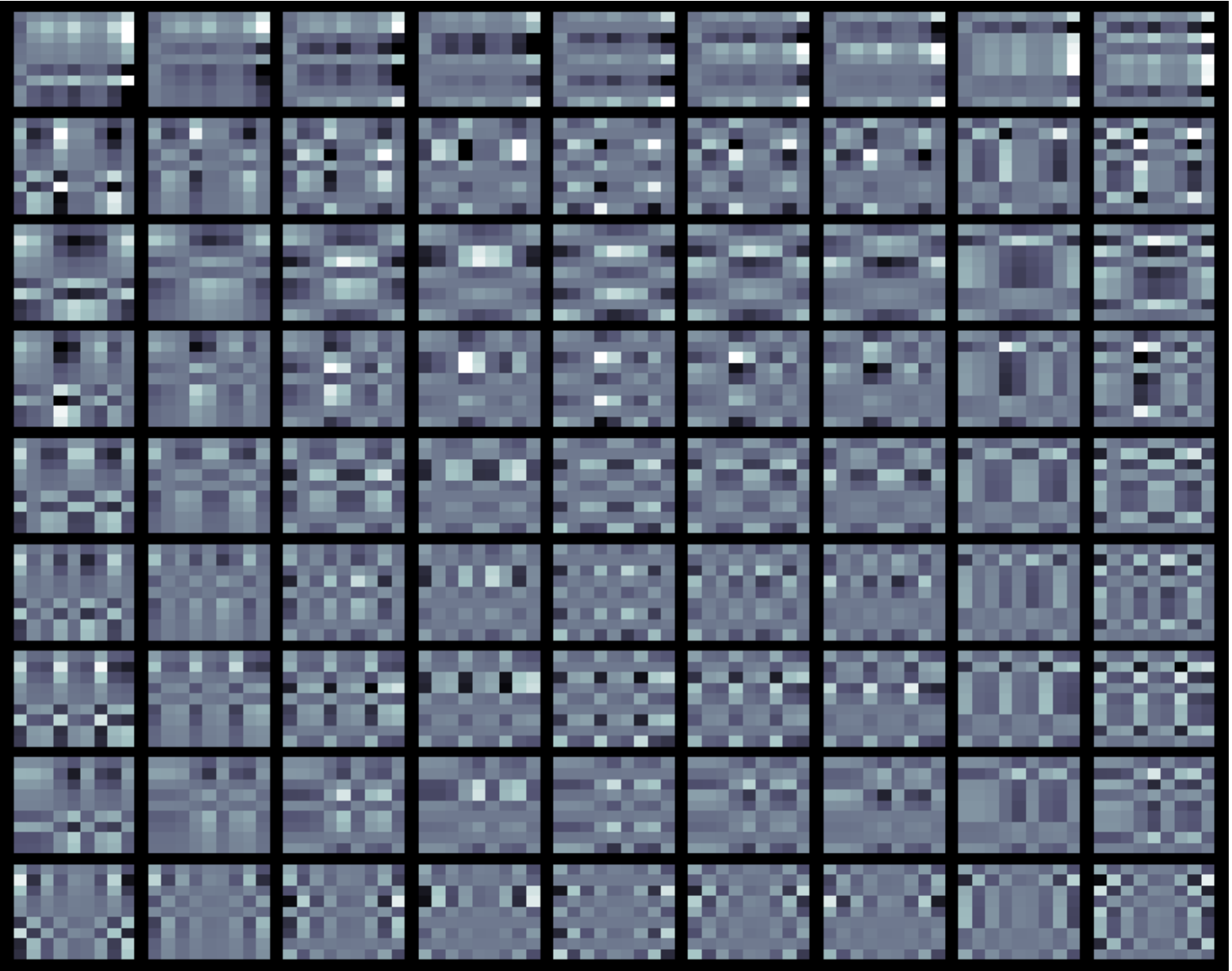}}
\subfigure[$\W_1\otimes\W_2$]{\includegraphics[scale =0.23,clip=true]{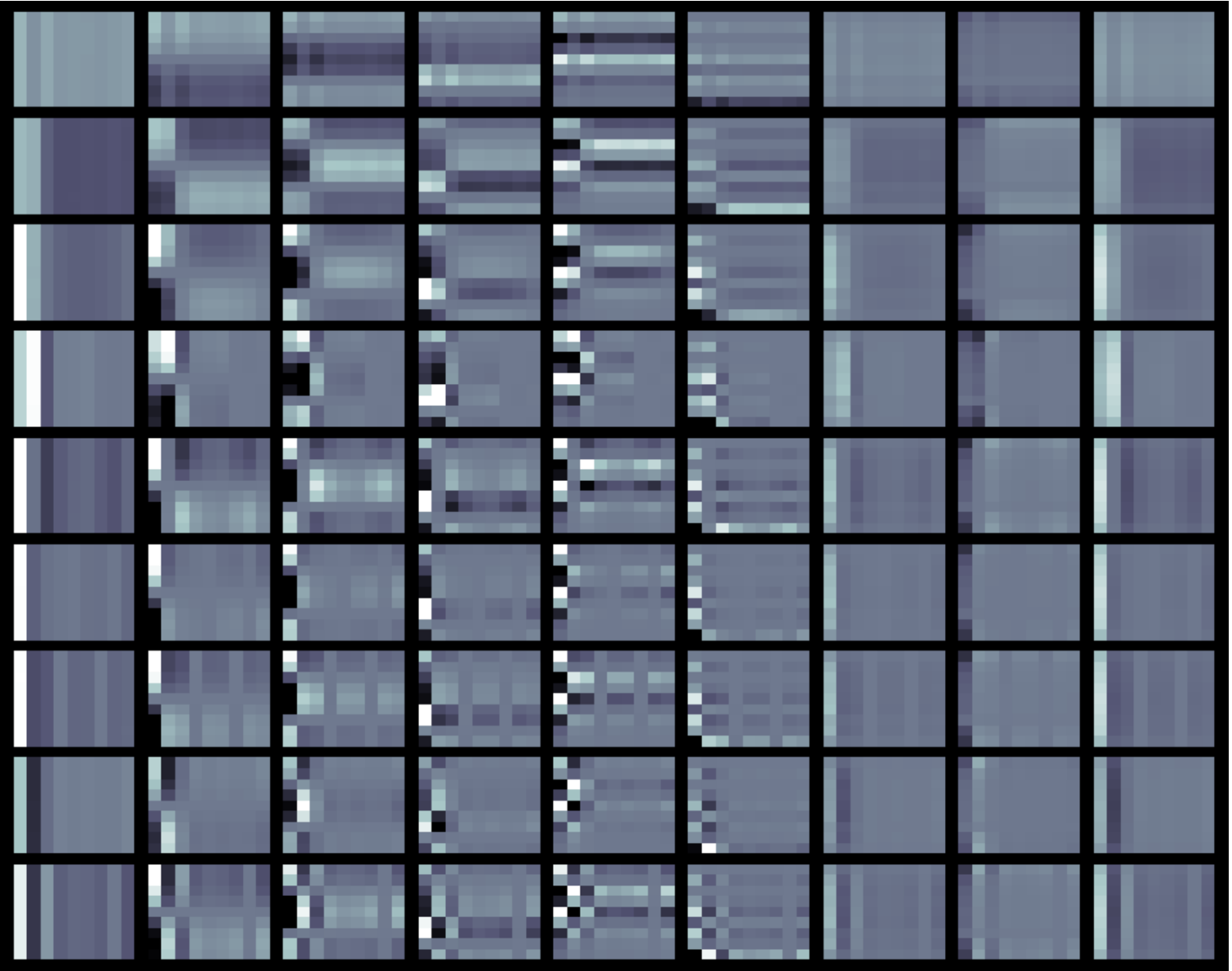}}
\subfigure[$\W_2\otimes\W_3$]{\includegraphics[scale =0.23,clip=true]{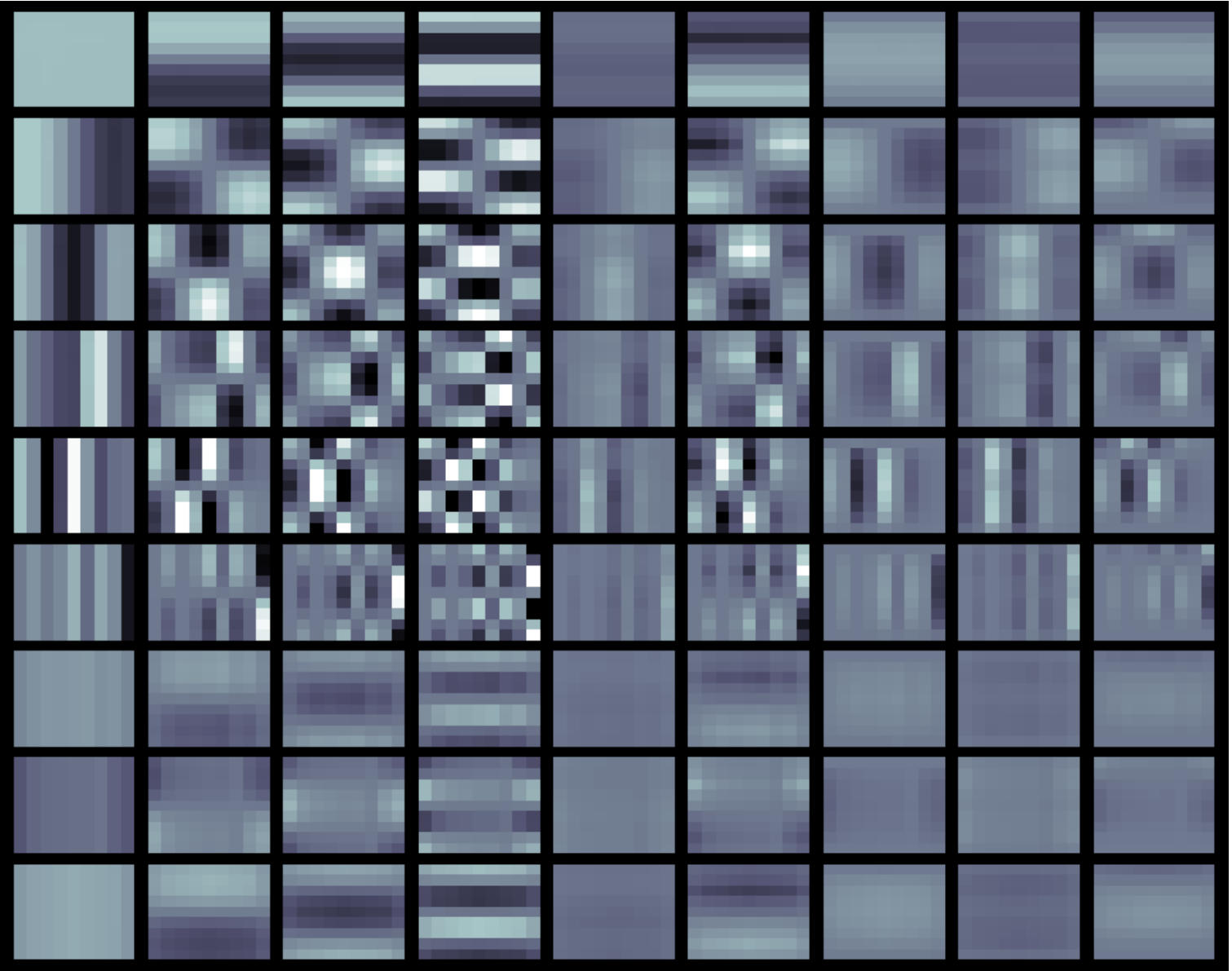}}
\subfigure[$\W_1\otimes\W_3$]{\includegraphics[scale =0.23,clip=true]{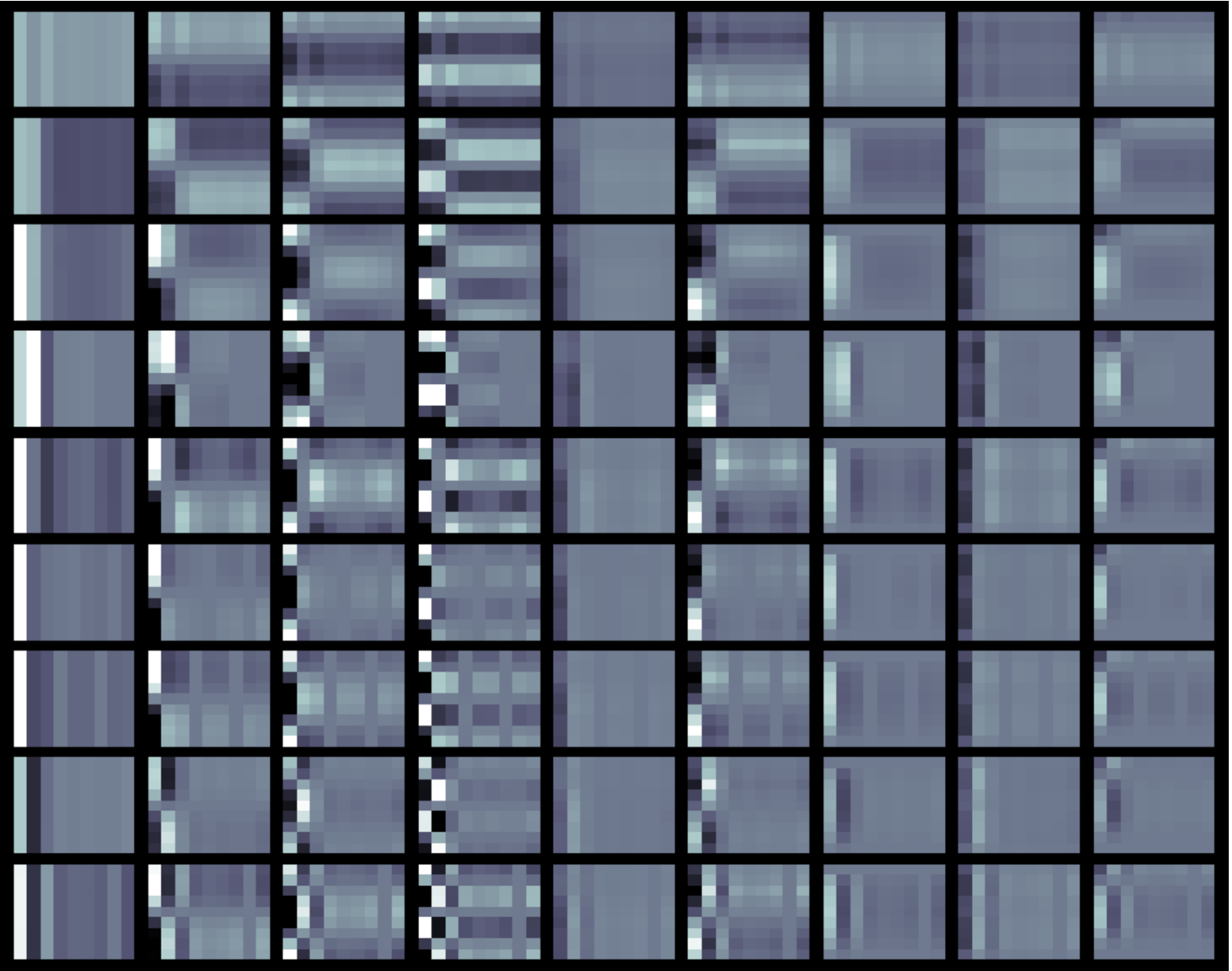}}
\caption{Visualization of  learned dictionaries for SMDS-Net  when $\{\D_j, \C_j, \W_j\}_{j=1,2,3}\in \Rbb^{9\times 9}$.} 
\label{fig:dict2}
 \end{figure*}

\subsubsection{Study of Dictionary Learning} Here, we conduct an ablation study to show the importance of dictionary learning of SMDS-Net. As for the case without dictionary learning, we direct set $\D_j$, $\W_j$, and $\C_j$ as DCT basis rather than parameters to be learned. Fig.~\ref{fig:dict} presents the effect of dictionary learning. It can be seen that there is a significant improvement in using dictionary learning because it can increase the ability of the network to model and adapt the data. Fig.~\ref{fig:dict2} visualizes the learned dictionaries which are constructed by the Kronecker product of $\D$, $\W$ and $\C$ along arbitrary dimensions. It can be seen that, on the one hand, there are  significant differences between the initialized DCT basis and the learned DCT basis, which means that the network learns to fit the data. On the other hand, there are differences between dictionaries, see (b)-(d), (e)-(g), (h)-(j), which proves that the learned dictionaries can capture the structural features of all dimensions.   

 \subsubsection{Study of Key Parameters}
 In general, the number of unfolding ($K$), the dictionary number ($[M_1,M_2,M_3]$) and the cube size $([I_1, I_2, I_3])$ are three factors influencing the denoising effectiveness  of SMDS-Net.  Here, we conduct a comprehensive study on these key parameters.  We add  additive white Gaussian  noise (AWGN) to every band whose  stand deviation $\sigma$ is set between 0 and 95. The averaged denoising performance  is reported.

\textbf{Impact of Dictionary Number}  We measure the influence of the dictionary number   in Table~\ref{tab:dic}.  For simplicity, we set $M_1=M_2=M_3$ and choose the value from the set of $\{5,7,9,11,13\}$. As shown in  Table~\ref{tab:dic}, using larger dictionaries allows for better denoising performance at the cost of more computation and parameters to be learned. Moreover, increasing the dictionary number from  $[5,5,5]$ to  $[13,13,13]$, we only  measure a difference of 0.15dB. Therefore, the dictionary number is suggested to set as $[9,9,9]$ considering a  trade-off between speed and the number of parameters. 
\begin{table}[thbp]
\caption{The influence of the  dictionary number.}\label{tab:dic}
\centering
\begin{tabular}{cccccccc|c|c|c|c|c|c|c|c|c|c|c|}
\hline
$[M_1,M_2,M_3]$&[5, 5, 5]&[7, 7, 7]&[9, 9, 9]&[11, 11, 11]&[13, 13, 13]\\
\hline
PSNR&39.21& 39.31&39.34&\textbf{39.43}&39.36\\
\hline
SSIM&0.9583&0.9576&0.9593&\textbf{0.9595}&0.9586\\
\hline
SAM&0.0307&0.0306&\textbf{0.0303}& 0.0305&0.0304\\
\hline
\#Parameters&1155&2625&5103&8877&14235\\
\hline
\end{tabular}
\end{table}

\begin{table}[thbp]
\caption{The influence of unfolding iteration $K$.}\label{tab:kimpact}
\centering
\begin{tabular}{ccccccccccc}
\hline
$K$&3&6&9&12&15\\
\hline
PSNR& 39.21&39.34& 39.33&\textbf{39.41}& 39.15\\
\hline
SSIM&0.9589&\textbf{0.9593}&\textbf{0.9593}&0.9587&0.9545\\
\hline
SAM&0.0305&\textbf{0.0303}&0.0306&\textbf{0.0303}&0.0304\\
\hline
\#Parameters&2916&5103&7290&9477&11664\\
\hline
\end{tabular}
\end{table}

\textbf{Impact of Unfolded Iteration $K$:}  $K$ indexes the number of unfolding and also directly controls the depth of the network.   Table~\ref{tab:kimpact} presents the denoising result with respect to different $K$.   SMDS-Net takes
the advantages of backpropagation and forward propagation to establish the relationship between current iteration and historical iterations which enforces the network to simultaneously takes historical information and current information into consideration for parameter update. This makes $K$ have insignificant influence on denoising performance as shown in Table~\ref{tab:kimpact}, implying layer number of SMDS-Net is easy to choose. When setting $K=6$, the number of parameters is 5103 which is noticeably less than that of QRNN3D~\cite{Wei2020} and HSI-SDeCNN~\cite{Maffei2020}, i.e., over 100 thousand. Small number of parameters also reduce the requirement of training samples size.  This attributes to the adoption of spectral low-rank projection and spatial multidimensional sparse modeling. 


\begin{table}[thbp]
\caption{The influence of   cube size.}\label{tab:cube}
\centering
\begin{tabular}{cccccccc|c|c|c|c|c|c|c|c|c|c|c|}
\hline
$[I_1,I_2,I_3]$&[3, 3, 3]&[5, 5, 5]&[7, 7, 7]&[9, 9, 9]&[11, 11, 11]\\
\hline
PSNR&   37.05&38.78&  39.13& \textbf{39.34}&    \textbf{39.34}\\
\hline
SSIM&0.9220&0.9537&0.9575&0.9593& \textbf{0.9595}\\
\hline
SAM&0.0358&0.0316&0.0311&\textbf{0.0303}&0.0305\\
\hline
\#Parameters&4617&4779&4941&5103&5265\\
\hline
\end{tabular}
\end{table}

\textbf{Study of Cube Size:}  Here, we analyze the influence of cube size $[I_1, I_2, I_3]$ on HSI denoising.  As the experiment in the study of dictionary number,  $I_1$, $I_2$ and $I_3$ are set the same.  By changing  the size from 3 to 11 with an interval of 2, Table~\ref{tab:cube} presents the denoising ability of SMDS-Net. Larger cube size is able to include more spatial nearby pixels for representation, therefore leading to better denoising performance especially in highly noisy cases.   From this experiment, it is recommended that the cube size is set as $[9,9,9]$, balancing the denoising performance and the number of parameters.
\section{Conclusion} \label{sec:con}
In this paper, we introduce a model guided deep neural network, i.e., SMDS-Net for HSI denoising. 
SMDS-Net simultaneously considers the spectral-spatial correlation, spectral low-rankness, and spatial sparsity priors in a unified end-to-end network.  Experimental results show  SMDS-Net provides the state-of-the-art denoising performance with strong learning capacity and higher interpretability, highlighting the benefit of equipping DL with model priors.   In the future, we will integrate the nonlocal similarity prior of HSIs into the network for more powerful denoising performance.

\appendices
\bibliographystyle{IEEEtran}
\bibliography{IEEEabrv,hsi-tracking-refs2}
\end{document}